\documentclass[braunschweig,print]{thesis}

\title{Investigations of small-scale magnetic features on the solar surface}
\author{Tino L. Riethm\"uller}
\town{Dingelst\"adt/Eichsfeld}

\refereea{Prof.~Dr.~S.~K.~Solanki}
\refereeb{Prof.~Dr.~K.-H.~Gla\ss{}meier}
\submitteddate{4.~Februar 2013}
\submittedyear{2013}
\examinationdate{3. Mai 2013}
\publicationyear{2013}
\isbn{978-3-942171-73-1}

\usepackage{natbib}
\usepackage{graphicx}
\usepackage{txfonts}
\usepackage{calc}
\usepackage{color}
\usepackage{rotating}
\usepackage[OT2, T1]{fontenc}
\usepackage[russian, USenglish]{babel} 
\usepackage{pdflscape}
\usepackage{makeidx}

\usepackage[breaklinks,pagebackref=true]{hyperref}  
\hypersetup{
    unicode=false,                        
    pdftoolbar=true,                      
    pdfmenubar=true,                      
    pdffitwindow=false,                   
    pdfstartview={FitH},                  
    pdftitle={Investigations of small-scale magnetic features on the solar surface},    
    pdfauthor={Tino~L.~Riethm\"uller},    
    pdfsubject={PhD Thesis},              
    pdfcreator={Tino~L.~Riethm\"uller},   
    pdfproducer={Tino~L.~Riethm\"uller},  
    pdfkeywords={Sunspots} {Quiet Sun} {Spectropolarimetry} {SUNRISE}, 
    pdfnewwindow=true,                    
    colorlinks=true,                      
    linkcolor=red,                        
    citecolor=blue,                       
    filecolor=green,                      
    urlcolor=cyan,                        
    citebordercolor={0 1 0}, 
    linktocpage=false,
    raiselinks=false
}

\newcommand{\Index}[1]{#1\index{#1}}
\newcommand{\bfit}[1]{\textbf{\textit{#1}}}

\makeindex

\begin{document}
\begin{sloppypar} 

\newcommand{\hinode}{\textsc{\Index{Hinode}}}
\newcommand{\sunrise}{\textsc{\Index{Sunrise}}}
\newcommand{\solarC}{\textsc{\Index{Solar-C}}}
\newcommand{\themis}{\textsc{\Index{Themis}}}
\newcommand\arcsec{\mbox{$^{\prime\prime}$}}
\newcommand\arcmin{\mbox{$^{\prime}$}}
\newcommand{\carcsec}{$\mbox{.\hspace{-0.5ex}}^{\prime\prime}$}
\newcommand{\carcmin}{$\mbox{.\hspace{-0.5ex}}^{\prime}$}

\maketitle

\chapter*{Vorver\"offentlichungen der Dissertation}

Teilergebnisse aus dieser Arbeit wurden mit Genehmigung der Fakult\"at f\"ur Elektrotechnik,
Informationstechnik, Physik, vertreten durch den Betreuer der Arbeit, in folgenden
Beitr\"agen vorab ver\"offentlicht:

\begin{itemize}

\item T.~L.~Riethm\"uller, S.~K.~Solanki, \& A.~Lagg,
\textit{Stratifications of Sunspot Umbral Dots from Inversion of Stokes Profiles Recorded by Hinode},
Astrophysical Journal Letters, 678, 157 (2008)

\item T.~L.~Riethm\"uller, S.~K.~Solanki, V.~Zakharov, \& A.~Gandorfer,
\textit{Brightness, distribution, and evolution of sunspot umbral dots},
Astronomy \& Astrophysics, 492, 233 (2008)

\item T.~L.~Riethm\"uller, S.~K.~Solanki, V.~Mart\'{\i}nez Pillet, J.~Hirzberger, A.~Feller, J.~A.~Bonet, N.~Bello Gonz\'alez, M.~Franz, M.~Sch\"ussler, P.~Barthol, T.~Berkefeld, J.~C.~del Toro Iniesta, V.~Domingo, A.~Gandorfer, M.~Kn\"olker, \& W.~Schmidt,
\textit{Bright Points in the Quiet Sun as Observed in the Visible and Near-UV by the Balloon-Borne Observatory \sunrise{}},
Astrophysical Journal Letters, 723, 169 (2010)

\item T.~L.~Riethm\"uller, S.~K.~Solanki, M.~van Noort, \& S.~K.~Tiwari,
\textit{Vertical flows and mass flux balance of sunspot umbral dots},
eingereicht bei Astronomy \& Astrophysics Letters

\end{itemize}

\begingroup
\hypersetup{linkcolor=black} 
\tableofcontents
\endgroup

\chapter*{Summary\markboth{Summary}{Summary}}
\addcontentsline{toc}{chapter}{Summary}

Solar activity is controlled by the magnetic field, which also causes the variability
of the solar irradiance that in turn is thought to influence the climate on Earth.
The magnetic field manifests itself in the form of structures of largely different sizes,
starting with sunspots ($\sim$30000~km), pores ($\sim$5000~km), and micropores ($\sim$1000~km) through to
bright points and umbral dots (both $\sim$200~km). The smallest known magnetic features are found
to play an important role in the dynamics and energetics of the solar atmosphere.
This thesis concentrates on two types of such small-scale magnetic elements: The first part
studies the properties of umbral dots, dot-like bright features in the dark umbra of a
sunspot. The obtained umbral dot properties provide a remarkable confirmation of the
magneto-hydrodynamical simulation results of \citet{Schuessler2006}. Observations as well as
simulations show that umbral dots differ from their surroundings mainly in the lowest
photospheric layers, where the temperature is enhanced and the magnetic field is weakened.
In addition, the interior of the umbral dots displays strong upflow velocities which are surrounded
by weak downflows. This qualitative agreement further strengthens the interpretation of umbral dots as
localized columns of overturning convection. The second part of the thesis investigates
bright points, which are small-scale brightness enhancements in the darker intergranular lanes
of the quiet Sun produced by magnetic flux concentrations. Observational data obtained
by the \sunrise{} mission, having the highest resolution reached for quiet-Sun magnetic
field measurements, are used in this thesis. An important part of the work underlying this thesis
was the development of the \sunrise{} Filter Imager (SuFI) software, the reduction of the SuFI data
after the flight, the development of fundamental parts of the software for the Instrument Control Unit,
and the conceptual design of the Data Storage Subsystem of the \sunrise{} observatory. With the help of
the unique SuFI data, for the first time contrasts of bright points in the important ultraviolet
spectral range are determined (this spectral range is of particular relevance for the Sun's influence
on chemistry of the stratosphere). A comparison of observational data with magneto-hydrodynamical
simulations revealed a close correspondence, but only after effects due to the limited spectral and
spatial resolution were carefully included. 98\% of the synthetic bright points are found to be
associated with a nearly vertical kilo-Gauss field. A small fraction of the observed
bright points with strong polarization signals (most likely corresponding to network elements)
cannot be found in the analyzed set of simulations. Larger and deeper computational
boxes to include supergranules are suggested for more realistic bright point simulations.

\chapter{Introduction}\label{Introduction}

The Sun as the central star of our solar system is the main energy source from outside Earth
and provides all the energy that is necessary to keep the Earth at a temperature needed
for higher life. Besides the energetically not so important solar wind and \Index{neutrino flux},
the main part of the solar energy is provided in the form of electromagnetic radiation.
Thus, variations of the \Index{insolation} were already early assumed to influence climate
on Earth. The Serbian mathematician Milutin Milankovitch established a theory now
named after him which partly refers cyclic climate variations, in particular the sequence
of \Index{ice age}s to periodic changes in the Earth's orbit and rotation axis \citep{Milankovitch1941}.
He considered the collective effects of the precession of the Earth's axis with a period
of about 23000 years, the variations in axial tilt (41000 year cycle), and variations in
eccentricity (100000 year cycle), calculated variations of the insolation in the range of 5-10\%,
and held this responsible for the occurrence of the ice ages. In the 1970s, the \Index{Milankovitch cycle}s
were confirmed by studies of deep-sea cores which allowed deducing seawater
temperatures influenced by glacial periods over about the past 500000 years
\citep{Hays1976,Berger1977}.

From the theory of stellar evolution it is known that the \Index{luminosity} of the Sun
increased during its lifetime of 4.6 billion years by about 39\% \citep{Stix2002},
but on smaller timescales a constant Sun was assumed for a long time. The
\Index{total solar irradiance} (\Index{TSI}), i.e. the spectrally integrated solar radiation measured
from outside the absorbing terrestrial atmosphere at a distance of 1~AU from
the Sun, was hence named {\bf \Index{solar constant}}. With the launch of
the \Index{Earth Radiation Experiment} onboard the \Index{Nimbus~7} satellite in 1978
\citep{Hoyt1992}, followed by \Index{ACRIM}~{\sc I} on the \Index{Solar Maximum Mission}
\citep{Willson1981}, the \Index{Earth Radiation Budget Experiment} onboard the
\Index{Earth Radiation Budget Satellite} \citep{Luther1986}, ACRIM~{\sc II} on the
\Index{Upper Atmosphere Research Satellite} \citep{Willson1994}, the \Index{Solar Variability Instrument}
on the \Index{European Retrievable Carrier} \citep{Crommelynck1993}, \Index{VIRGO} on \Index{SOHO}
\citep{Froehlich1997}, ACRIM~{\sc III} on ACRIMSat \citep{Willson2001}, and the
\Index{Total Irradiance Monitor} onboard the \Index{SORCE satellite} \citep{Kopp2005}, space-borne
radiometers are now operating and measuring the solar irradiance accurately for
more than 34 years.

Such measurements revealed that the total solar irradiance varies by about 0.1\%
coincidentally with the 11-year \Index{activity cycle} of the Sun \citep[e.g.][]{Froehlich2011,Ball2012}.
Surprisingly, the \Index{total solar irradiance} is highest if the Sun is most active \citep{Willson1988}.
This contrasts with the observation that on solar rotational timescales (a month or less)
the total irradiance of the Sun is reduced when dark \Index{sunspot}s and \Index{pore}s are present on the
solar disk. The reduced \Index{luminosity} is overcompensated by the \Index{bright point}s (small-scale
bright features in the dark interspaces of the granulation) whose number density is increased
during the activity maximum \citep{Domingo2009}. Such bright points can be found in
high resolution images of the solar surface at all places where the \Index{magnetic flux} is
concentrated into \Index{kilo-Gauss element}s \citep{Stenflo1973,Berger2001,Ishikawa2007}. Not much
is known about the \Index{total solar irradiance} variations on timescales of centuries from direct
observations, but models provide heavily diverging results, e.g. \citet{Krivova2007}.

Technically more challenging than the determination of the \Index{total solar irradiance} are measurements
of the spectral distribution of the solar radiation, since the needed detectors tend to degrade
in time, which is difficult to calibrate. Although the \Index{TSI} varies only weakly over the
\Index{solar cycle}, such measurements revealed that the largest part of the variations
are produced at wavelengths shorter than 4000~\AA, so that the irradiance in the
\Index{ultraviolet} (\Index{UV}) can vary by up to a factor of two over the \Index{solar cycle}, e.g. at the
wavelength of \Index{Lyman~$\mathrm{\alpha}$} \citep{Krivova2006,Harder2009}. High-resolution
observations of \Index{bright point}s with the stratospheric observatory \sunrise{}
\citep{Solanki2010,Barthol2011} showed that the \Index{bright point} contrasts are particularly
high in the \Index{UV} (see chapter~\ref{Bp1Chapter}), so that the radiative properties of the
\Index{bright point}s play an important role in influencing irradiance variations and hence
possibly the \Index{terrestrial climate}. Variations in the radiative UV flux can influence
the chemistry of the \Index{stratosphere}, which can propagate into the \Index{troposphere} and
finally change the climate \citep{London1994,Larkin2000,Haigh2010}. Correlations
between the solar irradiance variations and climate indicators suggest a causal relation
\citep[e.g.][]{Damon1994,Bond2001}. The most noted indication for such a solar forcing of
climate change is the so-called \Index{Maunder minimum} between 1645-1715, where the number of \Index{sunspot}s
was extremely low and exceptionally cold winters were observed in Europe and North America
\citep{Eddy1976,Bradley1993}.

One part of this thesis is a small contribution to a better understanding of these
photospheric bright points, while the other part addresses a further type of small-scale
magnetic structures, the \Index{umbral dot}s, i.e. small bright features inside the dark umbra
of sunspots. Umbral dots provide only a negligibly small contribution to the \Index{total solar irradiance}
and hence do not have a noteworthy influence on the \Index{terrestrial climate}, but they are
important for understanding the subsurface energy transport in \Index{sunspot}s. Outside
sunspots and pores the energy is transported to the solar surface by \Index{convection}, which
manifests itself in the \Index{granulation} pattern that is typical for images of the quiet \Index{photosphere}.
The strong and nearly vertical magnetic field of an \Index{umbra} suppresses convective processes
\citep{Biermann1941}, however, the observed umbral brightness is too high for a
completely inhibited energy transport. According to \citet{Adjabshirzadeh1983},
37\% of the radiative umbral flux is provided by umbral dots. Magnetoconvective processes
in umbral fine structure, such as umbral dots and \Index{light bridge}s, are assumed to be largely
responsible for the umbral energy transport \citep{Weiss1990,Weiss2002}.

Physical basics, phenomena, and techniques that are important for the understanding
of this thesis, are treated in chapter~\ref{Basics}. Chapter~\ref{Review} gives
an overview of the current state of research in the considered fields.
Chapter~\ref{Instrumentation} describes the instrumentation which was used for the
observations of the following five chapters. The focus is laid on the description of the
\sunrise{} Filter Imager, because the software engineering of this instrument was
a significant part of my work in the framework of this thesis. In chapter~\ref{Ud1Chapter}
\citep[published as][]{Riethmueller2008d}, I analyze a time series of a sunspot
that was observed with the \Index{Swedish Solar Telescope} on the Canary Island La Palma.
The time series contains thousands of umbral dots, whose analysis yields statistics of
lifetimes, sizes, horizontal velocities, peak intensities, and distances travelled over
their lifetimes. It is the first extensive statistical umbral dot study of a
diffraction-limited time series obtained with an 1-m telescope. The vertical temperature,
velocity, and magnetic field structure of umbral dots, as observed with the
\Index{spectropolarimeter} onboard the \hinode{} satellite, is analyzed in
chapter~\ref{Ud2Chapter} \citep[published as][]{Riethmueller2008c}. For the
first time, full Stokes profiles could be inverted for more than two vertical nodes,
so that the results could be compared in detail with state-of-the-art magneto-hydrodynamical
simulations. A close correspondence with the simulations was found.
Chapter~\ref{Ud3Chapter} refines the analysis of the same \hinode{} data set
by applying a new, considerably improved \Index{Stokes inversion} technique that greatly
reduces the effect of the spatial \Index{point spread function} of the telescope. For the
first time, systematic downflows in the close vicinity of umbral dots could be found in the
observations. The results also showed rather well balanced up- and downflow \Index{mass flux}es.
In chapter~\ref{Bp1Chapter} \citep[published as][]{Riethmueller2010}, I present the
first high resolution observations of the quiet Sun in the near-ultraviolet as observed with
the balloon-borne solar telescope \sunrise{} and I analyze brightness, velocity, and
\Index{polarization} of \Index{bright point}s. The observed bright points are compared with
magneto-hydrodynamical simulations in chapter~\ref{Bp2Chapter} from which new insights into
the nature of bright points are obtained. Finally, chapter~\ref{Outlook} gives an outlook
on how the presented studies on small-scale magnetic features could be usefully continued.

\chapter{Relevant basic physics}\label{Basics}

The fundamentals, this work relies on, are explained in this chapter. First,
the different layers of the solar atmosphere are presented and the solar
photosphere is introduced. Examples of the relevant solar surface phenomena
are given and the terms used in this thesis are described. Two important effects
employed to measure velocities and magnetic fields are explained, the \Index{Doppler effect}
and the \Index{Zeeman effect}. After an introduction to the \Index{Stokes formalism} of
describing the \Index{polarization} of light, the topic of \Index{radiative transfer} in the
solar atmosphere is raised and so-called inversion techniques are presented.
Inversion is a modern analysis method that has got more and more important
within the last years because it helps in retrieving all the
significant physical quantities of the solar surface from data that contain
spectral as well as polarimetric information. The basic equations used for
magneto-hydrodynamical simulations are explained in a further section and
finally, the \Index{diffraction limit} is mentioned since it limits the spatial
resolution of every observation with a telescope.

\section{Phenomena of the solar surface}\label{Phenomena}

The Sun is composed of a \Index{plasma}, i.e. matter containing free electrons and free ions
which make the plasma to an electrical conductor. Examples of plasmas are the gas of
the flame of a candle, the \Index{interstellar medium}, or the solar gas. In the case of rigid
bodies like planets and moons, it is clear what their surface is. It is not so clear
for the Sun because the density of the solar plasma decreases continuously with the
radial distance from the center of the Sun, but there is no clear phase transition.

\begin{figure*}
\centering
\includegraphics*[width=\textwidth]{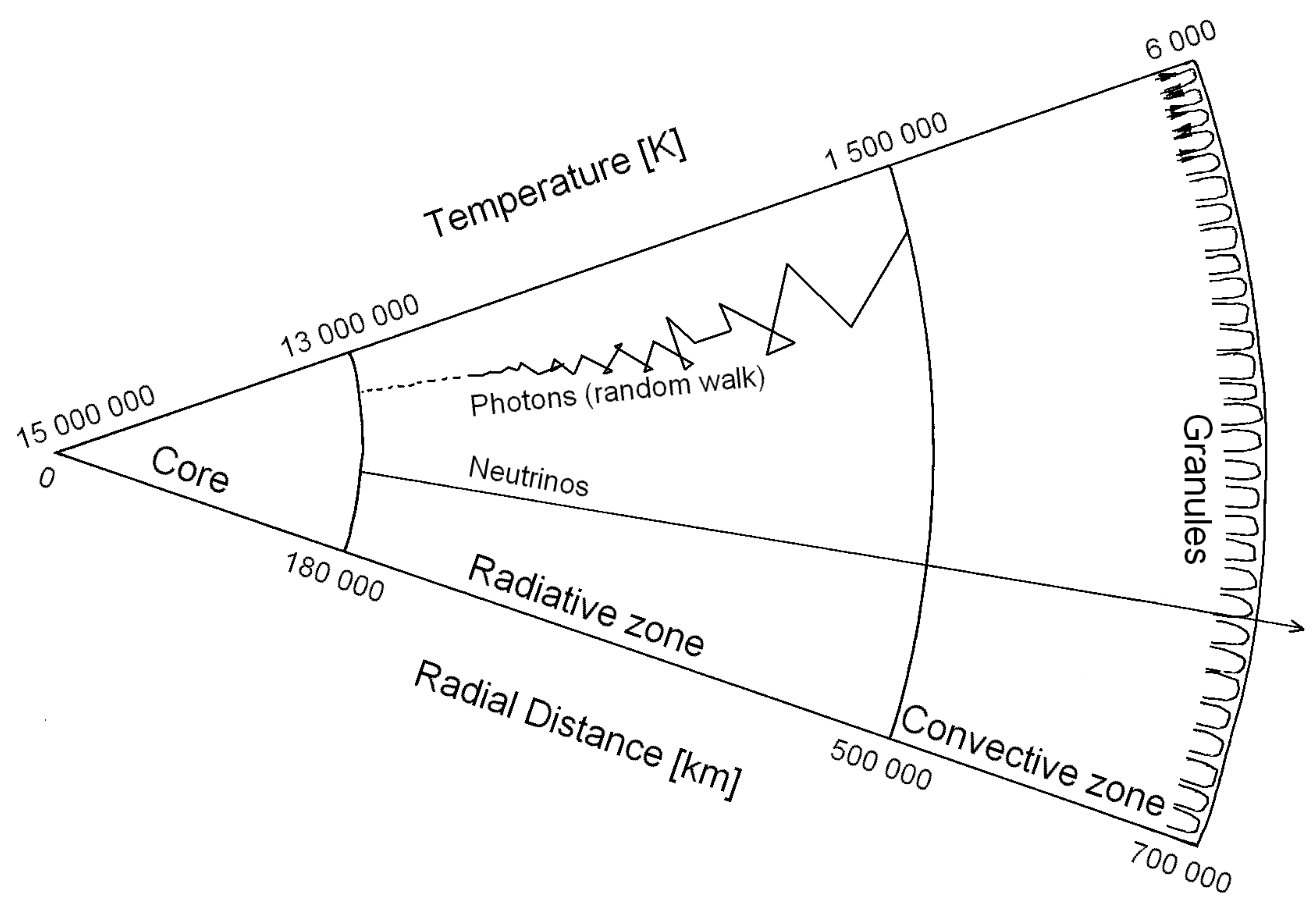}
\caption{Internal structure of the Sun. The energy of the Sun is generated in the
core by \Index{nuclear fusion} and then transported by radiation, later by \Index{convection} to
the solar surface. The temperature decreases from 15 million Kelvin in the core to
6000 Kelvin at the top of the \Index{convective zone}.}
\label{FigSolarInterior}
\end{figure*}

The Sun emits particles (mainly \Index{neutrino}s, \Index{electron}s, \Index{proton}s, and \Index{neutron}s) and
\Index{electromagnetic wave}s. Integrated over all wavelengths of the \Index{electromagnetic spectrum},
a total solar radiation of 1361-1363 watts per square meter (the so-called \Index{solar constant}) is
measured from near-Earth space \citep{Ball2012}. The surface of the Earth is hit by significantly less
radiation because the terrestrial atmosphere is not transparent for all wavelengths.
The total solar irradiation is $3.8\times 10^{26}$ watt. This huge energy flux originates
from the interior of the Sun by \Index{nuclear fusion} of \Index{hydrogen} to \Index{helium}. This \Index{fusion zone}
is also called the core of the Sun and has a temperature of up to 15 million Kelvin
and a radius of about 180000 kilometers (Fig.~\ref{FigSolarInterior}).

The \Index{nuclear fusion} creates, among others, \Index{photon}s and \Index{neutrino}s. The neutrinos
show hardly any interaction with matter so that they can escape the Sun practically
unhindered. The motion of the photons is disturbed all the time by collisions
with other \Index{plasma} particles due to the extremely high plasma density.
The photons need up to one million years to leave the Sun. Their \Index{mean free path}
is estimated to be less than a millimeter near the core. The \Index{mean free path} of the
photons increases as the density and temperature decrease on the outside. The
energy transport is completely dominated by radiation. This changes at a distance of
500000 kilometers from the center of the Sun. The \Index{radiation zone} ends there and
the \Index{convective zone} of about 200000 kilometers thickness starts. In this zone,
the energy is transported by \Index{convection}, i.e. plumes of hot gas rise, are cooled,
the cooler plasma descends, and the cycle starts again. In the lower \Index{photosphere},
the \Index{convection} is visible as the \Index{granulation} pattern shown in Fig.~\ref{FigSolarZoo2}.
The bright \Index{granule}s, with a typical diameter of 1000 kilometers, are regions of rising
hot plasma, the dark regions between the granules (the \Index{intergranular lane}s) are regions
of downflowing cold \Index{plasma}.

On top of the \Index{convective zone}, i.e. at a distance of almost 700000 kilometers
from the solar center, the \Index{plasma} density is so low and hence the \Index{mean free path} of
the \Index{photon}s is so large, that they can escape from the Sun. The exact position of
this depends on the wavelength of the photons. Almost all the visible light originates
from an approximately 500 kilometers thick layer that is called the \Index{photosphere}.
The visible \Index{continuum} is formed near the bottom of the \Index{photosphere} which is hence
called the \Index{solar surface}. If more precision is needed, often the \Index{solar surface}
is defined as the position where light from the \Index{continuum} at 500 nanometer reaches
\Index{optical depth} unity (see section~\ref{RadiativeTransfer}).

\begin{figure*}
\centering
\includegraphics*[width=\textwidth]{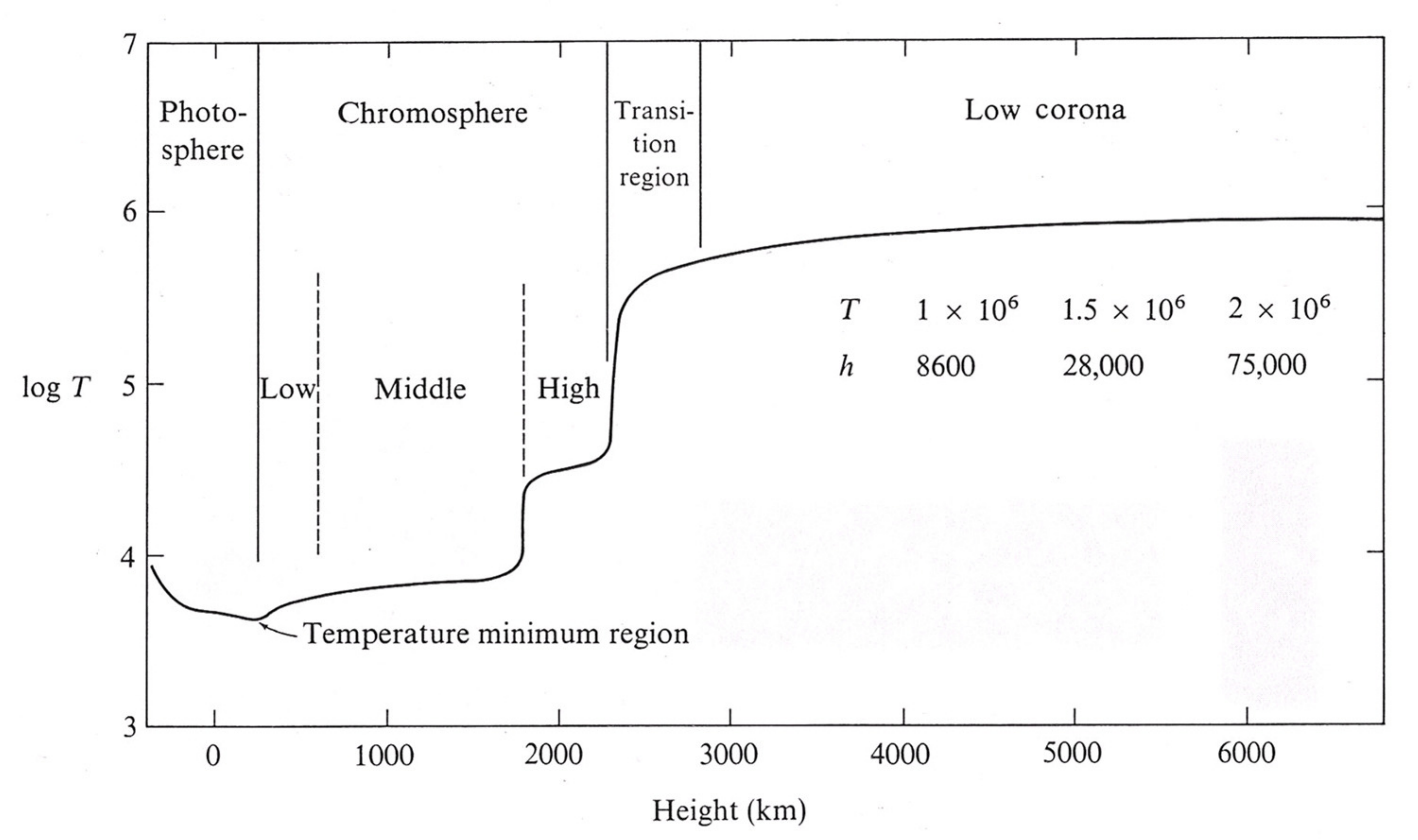}
\caption{The solar atmospheric layers and their temperatures. From \citet{Athay1976}.}
\label{FigTempVsHeight}
\end{figure*}

In classical \Index{plane-parallel} models of the solar atmosphere, the solar \Index{plasma} reaches
its \Index{temperature minimum} of roughly 4200 Kelvin (Fig.~\ref{FigTempVsHeight})
at the upper end of the \Index{photosphere}. Surprisingly, the temperature increases again up
to 50000 Kelvin in the adjacent atmospheric layer, the \Index{chromosphere}, while the \Index{plasma}
density continues to decrease. The chromosphere is about 2000 kilometers thick and is
continued by the thin \Index{transition region}, where the temperature suddenly escalates.
Temperatures between one and two million Kelvin are measured in the overlying \Index{corona}.
The \Index{corona} does not have a sharp outer boundary, but smoothly transits into the
\Index{solar wind}.

\begin{figure*}
\centering
\includegraphics*[width=\textwidth]{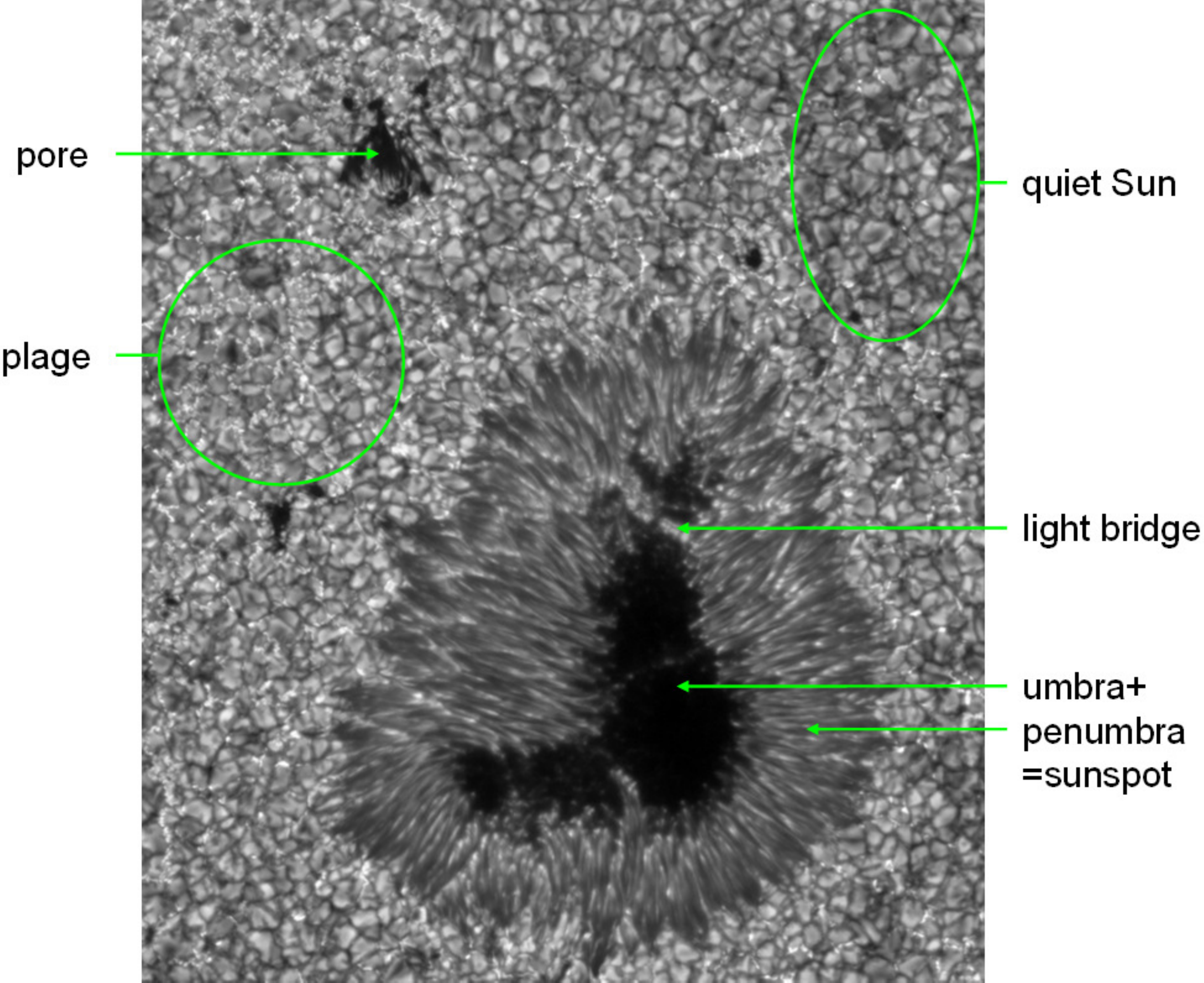}
\caption{The solar zoo of the photosphere as observed with the Solar Optical Telescope onboard the \hinode{}
satellite (downloaded from the European Hinode Science Data Center at Oslo University).}
\label{FigSolarZoo}
\end{figure*}

\begin{figure*}
\centering
\includegraphics*[width=\textwidth]{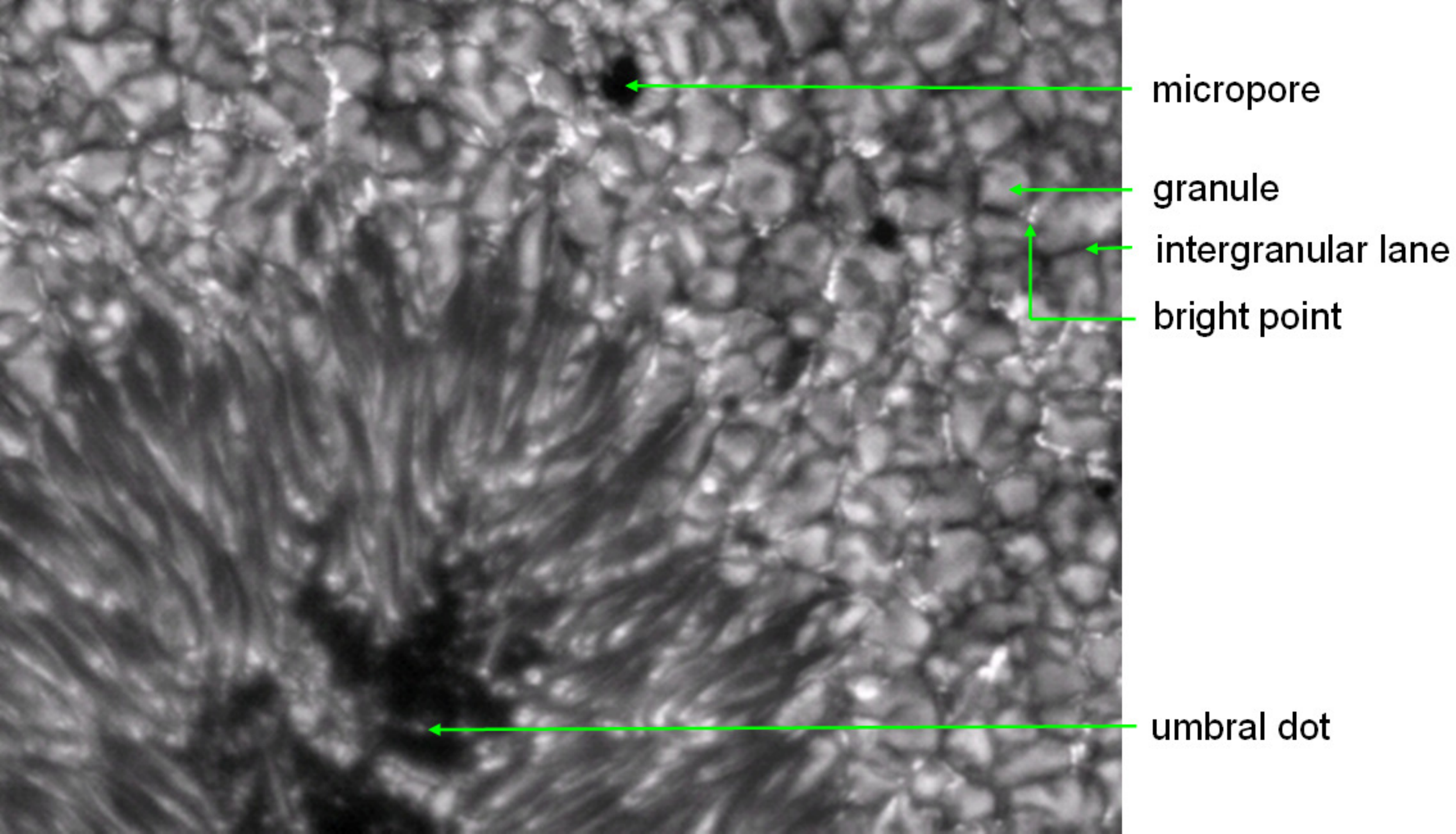}
\caption{The tiny features of the photospheric \Index{solar zoo}. }
\label{FigSolarZoo2}
\end{figure*}

This work studies the \Index{electromagnetic radiation} that originates from the \Index{photosphere}.
A typical photospheric observation can be seen in Fig.~\ref{FigSolarZoo}. It was
acquired with the space-borne observatory \hinode{} at a wavelength of 430 nanometers
on 2007 May 2. The field of view is $60\times70~\mathrm{Mm^2}$ and shows a fully
developed \Index{sunspot} and some of its vicinity.

The dark region within a \Index{sunspot} is named the {\bf \Index{umbra}}. The outer, slightly brighter region,
that consists of many radially oriented filaments, is called the {\bf \Index{penumbra}}. A dark region
without a penumbra is named a {\bf \Index{pore}}. On average, pores are smaller than sunspots. If
a pore has a size of only roughly a granule then it is called a {\bf \Index{micropore}}. Sometimes
a sunspot consists of several umbrae that are separated by {\bf \Index{light bridge}s}, bright lanes
connecting two parts of the penumbra. Outside spots and pores, one can see the \Index{granulation}
in which the \Index{convection} is manifested as a comblike structure. The dark regions in between
the granules are named {\bf \Index{intergranular lane}s}, see the enlarged image in Fig.~\ref{FigSolarZoo2}.
Sometimes one can find small roundish brightness enhancements in the darker intergranular
lanes - these are {\bf \Index{bright point}s}. If the granulation pattern contains only a few or
hardly any bright points, we call it a {\bf quiet-Sun} region. In contrast to that, it can
happen that the intergranular lanes are strongly filled with bright points, then we refer to
it as a {\bf \Index{plage}} region. Plage regions are found very often in the vicinity of sunspots.
Also the dark pores and umbrae can contain small roundish brightness enhancements called
{\bf \Index{umbral dot}s}.

\section{Doppler effect}

In the lower \Index{photosphere}, density and temperature of the solar \Index{plasma} are sufficiently
decreased so that the \Index{photon}s coming from the \Index{solar interior} can escape the Sun
(see section~\ref{Phenomena}). Some photons interact again with the photospheric gas.
Besides electrons, the gas consists of neutral atoms, ions, and molecules. Each of these
particles possesses various \Index{energy level}s. If a photon's wavelength corresponds
to the energy difference between two such levels, the photon can be absorbed by the
atom, ion, or molecule. The same particle can later emit a photon of the same energy
if it does not collide and get de-excited in that way first. Thus, the many thousand
absorption lines in the \Index{electromagnetic spectrum} of the  Sun, also known as \Index{Fraunhofer line}s,
are formed (see also section~\ref{LineFormation}).

Because of the up- and downflows in the \Index{convective zone}, the gas of the \Index{photosphere}
is also in motion. If a moving gas cloud emits photons, they exhibit the optical \Index{Doppler effect},
i.e. a \Index{spectral line} is shifted. Since typical photospheric velocities are on the
order of only a few km/s, it is sufficient to consider the non-relativistic \Index{Doppler shift} of:
\begin{equation}\label{Eq_DopplerShift}
\Delta\lambda=\lambda-\lambda_0=\lambda_0\frac{v_{\rm{LOS}}}{c} ,
\end{equation}
where $\lambda_0$ is the wavelength at rest, $c$ is the speed of light, and $v_{\rm{LOS}}$
is the \Index{line-of-sight} (\Index{LOS}) component (i.e. the component in the direction of the observer)
of the cloud velocity. In this thesis, the sign of the velocity is always defined such that
negative velocities are upflows (\Index{blueshift}s). More details about the \Index{Doppler effect} can be
found, e.g., in \citet{Grimsehl1988b} and \citet{BergmannSchaeferBd3}.

\begin{figure*}
\centering
\includegraphics*[width=\textwidth]{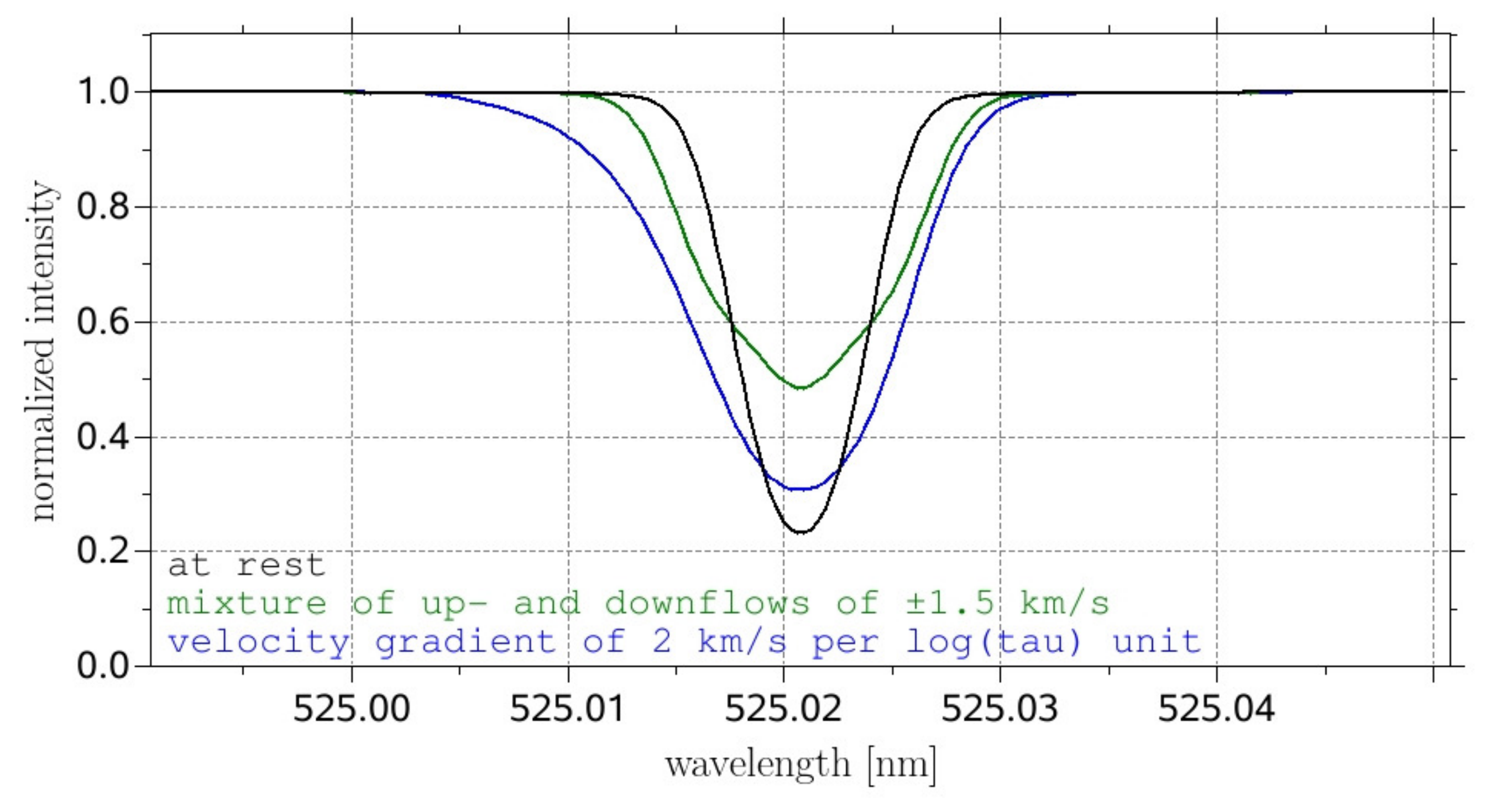}
\caption{The influence of the velocity on a spectral line. A mixture of moderate, height independent up- and downflows
of equal strength within the \Index{resolution element} broadens a spectral line. A velocity gradient leads
to an asymmetric profile.}
\label{FigDoppler2}
\end{figure*}

The temperature of the gas particles leads to a \Index{Maxwell-Boltzmann distribution} for their
velocities. Since the \Index{Doppler effect} acts also microscopically, the effect causes a broadening
of the \Index{spectral line}s \citep{SobelMan1973}. If there are upflows and downflows close to each other
within the \Index{resolution element} (or on top of each other along the line-of-sight), then the blue-
and \Index{redshift}s are superimposed macroscopically and can cause an additional \Index{line broadening} as shown
by the green line of Fig.~\ref{FigDoppler2} for a two-component atmosphere having height independent
velocities of $\pm$1.5~km\,s$^{-1}$ using the example of the neutral iron line at 525.02~nm. Since both
components are equally strong, the \Index{spectral profile} remains symmetric. Velocity gradients
along the line-of-sight lead to asymmetric profiles. The blue line of Fig.~\ref{FigDoppler2}
exhibits the profile of a one-component atmosphere having a gradient of 2~km\,s$^{-1}$ per
$\log(\tau)$ unit\footnote{The synthetic profiles of Figs.~\ref{FigDoppler2},
\ref{FigZeeman1}, \ref{FigZeeman2}, \ref{FigZeeman3}, \ref{FigZeeman4} and \ref{FigZeeman5}
are calculated with the \Index{STOPRO} routines \citep{Solanki1987} using a standard \Index{Kurucz atmosphere}
at 5750~K \citep{Kurucz1993}.}.

In the following section, the \Index{Zeeman effect} is introduced, which can also lead to a \Index{line broadening}.
The question arises how to separate the various reasons for \Index{line broadening} when observational data
are analyzed.

\section{Zeeman effect}\label{ZeemanEffect}

\begin{figure*}
\centering
\includegraphics*[width=\textwidth]{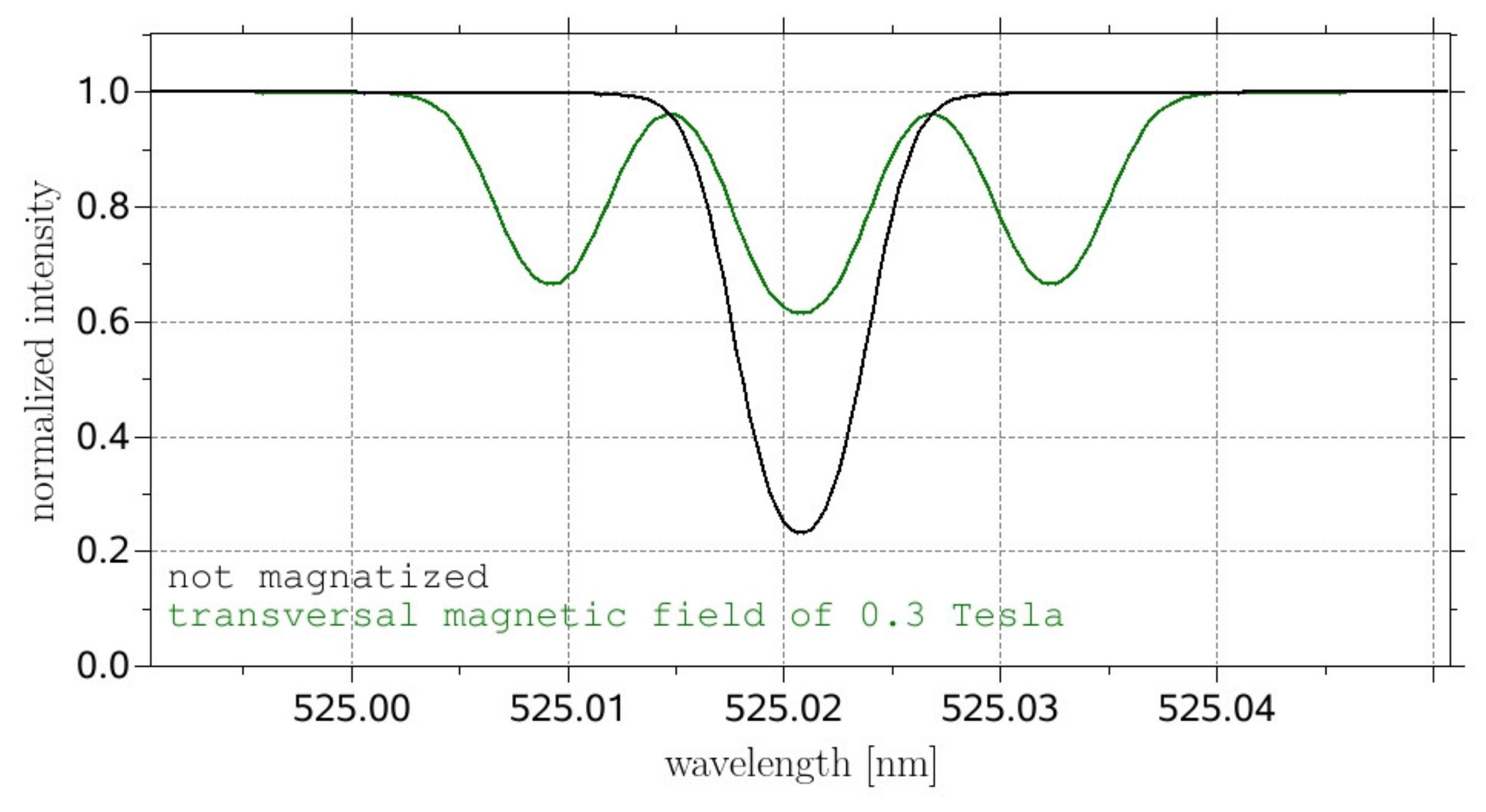}
\caption{The influence of a strong, height independent transversal magnetic field on a spectral line.
Due to the \Index{Zeeman effect}, the line is split up into a \Index{Lorentz triplet}.}
\label{FigZeeman1}
\end{figure*}

The \Index{Zeeman effect} gives us the possibility to measure the strength and the orientation of magnetic fields
on the solar surface. In 1896, the Dutch physicist Pieter Zeeman could confirm the prediction of his
colleague Hendrik Antoon Lorentz that many \Index{spectral line}s split into three components in the presence of
a magnetic field \citep{Zeeman1897a,Zeeman1897b,Zeeman1897c,Zeeman1897d}. Fig.~\ref{FigZeeman1} gives an
example, again for the Fe\,{\sc i} line\footnote{Fe\,{\sc i} means neutral iron, Fe\,{\sc ii} is singly
ionized iron, Fe\,{\sc iii} doubly ionized iron, etc.} at 525.02~nm. The black line shows the spectral line
in the absence of a magnetic field, the green line for a field of 0.3~Tesla (3000~Gau\ss{}), a typical
field strength in \Index{sunspot}s.

A \Index{spectral line} is formed by the transition of electrons between two \Index{energy level}s. The \Index{quantum number}s
$L$ (orbital angular momentum), $S$ (spin angular momentum), and $J$ (total angular momentum) describe
the quantum-mechanical state of such \Index{energy level}s \citep[see e.g.][]{Schwabl1992}. Without any magnetic
field, the energy of an atomic level only depends on the total angular momentum, i.e. on $J$. In the
presence of a magnetic field it is:
\begin{equation}\label{Eq_EnergySplit}
E_{J,M}=E_{J}+\frac{\hbar{}e}{2m_e}gMB ,
\end{equation}
i.e. the magnetic field splits the \Index{energy level} ($L$,$S$,$J$) into $2J+1$ sublevels of slightly different
energies. These sublevels are described by the magnetic \Index{quantum number} $M=-J,..,0,..,+J$.
In Eq.~(\ref{Eq_EnergySplit}) $\hbar{}=h/2\pi$ means the reduced \Index{Planck constant}, $e$ is the \Index{elementary charge},
$m_e$ the \Index{electron mass}, $B$ the \Index{magnetic field} strength, and $g$ is the \Index{Land\'e factor} of
the energy level \citep{Lande1923,Uhlenbeck1925,Uhlenbeck1926}, calculated as:
\begin{equation}\label{Eq_LandeFactor}
g=\frac{3}{2}+\frac{S(S+1)-L(L+1)}{2J(J+1)} .
\end{equation}
In the case of $J=0$ we also have to set $g=0$.

A negligible \Index{magnetic moment} of the nucleus is assumed in the Eqs.~(\ref{Eq_EnergySplit})
and (\ref{Eq_LandeFactor}) as well as the validity of the \Index{LS coupling} (also called \Index{Russel-Saunders coupling})
and that the coupling between the magnetic field and the atom is small compared to
the \Index{spin-orbit interaction} (linear \Index{Zeeman effect}). In the case of very strong magnetic fields or
spectral lines of atoms with a high atomic number, other coupling schemes, the quadratic Zeeman
effect, or the \Index{Paschen-Back effect} have to be taken into account \citep[see, e.g.,][]{SobelMan1973}.

\begin{figure*}
\centering
\includegraphics*[width=9cm]{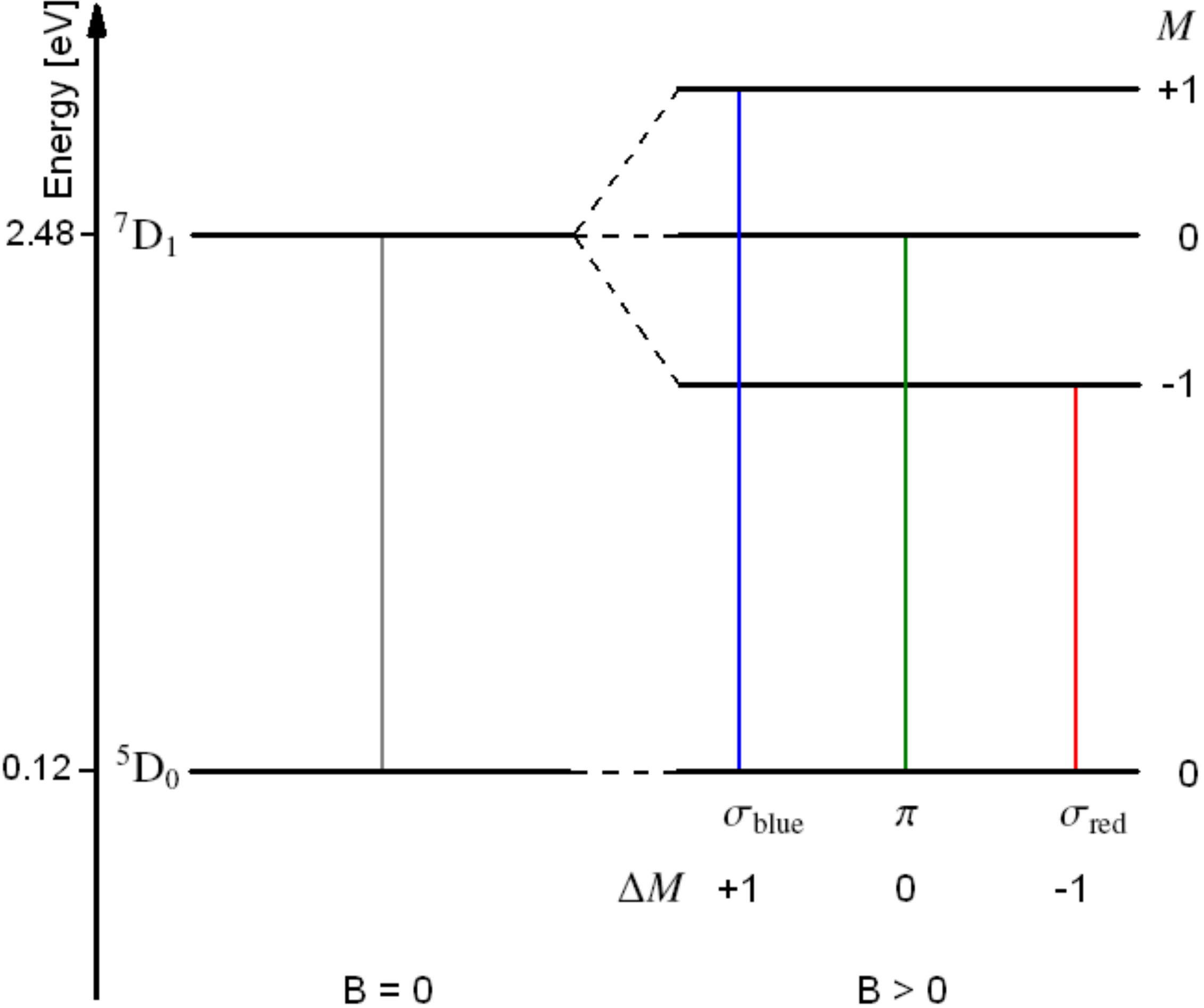}
\caption{The normal \Index{Zeeman effect} using the example of the neutral iron line at 525.02~nm.}
\label{FigNormalZeemanEffect}
\end{figure*}

The \Index{term scheme} in Fig.~\ref{FigNormalZeemanEffect} illustrates the \Index{Zeeman splitting} for the
iron line at 525.02~nm that is important for chapters~\ref{Bp1Chapter} and \ref{Bp2Chapter}.
A transition between the \Index{energy level}s $\mathrm{^{5}D_{0}\leftrightarrow{}^{7}D_{1}}$ forms
the \Index{spectral line}\footnote{The usual \Index{term symbol}s of the form
$\mathrm{^{2S+1}L_{J}}$ provide information about the \Index{quantum number}s $L$, $S$ and $J$. The
letters S,P,D,F, ... mean an orbital angular momentum corresponding to $L=0,1,2,3,...$
$\mathrm{{}^{7}D_{1}}$ means nothing else than $L=2, S=3, J=1$.}. Because of $J=0$, the
lower energy level $\mathrm{^{5}D_{0}}$ is not degenerated, i.e. it does not split in the
presence of a magnetic field. Thanks to $J=1$, the upper energy level splits into three
different sublevels with the magnetic quantum numbers $M=-1,0,+1$. A \Index{Lorentz triplet} is
formed, which is named the normal \Index{Zeeman effect}.

\begin{figure*}
\centering
\includegraphics*[width=9cm]{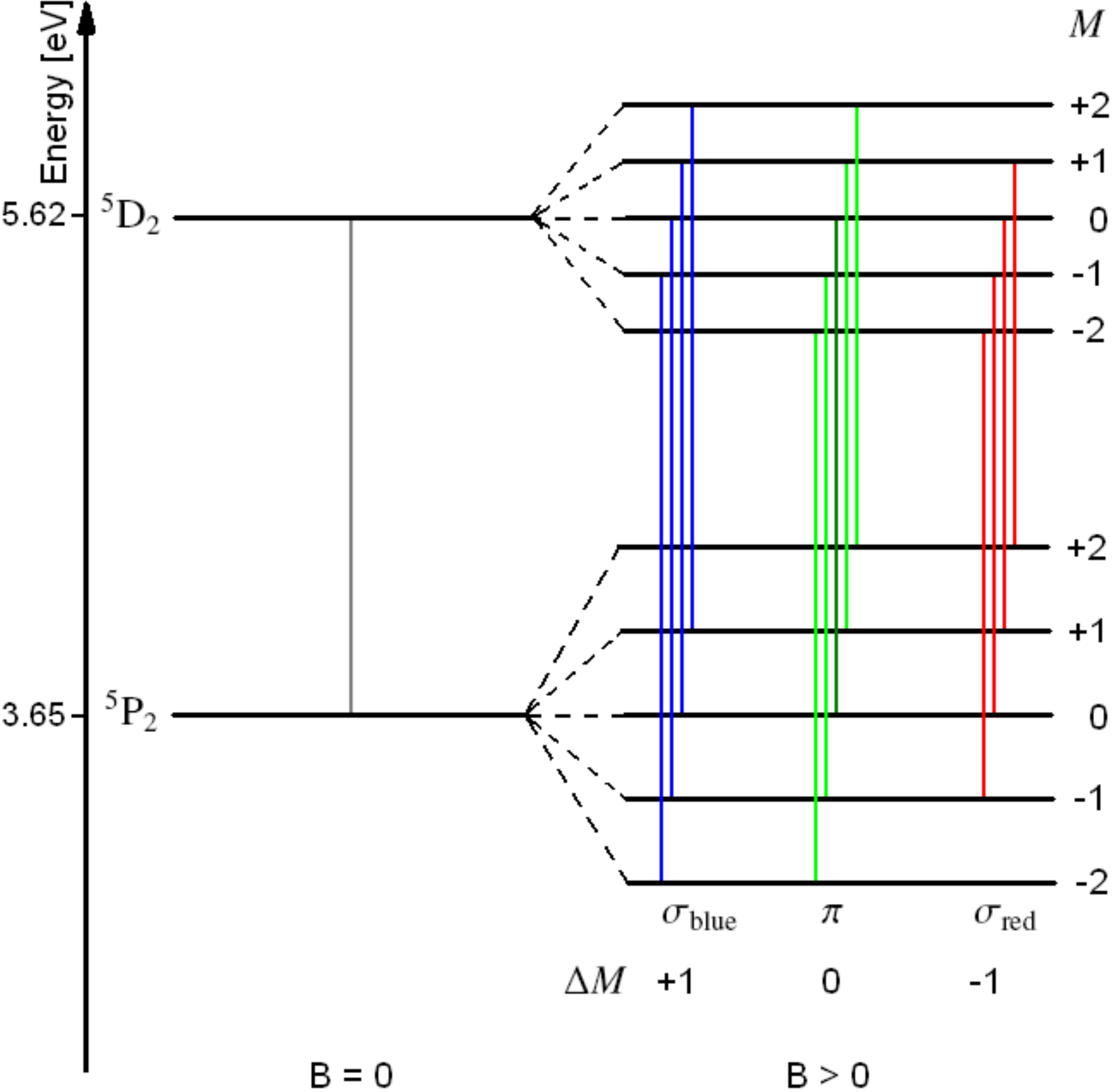}
\caption{The anomalous Zeeman effect using the example of the neutral iron line at 630.15~nm.}
\label{FigAnomalZeemanEffect}
\end{figure*}

A look at Eq.~(\ref{Eq_EnergySplit}) reveals that the formation of exactly three splitted lines
is not really ``normal'', but a particular case that only occurs if either one of the two
energy levels of the considered transition has $J=0$ (as shown in the example of
Fig.~\ref{FigNormalZeemanEffect}) or the \Index{Land\'e factor}s of the two levels are equal.
In the general case of the anomalous Zeeman effect, more than three components are observed.
For quantum-mechanical reasons not all transitions are allowed. The magnetic quantum number of the
two energy levels must not differ by more than one, i.e.:
\begin{equation}\label{Eq_SelectionRule}
\Delta M=M_{u}-M_{l}=0,\pm{}1 ,
\end{equation}
where the indices $l$ and $u$ mean the lower and upper energy level, respectively \citep[see, e.g.,][]{DelToroIniesta2003}.
The anomalous Zeeman effect is illustrated in Fig.~\ref{FigAnomalZeemanEffect} for the
$\mathrm{^{5}P_{2}\leftrightarrow{}^{5}D_{2}}$ transition of the iron line at 630.15~nm that
is important for chapters~\ref{Ud2Chapter} and \ref{Ud3Chapter}. Both energy levels have $J=2$ but differ in the
\Index{Land\'e factor}s, $1.8333$ and $1.5$, so that the line splits into 13 components. The transitions
with $\Delta M=0$ are called $\pi$~component\index{picomponent@$\pi$~component}s, the $\Delta M=\pm{}1$ transitions are the
$\sigma_{blue}$ and $\sigma_{red}$~components.

From Eq.~(\ref{Eq_EnergySplit}) follows for the Zeeman line splitting of the transition 
$l\leftrightarrow{}u$:
\begin{equation}\label{Eq_ZeemanSplit}
\Delta\lambda=\lambda-\lambda_0=\lambda_0\left[ \frac{1}{1+\frac{e B \lambda_0}{4 \pi c m_e}(g_u M_u - g_l M_l)}-1\right] ,
\end{equation}
where $\lambda_0$ is the reference wavelength of the non-magnetic case. Note that the strength
of the various components differs in the general case, which is not further considered here
\citep[see, e.g.][]{DelToroIniesta2003,LandiDeglInnocenti2004}.

In practice it is often sufficient to calculate the line splitting as a wavelength shift
between the center of gravity of the $\sigma$~components, $\lambda_{\sigma}$, to the
reference wavelength of the non-magnetic case, $\lambda_0$. It is:
\begin{equation}\label{Eq_ZeemanSplitCog}
|\Delta\lambda|=|\lambda_{\sigma}-\lambda_0|=\frac{eB\lambda_0^2}{4\pi c m_e}\cdot g_{eff} .
\end{equation}
The effective \Index{Land\'e factor} of the line is defined as \citep{Shenstone1929}:
\begin{equation}\label{Eq_EffectiveLandeFactor}
g_{eff}=\frac{1}{2}(g_u+g_l)+\frac{1}{4}(g_u-g_l)[J_u(J_u+1)-J_l(J_l+1)]
\end{equation}
and is a measurement for the sensitivity of the spectral line to Zeeman splitting.
More details about the Zeeman effect are given by
\citet{SobelMan1973,Schwabl1992,DelToroIniesta2003,LandiDeglInnocenti2004}.

The \Index{Land\'e factor} of the 525.02~nm line, $g=3$, is one of the biggest \Index{Land\'e factor}s
in the visible spectral range (only the Mn\,{\sc i} line at 407.03~nm, the V \,{\sc i} line
at 411.66~nm, and the Fe\,{\sc i} line at 422.45~nm have higher \Index{Land\'e factor}s) and
therefore makes the line very sensitive to the Zeeman effect, so that even weak magnetic
fields in the quiet Sun can be well measured.

\begin{figure*}
\centering
\includegraphics*[width=\textwidth]{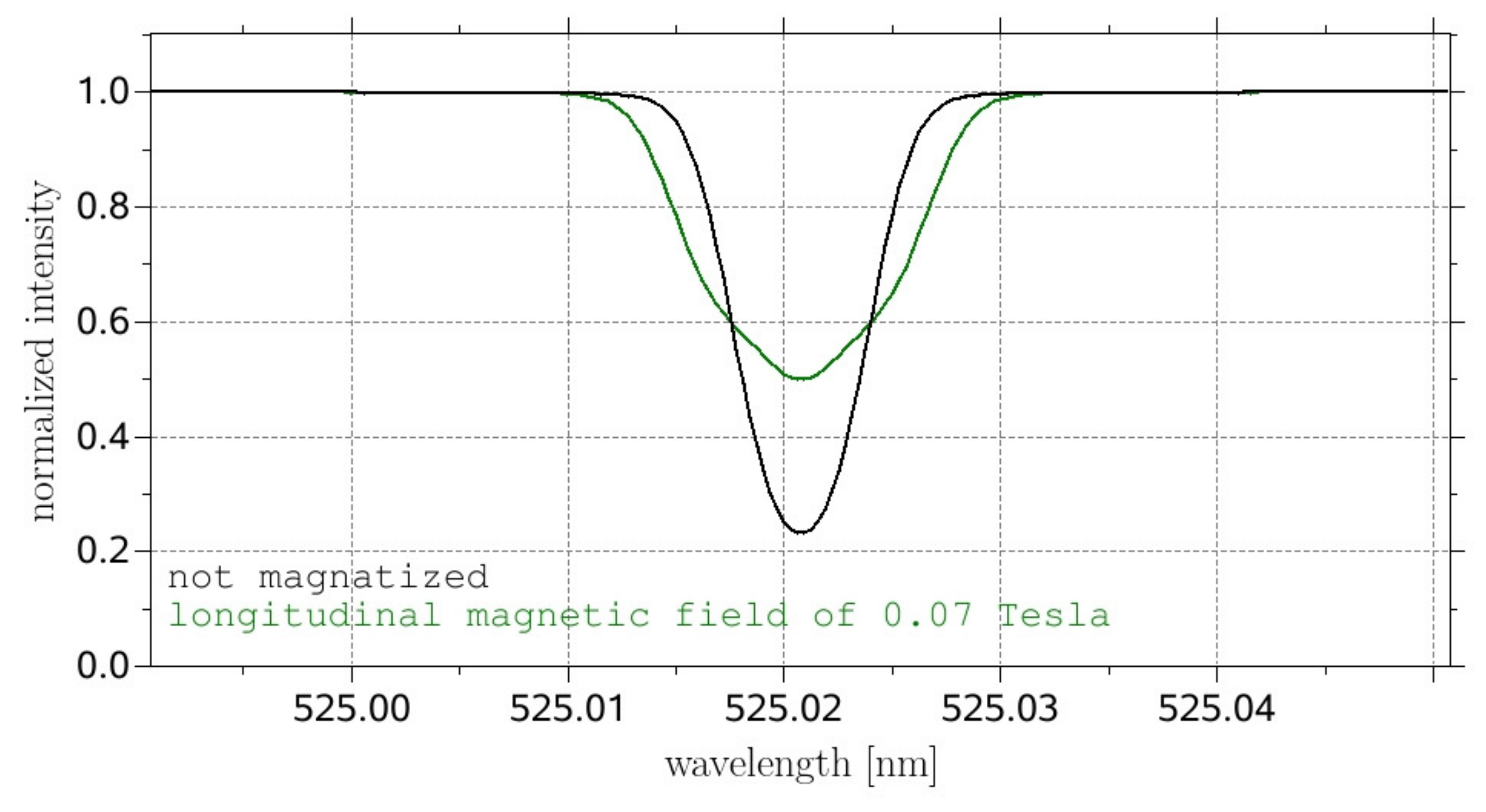}
\caption{The influence of a weak, height independent longitudinal magnetic field on a spectral line.
The field broadens the line in a quite similar way as the velocity effect illustrated in Fig.~\ref{FigDoppler2}.}
\label{FigZeeman2}
\end{figure*}

Eq.~(\ref{Eq_ZeemanSplitCog}) reveals that the splitting increases linearly with the magnetic field
strength and quadratically with the wavelength, i.e. the splitting is particularly conspicuous in the
infrared. Theoretically, the knowledge of the Zeeman splitting allows the determination of the
magnetic field strength, but in practice, typical solar surface field strengths lead very often to
Zeeman splittings on the order of the \Index{Doppler broadening} of the spectral lines. Fig.~\ref{FigZeeman2}
displays the weak splitting caused by a 0.07~T (700~G) magnetic field, which cannot be distinguished
from the \Index{line broadening} caused by velocity effects shown in Fig.~\ref{FigDoppler2} (green line).
Also the orientation of the field cannot be determined from the facts mentioned so far.

\begin{figure*}
\centering
\includegraphics*[width=\textwidth]{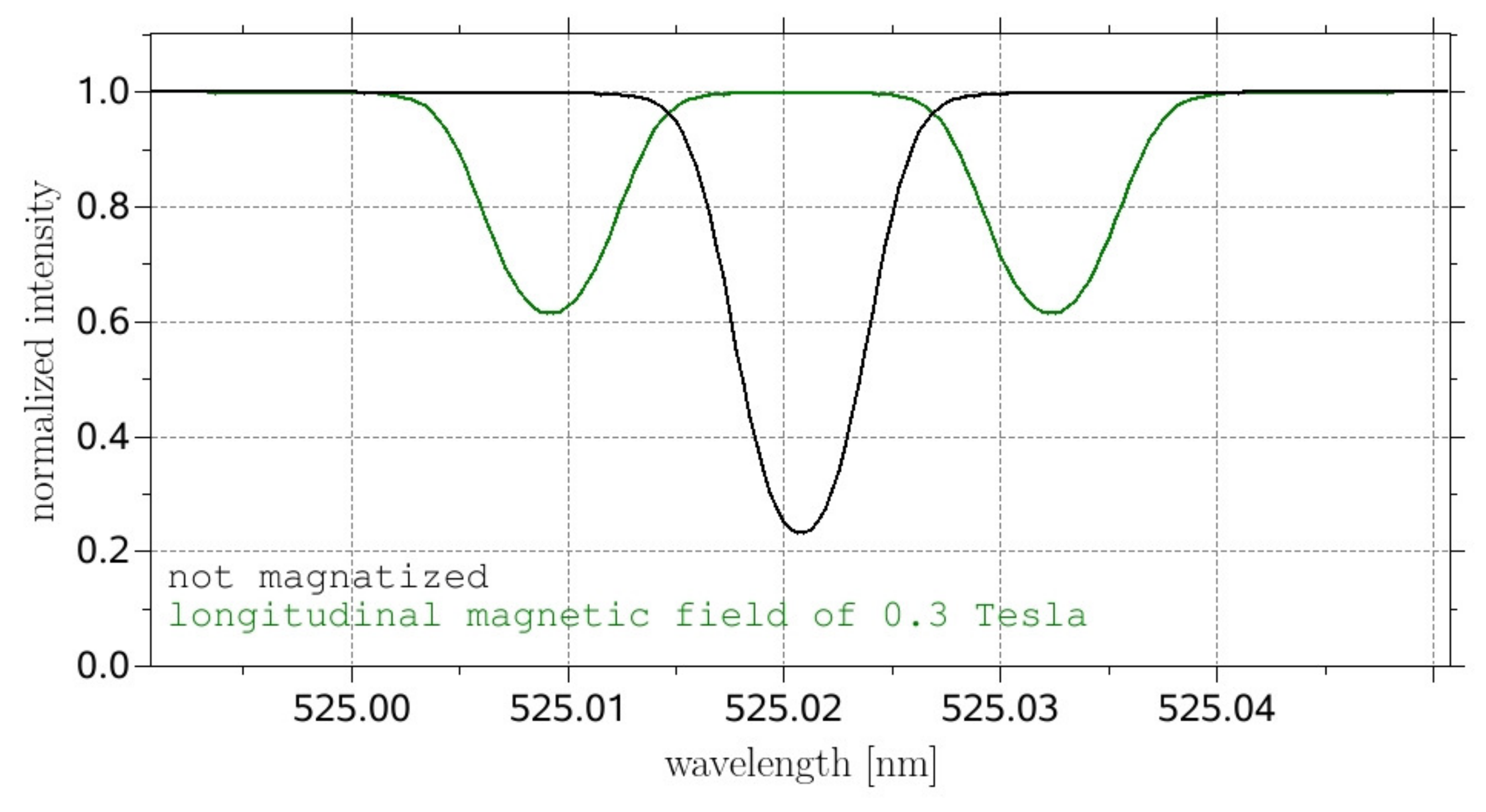}
\caption{The influence of a strong, height independent longitudinal magnetic field on a spectral line.
The $\pi$~component is not visible since the field is parallel to the line-of-sight.}
\label{FigZeeman3}
\end{figure*}

Fortunately, it has been shown that the components of the \Index{Zeeman splitting} are not only
wavelength shifted but are also polarized in the following way: If the magnetic field is
oriented perpendicular to the line-of-sight (\Index{transversal field}), the $\pi$~component\index{picomponent@$\pi$~component}s are
linearly polarized parallel to the magnetic field. The $\sigma$~components\index{sigmacomponent@$\sigma$~component} are also linearly
polarized, but perpendicular to the field. If the magnetic field is parallel to the line-of-sight
(longitudinal field), the $\sigma_{blue}$- and $\sigma_{red}$~components are oppositely
circularly polarized and the $\pi$~component\index{picomponent@$\pi$~component}s are missing completely (see Fig.~\ref{FigZeeman3}).
For arbitrary angles between the magnetic field vector and the line-of-sight, the light of
the \Index{Zeeman component}s is elliptically polarized.

Because velocity and temperature effects do not lead to a \Index{polarization} of the light,
these effects can be clearly distinguished from magnetic effects by measurements of the
polarization properties. The polarization of light shall therefore be treated in more detail
in the next section.

\section{Polarization of light}

\begin{figure*}
\centering
\includegraphics*[width=7cm]{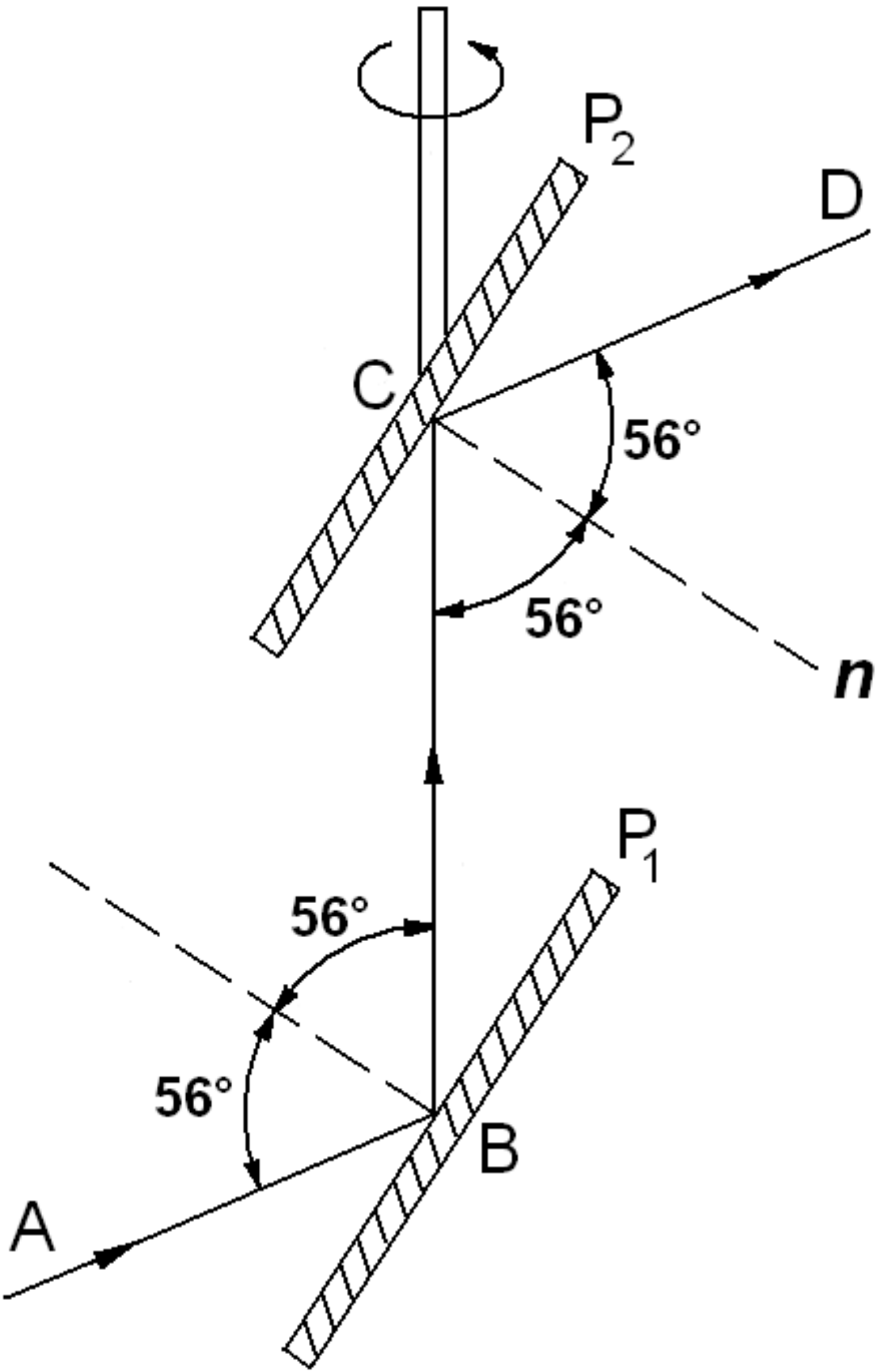}
\caption{Experiment to demonstrate the \Index{polarization} of light due to reflection on a plane surface
\citep[from][]{BergmannSchaeferBd3}.}
\label{FigLinearPolarizer}
\end{figure*}

In 1808, the French engineer and physicist \'Etienne Louis Malus discovered the broken symmetry
of a light beam around its direction of propagation if a glass plate reflects the beam \citep{Malus1809}.
Such an asymmetry of light was never observed before. Fig.~\ref{FigLinearPolarizer} shows a
simplified experimental setup. A common light ray coming from A is reflected by a glass plate $\rm{P_1}$
at point B. A second glass plate $\rm{P_2}$ makes the broken symmetry visible by reflecting the ray
again at point C into direction D. At the beginning, plate $\rm{P_1}$ is parallel to $\rm{P_2}$ and all
rays (i.e. AB, BC, CD) are in one plane. The experiment starts by rotating plate $\rm{P_2}$ around
the rotational axis BC, so that the ray CD leaves the plane while the rays AB and BC remain in the plane.
If the ray reflected by $\rm{P_1}$ would be fully symmetrical around its direction of propagation BC,
one would always observe a constant brightness for the ray CD for all rotation angles of $\rm{P_2}$.
This is not the case. One observes a maximal brightness for the rotation angles $\mathrm{0^\circ}$ and
$\mathrm{180^\circ}$, i.e. if the two plates $\rm{P_1}$ and $\rm{P_2}$ are parallel or antiparallel to
each other. A minimal brightness is observed for the rotation angles $\mathrm{90^\circ}$ and
$\mathrm{270^\circ}$. If the experiment is accomplished with monochromatic light and an incidence angle
of exactly $\mathrm{56^\circ}$, then the brightness goes down to zero. For a glass plate having a
\Index{refractive index} of 1.5, the so-called \Index{Brewster angle} is $\mathrm{56^\circ}$. The \Index{Brewster angle} is
defined as the incidence angle that leads to a reflected ray perpendicular to the refracted ray.

Nowadays we know that light exhibits properties of both waves and particles (\Index{wave-particle duality}).
The wave nature of light is described by \Index{electromagnetic wave}s with an electric field vector perpendicular
to the magnetic field vector and both vectors are perpendicular to the direction of propagation. Most
of the light sources are thermic emitters where the \Index{emission process}es of a huge amount of atoms are
superimposed. Even if a single electron of such a light source can be thought as a simple oscillator, each
electron oscillates independently, so that every oscillation direction is equiprobable. Such light is named
unpolarized light. The symmetry of unpolarized light around its direction of propagation is of a pure
statistical nature. If the atoms of a light source oscillate in only one direction, then the light is
linearly polarized. In contrast to the sensory organs of some insects, the human eye has very limited
capabilities to distinguish between different states of polarized light \citep[see][]{Haidinger1844}.

\begin{figure*}
\centering
\includegraphics*[width=9cm]{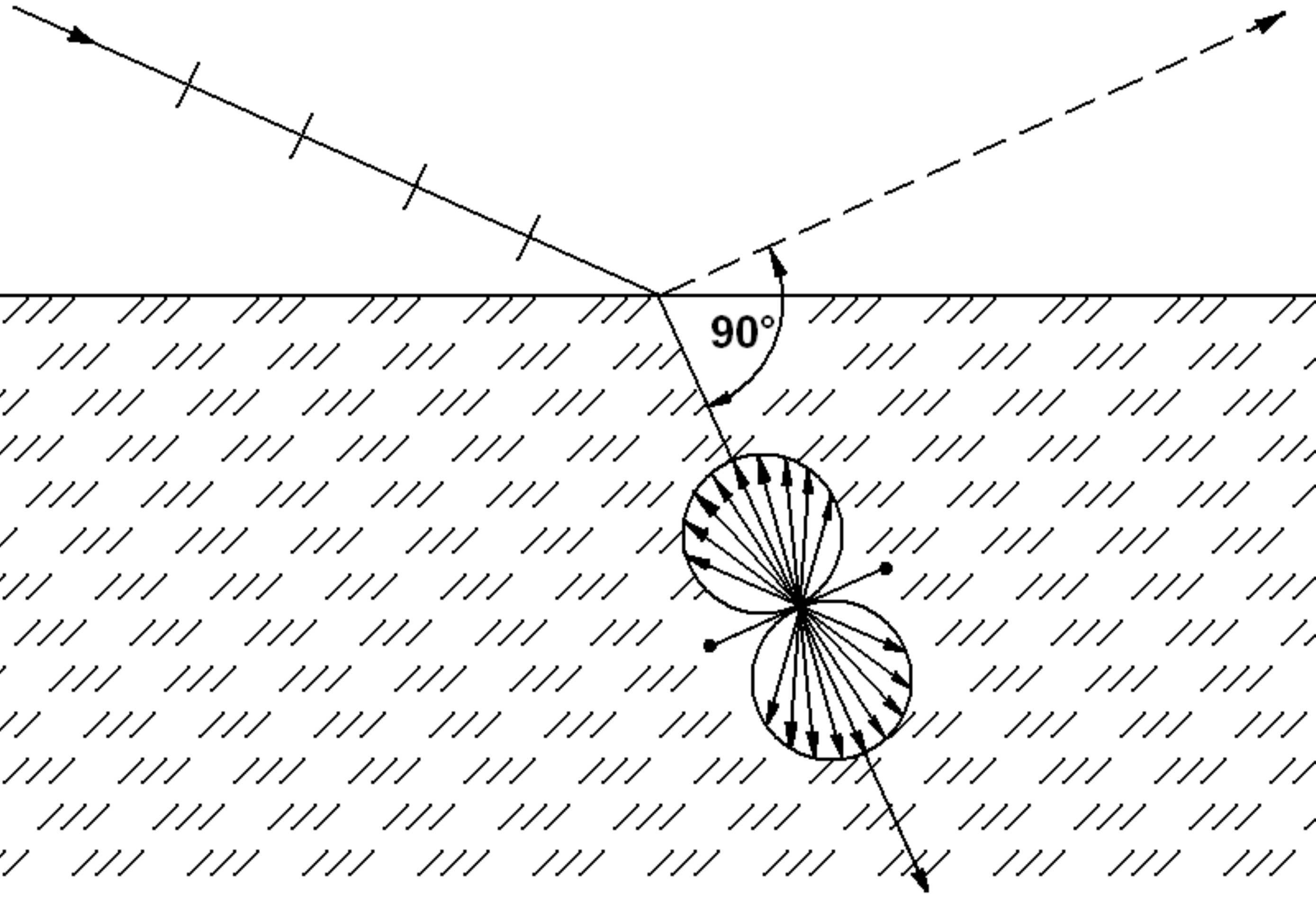}
\caption{Illustration of the reflection of linear polarized light under the \Index{Brewster angle}
by the radiation characteristic of a linear oscillating electron \citep[adapted from][]{BergmannSchaeferBd3}.}
\label{FigBrewster}
\end{figure*}

An apparatus creating linearly polarized light is called a \Index{linear polarizer}, an apparatus to detect linearly
polarized light is named an \Index{analyzer}. In the experiment of Fig.~\ref{FigLinearPolarizer}, the glass plate $\rm{P_1}$
serves as a \Index{linear polarizer} while plate $\rm{P_2}$ is the analyzer. What causes the creation of linearly polarized
light? As a start, one assumes monochromatic, linearly polarized light hitting the surface of a medium. As
indicated in Fig.~\ref{FigBrewster}, the electric field vector oscillates in the plane of the paper. The
electric field of the light source causes oscillations of the medium's electrons and each oscillating
electron is an electric dipole. A \Index{dipole} cannot radiate into the direction of its oscillation while the radiation
perpendicular to its oscillation direction is maximal. If the incident ray hits the surface at the \Index{Brewster angle},
there is no reflected ray, since it would lie in the oscillation direction of the \Index{dipole}s. If one assumes that
the electric field vector of the incident linearly polarized light oscillates perpendicular to the plane of the paper,
then, of course, there is a reflected ray, because this time it is perpendicular to the oscillation direction of the
\Index{dipole}s. Since unpolarized light can be considered as the superposition of linearly polarized light of all possible
oscillation directions, it is clear that unpolarized light is partly linearly polarized by a reflection at the
\Index{Brewster angle}.

\begin{figure*}
\centering
\includegraphics*[height=12cm]{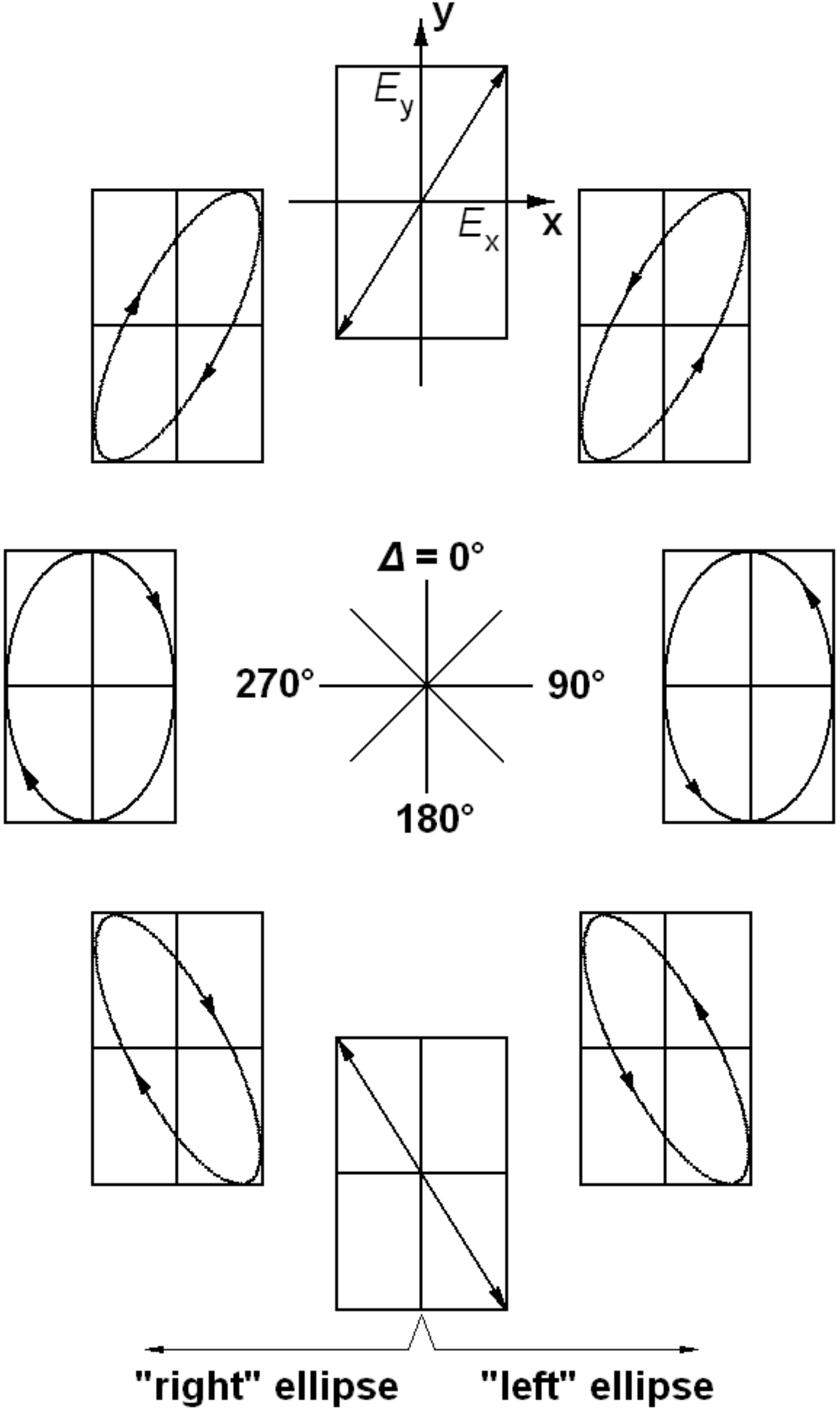}
\caption{Elliptical polarized light as the superposition of two linear polarized light beams \citep[adapted from][]{BergmannSchaeferBd3}.}
\label{FigEllipticalPolarization}
\end{figure*}

\begin{figure*}
\centering
\includegraphics*[width=\textwidth]{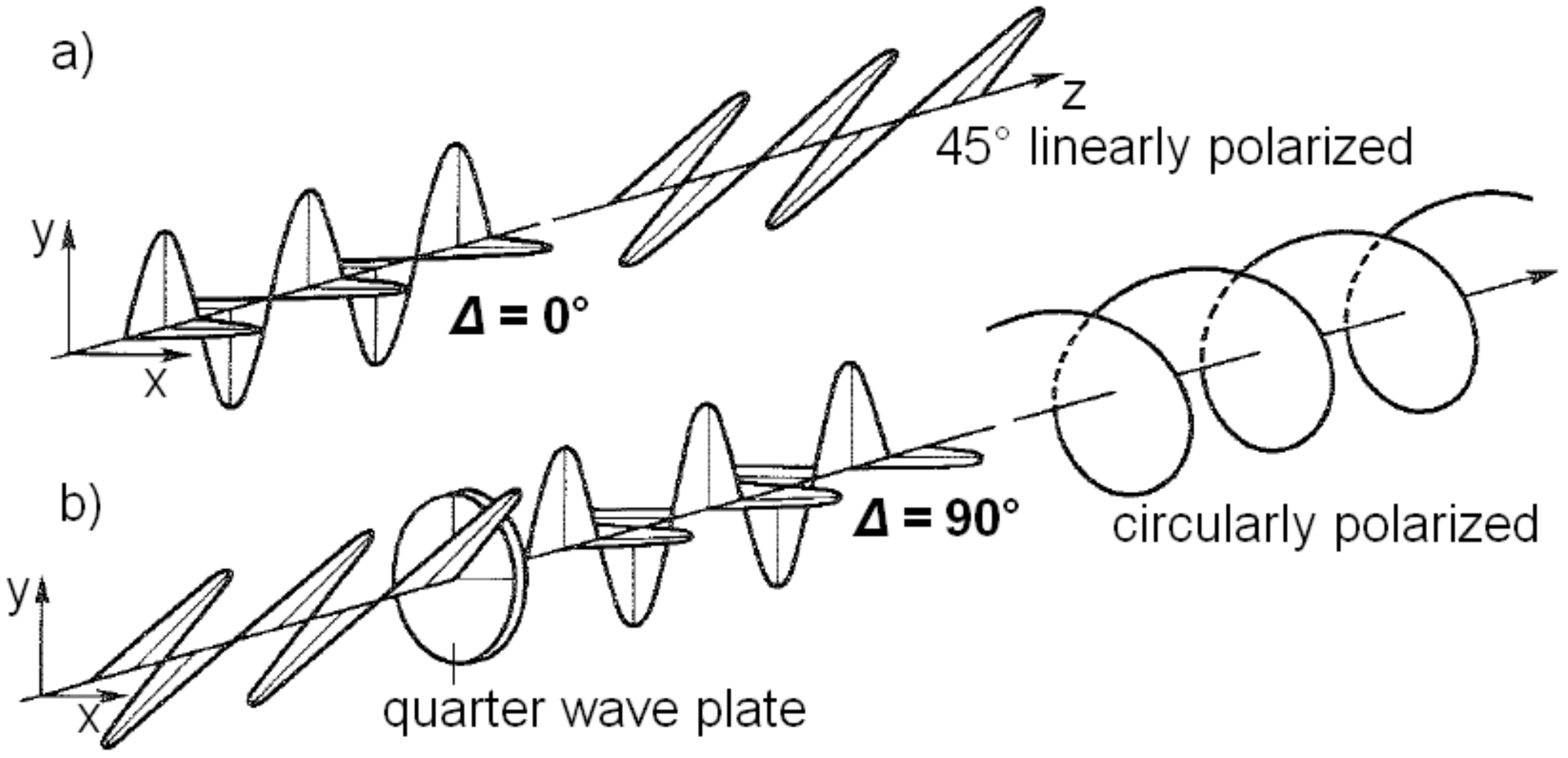}
\caption{The rotation of the polarization plane by a \Index{quarter-wave plate} \citep[from][]{MayerKuckuk}.}
\label{FigQuarterWavePlate}
\end{figure*}

The superposition of two monochromatic linearly polarized light rays whose oscillation directions are
perpendicular to each other, results in elliptical polarized light (think of Lissajous figures) depending
on the phase between the two. For the particular case of equal amplitudes $E_x = E_y$ and a phase shift
$\Delta$ between the two waves of  $90^{\circ}$ ($270^{\circ}$), the ellipse becomes a circle and the light
is called left (right) circularly polarized (see Fig.~\ref{FigEllipticalPolarization}). For phase shifts
of $0^{\circ}$ or $180^{\circ}$, the ellipse becomes a line and the light is again linearly polarized,
but with a rotated oscillation direction, which is illustrated in panel~a) of Fig.~\ref{FigQuarterWavePlate}
for the case of equal amplitudes.

Thin plates made of optically \Index{biaxial crystal}s (e.g. \Index{glimmer}) can be used to rotate the oscillation direction
of polarized light. A linearly polarized light ray hits such a \Index{retarder} plate at an angle of $45^{\circ}$
relative to the two axes of the plate, see Panel~b) of Fig.~\ref{FigQuarterWavePlate}. According to panel~a),
the incident light can be thought as a superposition of two linearly polarized light rays whose oscillation
directions are perpendicular to each other. Such a \Index{retarder} plate has different speeds of light along the
two axes of the plate, so that the two perpendicular oriented linearly polarized light rays get a phase
shift by passing the plate. In the considered case this phase shift is $90^{\circ}$, i.e. $\lambda / 4$
(therefore the \Index{retarder} is called a \Index{quarter-wave plate}). The incident ray oscillating in the x direction is
retarded by $\lambda / 4$ compared to the ray oscillating in the y direction, so that the x axis of the
\Index{quarter-wave plate} is named its \Index{slow axis} and the \Index{fast axis} is along the y direction. According to
Fig.~\ref{FigEllipticalPolarization}, the superposition of two linearly polarized light rays of equal
amplitudes and a phase shift of $90^{\circ}$ results in circularly polarized light. Thus, a \Index{quarter-wave plate}
can transform linearly polarized light into circularly polarized light and vice versa. Quarter-wave plates\index{quarter-wave plate}
are only rarely used in modern polarimeters. Nowadays, liquid crystal \Index{retarder}s are frequently used whose
\Index{retardance} (phase shift between the fast and slow axis) can be voltage-controlled.

\begin{figure*}
\centering
\includegraphics*[height=6cm]{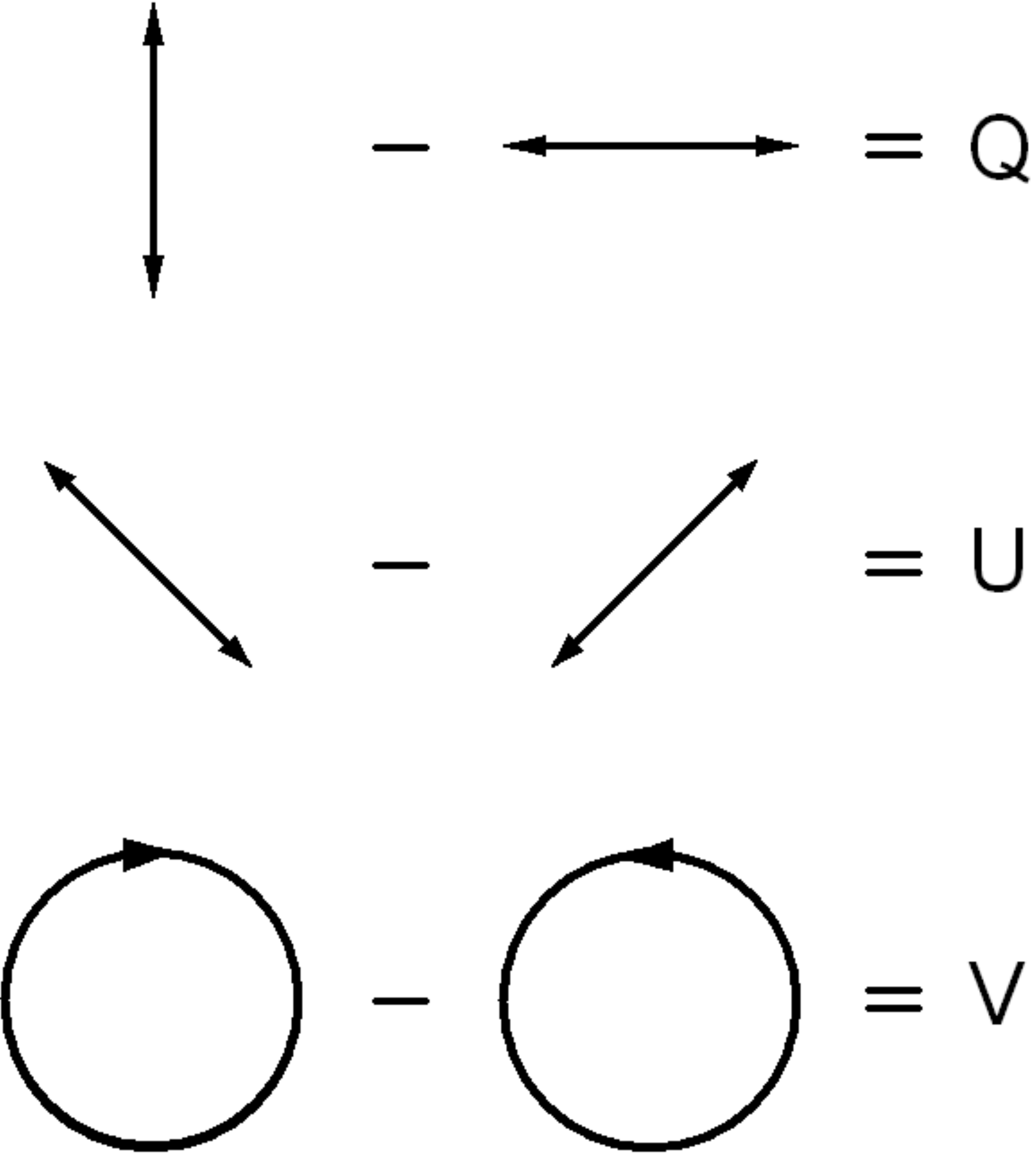}
\caption{Schematic representation of the definition of the \Index{Stokes parameter}s. The observer is assumed to
face the radiation source \citep[from][]{LandiDeglInnocenti2004}.}
\label{FigStokesDefinition}
\end{figure*}

The polarization state of light is sufficiently described by two quantities, the amplitude ratio
$E_x / E_y$ and the phase shift $\Delta$. This is reflected in the \Index{Jones formalism} which describes
completely polarized light and its manipulation by optical components, e.g. \Index{retarder}s or \Index{linear polarizer}s,
with the help of two-dimensional vectors and matrices \citep[see, e.g.,][]{Collet1992}. In solar physics
we practically never deal with completely polarized light but with a mixture of polarized and unpolarized
light. For such partly polarized light, the \Index{Stokes formalism} \citep{Stokes1852} has achieved
acceptance, in which the polarization state of the light is described by the four-dimensional \Index{Stokes vector}:
\begin{equation}\label{Eq_StokesVector1}
\vec{S}=\left( \begin{array}{c} I \\ Q \\ U \\ V \end{array} \right) .
\end{equation}
The symbols $I$, $Q$, $U$, and $V$ for the four \Index{Stokes parameter}s were introduced by \citet{Walker1954}
and are nowadays commonly used. Stokes~$I$ means just the intensity of the light, $Q$ is the intensity
difference measured with an ideal \Index{linear polarizer} at an angle of $0^{\circ}$ and $90^{\circ}$, respectively.
The \Index{linear polarizer} measurements at an angle of $45^{\circ}$ and $135^{\circ}$ lead to Stokes~$U$. Finally,
Stokes~$V$ is the intensity difference measured with a combination of an ideal \Index{quarter-wave plate}\footnote{i.e.
amongst others free of absorption} and an ideal \Index{linear polarizer} at an angle of $45^{\circ}$ and $135^{\circ}$,
respectively. Stokes~$V$ is hence the difference between the left and right circularly polarized part of the
light (see Fig.~\ref{FigStokesDefinition}). Beyond handling partially polarized light, the \Index{Stokes formalism}
also describes the possibility of determining the state of polarization by just intensity measurements. This
is important, since a detector is not able to measure polarization states directly, it can only measure
intensities and their differences.

The ratio of polarized and unpolarized light is expressed by the \Index{total polarization degree}
\begin{equation}\label{Eq_TotalPolGrad}
p=\sqrt{\left( \frac{Q}{I} \right)^2+\left( \frac{U}{I} \right)^2+\left( \frac{V}{I} \right)^2} ,
\end{equation}
which can have values from 0 to 1. Completely unpolarized light possesses $Q=U=V=0$ and hence $p=0$, while
completely polarized light holds $I^2=Q^2+U^2+V^2$, i.e. $p=1$. From the polarization properties of
the \Index{Zeeman component}s (see section~\ref{ZeemanEffect}) follows that Stokes~$V$ displays large signals
if vertical magnetic fields are observed at disk center (i.e. the longitudinal case where the field is
parallel to the line-of-sight). Large Stokes~$Q$ and $U$ signals are found if horizontal fields are observed
(transversal case where the field is perpendicular to the line-of-sight). Transverse fields are therefore
frequently described by the \Index{linear polarization degree}
\begin{equation}\label{Eq_LinPolGrad}
p_{lin}=\sqrt{\left( \frac{Q}{I} \right)^2+\left( \frac{U}{I} \right)^2}
\end{equation}
and longitudinal fields by the \Index{circular polarization degree}
\begin{equation}\label{Eq_CircPolGrad}
p_{circ}=\left| \frac{V}{I} \right| .
\end{equation}

\begin{figure*}
\centering
\includegraphics*[width=\textwidth]{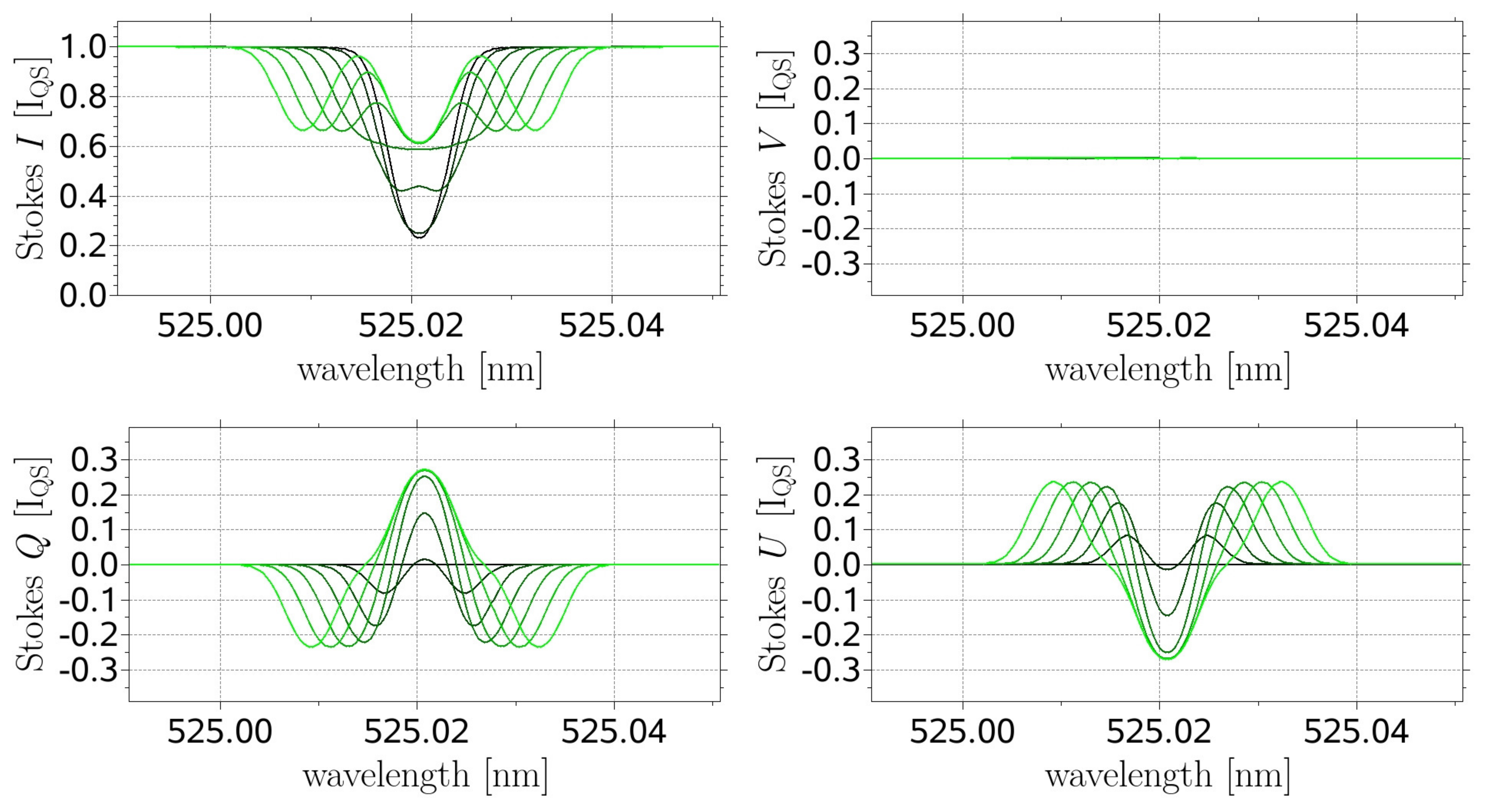}
\caption{The influence of a transversal magnetic field on the four Stokes profiles of a spectral line.
The height independent field increases from 0~G (black) to 3000~G (green) in steps of 500~G.}
\label{FigZeeman4}
\end{figure*}

\begin{figure*}
\centering
\includegraphics*[width=\textwidth]{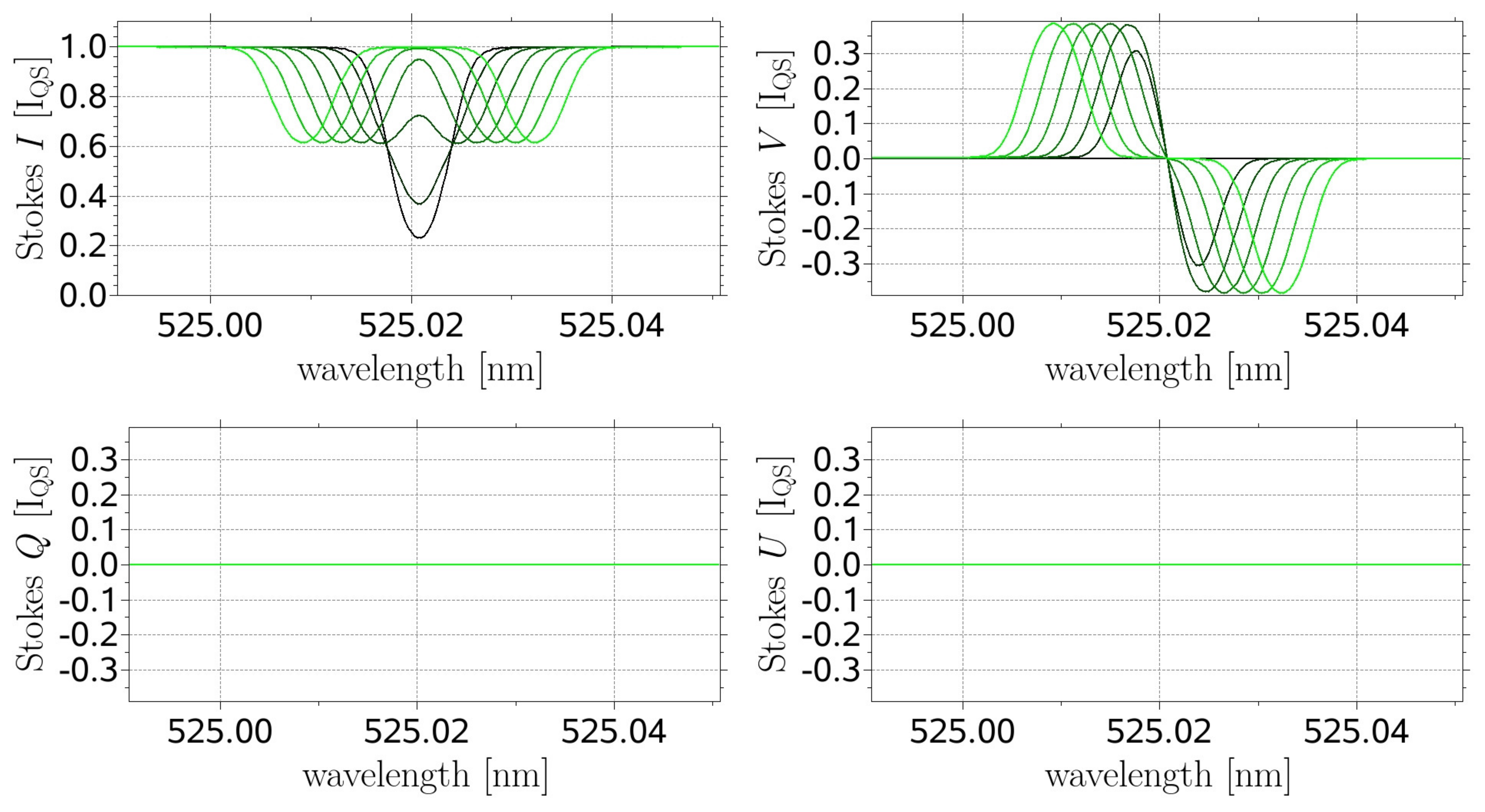}
\caption{The same as Fig.~\ref{FigZeeman4}, but for a longitudinal magnetic field.}
\label{FigZeeman5}
\end{figure*}

Fig.~\ref{FigZeeman4} shows Stokes profiles\footnote{normalized to the mean intensity of
quiet Sun, $\rm{I_{QS}}$} for magnetic fields perpendicular to the line-of-sight. The field strength
varies in steps of 0.05~T from 0 (black line) to 0.3~T (3000 G, green line), while the azimuthal
orientation of the field with respect to the Stokes~$Q$ and $U$ reference direction, $\mathrm{67.5^\circ}$,
is chosen such that Stokes~$Q$ and $U$ have the same amplitude but opposite polarities. The neutral
iron line at 525.02~nm serves again as an example. Fig.~\ref{FigZeeman5} displays the same as
Fig.~\ref{FigZeeman4} but this time the magnetic field is oriented towards the observer. The
Stokes~$V$ profiles in Fig.~\ref{FigZeeman5} nicely show the effect of the \Index{Zeeman saturation}.
The amplitude of Stokes~$V$ is proportional to the field strength in the case of weak longitudinal fields
(a few 100~G). For stronger fields, the $V$ amplitude increases only weakly, while the separation between
the red and the blue lobe of the $V$~profile is further proportional to the field strength.

All the magnetic (and velocity) fields we considered so far were constant over all atmospheric heights.
As a consequence, all Stokes~$I$, $Q$, and $U$ profiles were symmetrical and all $V$ profiles were
antisymmetrical. Asymmetrical Stokes profiles require gradients of the magnetic and/or velocity field.

If a light ray passes a system of optical components, the polarization properties of the light can
be influenced by the components, i.e. the incident \Index{Stokes vector} $\vec{S_{\rm{in}}}$ is transformed
into an outgoing \Index{Stokes vector} $\vec{S_{\rm{out}}}$. Such a transformation can be mathematically
described by a multiplication of the so-called M\"uller matrices\index{M\"uller matrix} by the incident \Index{Stokes vector}.
Imagine a ray to be transmitted through a \Index{retarder} and then through a \Index{linear polarizer}, which leads
to the following transformation:
\begin{equation}\label{Eq_MuellerCalculus}
\vec{S_{\rm{out}}}=M_{\rm{LP}} M_{\rm{Ret}} \vec{S_{\rm{in}}} ,
\end{equation}
where $M_{\rm{LP}}$ and $M_{\rm{Ret}}$ are the \Index{M\"uller matrix} of the \Index{linear polarizer} and the
\Index{retarder}. For an ideal \Index{linear polarizer}, whose transmission direction is identical with the reference
direction (e.g. the x axis), the \Index{M\"uller matrix} is:
\begin{equation}\label{Eq_MuellerLinearPolarizer}
M_{\rm{LP}}=\frac{1}{2}\left( \begin{array}{cccc} 1 & 1 & 0 & 0 \\
                                                  1 & 1 & 0 & 0 \\
                                                  0 & 0 & 0 & 0 \\
                                                  0 & 0 & 0 & 0 
                              \end{array} \right)
\end{equation}
and for an ideal \Index{retarder} having a \Index{retardance} of $\Delta$, whose fast axis is identical with the
reference direction, it is:
\begin{equation}\label{Eq_MuellerRetarder}
M_{\rm{Ret}}=\left( \begin{array}{cccc} 1 & 0 &  0            & 0            \\
                                        0 & 1 &  0            & 0            \\
                                        0 & 0 &  \cos{\Delta} & \sin{\Delta} \\
                                        0 & 0 & -\sin{\Delta} & \cos{\Delta} 
                    \end{array} \right) .
\end{equation}
If an optical component is rotated by the angle $\phi$, the \Index{M\"uller matrix} $M_{\rm{0}}$ is transformed
into:
\begin{equation}\label{Eq_MuellerRotation1}
M_{\rm{\phi}}=M_{\rm{Rot}}(-\phi) M_{\rm{0}} M_{\rm{Rot}}(\phi) ,
\end{equation}
where $M_{\rm{Rot}}(\phi)$ is the transformation matrix, which rotates the coordinate system by the angle $\phi$:
\begin{equation}\label{Eq_MuellerRotation2}
M_{\rm{Rot}}(\phi)=\left( \begin{array}{cccc} 1 & 0            & 0           & 0 \\
                                              0 &  \cos{2\phi} & \sin{2\phi} & 0 \\
                                              0 & -\sin{2\phi} & \cos{2\phi} & 0 \\
                                              0 & 0            & 0           & 1 
                          \end{array} \right) .
\end{equation}

Since every telescope and instrument to measure the state of polarization consists of many optical
components which change the polarization properties of the light, a way has to be found to disentangle
the \Index{instrumental polarization} from the solar polarization. This problem is solved by a \Index{polarimetric calibration}
that uses \Index{calibration optics} (often consisting of a rotatable \Index{linear polarizer} and a
rotatable \Index{quarter-wave plate}) in order to create various well-defined Stokes signals. The modification
of the well-known incident Stokes signals by the optical components between the \Index{calibration optics}
and the polarimeter are measured and used to determine the \Index{M\"uller matrix} that describes the
\Index{instrumental polarization}. This matrix is also called \Index{modulation matrix}, its inverse is named
\Index{demodulation matrix}. As part of the data reduction, the \Index{Stokes vector}s measured during the actual
observation (without the \Index{calibration optics}) are then multiplied by the \Index{demodulation matrix} so that
only the solar polarization remains. One problem of the method is that often the very first
components of the optical setup (e.g. primary mirror, entrance lens, coelostat mirrors) cannot be
calibrated because it is simply not feasible to mount \Index{calibration optics} of, e.g., 1-m diameter
in front of the telescope. Particularly, plane mirrors can considerably contribute to the \Index{instrumental polarization}
(remember the effect illustrated in Fig.~\ref{FigLinearPolarizer}). Often the part
of the telescope that is in front of the \Index{calibration optics} is theoretically modeled, so that
its part of the \Index{instrumental polarization} can also be removed.

\section{Radiative transfer}\label{RadiativeTransfer}

The bulk of our knowledge about the Sun is derived from the emitted \Index{electromagnetic radiation}.
If the radiation passes through the solar atmosphere, its properties change. This change depends
on the temperatures, densities, magnetic fields, velocities, etc. of the atmosphere.  The theory
of the \Index{radiative transfer} studies the interaction of the radiation with the atmospheric matter
in order to retrieve the physical properties of the atmosphere from observed Stokes profiles
\citep{Solanki1993,BellotRubio1998,Frutiger2000a,DelToroIniesta2003}.

\subsection{Fundamental terms}
\begin{figure*}
\centering
\includegraphics*[width=7cm]{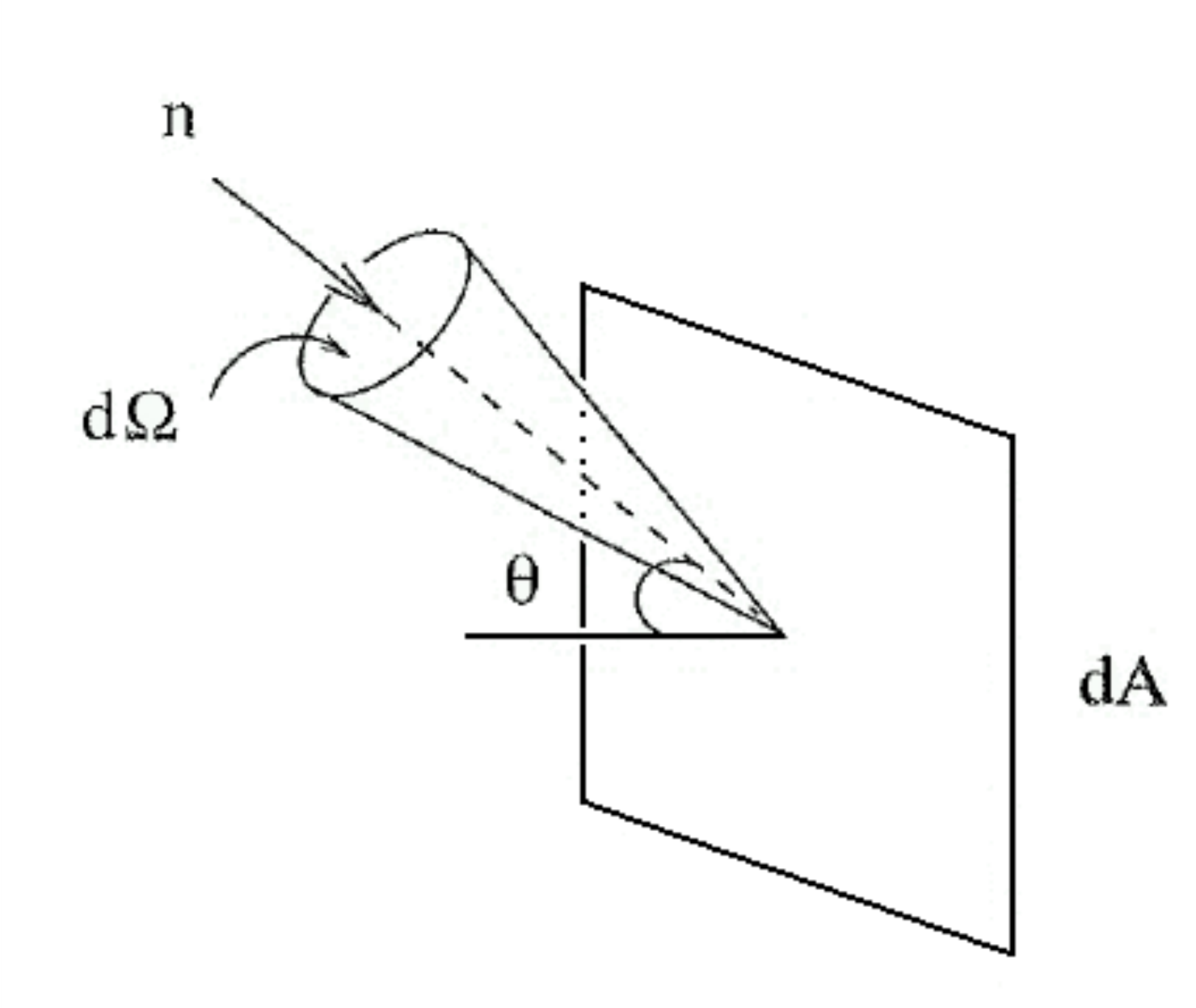}
\caption{Illustration of the \Index{specific intensity} \citep[from][]{Choudhuri2010}.}
\label{FigSpecificIntensity}
\end{figure*}

Let $dA$ be an infinitesimal small surface element at position ${\bf r}$ of the solar atmosphere
(see Fig.~\ref{FigSpecificIntensity}). Furthermore, let ${\bf n}$ be the unit vector in the direction
of the observer forming an angle $\theta$ with the surface normal of $dA$. The amount of radiation
$dE_{\rm \nu}\,d\nu$ passing through the surface element $dA$ from the solid angle $d\Omega$ in time
$dt$ in the frequency interval $\nu,\nu+d\nu$ is proportional to $d\Omega$, $d\nu$, $dt$, and to the
projected area $dA\cos\theta$:
\begin{equation}\label{Eq_SpecificIntensity}
dE_{\rm \nu}\,d\nu=I_{\rm \nu}\left(\bfit{r},t,\bfit{n}\right) dt\,d\Omega\,d\nu\,dA\cos\theta
\end{equation}
The proportionality factor $I_{\rm \nu}$ is named the {\bf \Index{specific intensity}} and is a function of the
position $\bfit{r}$, the time $t$, and the direction $\bfit{n}$. The index ${\rm \nu}$ used for various
quantities refers to the frequency dependence. If a body is in \Index{thermodynamic equilibrium}, i.e.
everywhere is the same temperature $T$, it emits \Index{black-body radiation} which is isotropic and has the
\Index{specific intensity}
\begin{equation}\label{Eq_PlanckFunction1}
B_{\rm \nu}\left(T\right)=\frac{2h\nu^3}{c^2}\frac{1}{\rm{e}^{{\it \frac{h\nu}{k_{\rm B}T}}}-1} ,
\end{equation}
where $k_{\rm B}$ is the \Index{Boltzmann constant} and $B_{\rm \nu}$ is the {\bf \Index{Planck function}}
\citep{Planck1900a,Planck1900b}. Black-body radiation\index{black-body radiation} is isotropic and temporally constant
if $T$ is constant and hence it depends only on temperature.

If radiation passes through matter, the \Index{specific intensity} can be changed by two effects. It can be
decreased if the radiation is absorbed by the matter or it can be increased if the matter emits
radiation. The absorption can be described by an \Index{absorption coefficient}, $\alpha_{\rm \nu}$, which is
proportional to the \Index{specific intensity}, i.e. the more radiation is present, the more radiation can be
absorbed by the matter. $1/\alpha_{\rm \nu}$ gives the \Index{mean free path} of the photons when passing
through the matter. The emission of radiation is described by an \Index{emission coefficient}, $j_{\rm \nu}$,
so that the change of the \Index{specific intensity} along the ray path is
\begin{equation}\label{Eq_RTE1}
\frac{{\it dI}_{\rm \nu}}{ds}=-\alpha_{\rm \nu}I_{\rm \nu}+j_{\rm \nu} .
\end{equation}
Eq.~(\ref{Eq_RTE1}) is called the {\bf R}adiative {\bf T}ransfer {\bf E}quation (\Index{RTE}). The ratio
of the \Index{emission coefficient} to the \Index{absorption coefficient}
\begin{equation}\label{Eq_SourceFunction1}
S_{\rm \nu}=\frac{j_{\rm \nu}}{\alpha_{\rm \nu}}
\end{equation}
is named the {\bf \Index{source function}}, $S_{\rm \nu}$, so that Eq.~(\ref{Eq_RTE1}) can also be written as:
\begin{equation}\label{Eq_RTE2}
\frac{{\it dI}_{\rm \nu}}{ds}=-\alpha_{\rm \nu}\left({\it I}_{\rm \nu}-S_{\rm \nu}\right) .
\end{equation}

In the case of non-emitting matter ($j_{\rm \nu}=0$), the \Index{radiative transfer equation}~(\ref{Eq_RTE1})
can be solved by a simple integration along the ray path:
\begin{equation}\label{Eq_RTE_Solution1}
{\it I}_{\rm \nu}\left(s\right)={\it I}_{\rm \nu}\left(s_0\right){\rm exp}\left(-\int\limits_{s_0}^s \alpha_{\rm \nu}(s^\prime)\,ds^\prime\right) .
\end{equation}
This suggests the introduction of the {\bf optical depth}, $\tau_{\rm \nu}$, as the proper coordinate for
\Index{radiative transfer} problems:
\begin{equation}\label{Eq_OpticalDepth1}
\tau_{\rm \nu}=\int\limits_{s_0}^s \alpha_{\rm \nu}(s^\prime)\,ds^\prime ,
\end{equation}
which is equivalent to:
\begin{equation}\label{Eq_OpticalDepth2}
\frac{d\tau_{\rm \nu}}{ds}=\alpha_{\rm \nu} .
\end{equation}
The emission-free solution of the \Index{radiative transfer equation} is therefore \citep{Choudhuri2010}:
\begin{equation}\label{Eq_RTE_Solution2}
{\it I}_{\rm \nu}\left(\tau_{\rm \nu}\right)={\it I}_{\rm \nu}\left(0\right){\rm e}^{-\tau_{\rm \nu}} .
\end{equation}
If matter only absorbs radiation but does not emit, then the \Index{specific intensity} of the radiation decreases
exponentially. Eq.~(\ref{Eq_RTE_Solution2}) makes clear why $\tau_{\rm \nu}$ is called optical
depth. If $\tau_{\rm \nu} \gg 1$ along a ray path through an object, then the light is strongly damped and
hence the object is called optically thick (opaque). Accordingly, in the case of $\tau_{\rm \nu} \ll 1$
the object is called optically thin (transparent).

\subsection{Spectral line formation}\label{LineFormation}
\begin{figure*}
\centering
\includegraphics*[width=9cm]{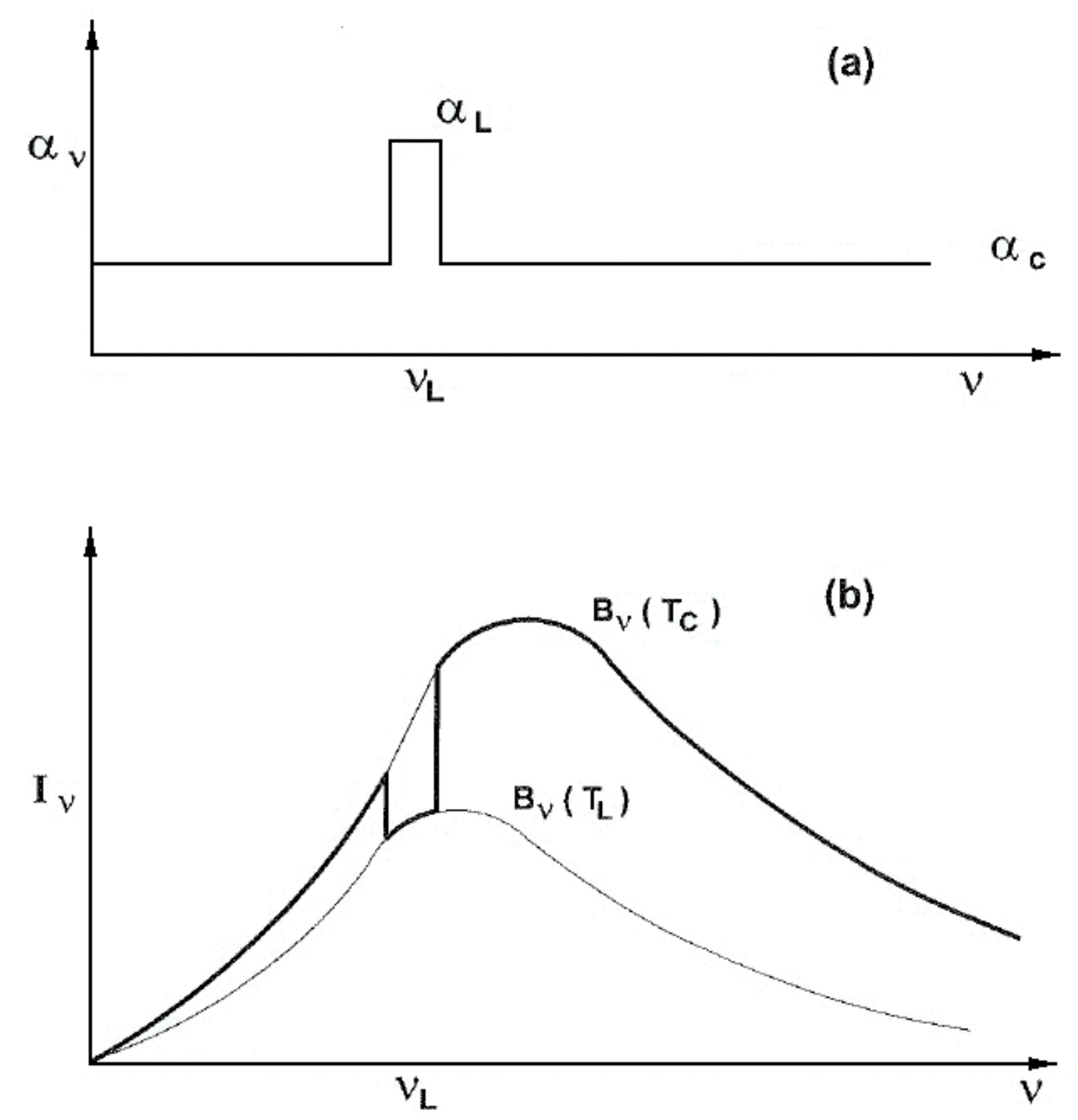}
\caption{Illustration of the formation of an absorption line \citep[from][]{Choudhuri2010}.}
\label{FigSpectralLineFormation}
\end{figure*}

Generally, the \Index{emission coefficient} is not zero and can, just as the \Index{absorption coefficient},
depend on the frequency and optical depth. (Only time-independent radiation fields are considered here.)
A stellar atmosphere whose \Index{absorption coefficient}, $\alpha_{\rm \nu}$, does not depend on frequency is
called a {\bf gray atmosphere}. In reality, gray atmospheres do not exist. They are just a
simplified model which although can be analytically solved \citep[see, e.g.,][]{Mihalas1970} but
cannot explain the formation of spectral lines. The formation of spectral lines can only be described
by non-gray atmospheres. The simplest frequency dependence of the \Index{absorption coefficient} is displayed
in panel~(a) of Fig.\ref{FigSpectralLineFormation}.
In a narrow interval around the frequency $\nu_{\rm L}$, the \Index{absorption coefficient} has an
increased value, $\alpha_{\rm L}$. Outside this interval it has always the value $\alpha_{\rm C}$.
(The frequency range outside spectral lines is also called {\bf continuum}.) It can be shown
that the \Index{specific intensity} at a frequency $\nu$ leaving a stellar atmosphere (e.g. observable
with a telescope from near-Earth space) is almost identical with the \Index{Planck function}
(\ref{Eq_PlanckFunction1}) at an atmospheric depth where the optical depth for that frequency
$\nu$ equals unity \citep{Choudhuri2010}. According to (\ref{Eq_OpticalDepth1}), the continuum
optical depth unity is reached at the depth $\alpha_{\rm C}^{-1}$ having the local temperature
$T_{\rm C}$. Hence, the \Index{specific intensity} for the continuum is $B_{\rm \nu}\left(T_{\rm C}\right)$.
Inside the spectral line, the optical depth unity is reached higher up in the atmosphere
at $\alpha_{\rm L}^{-1}$ relating to the somewhat lower temperature $T_{\rm L}$, which
corresponds to a \Index{black-body radiation}, $B_{\rm \nu}\left(T_{\rm L}\right)$. The superposition
of the two \Index{Planck function}s results in an absorption line as illustrated by panel~(b) of
Fig.\ref{FigSpectralLineFormation}.

\begin{figure*}
\centering
\includegraphics*[width=12cm]{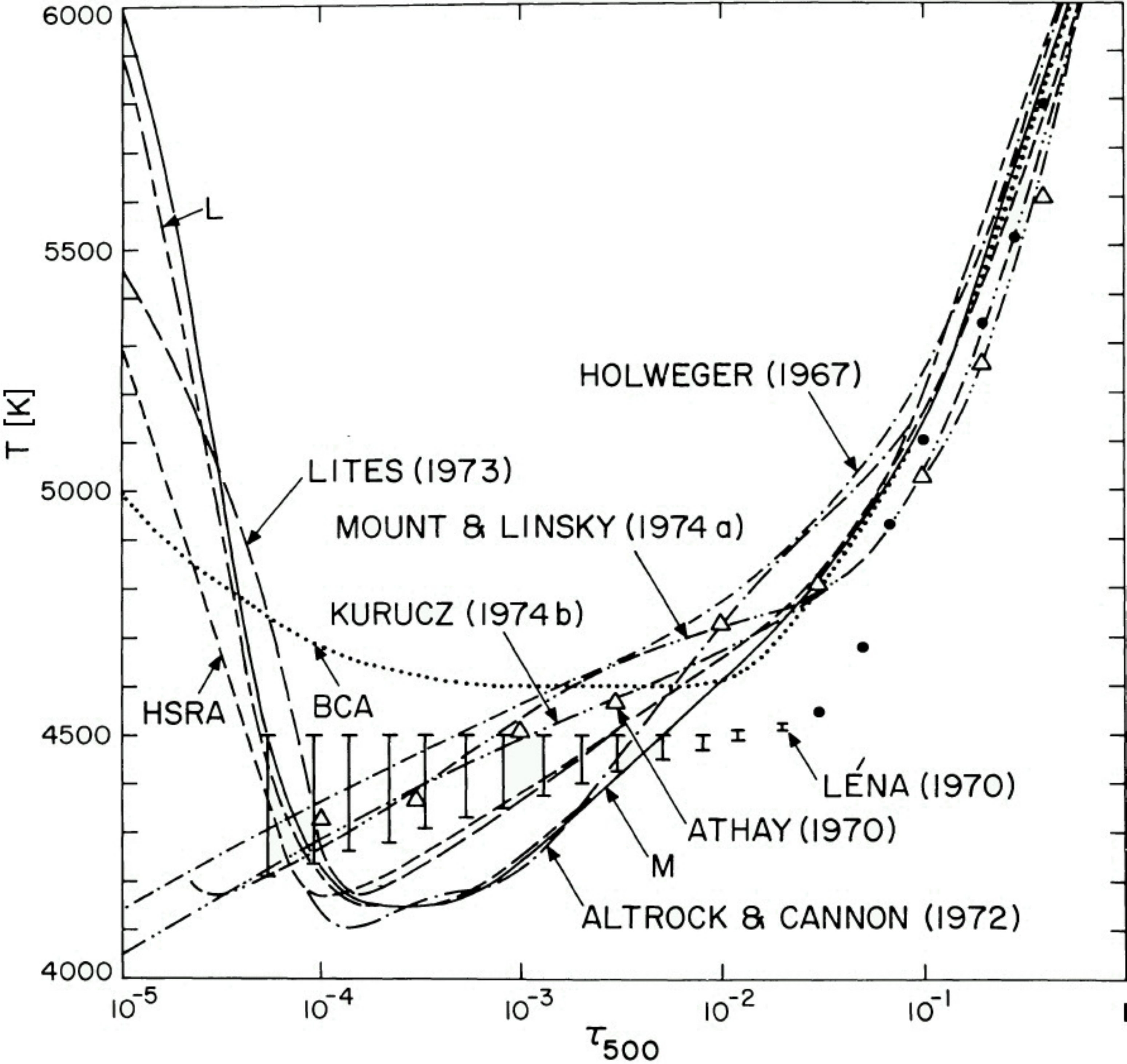}
\caption{Temperature as function of the continuum optical depth at 500~nm demonstrating the temperature minimum
for various model atmospheres of the Sun \citep[from][]{Vernazza1976}.}
\label{FigTemperatureMinimum}
\end{figure*}

The formation of spectral lines requires a frequency dependent \Index{absorption coefficient} as well as a
temperature gradient in the atmosphere. In the solar photosphere we find a decreasing temperature
with height which leads to the formation of absorption lines. At roughly the transition between photosphere
and chromosphere, the temperature reaches a minimum and increases again further outwards, see
Fig.~(\ref{FigTemperatureMinimum}). Above the photosphere the temperature gradient reverses its sign
so that emission lines are formed.

\subsection{Polarized \Index{radiative transfer}}
So far, it was just considered how a stellar atmosphere changes the \Index{specific intensity}, i.e. Stokes~$I$.
Not considered was the influence of the atmosphere on Stokes~$Q$, $U$, and $V$. As explained in
section~\ref{ZeemanEffect}, the interpretation of the full \Index{Stokes vector} is needed to disentangle
magnetic and temperature effects and to retrieve the orientation of the magnetic field and not only
its strength.

The propagation of electromagnetic radiation in a medium containing a homogeneous magnetic field
was formulated in the \Index{Stokes formalism} for the first time by \citet{Unno1956} for the normal Zeeman
effect and provided the first method for a retrieval of the orientation of the magnetic field.
Magneto-optical effects\index{magneto-optical effect} influence the Stokes profiles mainly in the core of the spectral lines,
which were included in the \Index{radiative transfer equation}s by \citet{Rachkovsky1962}. The anomalous
Zeeman effect was finally considered by \citet{Rachkovsky1967}.

Ignoring the index ${\rm \nu}$ expressing the frequency dependence, Eq.~(\ref{Eq_RTE1}) can be recast
in the generalized form:
\begin{equation}\label{Eq_UnnoRachkovsky1}
\frac{\textit{d}\bfit{I}}{ds}=-\bfit{K}\bfit{I}+\bfit{j} ,
\end{equation}
where $\bfit{I}$ is now the four-dimensional \Index{Stokes vector}
\begin{equation}\label{Eq_StokesVector2}
\bfit{I}=\left( \begin{array}{c} I \\ Q \\ U \\ V \end{array} \right) ,
\end{equation}
$\bfit{j}$ stands for the emission vector and $\bfit{K}$ for the total absorption matrix, which
consists of a continuum and a line part:
\begin{equation}\label{Eq_TotalAbsorptionMatrix}
\bfit{K}=\kappa_{\rm C}\left( \textbf{1}+\boldsymbol{\eta} \right) ,
\end{equation}
where $\kappa_{\rm C}$ denotes the continuum \Index{absorption coefficient}, $\textbf{1}$ is the
unity matrix, and
\begin{equation}\label{Eq_LineAbsorptionMatrix1}
\boldsymbol{\eta}=\left( \begin{array}{cccc} \eta_{\rm I} &  \eta_{\rm Q} &  \eta_{\rm U} &  \eta_{\rm V} \\
                                             \eta_{\rm Q} &  \eta_{\rm I} &  \rho_{\rm V} & -\rho_{\rm U} \\
                                             \eta_{\rm U} & -\rho_{\rm V} &  \eta_{\rm I} &  \rho_{\rm Q} \\
                                             \eta_{\rm V} &  \rho_{\rm U} & -\rho_{\rm Q} &  \eta_{\rm I} 
                         \end{array} \right)
\end{equation}
is the line absorption matrix with only seven independent elements, which meet the following equations:
\begin{equation}\label{Eq_UnnoRachkovsky2}
\begin{array}{cclcl} \eta_{\rm I} & = & \frac{1}{4} \Big[ 2\eta_{\rm\pi}\sin^2{\gamma} & + & (\eta_{\rm\sigma_{blue}}+\eta_{\rm\sigma_{red}}) \Big] \left( 1+\cos^2{\gamma} \right) 
                     \vspace{3mm}                                                                                                                                                   \\
                     \eta_{\rm Q} & = & \frac{1}{4} \Big[ 2\eta_{\rm\pi}               & - & (\eta_{\rm\sigma_{blue}}+\eta_{\rm\sigma_{red}}) \Big] \sin^2{\gamma} \sin{2\chi}      \\
                     \eta_{\rm U} & = & \frac{1}{4} \Big[ 2\eta_{\rm\pi}               & - & (\eta_{\rm\sigma_{blue}}+\eta_{\rm\sigma_{red}}) \Big] \sin^2{\gamma} \cos{2\chi}      \\
                     \eta_{\rm V} & = & \frac{1}{2} \Big[                              &   & (\eta_{\rm\sigma_{blue}}+\eta_{\rm\sigma_{red}}) \Big] \cos{\gamma}                    
                     \vspace{3mm}                                                                                                                                                   \\
                     \rho_{\rm Q} & = & \frac{1}{4} \Big[ 2\rho_{\rm\pi}               & - & (\rho_{\rm\sigma_{blue}}+\rho_{\rm\sigma_{red}}) \Big] \sin^2{\gamma} \sin{2\chi}      \\
                     \rho_{\rm Q} & = & \frac{1}{4} \Big[ 2\rho_{\rm\pi}               & - & (\rho_{\rm\sigma_{blue}}+\rho_{\rm\sigma_{red}}) \Big] \sin^2{\gamma} \cos{2\chi}      \\
                     \rho_{\rm V} & = & \frac{1}{2} \Big[                              &   & (\rho_{\rm\sigma_{blue}}+\rho_{\rm\sigma_{red}}) \Big] \cos{\gamma}
\end{array}
\end{equation}
The Eqs.~(\ref{Eq_UnnoRachkovsky1}) together with (\ref{Eq_UnnoRachkovsky2}) are usually named the
{\bf \Index{Unno-Rachkovsky equations}}. $\gamma$ and $\chi$ denote the inclination and azimuthal angle of
the magnetic field vector with respect to the line-of-sight. $\eta_{\rm\pi,\sigma_{blue},\sigma_{red}}$ and
$\rho_{\rm\pi,\sigma_{blue},\sigma_{red}}$ are the absorption\index{absorption profile} and \Index{dispersion profile}s at the
wavelength position of the $\pi$, $\sigma_{blue}$, and $\sigma_{red}$ \Index{Zeeman component}s. Details can
be found in \citet{BellotRubio1998}, \citet{Frutiger2000a}, \citet{DelToroIniesta2003}, or
\citet{LandiDeglInnocenti2004}.

For a better understanding, the line \Index{absorption matrix} can be decomposed in three matrices: 
\begin{equation}\label{Eq_LineAbsorptionMatrix2}
\boldsymbol{\eta}=\underbrace{ \left( \begin{array}{cccc} \eta_{\rm I} &  0            &  0            &  0            \\
                                                          0            &  \eta_{\rm I} &  0            &  0            \\
                                                          0            &  0            &  \eta_{\rm I} &  0            \\
                                                          0            &  0            &  0            &  \eta_{\rm I} 
                                      \end{array} \right)
                             }_{absorption}
                 +\underbrace{ \left( \begin{array}{cccc} 0            &  \eta_{\rm Q} &  \eta_{\rm U} &  \eta_{\rm V} \\
                                                          \eta_{\rm Q} &  0            &  0            &  0            \\
                                                          \eta_{\rm U} &  0            &  0            &  0            \\
                                                          \eta_{\rm V} &  0            &  0            &  0
                                      \end{array} \right)
                             }_{dichroism}
                 +\underbrace{ \left( \begin{array}{cccc} 0            &  0            &  0            &  0            \\
                                                          0            &  0            &  \rho_{\rm V} & -\rho_{\rm U} \\
                                                          0            & -\rho_{\rm V} &  0            &  \rho_{\rm Q} \\
                                                          0            &  \rho_{\rm U} & -\rho_{\rm Q} &  0
                                      \end{array} \right)
                             }_{dispersion}
\end{equation}
where the diagonal \Index{absorption matrix} only damps the \Index{specific intensity} of the light but does not
change its polarization. The \Index{dichroism matrix} describes the polarization dependent absorption.
Finally, the \Index{dispersion matrix} expresses the \Index{magneto-optical effect}s which can lead to a redistribution
of energy between the states of polarization. Circular \Index{birefringence} (different refractive indices\index{refractive index}
for left and right circularly polarized light) causes as {\bf \Index{Faraday effect}} a rotation of the
polarization plane of the linearly polarized part of the light \citep{Faraday1855}, while linear
\Index{birefringence} (different refractive indices\index{refractive index} for linearly polarized components of the light that
oscillate parallel and perpendicular to the magnetic field) leads as {\bf \Index{Voigt effect}} to a phase
shift between these components \citep{Wittmann1972}.

A system is not in \Index{thermodynamic equilibrium} if the temperature of the system is not unique, e.g. in the case
of the solar atmosphere. Nevertheless, there are situations in which the \Index{mean free path} of the gas particles
is small compared to the length scale over which considerable temperature changes occur. Such a situation
is named {\bf L}ocal {\bf T}hermodynamic {\bf E}quilibrium (\Index{LTE}) and is a valid description, e.g. for the
photosphere, which is considered in the following chapters. For the chromosphere and even more for the corona,
\Index{LTE} cannot be assumed. The velocities of the gas particles in an \Index{LTE} system are Maxwellian distributed,
the particle number of the various ionic species is determined by the \Index{Saha equation} \citep{Saha1920} and the
population numbers of the participating energy levels can be calculated with the help of the \Index{Boltzmann formula}
\citep[see, e.g.,][]{Unsoeld1968}.

The \Index{LTE} approximation simplifies the determination of the \Index{emission vector} to:
\begin{equation}\label{Eq_EmissionVector}
\bfit{j}=\kappa_{\rm C}B_{\rm \nu}(T)\left( \begin{array}{c} \eta_{\rm I}+1 \\ \eta_{\rm Q} \\ \eta_{\rm U} \\ \eta_{\rm V} \end{array} \right)
\end{equation}
and hence the \Index{radiative transfer equation} can be written in the form:
\begin{equation}\label{Eq_RTE3}
\frac{\textit{d}}{ds}\left( \begin{array}{c} I \\ Q \\ U \\ V \end{array} \right) = -\kappa_{\rm C}
\left( \begin{array}{cccc} \eta_{\rm I}+1 &  \eta_{\rm Q}   &  \eta_{\rm U}   &  \eta_{\rm V}   \\
                           \eta_{\rm Q}   &  \eta_{\rm I}+1 &  \rho_{\rm V}   & -\rho_{\rm U}   \\
                           \eta_{\rm U}   & -\rho_{\rm V}   &  \eta_{\rm I}+1 &  \rho_{\rm Q}   \\
                           \eta_{\rm V}   &  \rho_{\rm U}   & -\rho_{\rm Q}   &  \eta_{\rm I}+1   
\end{array} \right) \left( \begin{array}{c} I-B_{\rm \nu}(T) \\ Q \\ U \\ V \end{array} \right) .
\end{equation}

Eq.~(\ref{Eq_RTE3}) is a coupled system of differential equations, which can only be solved
analytically for some particular cases. For the general case, the \Index{RTE} has to be solved numerically.
Various numerical \Index{RTE code}s are available, for example the \Index{SIR} code \citep{RuizCobo1992,BellotRubio1998},
the \Index{STOPRO} code \citep{Solanki1987,Frutiger2000a} or the \Index{LILIA} code \citep{SocasNavarro2001}.

\subsection{Inversion of Stokes profiles}\label{StokesInversion}
In practice, the use of a numerical \Index{RTE code} starts with given stratifications of temperature,
magnetic field strength and orientation, LOS velocity (and eventually gas pressure) as function
of the optical depth for which the numerical solution of the \Index{radiative transfer equation}s
provides the synthetic \Index{Stokes parameter}s
\begin{equation}\label{Eq_SynStokes}
{\bfit I}^{\rm syn}(\lambda)=\left( \begin{array}{c} I^{\rm syn}(\lambda) \\ Q^{\rm syn}(\lambda) \\ U^{\rm syn}(\lambda) \\ V^{\rm syn}(\lambda) \end{array} \right)
\end{equation}
as a function of the wavelength. For example, the Figs.~\ref{FigDoppler2} and \ref{FigZeeman1} are calculated
with such {\bf \Index{forward calculation}s}, also called {\bf syntheses\index{synthesis}}.

In solar physics the inverse problem has often to be solved: An observation campaign provides
\Index{Stokes parameter}s at certain wavelength positions within the considered absorption line(s) with
the help of a \Index{spectropolarimeter}. The physical properties of the photosphere need to be retrieved
from the observational data. This process is called {\bf \Index{inversion}} and starts by making a first
guess of the atmospheric parameters, temperature, LOS velocity, magnetic field strength and orientation.
The numerical solution of the \Index{radiative transfer equation}s provides synthetic Stokes profiles
$\bfit{I}^{\rm syn}$, which are then compared with the observed Stokes profiles $\bfit{I}^{\rm obs}$.
Thus, a \Index{merit function}
\begin{equation}\label{Eq_MeritFunction}
\chi^2 = \sum_{i=1}^{4}{\sum_{j=1}^{N}{\left[ \frac{ I_{i}^{\rm syn}(\lambda_{j}) - I_{i}^{\rm obs}(\lambda_{j}) } {\sigma_{i}} \right]^2}}
\end{equation}
can be defined and minimized. (The index $i$ applies to the four \Index{Stokes parameter}s and index $j$ to
the wavelength positions. $\sigma_{i}$ denotes the uncertainty in the measurement of the \Index{Stokes parameter}
$i$.) The minimization of the \Index{merit function} is an iterative process in which the atmospheric parameters
are systematically modified \citep[mostly via the \Index{Levenberg-Marquardt algorithm}, see e.g.][]{Press2007}
until a best fit between synthetic and observed Stokes profiles is reached.

The described inversion technique is applied for each pixel of a one- or two-dimensional map of
the solar surface, i.e. the inversion of a pixel is completely independent of the neighboring pixels.
Usually, the mentioned first guess of the atmospheric parameters is taken from a model atmosphere
for the considered solar surface phenomenon (quiet Sun, plage, umbra) or from the inversion result
of a neighboring pixel.

Finally, it is pointed out that the underlying \Index{Zeeman effect} allows the determination of the magnetic
field's azimuthal direction $\psi$ not completely uniquely because of the ever-present
{\bf $\mathrm{180^\circ}$ ambiguity\index{ambiguity@$\mathrm{180^\circ}$ ambiguity}}. From the observed Stokes profiles one cannot distinguish
between $\psi$ and $\psi+180^\circ$ because both azimuthal angles lead to the same Stokes profile.
The inversion retrieves the inclination of the magnetic field vector with respect to the line-of-sight,
$\gamma$, and the azimuthal angle in the plane perpendicular to the line-of-sight with respect to the
Stokes~$Q$ and $U$ reference direction, $\psi$. Usually, the two solutions $(\gamma,\psi)$ and
$(\gamma,\psi+180^\circ)$ are transformed into a coordinate system with respect to the solar surface
normal (\Index{local reference frame}) and often a decision can then be taken which of the two possibilities
is the physically meaningful solution (e.g. the orientation of sunspot fields is roughly known).
An overview of various algorithms to solve the $\mathrm{180^\circ}$ ambiguity can be found in
\citet{Metcalf2006}.

\section{Magneto-hydrodynamical simulations}\label{MHD}

The previous sections \ref{Phenomena}-\ref{RadiativeTransfer} concentrated on such aspects of solar physics
which are of fundamental importance for an observer of the solar photosphere. Most of the present knowledge
about the Sun is retrieved from observations and analyses of the electromagnetic radiation emitted by the
Sun towards Earth and analyzed by our instruments from ground, from the \Index{stratosphere}, or from space.
The temporally and spatially highest resolution images stem from light of the photospheric absorption
lines in the spectral range 200$-$1600~nm (from the near ultraviolet over the visible to the near infrared).
The light of these \Index{absorption line}s and the associated \Index{continuum} is emitted by a very thin (compared to the
solar diameter) layer of the solar atmosphere of only a few hundred kilometers thickness. The retrieval
of information about the deeper layers has proved to be difficult. Even if \Index{helioseismology} tries to
obtain information about the solar interior by measuring solar \Index{acoustic oscillation}s \citep{Gizon2005,Gizon2010},
this promising technique cannot yet reach the spatial nor temporal resolution needed to understand the
fundamental processes of the interaction between plasma motions and magnetic fields. Also, the theoretical
possibility of learning more about the solar core with the help of large \Index{neutrino telescope}s that
collect \Index{neutrino}s moving practically unhindered from the solar interior towards Earth, is in its infancy
\citep{Waxman2007,Suzuki2009}.

One way out of the dilemma are magneto-hydrodynamical simulations. Magneto-hydrodynamics (\Index{MHD}) is
an important branch of \Index{plasma physics}, that describes the evolution of a plasma in a \Index{macroscopic approximation}.
A \Index{plasma} is always assumed to be \Index{quasi-neutral}, i.e. the negative charges of
the electrons are roughly compensated by the positive charges of the ions. Charge separation\index{charge separation}
can be neglected for spatial and temporal scales much larger than the \Index{Debye length},
$\lambda_D = \sqrt{\epsilon_{0}k_{\rm B}T/(e^{2}n_e)}$ \citep{Debye1923}, and the inverse \Index{plasma frequency},
$\omega_{P}^{-1}=\sqrt{\epsilon_0{}m_e/(e^{2}n_e)}$ \citep{Kippenhahn1975}, respectively,
where $\epsilon_0$ is the \Index{dielectric constant}, $k_{\rm B}$ the \Index{Boltzmann constant}, $T$ the plasma temperature,
$e$ the \Index{elementary charge}, $n_e$ the \Index{electron number density}, and $m_e$ the \Index{electron mass}.
Furthermore, the thermal energy of the plasma particles must not exceed the volume charge effects
so that a collective behavior of the plasma particles is possible, i.e. the \Index{Debye sphere} has to
contain many electrons, $4\pi\lambda_D^3n_e/3 \gg 1$ \citep{Cap1994}. These conditions are mostly
fulfilled in the \Index{convective zone} and in the lower solar atmosphere, e.g. typical photospheric values,
$T=5000\,\mathrm{K}$ and $n_e=10^{20}\,\mathrm{m}^{-3}$, lead to $\lambda_D = 5 \cdot 10^{-7}\,\mathrm{m}$,
$\omega_{P}^{-1}=2 \cdot 10^{-12}\,\mathrm{s}$, and $4\pi\lambda_D^3n_e/3=49$.

\subsection{Hydrodynamics}
In the absence of electromagnetic phenomena, a plasma behaves like a neutral fluid. The dynamic properties
of neutral gases and fluids can be described by hydrodynamic equations. In \Index{hydrodynamics} we distinguish
between two kinds of time derivatives of dynamic quantities. The partial derivative of a scalar quantity
$Q(\vec{x},t)$ with respect to the time at a fixed position, $\vec{x}$, is called {\bf \Index{Eulerian derivative}}
and is symbolized, as usual, by $\partial / \partial t$.
\begin{equation}\label{Eq_EulerianDerivative}
\frac{\partial Q}{\partial t} = \lim_{\delta t \to 0} \frac{Q(\vec{x},t+\delta t)-Q(\vec{x},t)}{\delta t} .
\end{equation}
In contrast, $d / dt$ symbolizes the {\bf \Index{Lagrangian derivative}} (also named {\bf \Index{material derivative}})
and means the temporal variation of a quantity if one moves with the fluid element having the velocity
$\vec{v}$.
\begin{equation}\label{Eq_LagrangianDerivative}
\frac{dQ}{dt} = \lim_{\delta t \to 0} \frac{Q(\vec{x}+\vec{v}\delta t,t+\delta t)-Q(\vec{x},t)}{\delta t} .
\end{equation}
A first order Taylor expansion of $Q(\vec{x}+\vec{v}\delta t,t+\delta t)$ provides
\begin{equation}\label{Eq_TaylorExpansion}
Q(\vec{x}+\vec{v}\delta t,t+\delta t) = Q(\vec{x},t) + \vec{v}\delta t \nabla Q + \delta t \frac{\partial Q}{\partial t}
\end{equation}
and inserted into Eq.~(\ref{Eq_LagrangianDerivative}) it results in
\begin{equation}\label{Eq_EulerLagrangeScalar}
\frac{dQ}{dt} = \frac{\partial Q}{\partial t} + \vec{v} \cdot \nabla Q\,,
\end{equation}
which provides a relation between the Eulerian and Lagrangian derivative\footnote{Differential
operators, e.g. $\nabla$ or $\Delta$ and some of the related calculation rules are explained in
appendix~\ref{Appendix_A}.}. If $\vec{A}$ is a vector field ($\vec{A}=\{A_i\}$), Eq.~(\ref{Eq_EulerLagrangeScalar})
becomes in components
\begin{equation}\label{Eq_EulerLagrangeVector}
\frac{dA_i}{dt} = \frac{\partial A_i}{\partial t} + \sum_{j}{v_j \nabla_j A_i} = \frac{\partial A_i}{\partial t} + (\vec{v} \cdot \nabla) A_i\,.
\end{equation}

In \Index{hydrodynamics}, the conservation of mass is expressed by the {\bf \Index{continuity equation}}
\begin{equation}\label{Eq_Continuity1}
\frac{\partial \rho}{\partial t} + \nabla \cdot (\rho \vec{v}) = 0 ,
\end{equation}
where $\rho$ stands for the mass density of the fluid. With the help of the product rule for the
divergence operator and use of Eq.~(\ref{Eq_EulerLagrangeScalar}), the \Index{continuity equation} can be
written as
\begin{equation}\label{Eq_Continuity2}
\frac{d \rho}{dt} + \rho \nabla \cdot \vec{v} = 0\,.
\end{equation}
A fluid is incompressible\index{incompressible fluid} ($d\rho/dt = 0$) if and only if it has a source-free velocity
field ($\nabla \cdot \vec{v} = 0$).

If we consider \Index{incompressible fluid}s and neglect inner friction (\Index{viscosity}),
the \Index{equation of motion} for the velocity vector is the {\bf \Index{Euler equation}}
\begin{equation}\label{Eq_Euler}
\rho \frac{d\vec{v}}{dt} = \rho \left( \frac{\partial \vec{v}}{\partial t} + (\vec{v} \cdot \nabla) \vec{v} \right) = - \nabla p + \vec{f}\,,
\end{equation}
which extends to the {\bf \Index{Navier-Stokes equation}}
\begin{equation}\label{Eq_NavierStokes1}
\rho \frac{d\vec{v}}{dt} = \rho \left( \frac{\partial \vec{v}}{\partial t} + (\vec{v} \cdot \nabla) \vec{v} \right) = - \nabla p + \vec{f} + \nabla \hat{\tau}
\end{equation}
if \Index{viscosity} and \Index{compressibility} of the fluid are also considered. $p$ denotes the gas pressure, $\vec{f}$
the force density (unit ${\rm N/m^3}$) which acts on the fluid particles, e.g. the gravitational force density
$\rho\vec{g}$, and $\hat{\tau}$ is the \Index{viscous stress tensor}. For \Index{incompressible fluid}s, the viscous term
simplifies to $\nabla \hat{\tau} = \mu\Delta \vec{v}$, where $\mu$ is the dynamic \Index{viscosity} (unit ${\rm Ns/m^2}$).
More details can be found in the relevant textbooks, e.g. \citet{Fluegge1959,Landau1974,Greiner1991a,Choudhuri1998}.

\subsection{Magneto-hydrodynamics}
As the name already expresses, in \Index{magneto-hydrodynamics} the afore mentioned hydrodynamical equations
are generalized by combining them with the electrodynamical equations in order to come to a description
of fluids which are good electrical conductors. The \Index{Maxwell equation}s in their usual differential form
are written as:
\begin{equation}\label{Eq_Maxwella1}
\nabla \cdot \vec{D} = \rho_q
\end{equation}
\begin{equation}\label{Eq_Maxwella2}
\nabla \cdot \vec{B} = 0
\end{equation}
\begin{equation}\label{Eq_Maxwella3}
\nabla \times \vec{E} = -\frac{\partial \vec{B}}{\partial t}
\end{equation}
\begin{equation}\label{Eq_Maxwella4}
\nabla \times \vec{H} = \vec{j} + \frac{\partial \vec{D}}{\partial t}
\end{equation}
and are complemented by Ohm's law \citep[see, e.g.,][]{Greiner1991b,Grimsehl1988a}:
\begin{equation}\label{Eq_OhmsLawa}
\vec{j} = \sigma \vec{E} .
\end{equation}
$\vec{E}$ stands for the \Index{electric field} (unit ${\rm V/m}$), $\vec{D}$ for the \Index{dielectric displacement field}
(unit ${\rm As/m^2}$), $\vec{H}$ for the \Index{magnetic field}
(unit ${\rm A/m}$), $\vec{B}$ for the \Index{magnetic flux density} (also called \Index{magnetic induction},
unit ${\rm Vs/m^2=Tesla=10^4}$\,Gau\ss{}), $\rho_q$ for the \Index{charge density}  (unit
${\rm As/m^3}$), $\vec{j}$ for the \Index{current density} (unit ${\rm A/m^2}$), and $\sigma$ for the
\Index{electrical conductivity} (unit ${\rm 1/(\Omega m)}$).

A distinction between $\vec{E}$ and $\vec{D}$ or $\vec{B}$ and $\vec{H}$, respectively, makes only
sense if one can distinguish between charges and currents in the conductor and its surrounding medium.
In the case of plasmas this does not play a role, so that we can write $\vec{D}=\epsilon_0\vec{E}$ and
$\vec{B}=\mu_0\vec{H}$, where $\epsilon_0=8.8542 \cdot 10^{-12}\,{\rm As/(Vm)}$ is the \Index{dielectric constant}
(\Index{permittivity} of vacuum) and $\mu_0=1.2566 \cdot 10^{-6}\,{\rm Vs/(Am)}$ is the \Index{magnetic constant}
(\Index{permeability} of vacuum). Therefore in astrophysics the \Index{magnetic flux density}, $\vec{B}$,
is often sloppily called the \Index{magnetic field}. Since the plasma is generally not at rest, but moves
with a velocity, $\vec{v}$, the electrostatic force, $q\vec{E}$, acting on a particle of electric charge $q$,
has to be extended by the \Index{Lorentz force}, $q(\vec{v}\times\vec{B})$. The basic suppositions
of the \Index{MHD approximation} are \citep{Kippenhahn1975,Priest1982}:
\begin{itemize}
   \item All plasma velocities are small compared to the speed of light:
   \begin{equation}\label{Eq_MhdSuppos1}
   \frac{v}{c} \ll 1\,.
   \end{equation}
   Additionally, each temporal variation of a field quantity shall be small in the sense that
   \begin{equation}\label{Eq_MhdSuppos2}
   \frac{l_0/t_0}{c} \ll 1\,,
   \end{equation}
   where $l_0$ and $t_0$ are characteristic spatial and temporal scales on which a field quantity changes.
   \item The plasma's \Index{electrical conductivity} has to be high, so that strong electric fields cannot
   occur:
   \begin{equation}\label{Eq_MhdSuppos3}
   \frac{E_0}{cB_0} \ll 1\,,
   \end{equation}
   where $E_0$ and $B_0$ are characteristic values of the electric and magnetic field strength.
\end{itemize}

An order of magnitude estimate of the ratio of the second term on the right-hand side of
Eq.~\ref{Eq_Maxwella4} (\Index{displacement current}) to the left-hand side of Eq.~\ref{Eq_Maxwella4} leads to:
\begin{equation}\label{Eq_MhdSuppos4}
\frac{\frac{\partial \vec{E}}{\partial t}}{c^2 \nabla \times \vec{B}} \approx \frac{\frac{E_0}{t_0}}{c^2 \frac{B_0}{l_0}} = \frac{E_0}{cB_0} \cdot \frac{l_0/t_0}{c} \ll 1\,,
\end{equation}
so that \Index{displacement current}s can be neglected \citep[see also][]{Cap1994,Choudhuri1998}.
Hence, the Eqs.(\ref{Eq_Maxwella1})-(\ref{Eq_OhmsLawa}) can be simplified:
\begin{equation}\label{Eq_Maxwellb1}
\nabla \cdot \vec{E} = \frac{\rho_q}{\epsilon_0}
\end{equation}
\begin{equation}\label{Eq_Maxwellb2}
\nabla \cdot \vec{B} = 0
\end{equation}
\begin{equation}\label{Eq_Maxwellb3}
\nabla \times \vec{E} = -\frac{\partial \vec{B}}{\partial t}
\end{equation}
\begin{equation}\label{Eq_Maxwellb4}
\nabla \times \vec{B} = \mu_0 \vec{j}
\end{equation}
\begin{equation}\label{Eq_OhmsLawb}
\vec{j} = \sigma (\vec{E} + \vec{v}\times\vec{B})\,.
\end{equation}

Eq.~(\ref{Eq_OhmsLawb}) inserted into Eq.~(\ref{Eq_Maxwellb4}) results in
\begin{equation}\label{Eq_ElectricField}
\vec{E} = \eta \nabla \times \vec{B} - \vec{v} \times \vec{B}\,,
\end{equation}
where $\eta=1/(\mu_0 \sigma)$ denotes the \Index{magnetic diffusivity}. Therefore, the \Index{electric field} vector
is not an independent dynamic variable of \Index{magneto-hydrodynamics}, but it can be calculated from the
\Index{magnetic field} $\vec{B}$ and the velocity $\vec{v}$. The only one new dynamic variable is the
magnetic field vector, $\vec{B}$, whose temporal evolution has to be described by a further equation.
Thereto, Eq.~(\ref{Eq_ElectricField}) is inserted into Eq.(\ref{Eq_Maxwellb3}) and assumed that the
\Index{electrical conductivity}, $\sigma$, is spatially constant. With the help of the operator identity~(\ref{Eq_OpIdent8})
and Eq.~(\ref{Eq_Maxwellb2}), we can derive the {\bf \Index{induction equation}}
\begin{equation}\label{Eq_Induction}
\frac{\partial \vec{B}}{\partial t} = \nabla \times (\vec{v} \times \vec{B}) + \eta \Delta \vec{B}\,.
\end{equation}

Besides this new equation for the magnetic field, the above mentioned hydrodynamical equations
(\ref{Eq_Continuity1}) and (\ref{Eq_NavierStokes1}) are still valid but have to
be extended by the magnetic field. The \Index{continuity equation}~(\ref{Eq_Continuity1}) remains but the
\Index{Navier-Stokes equation}~(\ref{Eq_NavierStokes1}) has to be extended by the \Index{magnetic force density}
$\vec{j} \times \vec{B}$ \citep[see, e.g.,][]{Gerthsen1992,Grimsehl1988a}
\begin{equation}\label{Eq_NavierStokes2}
\rho \left( \frac{\partial \vec{v}}{\partial t} + (\vec{v} \cdot \nabla) \vec{v} \right) = - \nabla p + \vec{j}\times\vec{B} + \rho\vec{g} + \nabla \hat{\tau}\,,
\end{equation}
which, after using Eqs.(\ref{Eq_Maxwellb4}) and (\ref{Eq_OpIdent9}), results in the following MHD \Index{equation of motion}:
\begin{equation}\label{Eq_NavierStokes3}
\rho \left( \frac{\partial \vec{v}}{\partial t} + (\vec{v} \cdot \nabla) \vec{v} \right) = - \nabla \left( p + \frac{B^2}{2\mu_0}\right) + \frac{(\vec{B}\cdot\nabla)\vec{B}}{\mu_0} + \rho\vec{g} + \nabla \hat{\tau}\,.
\end{equation}
The presence of a magnetic field leads to an additional \Index{magnetic pressure} of the form $B^2/(2\mu_0)$
(in the cgs system: $B^2/(8\pi)$). The ratio of the thermal\index{thermal pressure} to the magnetic pressure\index{plasma beta@plasma $\beta$}
\begin{equation}\label{Eq_PlasmaBeta}
\beta = \frac{\mathrm{thermal~pressure}}{\mathrm{magnetic~pressure}} = \frac{2 \mu_0 p}{B^2}
\end{equation}
plays an important role in the investigation of magneto-hydrodynamical phenomena. In the case
$\beta \gg 1$, e.g. in the \Index{convective zone}, the gas plays the important role and forces the magnetic
field lines to follow the gas motions. On the other hand, in the case $\beta \ll 1$, e.g. in the solar
corona, the magnetic field dominates the dynamic behavior of the plasma. The plasma particles have to
follow the magnetic field. $(\vec{B}\cdot\nabla)\vec{B}/\mu_0$ is the second term of
Eq.~(\ref{Eq_NavierStokes3}) which is of a magnetic nature. It is a \Index{magnetic tension} along the field lines.

The conservation of energy can be expressed in terms of the total energy per unit volume,
$e$ (unit ${\rm J/m^3}$), which is the sum of the internal energy density, the kinetic energy density,
and the magnetic energy density: $e = e_{\rm int} + \rho v^2/2 + B^2/2\mu_0$. The time derivative of this
energy balance as well as the use of the \Index{induction equation} (\ref{Eq_Induction}) and the \Index{equation of motion}
(\ref{Eq_NavierStokes3}) lead to the MHD \Index{energy equation} \citep{Voegler2003,Voegler2005b}:
\begin{equation}\label{Eq_Energy}
\frac{\partial e}{\partial t} + \nabla \cdot \left[ \vec{v} \left( e+p+\frac{|\vec{B}|^2}{2\mu_0} \right) - \frac{1}{\mu_0}\vec{B}(\vec{v}\cdot\vec{B}) \right] = \\
\frac{1}{\mu_0} \nabla \cdot (\vec{B} \times \eta \nabla \times \vec{B}) + \nabla \cdot (\vec{v} \cdot \hat{\tau}) + \nabla \cdot (K \nabla T) + \rho(\vec{g}\cdot\vec{v}) + Q_{\rm rad}\,,
\end{equation}
where $K$ stands for the \Index{thermal conductivity} (unit ${\rm W/(mK)}$) and $T$ for the temperature. $Q_{\rm rad}$
takes into account radiation heating and cooling processes and requires \Index{radiative transfer} calculations.

The relations between the thermodynamic quantities of the plasma are described by the equations of state\index{equation of state},
e.g. in the form $T=T(\rho,e_{\rm int})$ and $p=p(\rho,e_{\rm int})$ and complete the system of
\Index{MHD equation}s~(\ref{Eq_Continuity1}), (\ref{Eq_Induction}), (\ref{Eq_NavierStokes3}), (\ref{Eq_Energy}).
Derivations and examples of physical applications of the MHD equations can be found in, e.g.,
\citet{Priest1982,Cap1994,Choudhuri1998}.

\subsection{Numerical codes}
Subject of this study is the investigation of solar photospheric phenomena. Such phenomena often
have their origin in the \Index{convective zone} or are determined by the interplay of the convection
zone with the photosphere. Since the \Index{convective zone} eludes a direct observation, the development
of numerical codes is pursued, which simulate the behavior of the plasma in the \Index{convective zone} and
\Index{photosphere} by solving the \Index{MHD equation}s. If the simulation results agree well with the photospheric
observations, there is hope that the simulations can also reproduce the behavior of the \Index{convective zone}.
Numerical \Index{MHD code}s do not only give the physicists the possibility to look below the solar surface,
but they also provide results at an arbitrarily high temporal, spatial, and spectral resolution as well
as free of \Index{stray light} and noise.

In the transition from the convective zone to the photosphere, the plasma density is sufficiently decreased,
so that the plasma becomes transparent for wavelengths in the visible spectral range. The convective energy
transport changes into a radiative energy transport. The irradiance emitted by the Sun as well as the
temperature profile of the photosphere depend on the energy exchange between gas and radiation. Therefore,
the \Index{energy equation} of a numerical code simulating the photosphere has to comprise a radiative term, i.e.
the \Index{radiative transfer equation} has also to be solved numerically. Such a code is called a radiation \Index{MHD code}.

Examples of numerical radiation \Index{MHD code}s are the \Index{MURaM} code\footnote{I calculated the
MHD simulations analyzed in chapter~\ref{Bp2Chapter} with the MURaM code.} \citep{Voegler2003,Voegler2005b},
the \Index{CO$^5$BOLD} code \citep{Freytag2008,Freytag2012}, or the Stagger code
\citep{Nordlund1995,Galsgaard1996}. A comparison of the results of these three codes can
be found in \citet{Beeck2012}.

\section{Miscellaneous}
 
\subsection{Arcseconds}

A solar physicist says, an object of the solar surface has a size of one \Index{arcsecond}, if the object
has an \Index{angular resolution} of $1/3600^{\circ}$ when observed from Earth. The conversion between
arcsecond and kilometer depends on the distance between Earth and Sun, which varies in the course
of a year. \citetext{Algorithms to calculate the Earth-Sun distance can be found in \citealt{Meeus1991},
source codes of an implementation using the programming language C++ are part of \citealt{Hirzberger2009}.}
Hence, an \Index{arcsecond} corresponds to 713\,km~-~737\,km on the solar surface. Averaged over the year, one
finds: 1\arcsec{} = 725\,km. When \sunrise{} observed the data used in chapter~\ref{Bp1Chapter} and
\ref{Bp2Chapter}, the relation was: 1\arcsec{} = 736.4\,km.

\begin{figure*}
\centering
\includegraphics*[width=9cm]{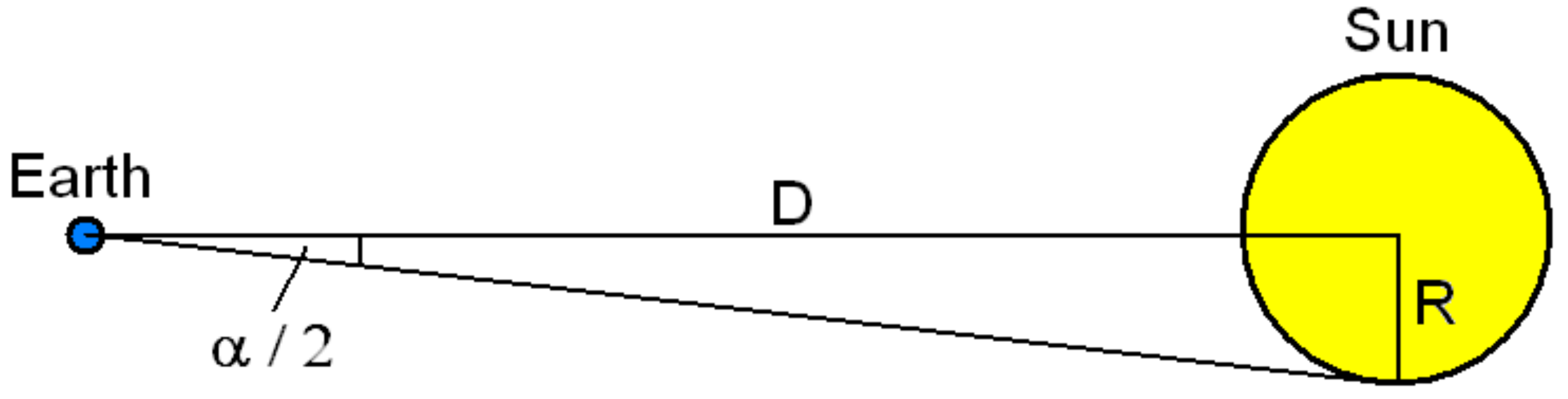}
\caption{Viewing angle of the Sun as observed from Earth.}
\label{FigArcsecond}
\end{figure*}

The Sun has a radius of R~=~696\,000\,km and the mean distance to Earth is D~=~149\,600\,000\,km
(see Fig.~\ref{FigArcsecond}). If observed from Earth, the solar diameter, $\alpha$, is in arcseconds:
$\tan{(\alpha/2)} = R/D$, hence $\alpha = 2 \arctan{(R/D)} = 0.533^{\circ} = 32\arcmin = 1919\arcsec$.

\subsection{Heliocentric angle}

\begin{figure*}
\centering
\includegraphics*[width=\textwidth]{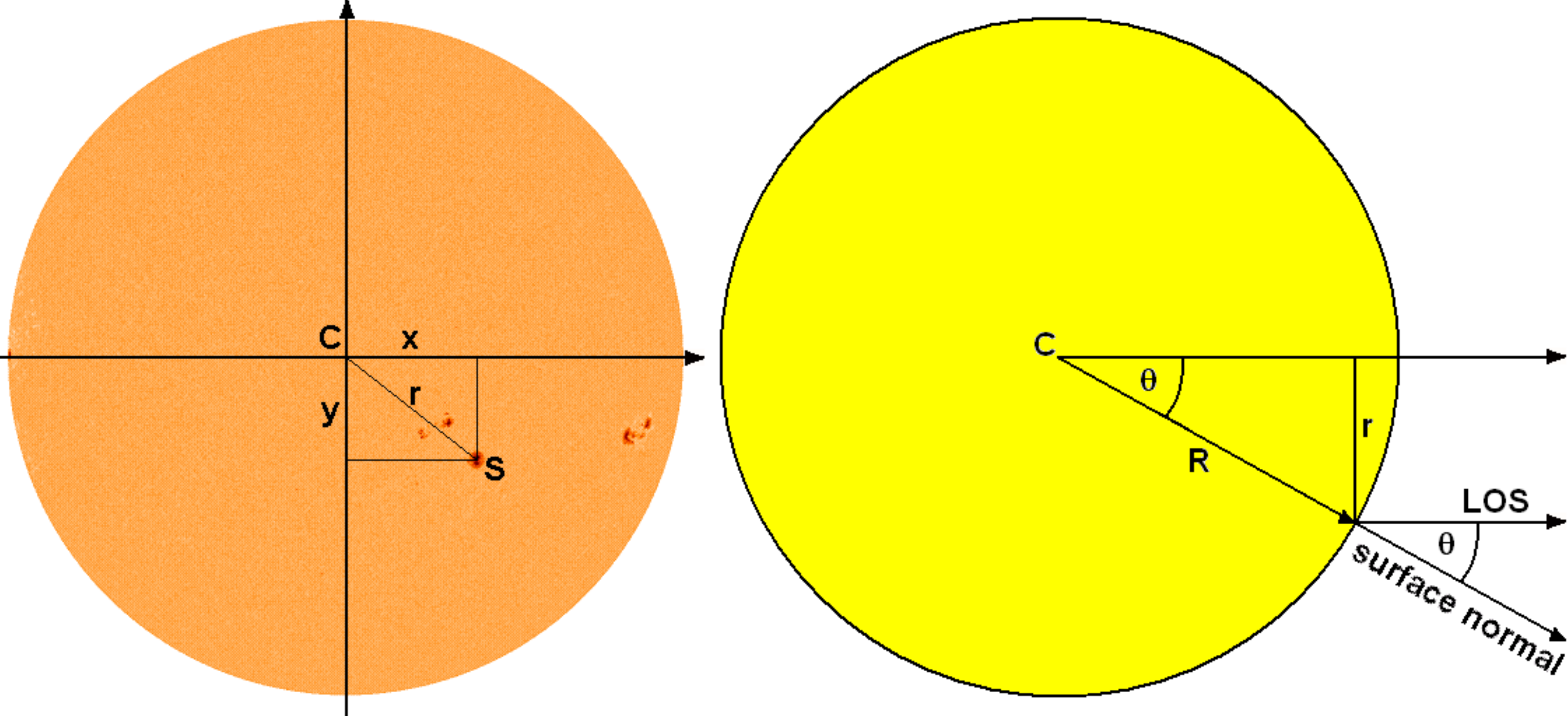}
\caption{Definition of the heliocentric angle $\theta$ as angle between the line-of-sight (LOS) and the surface normal.}
\label{FigHeliocentricAngle}
\end{figure*}

Since the Sun is a sphere, we only look perpendicularly through the solar atmosphere if the observed
phenomenon is located in the center of the solar disk. In the case of off-center observations, the
line-of-sight is inclined with respect to the solar surface normal. This leads to \Index{projection effect}s
which are more pronounced close to the solar limb. Often the \Index{heliocentric angle} of the observed region,
i.e. the angle between the \Index{line-of-sight} (direction to the observer) and the surface normal, is given
to quantitatively capture the strength of such \Index{projection effect}s.
The left panel of Fig.~\ref{FigHeliocentricAngle} exhibits the Sun as it was observed on September 9, 2004.
The sunspot S of active region NOAA~10667 (extensively analyzed in chapter~\ref{Ud1Chapter}) was
two days after our observation at the \Index{Swedish Solar Telescope} at the position $x=479\arcsec,~y=-280\arcsec$.
The distance between S and disk center was therefore $r=\sqrt{x^2+y^2}=555\arcsec$. The right panel of 
Fig.~\ref{FigHeliocentricAngle} shows a cut through the plane spanned by $r$ and the LOS, with the observer
being on the right side. The \Index{heliocentric angle} can be calculated as follows:
\begin{equation}\label{Eq_HeliocentricAngle}
\sin{\theta} = \frac{\sqrt{x^2+y^2}}{R} ,
\end{equation}
where $R=1919\arcsec/2$ denotes the solar radius in arcseconds. Often the value $\mu=\cos{\theta}$
is given.

\subsection{Diffraction limit}

If light enters a telescope, the wavefront is restricted, mostly by the size of the main mirror or,
in case of a refractor, by the size of the entrance lens. Hence, the light is diffracted at the border
of the entrance aperture, i.e. a point-like object (e.g. a star) is imaged by a telescope as an
\Index{Airy disk} instead of a point. If two nearby point sources are observed, the two Airy disks are
superimposed and cannot be separated if their distance is sufficiently small. According to the so-called
{\bf \Index{Rayleigh criterion}}, two point sources can be separated if their distance is at least the radius
of the Airy disk, i.e. if the first minimum of the one Airy disk is at the position of the maximum
of the other \Index{Airy disk}. The corresponding \Index{angular resolution} is:
\begin{equation}\label{Eq_RayleighCriterion}
\sin{\alpha} = 1.22 \frac{\lambda}{D} ,
\end{equation}
where $\lambda$ stands for the observed wavelength and $D$ for the diameter of the telescope's
\Index{entrance aperture}. Hence, the \Index{spatial resolution} of a telescope is better for larger telescopes
and for shorter wavelengths. At a wavelength of 6301.5\,\AA{}ngstr\"om\footnote{1\,\AA{} = 0.1\,nm
= $10^{-10}$\,m}, the 50-cm telescope SOT onboard the \hinode{} satellite (see chapters~\ref{Ud2Chapter} and \ref{Ud3Chapter})
has a \Index{diffraction limit} of $\alpha=\arcsin{(1.22\times 6301.5\times 10^{-10}\,\mathrm{m}\,/\,0.5\,\mathrm{m})}
= 0.0000979^{\circ} = 0$\carcsec{}$35 = 255$\,km. For the shortest wavelength of \sunrise{}
(see chapters~\ref{Bp1Chapter} and \ref{Bp2Chapter}), the \Index{diffraction limit} is
$\alpha=\arcsin{(1.22\times 2140\times 10^{-10}\,\mathrm{m}\,/\,1.0\,\mathrm{m})} = 0$\carcsec{}$05 = 40$\,km.

\chapter{Review}\label{Review}

After the outline, in Chapter~\ref{Basics}, of the physical basics essential for the understanding
of this thesis, this chapter reviews the historical development of research on small-scale
magnetic features. An overview of the most important umbral dot studies is given first, followed by the
papers on bright points. Completeness cannot be claimed owing to the large number of publications on these
topics. Instead the review is restricted to such aspects that are relevant for this thesis.

\section{Umbral dots}\index{umbral dot}

\begin{figure*}
\centering
\includegraphics*[width=\textwidth]{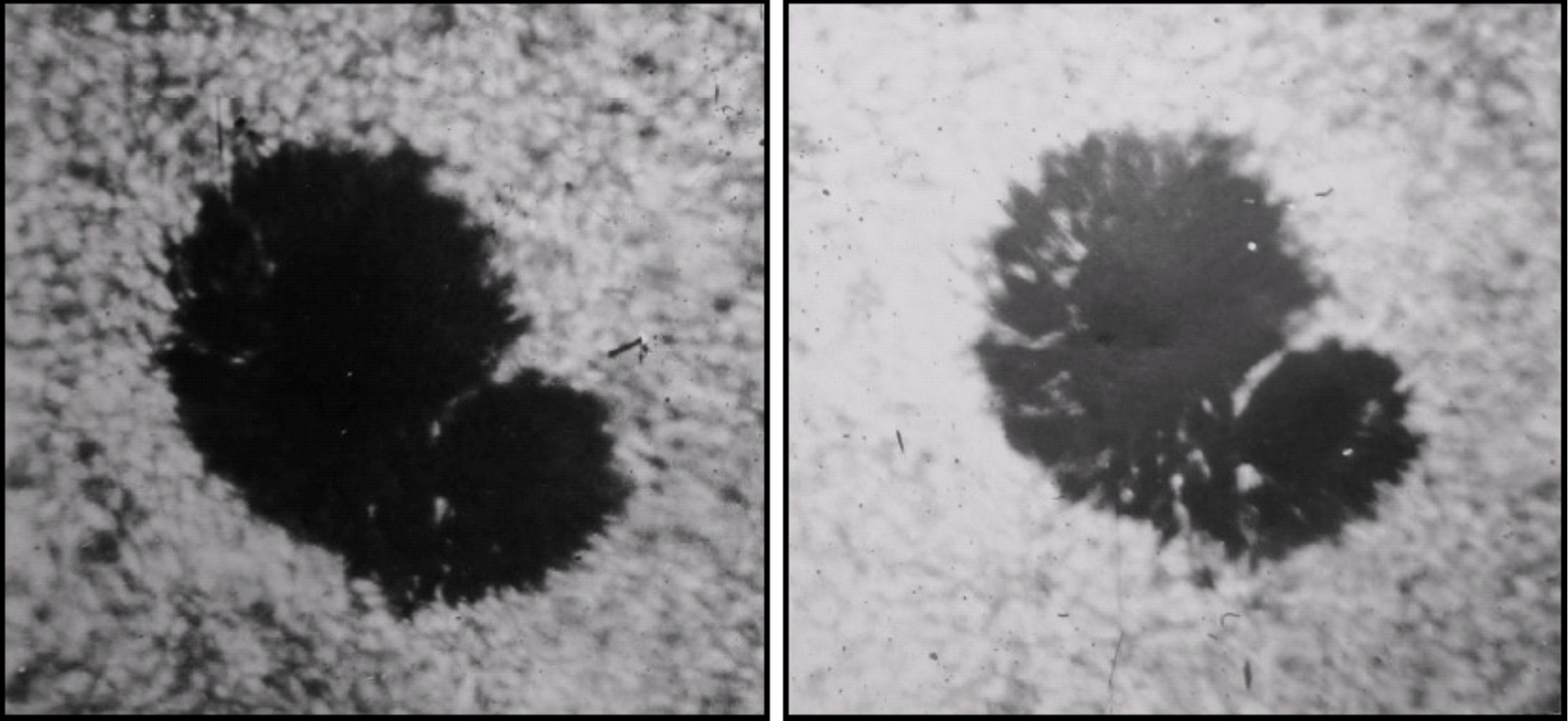}
\caption{Bright features visible in the dark umbra of a sunspot observed by the Jesuit Stanislas Chevalier in 1907
with a 40-cm refractor at the ground-based observatory of Z\^o-S\`e close to Shanghai \citep{Chevalier1916a}.
The solar scene was photographed twice with a different exposure time.}
\label{FigChevalierUDs}
\end{figure*}

Bright sub-structures inside a sunspot umbra were observed for the first time by Father Stanislas Chevalier
in 1907 \citep[see][and also Fig.~\ref{FigChevalierUDs}]{Chevalier1914a,Chevalier1916a}.
The leader of the Jesuitian observatory near Shanghai discovered with a 40-cm refractor a small-scale
granulation-like pattern in the \Index{umbra} which appeared more coarse than the quiet-Sun granulation \citep{Chevalier1916b}.
In 1941, Ludwig Biermann stated, that strong magnetic fields in sunspots suppress the convection by
\Index{eddy-current braking} \citep{Biermann1941}. In a sunspot, the energy can only be transported by radiation
but not by convection. In 1950, Georg Thiessen confirmed the \Index{umbral granulation} observed by Chevalier
with the 60-cm refractor of the observatory of Hamburg-Bergedorf \citep{Thiessen1950}. He found
a mean diameter of an umbral granule of 1\arcsec{}, while the size of a typical photospheric \Index{granule}
was 1\carcsec{}3. Sometimes the umbra did not show a \Index{granulation} pattern, instead he found bright dots
inside the umbra of only 0\carcsec{}3. In 1959, Bray \& Loughhead determined for the first time
the lifetime of umbral granules and found values between 15 and 120 minutes \citep{Bray1959,Loughhead1960}.
Since their telescope, having an aperture of 12.7~cm, had only a limited spatial resolution and
the \Index{seeing}\footnote{Wavefront aberrations caused by turbulence in the terrestrial atmosphere are called "seeing".}
was probably not optimal, they only found large values of 2\carcsec{}3 and 2\carcsec{}9 for the mean size
of granules in the umbra and quiet Sun, respectively.

\begin{figure*}
\centering
\includegraphics*[width=\textwidth]{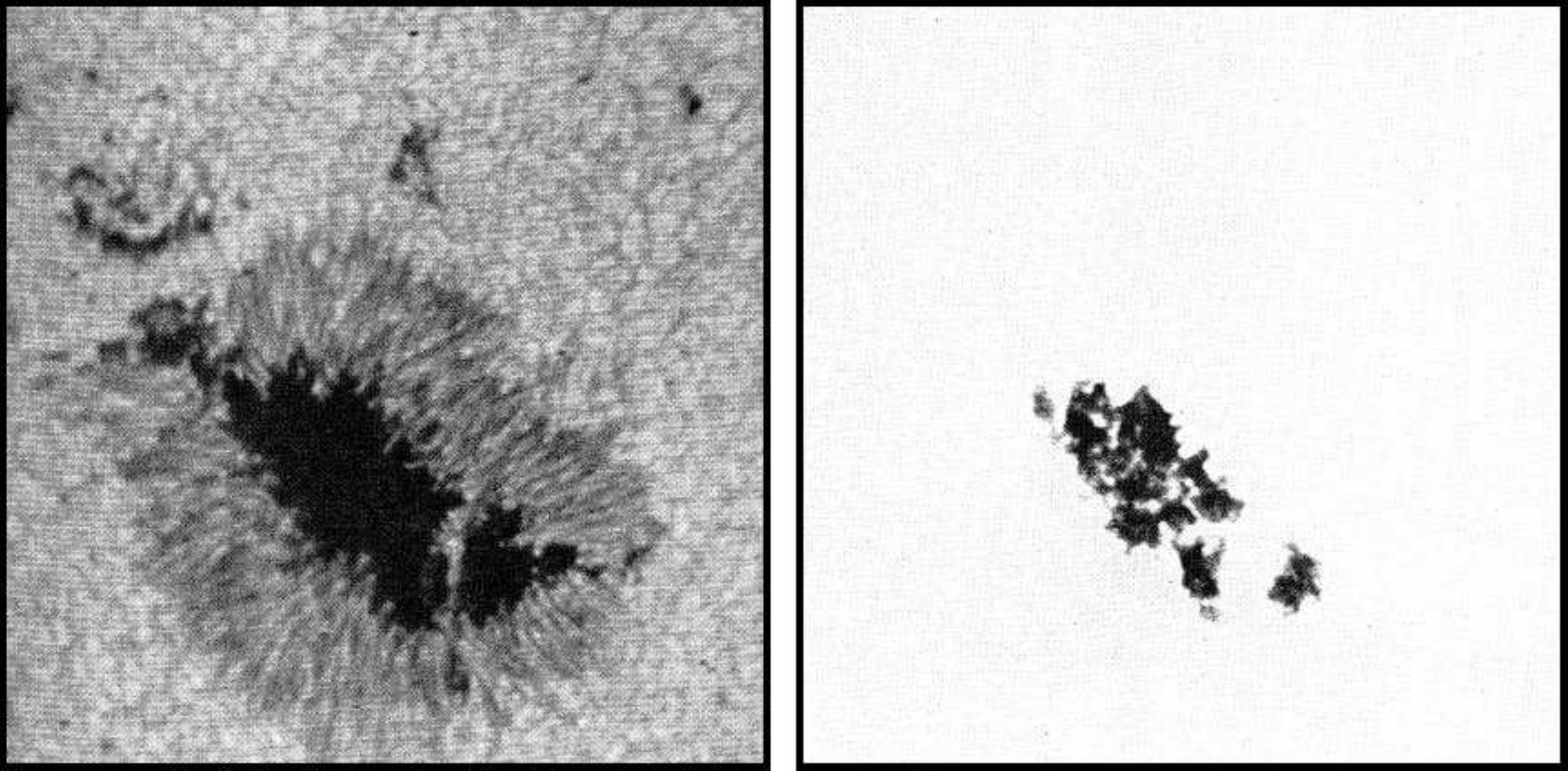}
\caption{Umbral dots observed with the balloon-borne 30-cm telescope Stratoscope~I in 1959 \citep{Danielson1964}.
The solar scene was photographed twice with a different exposure time.}
\label{FigDanielsonUDs}
\end{figure*}

The early observations were made from ground and without using \Index{adaptive optics}, hence they suffered
from variable seeing conditions. The situation changed with the launches of the balloon-borne 30-cm telescope
\Index{Stratoscope~I} in the years 1957-1959 \citep{Rogerson1958,Danielson1961}. The flight on September 24, 1959
concentrated on umbral observations. \citet{Danielson1964} introduced the name \Index{umbral dot} (UD) for the
small-scale bright dots inside umbrae and argued that the so-called \Index{umbral granulation} does not really
exist but is only the result of observations of UD groups at insufficient spatial resolution or under
poor seeing conditions, respectively (see Fig.~\ref{FigDanielsonUDs}). He could only limit the UD lifetimes
roughly between 4 and 50 minutes owing to the small number of observations. This range of UD lifetimes was
consistent with the lifetimes of umbral granules found by \citet{Loughhead1960}. Observations with the 30-cm
telescope of the Sacramento Peak Observatory confirmed that the bright umbral sub-structures are not closed
patterns like in the quiet Sun but rather isolated emission dots \citep{Beckers1968b}.

\begin{figure*}
\centering
\includegraphics*[width=10cm]{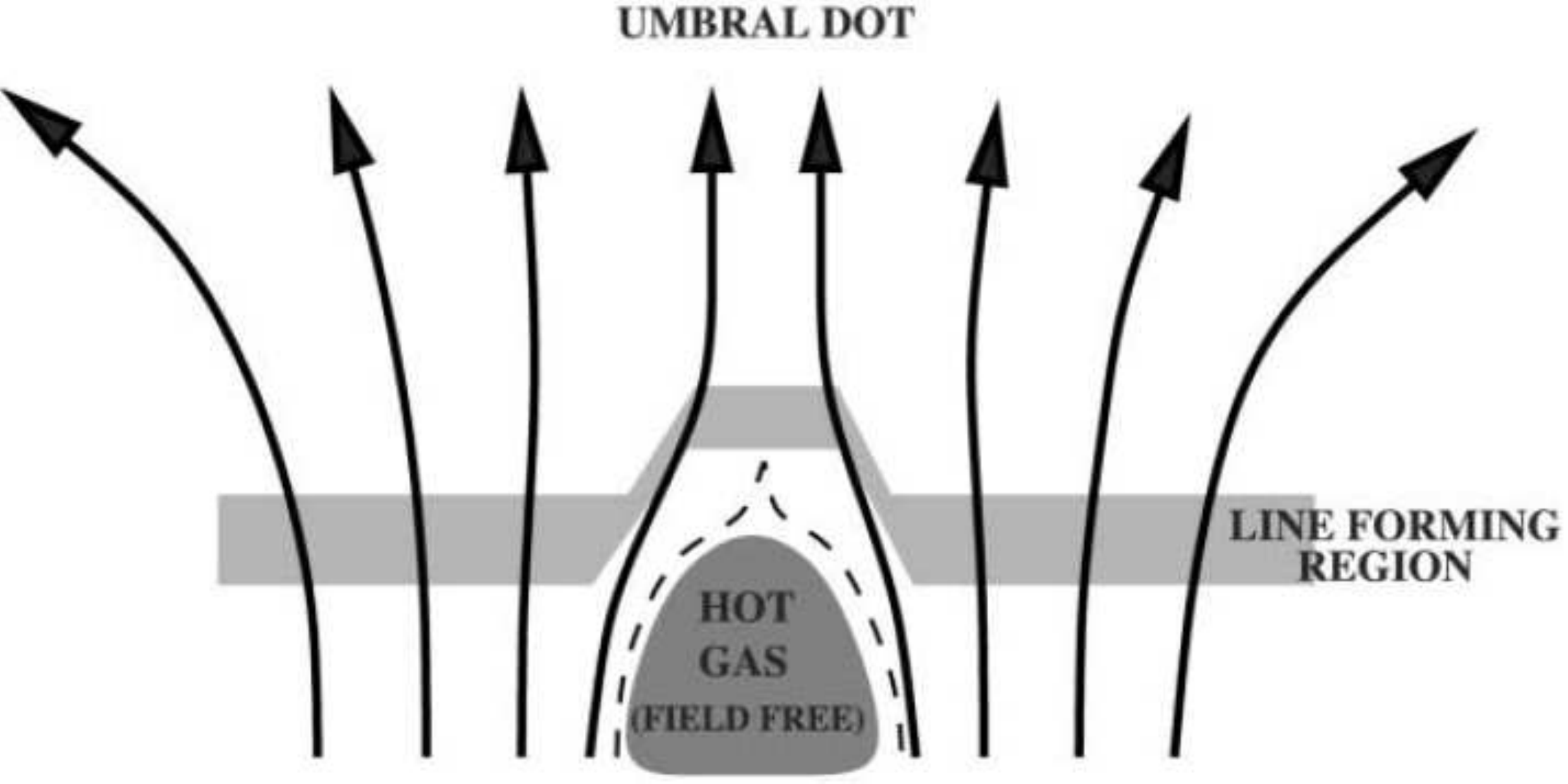}
\caption{Sketch of an umbral dot as an intrusion of field-free hot gas in between a bundle of thin
magnetic flux ropes of the \Index{cluster model} of sunspots. The increased temperature of the UD shifts the
line forming region (shaded) to higher layers. From \citet{SocasNavarro2004}.}
\label{FigUDCartoon}
\end{figure*}

Spectroscopical observation methods finally provided information about the magnetic field strength
and LOS velocity of the UDs. \citet{Kneer1973} found a field weakening from 2600~G in the umbral
vicinity to 1200~G in the observed UD. Considerably lower field weakenings of only 5-20\% are reported,
e.g., by \citet{Buurman1973, Adjabshirzadeh1983,Pahlke1990,Schmidt1994,Tritschler1997}. On the contrary,
other authors did not find indications of magnetic field weakenings in UDs
\citep[see, e.g.,][]{Zwaan1985,Lites1989,Lites1991}, but used only Stokes~$I$ observations of single spectral
lines, which limited the accuracy of their results. \citet{Schmidt1994} pointed out, that a field
weakening of 5-20\% can also be explained by the fact that the magnetic field in a sunspot decreases with height
and that the visible surface of an UD is enhanced compared to its vicinity, see Fig.~\ref{FigUDCartoon}.
According to \citet{Parker1979} and \citet{Choudhuri1986}, the UDs are seen as thin columns of field-free
hot gas penetrating the cluster of small magnetic \Index{flux tube}s that form the sub-photospheric structure of a sunspot.
In addition to the \Index{cluster model}, a second \Index{sunspot model} exists. In the \Index{monolithic model}, sunspots are
homogeneous even below the solar surface. Numerical simulations revealed that magneto-convective processes
in such strongly magnetized plasmas can lead to spatially modulated oscillations, that possibly can be
observed as UDs \citep{Weiss1990,Hurlburt1996}.

A similar heterogeneous picture was found for the LOS velocities of UDs. While \citet{Kneer1973} and
\citet{Pahlke1990} determined upflow velocities in the range 1-3~km~s$^{-1}$, \citet{Rimmele1997},
\citet{Hartkorn2003} and \citet{SocasNavarro2004} found only small blueshifts of 50-300~m~s$^{-1}$.
\citet{Zwaan1985}, \citet{Schmidt1994} and \citet{Wiehr1994} could not retrieve any vertical plasma
flows in the UDs relative to their surroundings. Besides insufficient spatial resolution, there are
further reasons for the diversity of the results: On the one hand, the observations were done in
various spectral lines, which are formed in different atmospheric heights. Since the field lines
of the umbral magnetic field show a \Index{canopy}-like structure above the UD (see below), the measured field
weakening (and LOS velocity) depends strongly on the formation height of the used spectral line.
On the other hand, the observations could be influenced by varying atmospheric seeing conditions
and different levels of \Index{stray light} contamination.

Among the most comprehensive UD analyses are those of \citet{Sobotka1997a,Sobotka1997b}.
With the Swedish 50-cm telescope (\Index{SVST}), an umbra could be observed for 4.5 hours under outstanding
seeing conditions. The analysis of 662 observed UDs revealed monotonically decreasing histograms for
UD lifetime and effective diameter. On average, an UD lived 13.8~minutes and had a diameter of 0\carcsec{}42.
The UDs moved horizontally with velocities in the range of 0-1000~m~s$^{-1}$, where the UD speeds were
slightly grouped at 100~m~s$^{-1}$ and 400~m~s$^{-1}$. Large and long-lived UDs were preferentially
found near the umbral border. \citet{Sobotka2005} repeated the determination of the UD diameters with
the now upgraded 1-m Swedish telescope (SST) and found a maximum in the histogram at 0\carcsec{}23.
While more recent UD studies only investigated individual UD properties and did not determine, e.g.,
UD trajectories \citep{Tritschler2002,Hartkorn2003,Sobotka2005}, chapter~\ref{Ud1Chapter} extends the work
of Sobotka et al. by analyzing a multitude of UD properties (e.g. lifetimes, diameters, horizontal velocities,
peak intensities) of an almost two-hour SST time series of photometric sunspot data. For the first time,
the observations with a 1-m telescope were brought very close to the theoretical \Index{diffraction limit} of
0\carcsec{}18 at the observing wavelength with the help of a modern \Index{image reconstruction} technique
\citep[\Index{Multi-Frame Blind Deconvolution},][]{Loefdahl2002}.

\begin{figure*}
\centering
\includegraphics*[width=10cm]{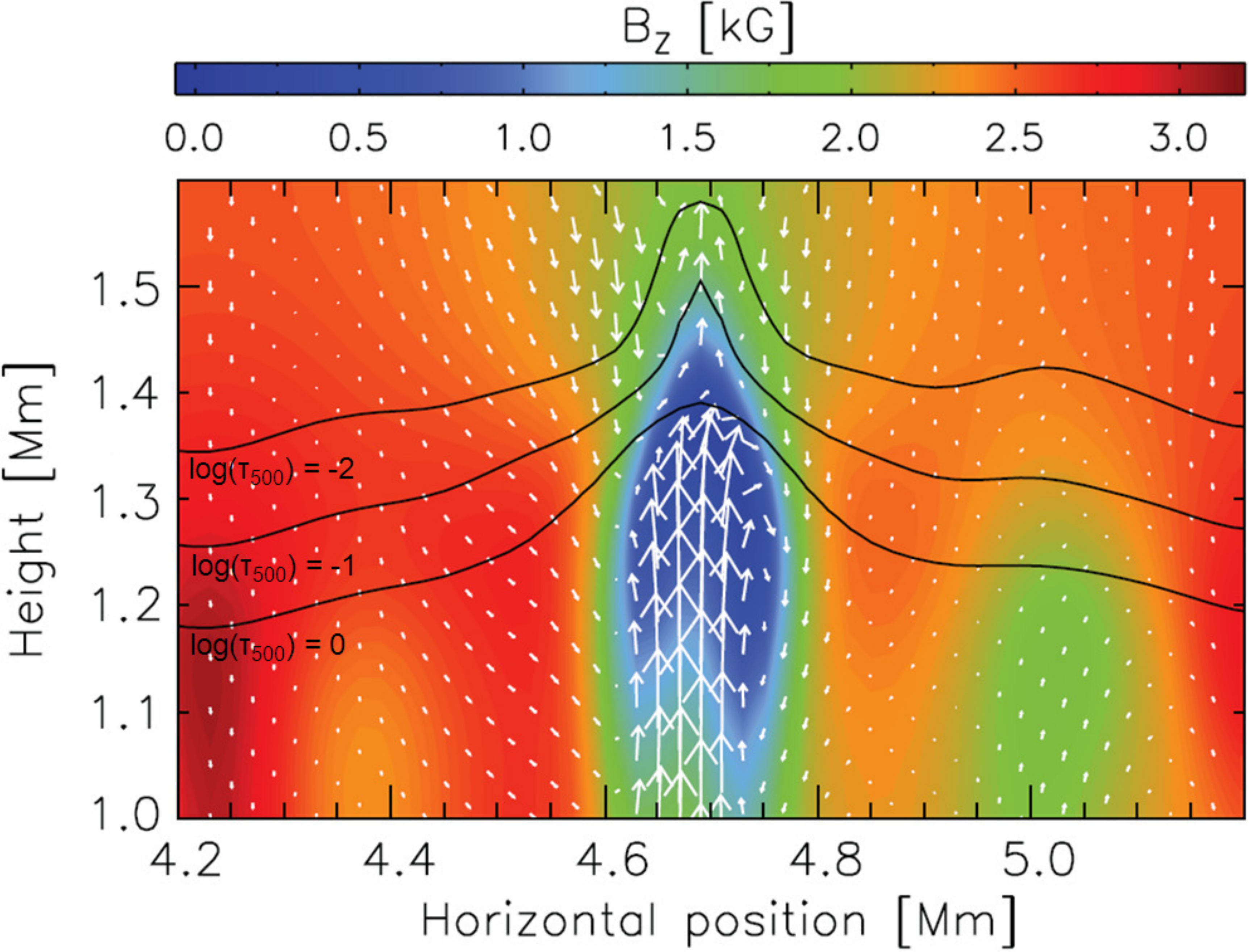}
\caption{Vertical cut trough an UD as simulated by \citet{Schuessler2006}. Colors indicate
magnetic field strength, and the arrows represent (projected) velocity vectors. The longest
arrow corresponds to a velocity of 2.7~km~s$^{-1}$. The black lines indicate levels of constant optical
depth $\log(\tau_{500}) = 0$, $-1$, and $-2$ (bottom to top). The upper part of the plume has developed a
cusplike shape with a strongly decelerated upflow and a weak magnetic field.}
\label{FigUDVertCut}
\end{figure*}

A substantial step in the understanding of UDs was taken with the help of MHD simulations.
The realistic simulation of three-dimensional radiative magneto-convection of an umbra by
\citet{Schuessler2006} yielded numerous UDs as a natural consequence of the convection in a strong, initially
monolithic, magnetic field. Upflows of hot gas are present inside the synthetic UDs and narrow downflow
channels surround the UDs. Close to the visible surface, the magnetic field of the UDs is significantly
weakened (see Fig.~\ref{FigUDVertCut}). Most of the UDs are slightly elongated in the horizontal direction
and show a central \Index{dark lane}. At least for relatively large UDs, the \Index{dark lane}s predicted by the simulations
were found in the observations of \citet{Bharti2007} and \citet{Rimmele2008}.

Another important milestone in the investigation of UDs was the introduction of \Index{inversion} techniques for
analyzing Stokes profiles (see section~\ref{StokesInversion}). Older analyses of spectroscopic data
retrieved the magnetic field strengths and LOS velocities directly from the splitting, broadening, and
shift of the Stokes profiles, in many cases only from the Stokes~$I$ profiles. With the advent of improved
instruments that are able to observe the full \Index{Stokes vector} at sufficiently high spectral sampling and
resolution, the full \Index{radiative transfer equation} can be solved numerically, so that the inversions
can now retrieve the full magnetic field vector, the LOS velocity, as well as the temperature and their
height dependence. A pioneering work in the field of UD analyses was the study of \citet{SocasNavarro2004},
in which eight UDs were observed spectropolarimetrically with the 50-cm telescope \Index{SVST}. The inversion of the
full \Index{Stokes vector} in the spectral range 6301\AA-6304\AA{} allowed the determination of gradients for
temperature, magnetic field, and LOS velocity. The UDs showed a temperature enhancements of 1000~K, a field
weakening of 500~G, and upflows of 100~m~s$^{-1}$. The influence of atmospheric seeing and \Index{stray light}
degraded the data significantly, so that the physical quantities could only be retrieved for two $\log(\tau)$
nodes (i.e. gradients). Chapter~\ref{Ud2Chapter} is a continuation of the work of \citet{SocasNavarro2004}.
The Spectropolarimeter onboard the 50-cm space telescope \hinode{} recorded sunspot data which are
seeing-free and almost free of stray light. 51 UDs were statistically analyzed and, for the first time,
full Stokes profiles could inverted at four $\log(\tau)$ nodes which led to an observational confirmation
of essential aspects of the simulations done by \citet{Schuessler2006}.

\section{Photospheric bright points}

\begin{figure*}
\centering
\includegraphics*[width=\textwidth]{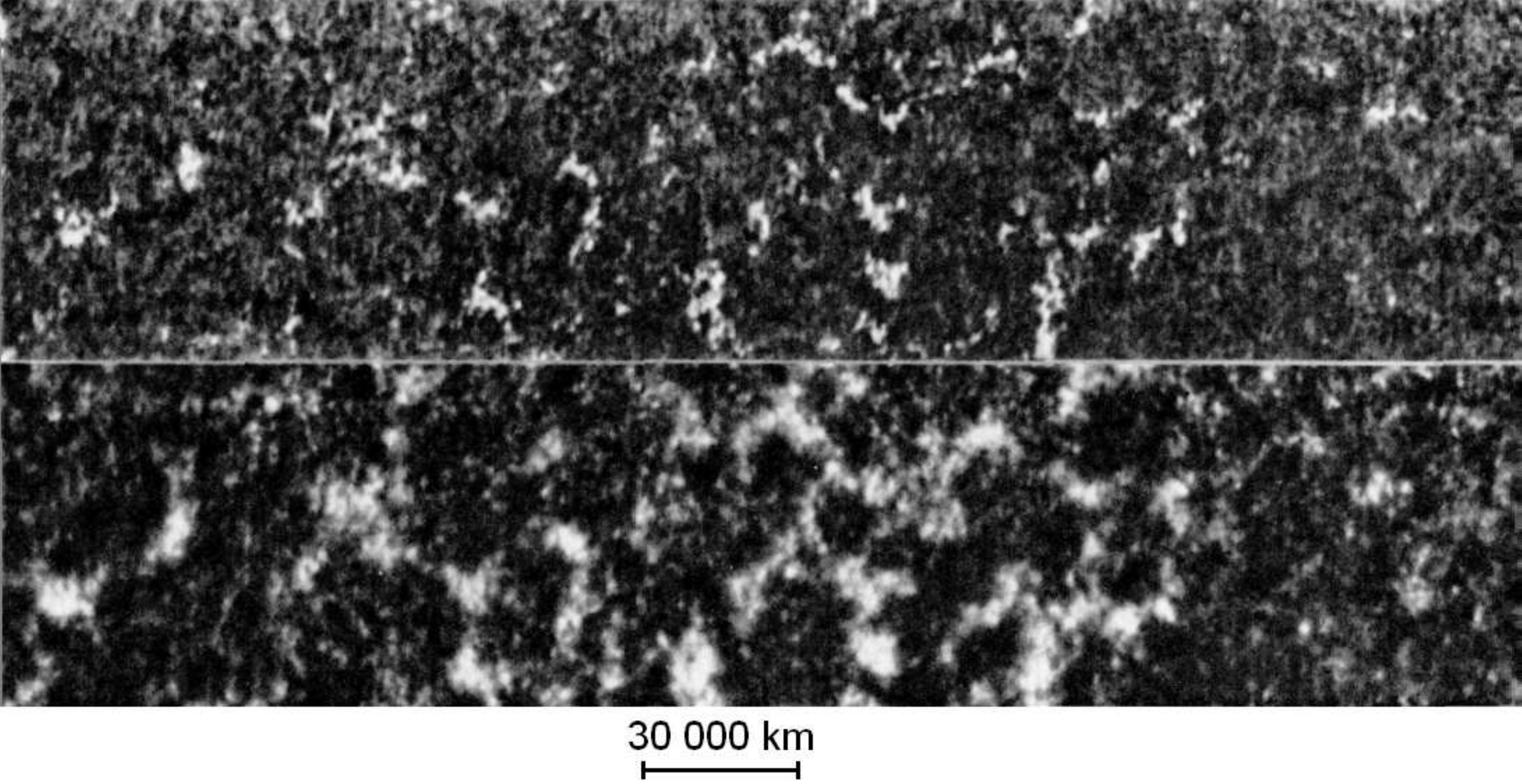}
\caption{Simultaneous observations of the photospheric network in the CN band at 388~nm (top panel)
and the \Index{chromospheric network} in the Ca\,{\sc ii}~K line at 393~nm (bottom panel). The supergranular cells
are visible due to bright network elements located at the cell boundaries and have a typical size of
30~Mm. From \citet{Liu1971}.}
\label{FigNetwork}
\end{figure*}

The English physicist Avril Sykes, nee Hart, carried out spectroscopical investigations of the
solar rotation and discovered an irregular distribution of velocity fluctuations \citep{Hart1956}.
In such velocity fluctuations, \citet{Leighton1962} found a cellular pattern with a mean cell
distance of 30~Mm. The cells were uniformly distributed over the entire Sun and they called the
cellular pattern \Index{supergranulation}. Finally, \citet{Simon1964} found a coincidence between the
supergranular edges and the \Index{chromospheric network} that was already known from Ca\,{\sc ii} and
H$\alpha$ images (see Fig.~\ref{FigNetwork}). A typical lifetime on the order of one day was found
for both, the supergranules and the \Index{chromospheric network}
\citep{Simon1964,Janssens1970,Rogers1970,Worden1976,Duvall1980,Wang1989,Shine2000}.

\citet{Dunn1973} observed the \Index{chromospheric network} of the Sun with the 60-cm telescope of the
Sacramento Peak Observatory using a narrowband (0.25~\AA{}) H$\alpha$ filter. By tuning the filter
from the core of the line to the wing, one looks at deeper and deeper layers of the solar atmosphere.
Under excellent seeing conditions, Dunn \& Zirker saw at 2~\AA{} distance from the core of the H$\alpha$
line a pattern of grains roughly 0\carcsec{}25 in size which were often interconnected into chains or
crinkles. The grains and crinkles of this bright photospheric pattern named \Index{filigree} were mostly
located in the dark \Index{intergranular lane}s.

\begin{figure*}
\centering
\includegraphics*[width=11cm]{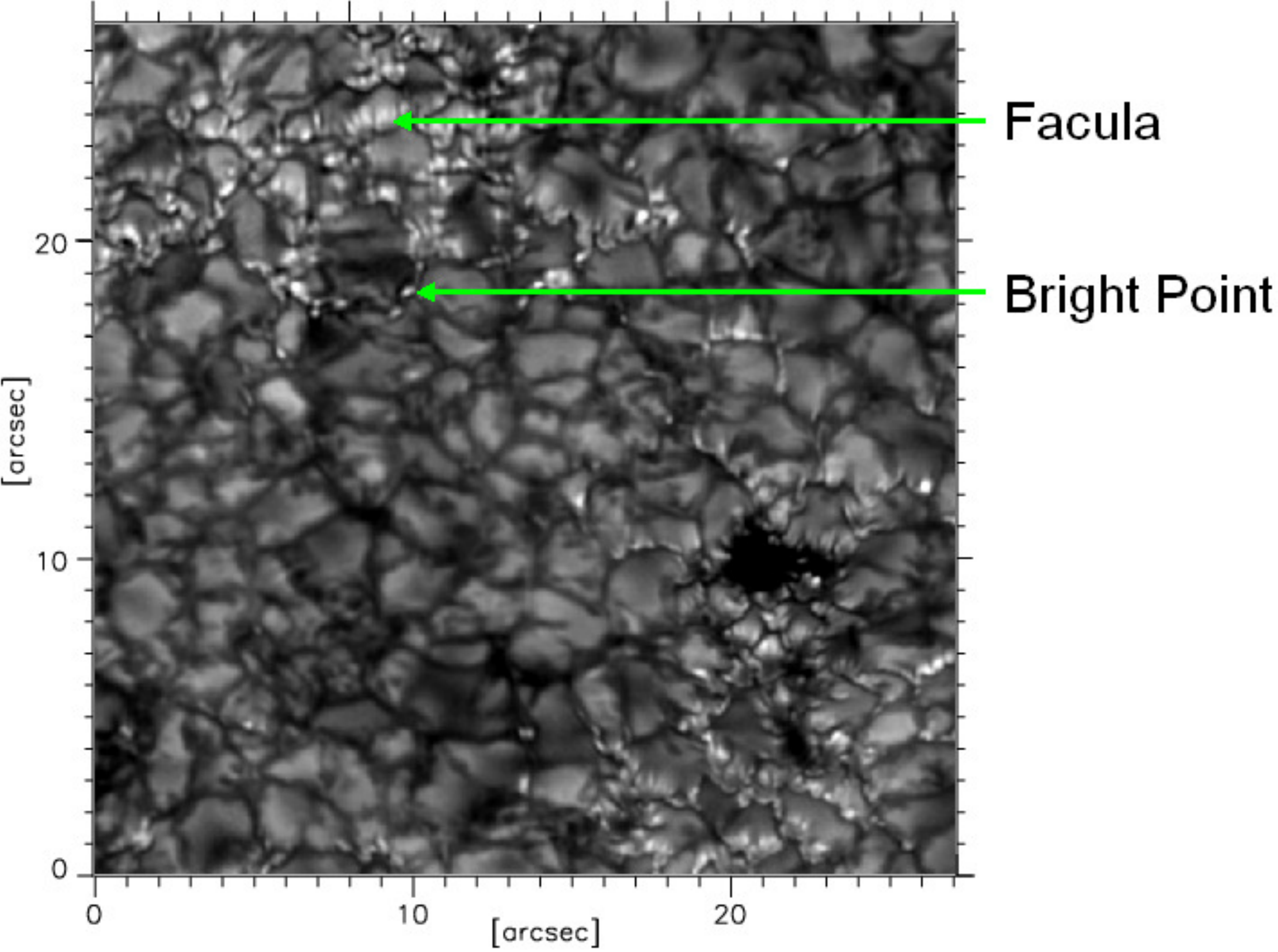}
\caption{G-Band image showing various \Index{faculae} and \Index{bright point}s as recorded at the 1m Swedish Solar Telescope
in September 2004. Heliocentric angle was $\mu=\cos{\theta}=0.9$. The disk center direction is downwards.
Adapted from \citet{Kobel2009}.}
\label{FigFacula}
\end{figure*}

Small bright features that are preferentially visible near the solar limb are called {\bf \Index{faculae}}
(see Fig.~\ref{FigFacula}) and were already observed by \citet{Chevalier1914b}. \citet{tenBruggencate1940}
determined facular sizes of 1\arcsec{}-2\arcsec{} and a mean lifetime of about one hour. Only
\citet{Mehltretter1974} was able to observe disk center faculae in his narrowband Ca\,{\sc ii}~K
images (again with the 60-cm telescope of the Sacramento Peak Observatory). Under exceptionally good
seeing conditions, he could resolve some of the faculae into several \Index{bright point}s (BPs) about
0\carcsec{}25 in size. The BPs were clearly found in the dark \Index{intergranular lane}s.
Simultaneous observations in Ca\,{\sc ii}~K and H$\alpha$ revealed that the BPs and the \Index{filigree}
observed by \citet{Dunn1973} are the same phenomenon. From a comparison between his solar images
and a Kitt Peak \Index{magnetogram}, Mehltretter proposed magnetic flux concentrations in the photosphere
to explain the facular points.

Since the discovery of magnetic fields in sunspots \citep{Hale1908}, magnetic flux concentrations were also
searched for outside spots and pores. Hale determined in his first successful measurements field strength in
the range 200-300~G \citep{Hale1922}. The invention of the photoelectric \Index{magnetograph} \citep{Babcock1953a}
increased significantly the sensitivity of the magnetic field measurements and decreased the lower
limit of the measurements to 1-200~G at a spatial resolution of 5\arcsec{}-20\arcsec{}. In the 1960s
and beginning of the 1970s, the resolution of the magnetic field measurements could be partly improved to
1\arcsec{}, but the results were still contradictory. Field strengths of a few 100~G were found by
\citet{Sheeley1967,Grigorjev1969,Abdusamatov1969}, while \citet{Beckers1968a} and \citet{Simon1974}
reported kilo-Gauss fields. The limited spatial resolution of these measurements made it difficult
to distinguish between strong field features that were much smaller than the resolution element and
weak fields homogeneously distributed over the \Index{resolution element}. Also \Index{dipole}-like fields within the
resolution element could not be detected since opposite polarities in the Stokes~$V$ signal cancel
each other.

The situation changed with the introduction of the \Index{line ratio method}. \citet{Stenflo1973} used the
multi-channel \Index{magnetograph} at the Kitt Peak Observatory to measure Stokes~$I$ and $V$ profiles
simultaneously for the two Fe\,{\sc i} lines at 5247.0~\AA{} ($g=2$) and 5250.2~\AA{} ($g=3$).
Both lines are formed in practically the same manner in the solar atmosphere and in the same
atmospheric height range since they only differ in their effective \Index{Land\'e factor}s but have very
similar properties otherwise (almost identical \Index{oscillator strength}s and \Index{excitation potential}s for
the lower level, equal spin-orbit coupling energies for the lower and upper level), so that
temperature and velocity in the solar atmosphere affect the lines identically. From the
different \Index{Zeeman saturation}s of the two lines, Stenflo could retrieve field strengths, which
were, for the first time, independent of the \Index{spatial resolution} and of an underlying model.
For network elements of the quiet Sun, he determined field strengths around 2000~G and
characteristic sizes of the magnetic features in the range 100-300~km.

The kilo-Gauss fields found by \citet{Stenflo1973} were confirmed several times by other authors.
\citet{Harvey1975} used the very Zeeman sensitive Fe\,{\sc i} line at 15648~\AA{} (see the quadratic
$\lambda$ dependence in Eq.~(\ref{Eq_ZeemanSplitCog})) and found field strengths between 1500~G and 2000~G
for magnetic elements outside active regions. \citet{Wiehr1978} extended Stenflo's \Index{line ratio method}
and used three spectral lines to determine field strengths of 1500-2200~G for Ca~K BPs and H$\alpha$
\Index{faculae}. Improvements in the spatial resolution made it possible to resolve BPs, so that the field
strengths predicted by the \Index{line ratio method} could be confirmed, for the first time by \citet{Keller1992}
with the help of \Index{speckle polarimetry} and later by several authors \citep{Berger2007,Ishikawa2007,Lagg2010}.

\begin{figure*}
\centering
\includegraphics*[width=8cm]{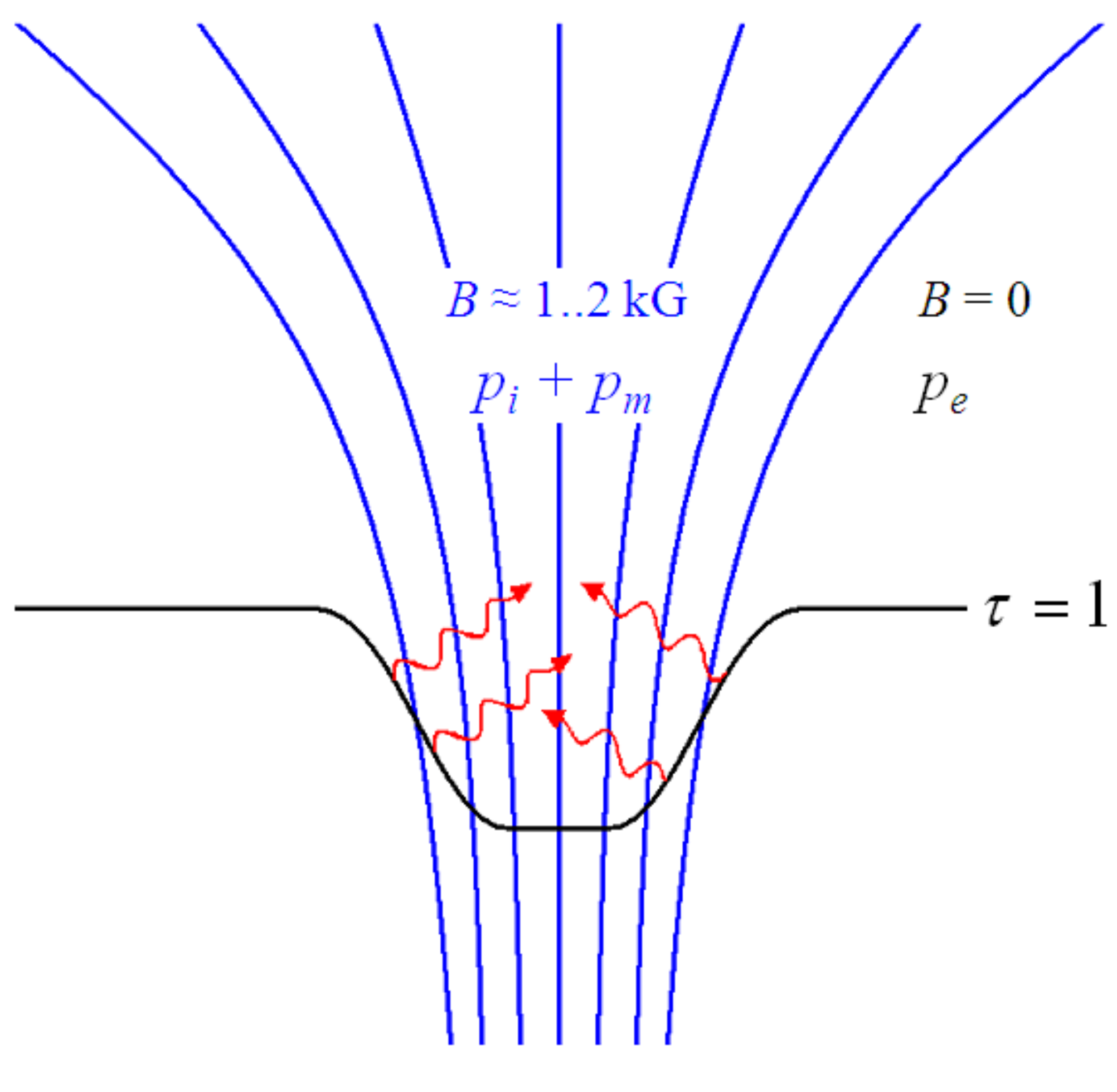}
\caption{Sketch of a \Index{bright point} as an expanding vertical \Index{flux tube}. The magnetic field lines are drawn
in blue color, the optical depth unity surface in black. The wavy red arrows illustrate the lateral
inflow of heat through the walls of the flux tube.}
\label{FigBPCartoon}
\end{figure*}

The new insights gained in the 1970s and 1980s led to a picture of the formation and the structure
of BPs that is largely still valid today: The energy produced by \Index{nuclear fusion} processes in the
core of the Sun is transported to the solar surface by radiation and in the outer zone by convection
(see Fig.~\ref{FigSolarInterior}). The outer part of the convective zone is the place of the
granulation cells, visible as the photospheric granulation pattern outside sunspots and pores.
The \Index{electrical conductivity} of the solar plasma is high so that the convective motions of the granulation
transport the quasi frozen-in magnetic field lines from the inner parts of the granules to their
outer regions, the \Index{intergranular lane}s \citep{Parker1963,Weiss1966,Tao1998}.
This flux expulsion process can concentrate the field in the intergranular lanes, in particular in
the vertices of the granulation cells \citep{Clark1967}, until the magnetic energy density is
roughly equal to the kinetic energy density. The plasma trapped between the field lines cools by
radiative heat losses at the solar surface \citep{Parker1978,GrossmannDoerth1998}. The density
increases and the plasma material flows down. Owing to the low density in higher layers, not enough
material is replenished from above. Additionally, the nearly vertical magnetic field suppresses
horizontal plasma motions. Thus, the intergranular regions of magnetic field are evacuated.
The higher pressure outside these regions compresses the evacuated magnetic regions. This effect
intensifies the inherent inhomogeneities and leads to a clustering of magnetic field lines so that
thin \Index{flux tube}s and elongated \Index{flux sheet}s are formed. Additionally, the interchange instability
can cause a fragmentation of the elongated flux sheets into smaller flux structures. The evacuation
of the flux tubes leads to a depression of the iso-$\tau$ planes, see Fig.~\ref{FigBPCartoon}.
The radiative losses are in balance with the lateral inflow of heat through the walls of the
depression. The inflowing heat makes the flux tubes hot and bright \citep{Spruit1976} so that
they can be observed as photospheric BPs \citep{Deinzer1984}. Near the solar limb, the
\Index{line-of-sight} is inclined relative to the flux tubes, so that the \Index{hot wall}s of the neighboring
granules are observed as \Index{faculae} \citep{Berger2007}. The flux tubes are mainly vertically oriented
due to \Index{magnetic buoyancy} \citep{Schuessler1986}. Because \Index{faculae} are sometimes observed close to
the disk center and BPs can also be detected near the limb, varying flux tube inclinations are
under discussion \citep{Kobel2009}. However, the angle at which the \Index{hot wall}s of a flux tube become
visible depends also on the thickness of the flux tube \citep[see, e.g.,][]{Solanki2006}. The external
gas pressure, $p_{\mathrm{e}}$, is balanced by the pressure inside the flux tube that consists of the
internal gas pressure, $p_{\mathrm{i}}$, and the magnetic pressure:
\begin{equation}\label{Eq_PressureBalance}
p_{\mathrm{e}}=p_{\mathrm{i}}+\frac{B^2}{2\mu_0} .
\end{equation}
Pressure balance\index{pressure balance} (Eq.~\ref{Eq_PressureBalance}) and magnetic flux conservation (Eq.~\ref{Eq_Maxwella2})
cause an expansion of the \Index{flux tube} with decreasing external gas pressure, i.e. with increasing height.

In the quiet Sun, such magnetic elements are concentrated at the borders of the \Index{supergranulation}
cells and form the network. At sufficiently high spatial resolution, magnetic flux elements can also be
observed in the interiors of supergranules where they are named internetwork elements. In active regions,
outside spots and pores, the number density of magnetic elements is considerably higher than in the quiet
Sun because of the higher mean flux density in active regions (see chapter~\ref{Bp2Chapter}). A large
fraction of the \Index{intergranular lane}s is filled with \Index{flux tube}s or sheets\index{flux sheet} and hence \Index{plage} regions are formed.

Since \citet{Muller1984}, photospheric BPs are preferentially observed in Fraunhofer's G~band,
a spectral range between 4295~\AA{} and 4315~\AA{} which is dominated by lines of the CH molecule.
The degree of dissociation of this molecule depends strongly on temperature and hence causes high
contrasts for the G~band BPs \citep{Rutten1999,Steiner2001,Schuessler2003}. The same is true for
other molecular bands, e.g. the violet CN~band at 3880~\AA{}. Magnetograph\index{magnetograph} observations are
typically done in the green and red spectral range. In contrast to this, imaging in the blue G~band
has some advantages: a) shorter wavelength and hence lower \Index{diffraction limit}, b) lower integration
times (broad wavelength band) and hence less degradation due to seeing effects, c) higher cadences,
and most notably d) fewer instrumental requirements. Almost all G~band observations have a higher
spatial resolution than comparable magnetographic observations, so that the field of \Index{proxy-magnetometry}
was established, a technique to identify magnetic features owing to their locally enhanced intensity.
The usefulness of proxy-magnetometry was confirmed by comparisons between simultaneous recordings of
G~band images and \Index{magnetogram}s \citep{Berger2001,Berger2007}. Further confirmation came from MHD simulations
which revealed BPs as radiative signatures of magnetic flux concentrations \citep{Schuessler2003,Shelyag2004}.
The converse argument is not valid since magnetic field concentrations do not always coincide with
an enhanced intensity. Very small magnetic elements can appear dark because their contrast is
smeared over the darker intergranular lanes \citep{Title1996}. Somewhat larger magnetic features
display neutral contrast or are darker than the mean quiet Sun \citep{Spruit1981,GrossmannDoerth1994}.
\citet{Ishikawa2007} studied a \Index{plage} region and reported on extended areas with high magnetic flux
(magnetic islands). BPs are preferentially located near the boundary of such islands.

The long history of BP studies is continued in chapter~\ref{Bp1Chapter} by analyzing quiet-Sun regions
at disk center observed by the balloon-borne 1-m telescope \sunrise{}. Broadband observations
in various spectral ranges of the near UV were achieved simultaneously with full Stokes measurements
of a \Index{magnetograph} operating in the visible spectral range. For the first time, BPs could be spatially
resolved for wavelengths below 350~nm. Brightness, velocity, and polarization degree of a few hundred
BPs are analyzed. \sunrise{} observations are compared with three-dimensional state-of-the-art MHD
simulations of several mean flux densities in chapter~\ref{Bp2Chapter}. Synthetic Stokes profiles are
calculated from the MHD data by solving the \Index{Unno-Rachkovsky equations} numerically, including for the first
time for the OH~band at 312~nm. While the comparison between observation and simulation of existing
studies mainly concentrated on intensity histograms, chapter~\ref{Bp2Chapter} considers many more observational
quantities (intensity at multiple wavelengths, LOS velocity, spectral line width, and polarization degree).

\chapter{Instrumentation}\label{Instrumentation}

The core of this thesis are the five studies that constitute the chapters~\ref{Ud1Chapter}-\ref{Bp2Chapter}.
The used observational data were recorded with three different telescopes: Observations with the ground-based
{\bf S}wedish-{\bf S}olar-{\bf T}elescope (SST) are analyzed in chapter~\ref{Ud1Chapter}.
Chapters~\ref{Ud2Chapter} and \ref{Ud3Chapter} utilize recordings of the space-borne Solar Optical Telescope (SOT)
onboard the \hinode{} satellite. Data of the balloon-borne observatory \sunrise{} are studied in the
chapters~\ref{Bp1Chapter} and \ref{Bp2Chapter}.

This chapter describes the used telescopes, optical setups, and instrumentation in more detail than possible
in a scientific publication. I spent a large part of my Ph.D. time with technical work for the \sunrise{}
observatory. Some examples are the entire software development of the \sunrise{} Filter Imager (SuFI), the
reduction of the SuFI data after the flight, the development of fundamental parts of the software for the
\Index{Instrument Control Unit} (\Index{ICU}), and the conceptual design of the \Index{Data Storage Subsystem} (\Index{DSS}). For this reason,
in the following the \sunrise{} topic is considerably more pronounced than the SST and SOT topics.

\section{The Swedish Solar Telescope}

\begin{figure*}
\centering
\includegraphics*[width=12cm]{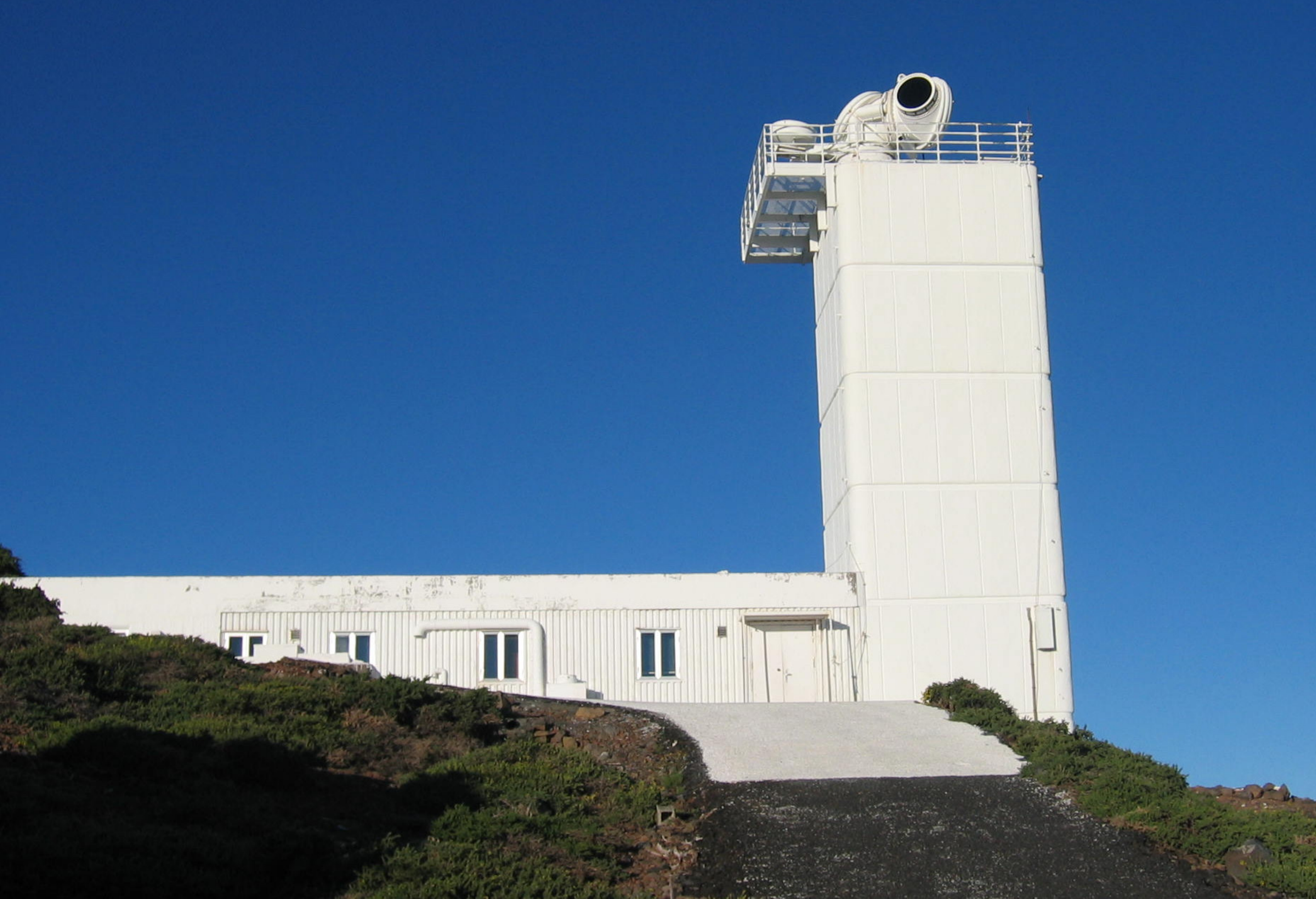}
\caption{The \Index{Swedish Solar Telescope} on the Canary Island La Palma.}
\label{FigSST}
\end{figure*}

The \Index{SST} is a ground-based telescope of the 1-m class \citep{Scharmer2002,Scharmer2003a}. It is
located on the Canary Island La Palma (Spain), close to the peak of the Roque de los Muchachos
(see Fig.~\ref{FigSST}). The low temperature variations of the Atlantic Ocean and the low air
pressure at an altitude of 2360~m above sea level make this one of the most suitable places
for solar observations worldwide. The telescope is run by the Institute for Solar Physics
of the Royal Swedish Academy of Sciences in Stockholm. With a clear aperture of 96~cm, the SST
was put into operation on 2002 May 21 and replaced the 47.5~cm \Index{Swedish Vacuum Solar Telescope} (\Index{SVST})
that was operated at the same location from 1985 to 2000 \citep{Scharmer1985}.

Most of the modern solar telescopes are reflectors having an on-axis design (e.g. the \sunrise{}
telescope described in section~\ref{SunriseInstr}), i.e. the primary mirror has a central hole
through which the light beam passes. The hole as well as the spiders needed to mount the secondary
mirror cause an obscuration which lowers the contrast of the images significantly. This constructional
shortcoming was avoided in the SST design by using a 1~m primary lens. The lens is also used as the
entrance window to the \Index{vacuum tower} in which the sunlight is re-directed with the help of two
plane mirrors to the optics laboratory located in the basement of the telescope building. The vacuum
prevents the deformation of the wavefront by air turbulence inside the tower. The SST has an
alt-azimuth mount. Chromatic aberrations caused by the primary lens are compensated by a
\Index{Schupmann corrector} \citep{Schupmann1899}. The focal length of the telescope is 20.3~m (determined
at a wavelength of 460~nm).

\begin{figure*}
\centering
\includegraphics*[width=\textwidth]{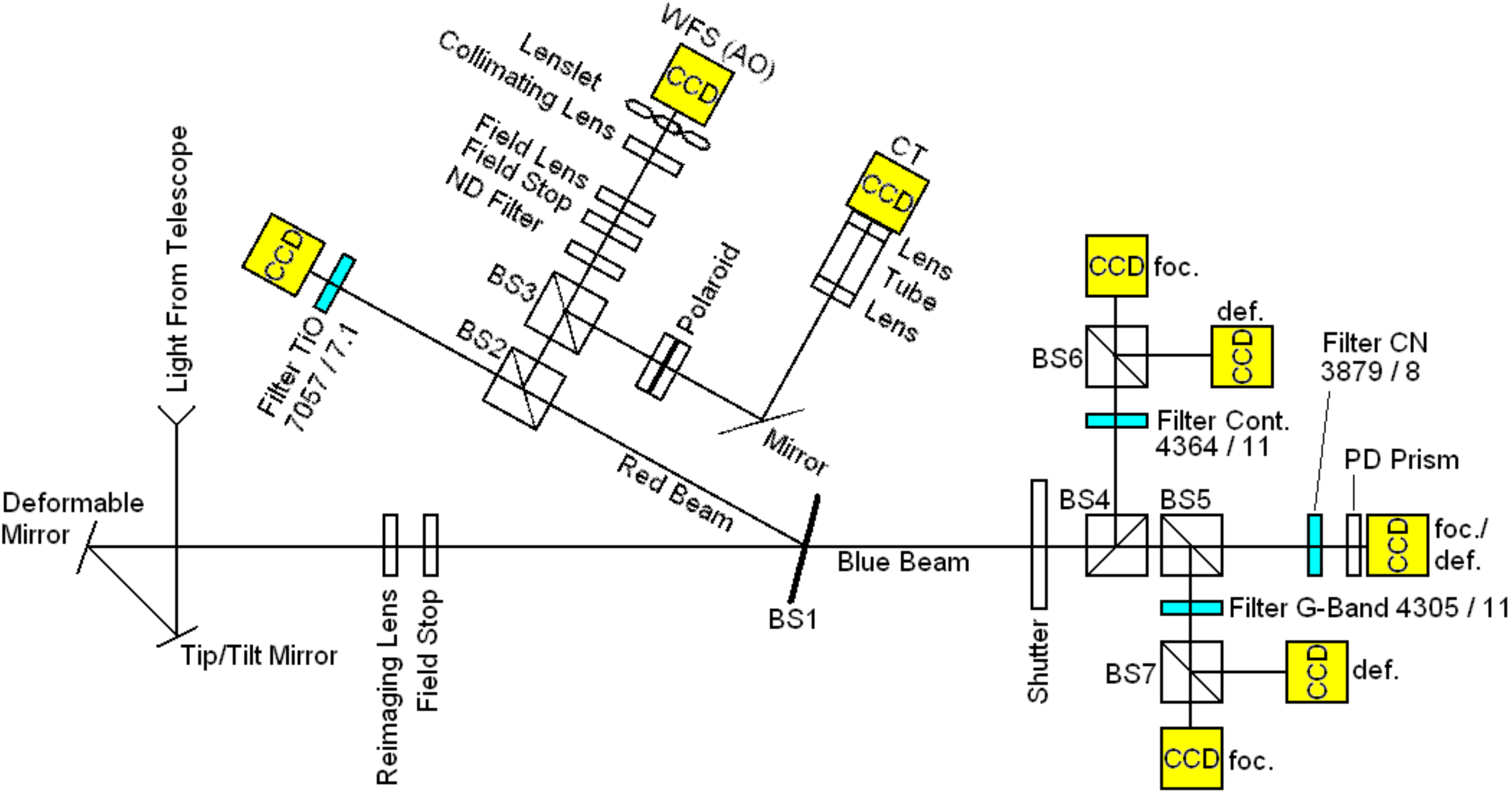}
\caption{Optical setup used on September 7, 2004 for imaging in the titanium oxide band head with the
Swedish Solar Telescope.}
\label{FigSetupSst}
\end{figure*}

For the study in chapter~\ref{Ud1Chapter}, the setup shown in Fig.~\ref{FigSetupSst} was used
by Vasily Zakharov for observations of the active region NOAA~10667 on 2004 September 7. The light
leaves the vacuum tank in the optics laboratory, is reflected by the tip/tilt and deformable mirror,
passes the reimaging lens and the field stop, and hits the first beam-splitter BS1. Wavelengths shorter
than 500~nm can pass the beam-splitter (blue beam) but longer ones are reflected (red beam). 90\% of
the red light can pass a second beam-splitter BS2 and hit the TiO 7057~\AA{} interference filter
having a width of 7.1~\AA{}. The following Kodak Megaplus CCD camera (9~$\mu$m pixel size) acquired
the data of chapter~\ref{Ud1Chapter}. The remaining 10\% of the red light are reflected by BS2 to
the \Index{adaptive optics} (AO) of the SST \citep{Scharmer2003b}. BS3 divides the beam once more between
\Index{correlation tracker} camera and \Index{wavefront sensor} camera. The correlation tracker controls the two
axes of the \Index{tip/tilt mirror} and hence compensates for the rotation of the Sun around its own poles
as well as for smaller image shifts due to turbulence in the terrestrial atmosphere, so that the
observed solar object is at a fixed position within the field of view.

\begin{figure*}
\centering
\includegraphics*[width=10cm]{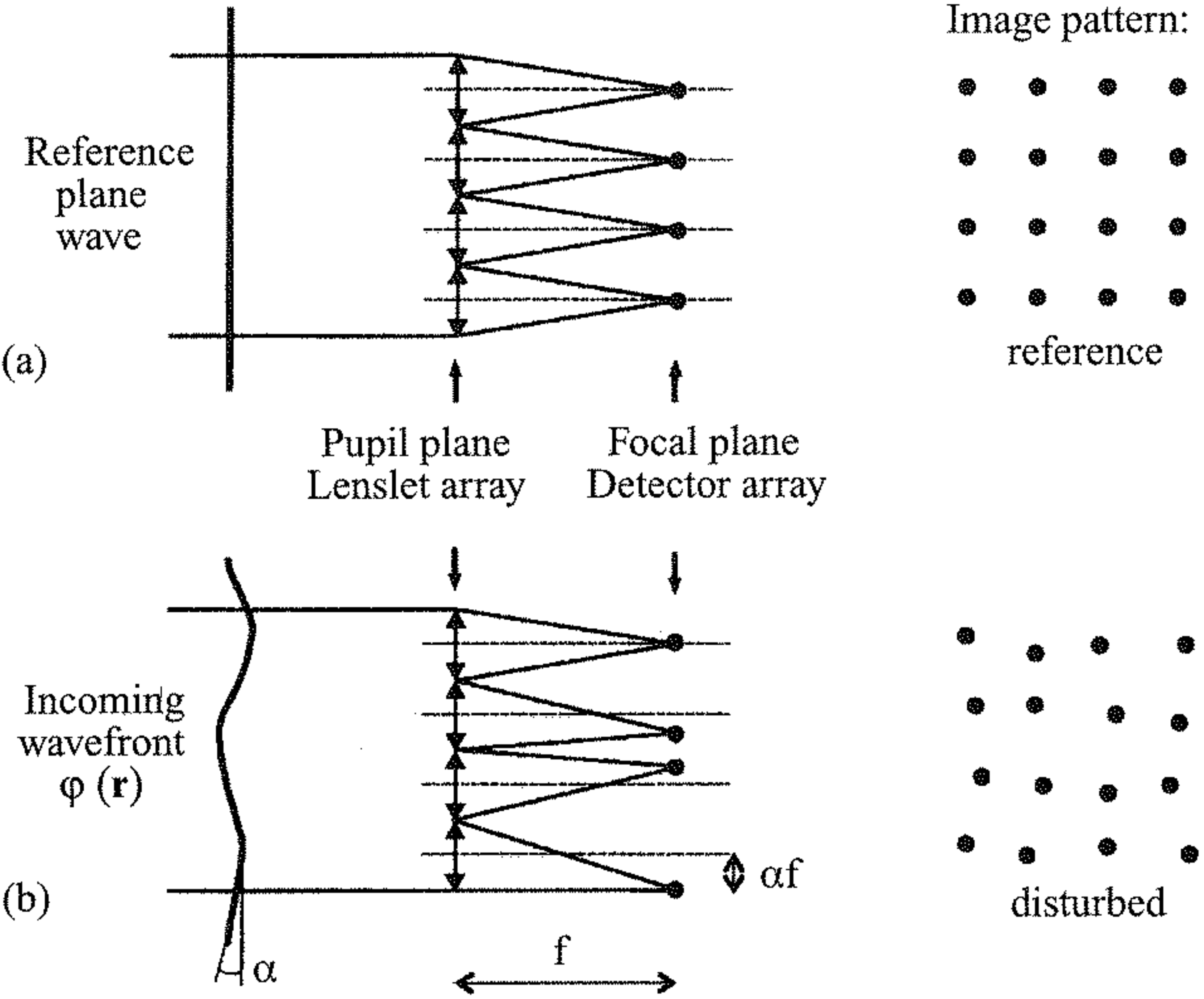}
\caption{Principle of the \Index{Shack-Hartmann wavefront sensor}. The example of a plane wave is given in panel (a),
a disturbed incoming wavefront in panel (b). From \citet{Roddier1999}.}
\label{FigShackHartmann}
\end{figure*}

The \Index{Shack-Hartmann wavefront sensor} \citep{Hartmann1900,Shack1971} provides the control signals
for the \Index{deformable mirror} which compensates for higher order aberrations. Fig.~\ref{FigShackHartmann}
demonstrates the main principle. The heart of the Shack-Hartmann \Index{wavefront sensor} at the SST is
a hexagonal microlens array (called lenslet array) with 37 elements placed in a conjugate pupil plane.
Each microlens forms an image of the source. The shifts of the images among each other are determined
via correlation functions and yield the gradients of the \Index{wavefront deformation}s. The deformable mirror
is also placed in a conjugate pupil plane and its 37 electrodes are now controlled such that the
wavefront deformations are compensated as well as possible.

The five CCD cameras of the blue beam are all synchronized by an external shutter (for exactly simultaneous
observations). A phase diversity configuration (see section~\ref{PD_prism}) in each case records data for the
violet CN~band, the G~band, and the blue continuum. The data of the blue beam are analyzed by
\citet{Zakharov2006} and \citet{Kobel2009}, but are ignored in this thesis. Therefore we skip details.

As a final remark it is mentioned, that the active region NOAA~10667 investigated in chapter~\ref{Ud1Chapter}
was also observed at the Italian-French \themis{} observatory. Spectropolarimetric observations were done
simultaneously in the six wavelength ranges 5139-5145~\AA{}, 5197-5203~\AA{}, 5247-5253~\AA{}, 5872-5879~\AA{},
6705-6709~\AA{}, and 7053-7059~\AA{} and first results are published in \citet{Arnaud2006} and \citet{Wenzel2010}.
In spite of a similar aperture of 90~cm, the effectively reached spatial resolution of the \themis{} data
is significantly lower than for the SST data due to the higher \Index{optical performance} of the SST.

\section{The spectropolarimeter aboard the \hinode{} satellite}

\begin{figure*}
\centering
\includegraphics*[width=12cm]{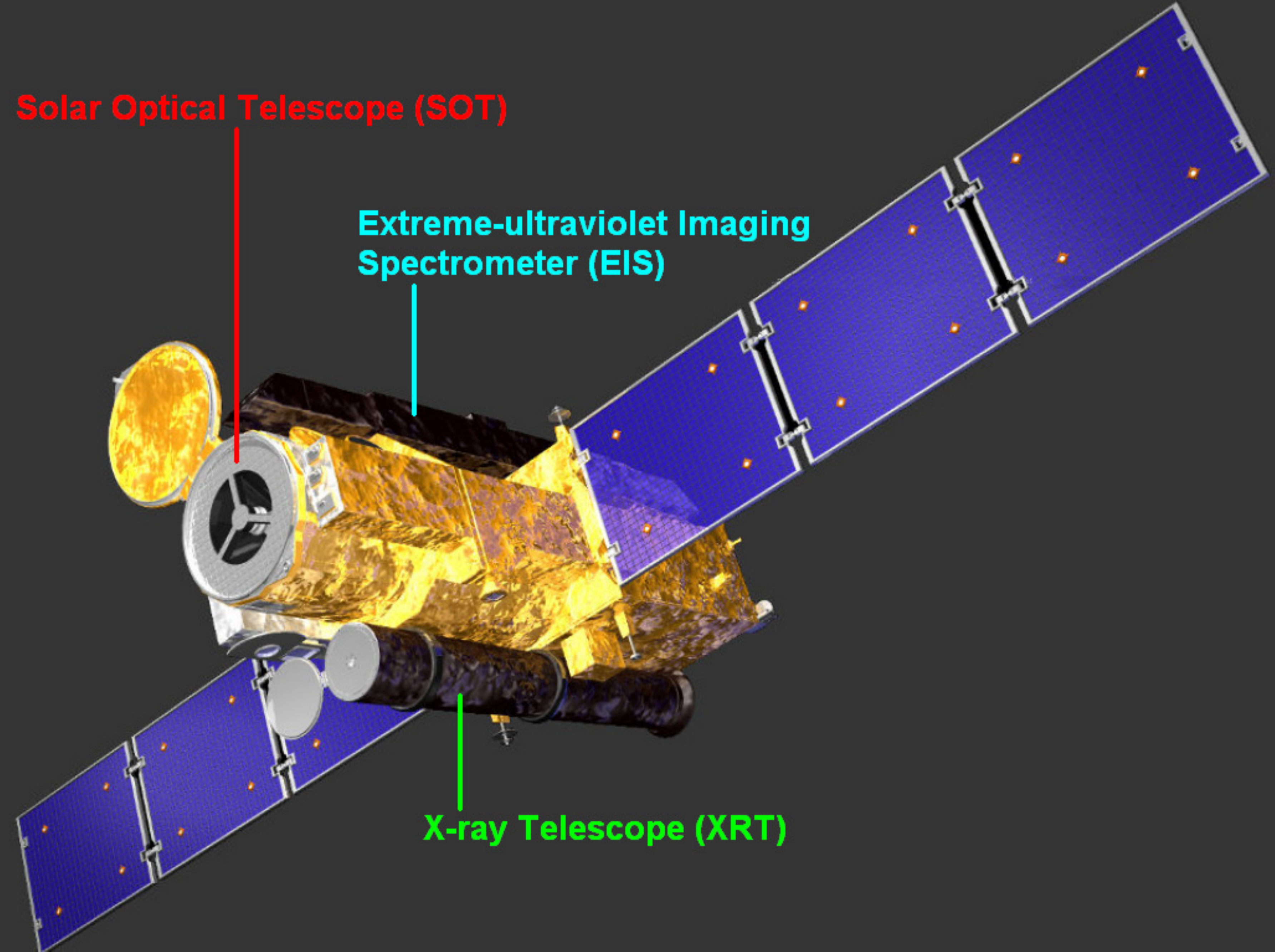}
\caption{The \hinode{} satellite carrying the \Index{Solar Optical Telescope}. Adapted from \citet{Shimizu2008}.}
\label{FigHinode}
\end{figure*}

The \hinode{} satellite (former name: \Index{Solar-B}) is a space-borne solar observatory which was developed
under the leadership of the Japanese space agency JAXA and is operated in cooperation with
American and European institutes \citep{Kosugi2007}. \hinode{} was launched on 2006 September 23 at
06:36 (JST) from the Kagoshima space center with a Japanese M-V rocket. At an altitude of about 600~km,
the \hinode{} satellite orbits the Earth and observes the Sun with three telescopes, see Fig.\ref{FigHinode}.
Corona and transition region are observed by the X-ray telescope (XRT) and the EUV imaging spectrometer
(EIS). Photosphere and chromosphere are the target of the \Index{Solar Optical Telescope} (\Index{SOT}).

\begin{figure*}
\centering
\includegraphics*[width=\textwidth]{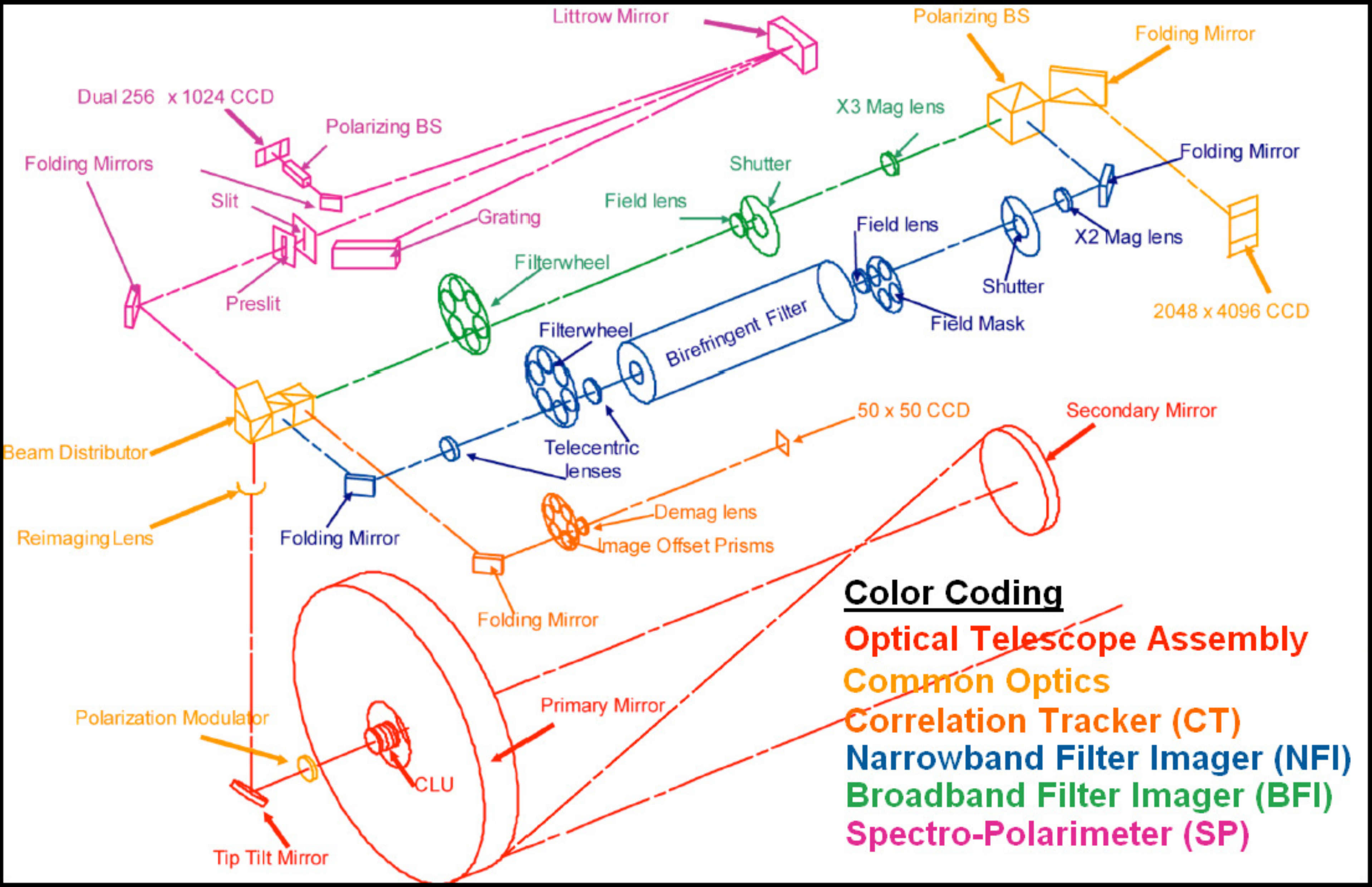}
\caption{Optical schematic of the \Index{Solar Optical Telescope} (\Index{SOT}) onboard the \hinode{} satellite and
its focal plane instrumentation. From \citet{Hurlburt2009}.}
\label{FigSOT}
\end{figure*}

The SOT is a Gregory-type on-axis telescope \citep{Suematsu2008,Tsuneta2008} whose design is schematically
drawn in Fig.~\ref{FigSOT}. The aperture of the SOT is 50~cm, the effective focal length is 4527~mm.
The sunlight is reflected by the primary and secondary mirror, passes the hole in the primary mirror,
and hits the Collimator Lens Unit (CLU). The collimated beam passes the Polarization Modulator Unit (PMU,
a \Index{retarder} plate rotating with 5/8~Hz), is reflected by the \Index{tip/tilt mirror}, and enters the beam
distributor which then distributes the light to the Correlation Tracker\index{correlation tracker} \citep[CT,][]{ShimizuEtAl2008}
and to the three scientific instruments, \Index{NFI}, \Index{BFI}, and \Index{SP}. While BFI (Broadband\footnote{In solar physics,
a filter is called to be broadband, if its pass band is much broader than the width of a spectral line.} Filter
Imager) records purely photometric images for five photospheric (3883, 4305, 4504, 5550, and 6684~\AA{})
and one chromospheric (3968~\AA{}) wavelength range, the NFI (Narrowband\footnote{The width of a narrowband
filter is smaller than the width of a spectral line.} Filter Imager) and SP (Spectropolarimeter) are two
vector \Index{magnetograph}s. NFI allows the observation of two-dimensional Stokes images with a high cadence but
only moderate spectral resolution of 90~m\AA{} (at 6300~\AA{}) with the help of a tunable \Index{Lyot filter}.
For photospheric \Index{magnetogram}s, the Fe\,{\sc i} lines at 5247.1, 5250.2, 5250.6, 6301.5, 6302.5~\AA{} or
the Ti\,{\sc i} 6303.8~\AA{} line can be scanned. Chromospheric \Index{magnetogram}s and \Index{Dopplergram}s can be
retrieved from scans of the Na\,{\sc i} 5896~\AA{} or Mg\,{\sc i} 5172.7~\AA{} line, photospheric Dopplergrams
from the Fe\,{\sc i} 5576~\AA{} line, and finally, the chromospheric H$\alpha$ 6563~\AA{} line can be scanned.
Since chapters~\ref{Ud2Chapter} and \ref{Ud3Chapter} only uses data of the spectropolarimeter, no further details
about the broad- and narrowband imagers are given here. The interested reader is referred to \citet{Hurlburt2009}.

The \Index{SP} is a slit spectrograph in Littrow configuration for the wavelength range 6300.8-6303.2~\AA{}
with a high spectral resolution of 30~m\AA{} at a low cadence \citep{Lites2001}.
An Echelle grating having 79 grooves/mm is used as the dispersive element and is operated in the 36th
order. A polarizing beam-splitter serves as \Index{analyzer} and creates two orthogonal linearly polarized
states, each of them imaged at one half of the CCD. The \Index{dual-beam configuration} allows a considerable
reduction of \Index{cross talk}s which can occur by residual motions or vibrations of the satellite
\citep{Lites1987,Skumanich1997}. The CCD camera is continuously exposed and read out
synchronously with the PMU: 16 spectra are recorded per PMU revolution. A prototype of the SP
was installed at the \Index{Dunn Solar Telescope} as the \Index{Diffraction Limited Spectro Polarimeter} (\Index{DLSP})
\citep{Sigwarth2001}.

The slit of the \Index{SP} is 0\carcsec{}16 wide and 164\arcsec{} long and is oriented parallel to the
solar North-South direction. It can be moved in the range of $\pm 164\arcsec{}$ along the solar
East-West direction with the help of a tiltable mirror, so that the solar surface can be scanned
in a field of view of maximal $164\arcsec{} \times 328\arcsec{}$. SP has four operating modes:
a normal map mode, a fast map mode, a dynamics mode, and a deep \Index{magnetogram} mode. Only data of the
normal map mode are studied in chapter~\ref{Ud2Chapter}, where the integration time per slit position
is 4.8~s (3 revolutions of the modulator) and a polarimetric accuracy of $10^{-3}$ can be reached.
The spatial sampling in the normal map mode is 0\carcsec{}16 $\times$ 0\carcsec{}16 and a 160\arcsec{}
wide map can be scanned in 83~min. The characteristics of the other three modes can be found in
\citet{Hurlburt2009}.

Compared to an imaging \Index{magnetograph} like \Index{NFI}, the big advantage of a slit spectrograph like \Index{SP} is its
high spectral resolution/sampling, so that a large number of wavelength points per spectral line is
achieved. As demonstrated in chapters~\ref{Ud2Chapter} and \ref{Ud3Chapter}, this allows not only
inversions that retrieve averaged physical quantities of the solar atmosphere, but also height-dependent
inversions that can retrieve stratifications of the atmospheric quantities.

\section{The balloon-borne observatory \sunrise{}}\label{SunriseInstr}

\sunrise{} is a balloon-borne solar observatory developed under the leadership of the
Max-Planck-Institut f\"ur Sonnensystemforschung (MPS) in Lindau/Eichsfeld in cooperation with the
Kiepenheuer-Institut f\"ur Sonnenphysik (KIS) in Freiburg/Breisgau, the High Altitude Observatory (HAO)
in Boulder/Colorado, the Instituto de Astrof\'{\i}sica de Canarias (IAC) in La Laguna/Tenerife,
the Instituto de Astrof\'{\i}sica de Andaluc\'{\i}a (IAA) in Granada, the Instituto Nacional
de T\'ecnica Aerospacial (INTA) in Madrid, the Grupo de Astronom\'{\i}a y Ciencias del Espacio (GACE)
in Valencia, and the Lockheed Martin Solar and Astrophysics Labotory (LMSAL) in Palo Alto/California.

\begin{figure*}
\centering
\includegraphics*[width=\textwidth]{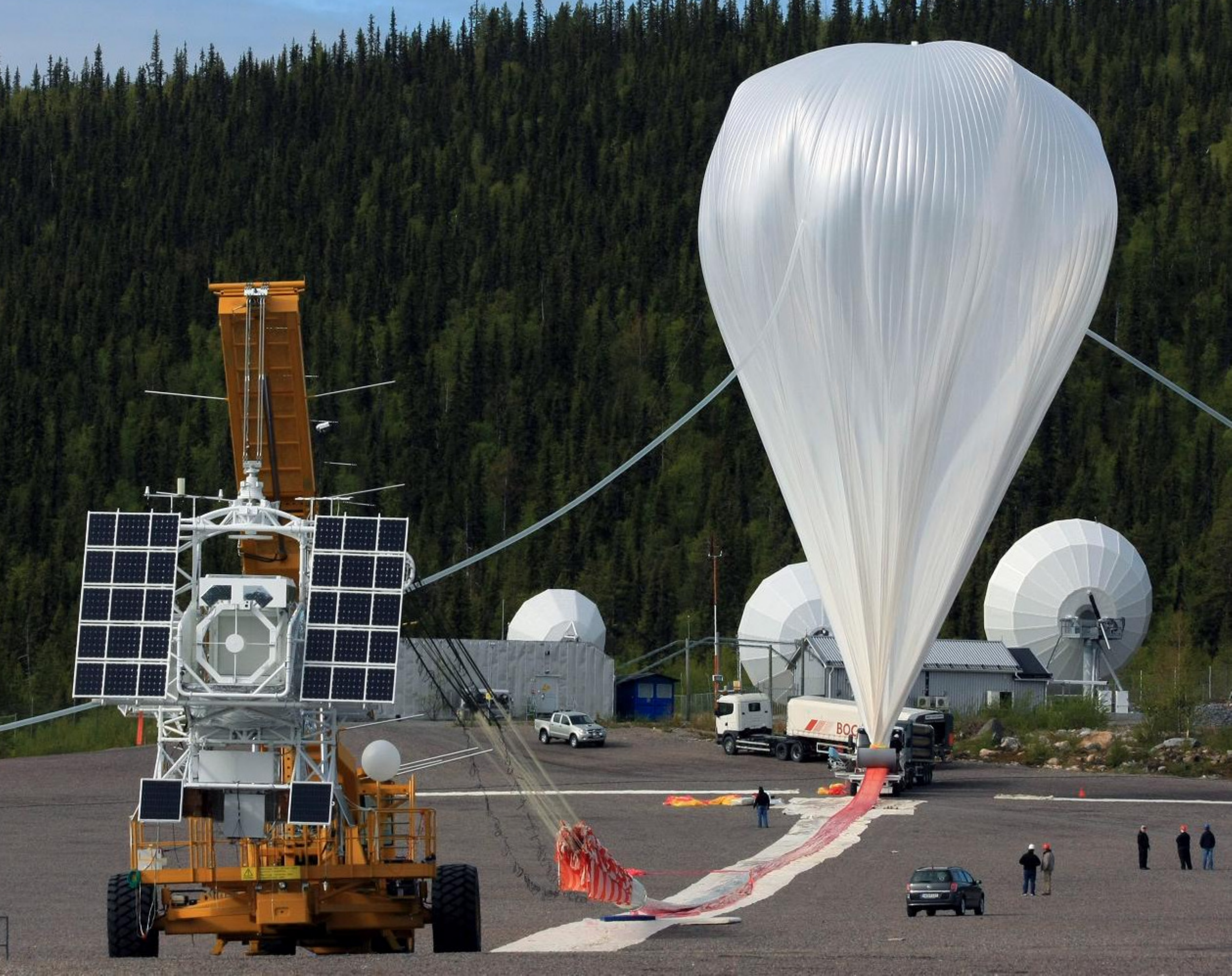}
\caption{The \sunrise{} observatory a few minutes before the launch of the first science flight on 2009 June 8.}
\label{FigSunriseBeforeLaunch}
\end{figure*}

The first scientific flight took place in June 2009 from Kiruna in Northern Sweden to Somerset Island
in Northern Canada (Fig.~\ref{FigSunriseBeforeLaunch}). The distance of more than 4000~km was covered
in 137 hours. The flight altitude of the balloon showed cyclical variations between day and night in the
range 34-37~km. At a mean northern latitude of $\mathrm{71^\circ}$, the Sun was always above the horizon
and could be observed 24 hours a day. Stratospheric balloon observations have two more advantages.
99\% of the air mass are below the balloon, so that disturbing seeing effects known from ground-based
observations practically do not exist. In addition, also wavelengths below 350~nm can be observed which
is not possible from ground because of the absorption of the \Index{UV} radiation in the terrestrial \Index{ozone layer}.
\sunrise{} was therefore designed for observations in the visible and near UV spectral range from the
very first. The Post Focus Instrumentation (PFI) contains two scientific instruments: \Index{SuFI}, a filter
imager for the near UV and \Index{IMaX}, an imaging \Index{magnetograph} for the Fe\,{\sc i} 5250.2~\AA{} line in
the visible (see below).

A complete description of \sunrise{} is not possible within the limits of this thesis. I only concentrate on
those topics that were part of my work at MPS in the recent years. Extensive information about technical
aspects of \sunrise{} can be found in \citet{Barthol2011} and the following \sunrise{} articles of same
special issue of Solar Physics \citep{Berkefeld2011,Gandorfer2011,MartinezPillet2011}. First scientific
results of the 2009 flight were published as special issue of The Astrophysical Journal Letters, amongst
others the study reproduced in this thesis as chapter~\ref{Bp1Chapter}. A good overview is given by the
first article of the special issue, \citet{Solanki2010}.

\subsection{Telescope}

\begin{figure*}
\centering
\includegraphics*[width=12cm]{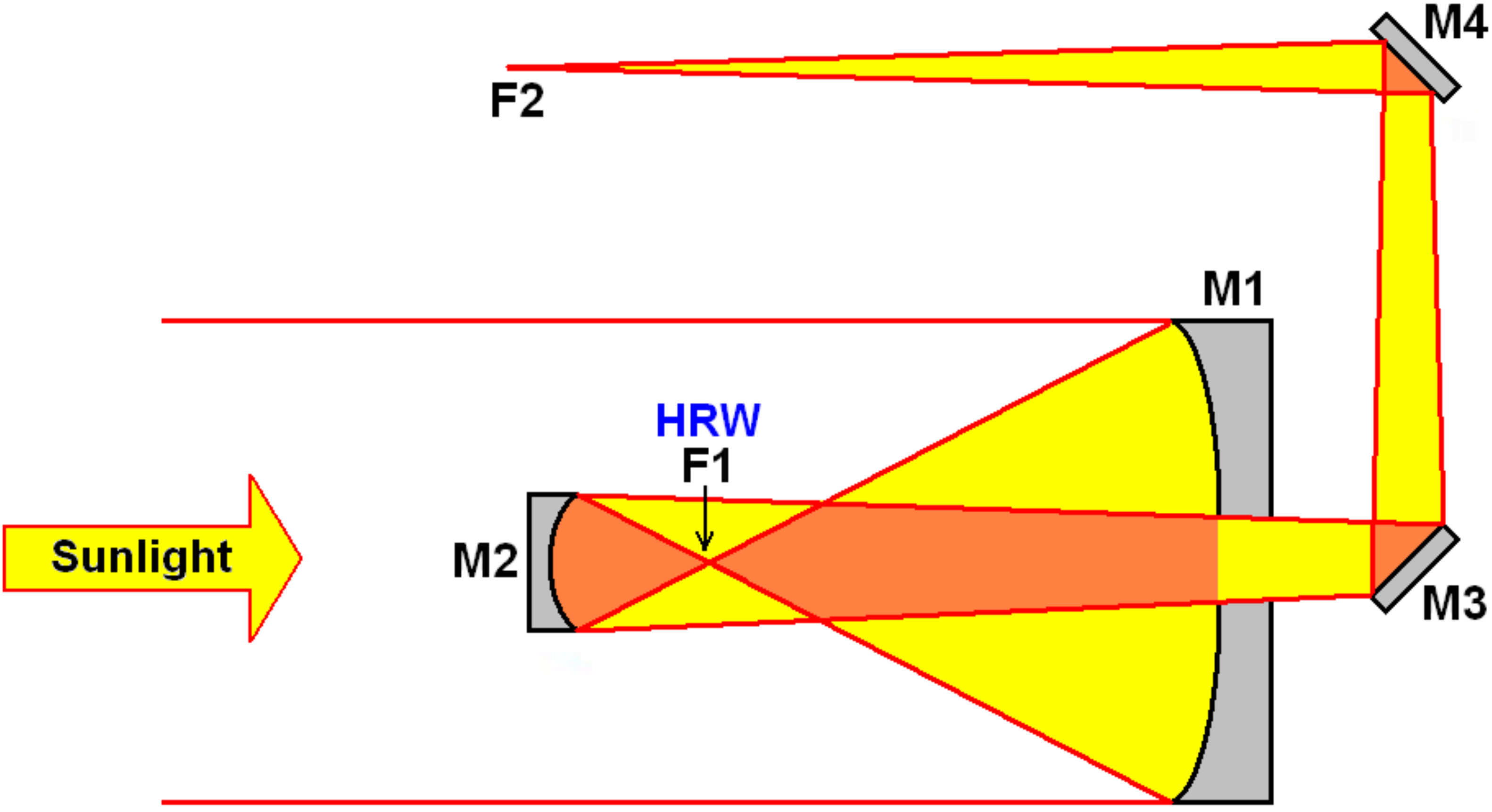}
\caption{Optical configuration of the Gregory telescope of \sunrise{}.}
\label{FigSunriseOpticalLayout}
\end{figure*}

Similar to \hinode{}/SOT, \sunrise{} is an on-axis Gregory-type telescope, whose optical configuration
can be seen in Fig.~\ref{FigSunriseOpticalLayout}. The parabolic primary mirror (M1) has a focal length of
2.42~m and a diameter of 1~m and makes \sunrise{} the largest solar observatory that ever left Earth's surface.
A \Index{Heat Rejection Wedge} (\Index{HRW}) is placed in the primary focus (F1) and reflects 99\% of the unwanted heat load
into cold space. Only the light passing a 2.8-mm wide hole in the HRW can hit the 24.5~cm elliptical secondary
mirror (M2). The effective focal length of the Gregory telescope is 24.2~m. The two plane mirrors M3 and M4
direct the light into the PFI placed on top of the telescope. The mirrors M2, M3, and M4 can be moved via
mechanisms, so that the telescope can be readjusted in flight, in particular it can be refocussed.

\subsection{Light distribution and image stabilization}
The {\bf I}mage {\bf S}tabilization and {\bf Li}ght {\bf D}istribution (\Index{ISLiD}) unit distributes
the light with the help of dichroic beam-splitters, according to wavelength, to the
two scientific instruments \Index{SuFI} (200-400~nm) and \Index{IMaX} (525~nm) as well as to the {\bf C}orrelating
{\bf W}avefront {\bf S}ensor (\Index{CWS}, 500~nm), so that the three camera-based instruments can be
operated simultaneously. ISLiD has to preserve the diffraction-limited performance and the polarization
information of the light (only important for IMaX). A magnifying optical system reimages the secondary
focus (F2) onto SuFI science focus, so that the effective focal length of SuFI is 121~m.

A motor-driven filter wheel placed in the secondary focus has a closed position for taking
dark current measurements, a pinhole position for checking the alignment between the three
camera-based systems and for internal calibrations of the CWS, and an open field position for
the normal solar observations. The \Index{CWS} is a six-element \Index{Shack-Hartmann wavefront sensor}, which
compensates for image motions in the range 1-30~Hz by controlling a fast \Index{tip/tilt mirror}
($\pm$ 46\arcsec{} range) and corrects for \Index{defocus} and \Index{coma} by moving the three axes of M2.
More details about the \Index{CWS} can be found in \citet{Berkefeld2011}, details about \Index{ISLiD} in
\citet{Gandorfer2011}.

\subsection{Software architecture}\label{SoftwareArchitecture}

\begin{figure*}
\centering
\includegraphics*[width=\textwidth]{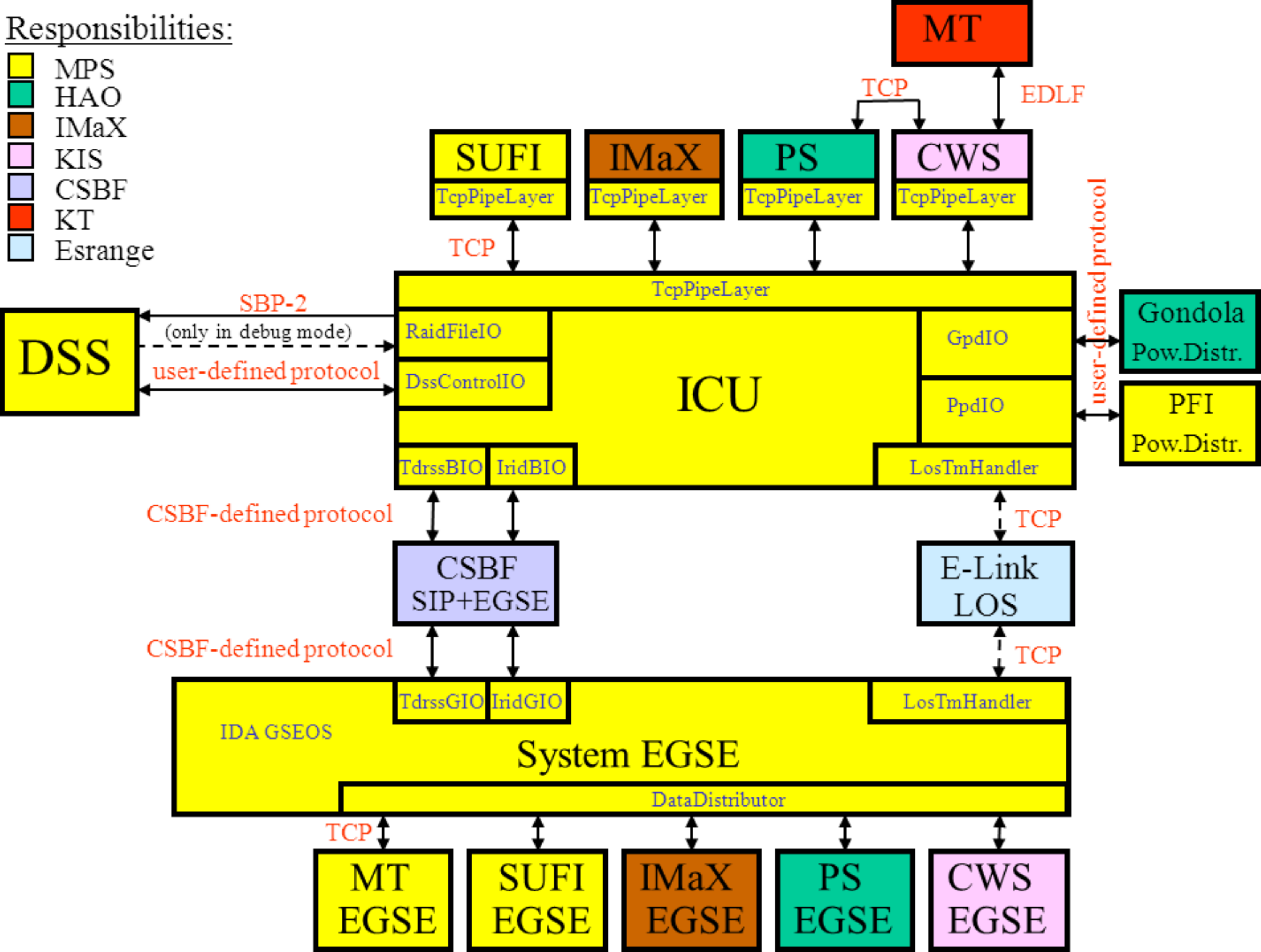}
\caption{Flight configuration of the \sunrise{} software architecture. From \citet{Riethmueller2006a}.}
\label{FigSwArchitecture}
\end{figure*}

The challenge of designing the hardware and software for the \sunrise{} observatory is posed by
the science requirements \citep{Schuessler2004} which had to be reconciled with the technical feasibilities,
in particular with respect to mass, power consumption, and telemetry bandwidth. A substantial design
driver was the fact, that the data rate provided by IMaX and SuFI in the observing modes (5-7~MBytes/s)
are orders of magnitudes higher than the available telemetry downlink rate of 6~kBit/s (gross).
The following conceptual design was chosen: Every camera-based instrument is equipped with an embedded PC
acquiring and pre-processing the camera data as well as controlling the mechanisms. Via a 100~MBit Ethernet,
each instrumental computer is connected with the {\bf I}nstrument {\bf C}ontrol {\bf U}nit\index{Instrument Control Unit} (\Index{ICU}), a central
computer also located in the gondola and connected to the ground station via telemetry channels, see
Fig.~\ref{FigSwArchitecture}. Each instrument transfers its image and \Index{housekeeping data} to the ICU,
which stores all data on a central {\bf D}ata {\bf S}torage {\bf S}ubsystem\index{Data Storage Subsystem} (\Index{DSS}) and, if telemetry bandwidth
is available, sends important parts of the data to the ground station.

The ground station receives the telemetry data centrally via the System \Index{EGSE} ({\bf E}lectrical {\bf G}round
{\bf S}upport {\bf E}quipment). The most important software component of the System EGSE is
a configurable ground support software named \Index{GSEOS}, which was developed by the {\bf I}nstitut f\"ur
{\bf Da}tentechnik und Kommunikationsnetze (IDA) in Braunschweig \citep{Reiche2009}. System relevant
housekeeping values are displayed on various screens and a color-code indicates out-of-range values.
The System EGSE also provides the possibility of transferring tele-commands from ground to the ICU.
Instrument-specific housekeeping displays and commanding capabilities are provided by separate Instrument
EGSEs which are connected to the \Index{DataDistributor} software via a TCP socket interface \citep{Riethmueller2006b}. 
The DataDistributor was developed by MPS and establishes a data connection to the GSEOS running on the
System EGSE \citep{Riethmueller2006c}. Since GSEOS is specialized in processing and displaying housekeeping
data but is not optimized for the processing of large amounts of image data, an additional \Index{Qt}\footnote{Qt is
a platform-independent C++ class library for the development of graphical user interfaces.}-based
SuFI EGSE software can be connected to the DataDistributor which is aimed at the analysis and processing
of solar images.

The most time-critical parts of the \Index{ICU} software due to high data rates are the receiving of image data
send by the instruments, the \Index{RAID} level~5 coding in software, and the data saving on the \Index{DSS}. Additionally,
the ICU has to perform various I/O tasks having uncritical data rates, e.g. sending of telemetry data,
receiving and execution of tele-commands, operating the gondola and PFI power distribution units,
acquisition of housekeeping data via the \Index{Sensor Interface Board} \citep{vonderWall2005}, etc.
The execution of one service must not block the software and hence inhibit the completion of other services.
At the same time the software has to be maintainable and upgradeable. Such requirements can only be
realized by a concurrent programming model.

Classical operating systems like Microsoft Windows or Linux provide concurrency in the form of threads
and processes. In the case of the ICU software, the telemetry services were put in separate
processes in order to decouple the telemetry processing from other software parts as completely as possible,
so that a possible fatal software error does not fully block the tele-commanding capabilities of the entire
system. The remaining tasks are distributed over several threads of a single process in order to guarantee
a maximal data exchange rate between the concurrent branches of the software.

A distributed multi-threaded software is error-prone and difficult to debug and hence a challenge even
for software engineers with many years of experience (think of \Index{deadlock}s, \Index{race condition}s,
\Index{thread synchronization}, etc.). Nevertheless, a notedly reliable software had to be developed which can run
without errors or interruptions for at least six days. The following software design tries to fulfill these
requirements:

1) From the software's point of view the interface between instrument and ICU is placed inside the
instrument software and not, as usual, at the Ethernet hardware. Since a symmetrical programming style for
the data sending and receiving software parts enables the lower transport layers of the software
(named \Index{TcpPipeLayer} in Fig.~\ref{FigSwArchitecture}) to be identical for the ICU and the instruments,
the know-how developed at MPS could be made accessible to the partner institutes in the form of
software libraries. This reduced their development effort and limited significantly the effort needed
to optimize the performance of the data transfer, because an instrument simulator provided as part
of the MPS software package could be used as benchmark and as application example of the libraries. 

Taking these advantages brought up two difficulties. On the one hand, the instruments are operated with
different operating system families and hardware architectures\footnote{Often the choice of the operating
system is not entirely free because, e.g., the camera driver is only available for a certain operating system.}.
This requires a careful treatment of byte order issues when transferring multi-byte numbers and it excludes
the usage of non-portable operating system functions. The native \Index{Application Programming Interface}s (\Index{API}s)
for the socket (network) interface and thread programming are, however, not portable between different operating
system families \citep{Stevens1997,Schmidt2000}. On the other hand, the \Index{TcpPipeLayer} could not be designed with
an object-oriented interface (which would have had some advantages, e.g. type safety), because not all instrument
suppliers used the object-oriented programming language C++.

2) Portability between different platforms\footnote{Platform is defined as a combination of a certain
hardware and a certain operating system.} as well as robustness and reliability of the software were
reached by the forceful usage of the \Index{Adaptive Communication Environment} (\Index{ACE}) library
\citep{Schmidt2001,Schmidt2002,Huston2003} which has been used very successfully on many platforms for
more than two decades. By the application of \Index{design pattern}s\footnote{Design patterns are well designed standard
solutions for frequently occurrent software problems.} \citep{Gamma2005} like Wrapper Facade,
Thread-Safe Interface, Acceptor/Connector, etc. \citep{Rising2000,Schmidt2000}, non-portable data structures
and \Index{API}s are encapsulated and hence object-oriented and type-safe class interfaces are provided by the
creation of reasonable abstractions.

The heart of the \Index{TcpPipeLayer} is a \Index{Thread Pool Reactor} \citep{Schmidt2000,Schmidt2002}, which is
preferentially responsible for the data receiving from the socket interface. A pool of 10 threads
makes the TcpPipeLayer responsive at all times, even if the ICU is receiving image and housekeeping
data simultaneously from multiple instruments. The data sending is realized with synchronous function
calls.

Access to a serial interface is mostly implemented in a separate thread which processes a queue
containing all write requests to the interface. Reading from the serial interface is done with
synchronous function calls according to a strict master-slave principle, i.e. data are only
expected to be read after a previous write access.

\begin{figure*}
\centering
\includegraphics*[width=13cm]{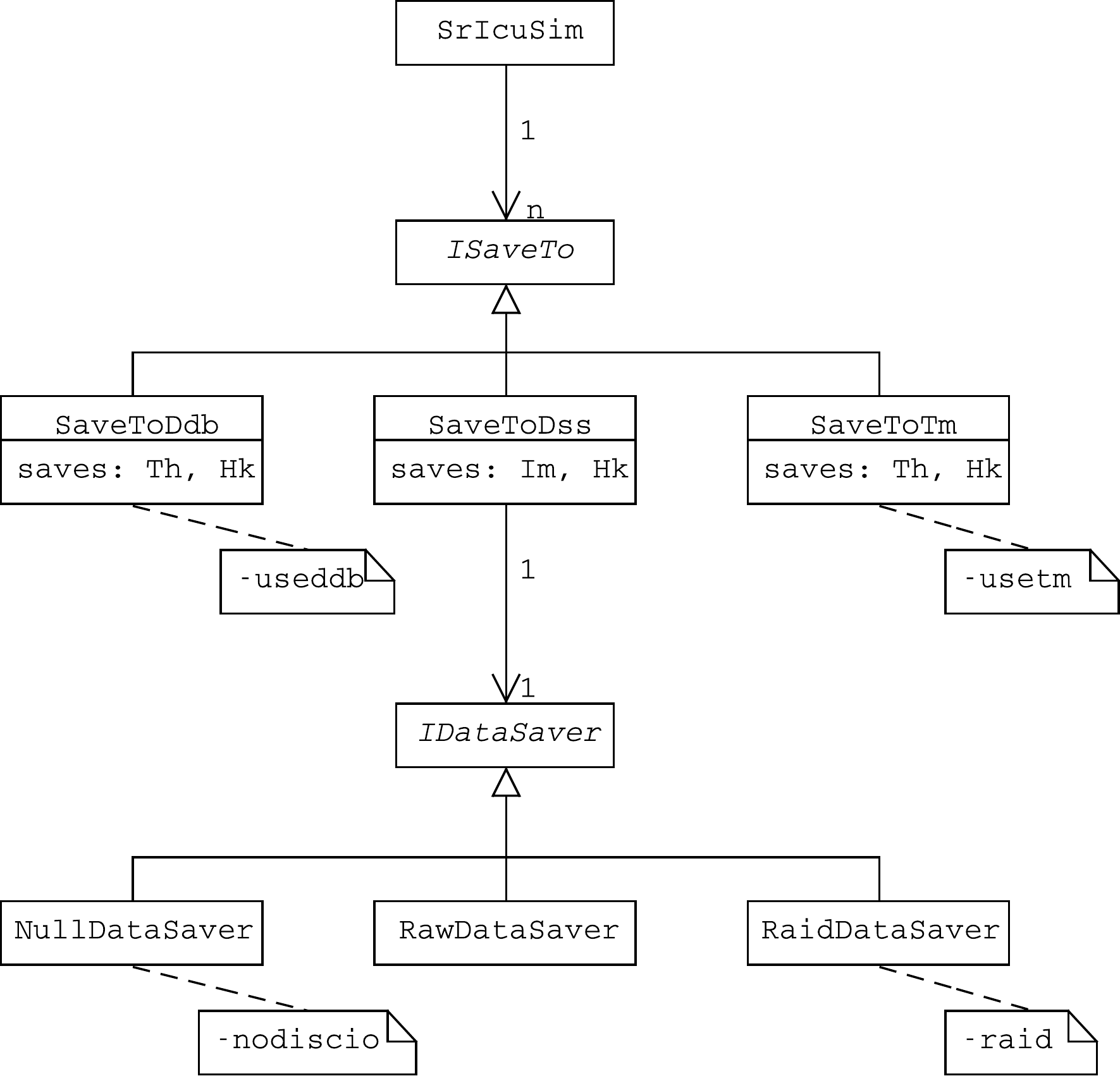}
\caption{Class diagram showing the various methods of data saving implemented in the ICU software.}
\label{FigIcuDataSaving}
\end{figure*}

3) An easy expandability of the software is reached by the forceful use of \Index{interface class}es
\citep{Meyers1995,Meyers1997}, which makes the application of a software component independent
of its implementation. Fig.~\ref{FigIcuDataSaving} shows as an example the interface class ISaveTo,
which uses the ICU to save data received from an instrument. The ICU flight software is identical
with the ICU simulator (SrIcuSim) and differs only in the usage of different command line options.
The ICU software contains always a SaveToDss object, which stores images (Im) and housekeeping data (Hk)
either on the DSS in a \Index{RAID} level~5 coded format (flight version, command line option -raid) or on a hard
drive in an uncoded format (important for the test and calibration phase, no command line option) or
the data are not stored at all (important for measurements of the data rate, command line option -nodiscio).
Housekeeping data and \Index{thumbnail}s (Th, small images suitable for transferring to the ground station) can be
optionally send to the ground station via a telemetry channel (command line option -usetm) or to the \Index{DataDistributor}
\citep[command line option -useddb, important for the software development phase, see also][]{Riethmueller2006a}.
An extension by a fourth alternative of data saving is easily possible because the ICU software knows
nothing about the concrete implementation but only calls member functions of the ISaveTo interface.

4) Run-time errors are difficult to identify, in particular in distributed concurrent software projects.
Therefor multiple platform-independent tools were developed that are specialized in the detection
of some typical run-time problems.

Data saving of large amounts of scientific data on the \Index{DSS} can only be implemented efficiently in
software if possible bottlenecks can be detected quickly and reliably, e.g. with the help of a \Index{profiler}.
A profiler tells the software developer which functions of a running software are called how many times
and how much CPU time they spend. In a multi-threaded software, the output of a profiler should be
separated by threads. If the software runs into a deadlock, the problem can be mostly identified
by outputting the \Index{call stack}\footnote{A call stack is a list of function names in the order they are
called by the software. The last list entry is the name of the currently running function.} of the
software, again separated by threads. Call stack logging and profiling are realized by the technique
of \Index{Method Call Interception} \citep[MCI,][]{Sayfan2005} which is explained below.

As a start, a tool named \Index{InsertMci} was developed which is called as a pre-built command for each
source file (similar to the C/C++ pre-processor tool). InsertMci inserts an additional source code
line in the format 'MCI(ClassName,MemberName);' at the entry point of each member function.
The name of the class and the member function are previously determined by the InsertMci tool.
At run-time, each entry point of a member function calls the constructor of an MCI object,
the corresponding destructor is called when the member function is left. If the \Index{profiler}
was registered at the MCI object, the time elapsed between constructor and destructor call can
be measured and the call counter can be incremented. In the case of the \Index{call stack logger}, class
and member function name are pushed onto a stack in the constructor and removed from the stack in
the destructor. Since the execution of the MCI constructors and destructors also need resources,
MCI can be completely ignored via a compiler option. There is also the possibility to switch off
MCI only for some time-critical member functions. The output of a call stack or profile can be
triggered from outside at any time by running a further small tool that plays the role of a
client and establishes a socket connection to the ICU software. Call stack logger and \Index{profiler}
are part of the ICU software and contain a server component which responds to the client request
by outputting the call stack or profile, respectively.

Finally, the \Index{memory checker} tool is mentioned which can detect the following memory management problems:
a) memory leaks, i.e. memory was allocated but never released, b) multiple-releases of memory, and c)
programming errors of the form new/delete[] or new[]/delete. Note that neither memory checker nor
MCI functions are allowed to use the \Index{ACE} library, otherwise a run-time analysis of the initialization
and finalization of ACE, including the Singletons managed by the ACE object manager, would not be
possible.

5) Finally, the robustness of the software was improved by the use of the C++ standard library
with its highly efficient container classes and algorithms \citep{Kuhlins2005} as well as by
ensuring const correctness \citep{Meyers1997}, i.e. whenever an object should not be modified,
this fact has to be communicated to the compiler and to other software engineers by using the C++ key
word ``{\tt const}''.

\subsection{Data storage}
The {\bf D}ata {\bf S}torage {\bf S}ubsystem\index{Data Storage Subsystem} (\Index{DSS}) is an onboard mass memory to store the
housekeeping and scientific image data persistently. The DSS consists of 48 hard disks (2.5\arcsec{})
having a capacity of 100~GBytes each. To save electrical power, only four of the 48 disks are
switched on. One of the four disks stores the parity bits of the other three disks (\Index{RAID} level~5),
so that the data can be completely recovered in the case of a single failure. Hence, the net capacity
of the DSS is 3.6~TBytes. The switch over from one four-disk-chain to the next requires a hot-pluggable
interface with a bandwidth of more than 100~MBit/s, e.g. USB or IEEE~1394\footnote{also called \Index{Firewire}}.
It was decided to use the \Index{IEEE~1394} interface because the operation of multiple hard disks using
this interface does not require an additional hub \citep[see also][]{Riethmueller2006d}.
Housekeeping data and commands are transferred between ICU and DSS via an RS422 interface
\citep{Tomasch2008}.

\begin{figure*}
\centering
\includegraphics*[height=8cm]{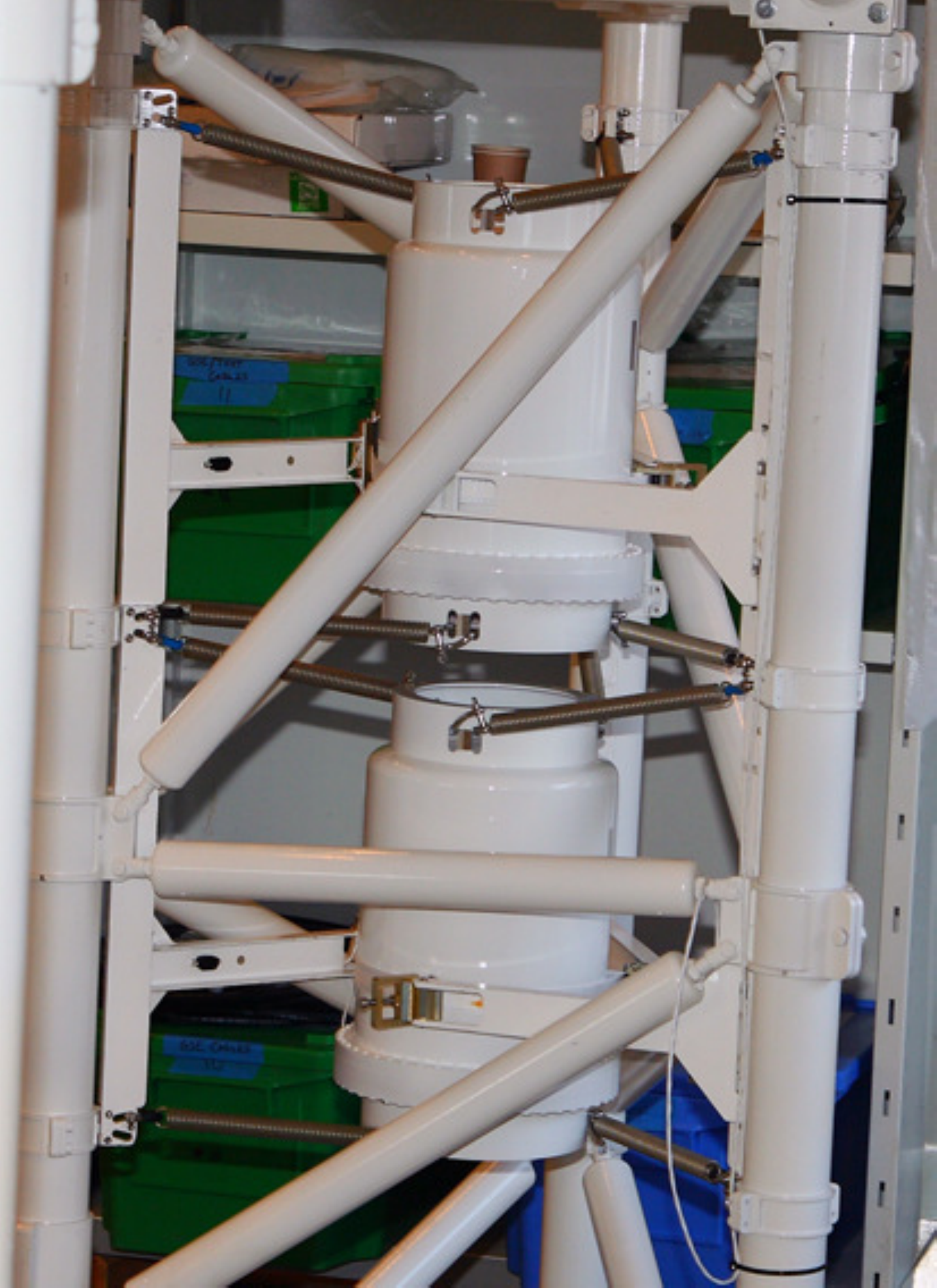}
\caption{The 2 pressurized vessels of the Data Storage Subsystem.}
\label{FigDSS}
\end{figure*}

The used hard disks need certain air pressure and humidity levels that cannot be found in
the \Index{stratosphere}. Therefore, a pressurized vessel was used, where the 48 disks were distributed
over two vessels for redundancy reasons. The vessels are mounted at a position within the gondola
structure that is as safe as possible during the landing, see Fig.~\ref{FigDSS}.

\subsection{Filter Imager}
The {\bf Su}nrise {\bf F}ilter {\bf I}mager (\Index{SuFI}) is a \Index{phase diversity} assisted broadband imager
for the near \Index{UV} which takes advantage of the observing conditions in the \Index{stratosphere}. SuFI
samples the solar \Index{photosphere} and \Index{chromosphere} in five different wavelength ranges. The shortest
wavelength, 214~nm, allows investigations of the upper photosphere at a theoretical \Index{diffraction limit}
of 0\carcsec{}05, corresponding to 40~km on the solar surface. Imaging the quiet photosphere
with the 312~nm or 388~nm filter provides intensity images that show high contrasts for \Index{bright point}s
because of the large number of molecular lines in these spectral ranges. In contrast,
the 300~nm filter allows photospheric observation without important molecular contributions.
The emission peaks on both sides of the core of the Ca\,{\sc ii}~H line at 397~nm are thermometers
measuring the temperature structure of the chromosphere.

SuFI consists of a filter wheel for the selection of the wavelength range, a shutter to limit
the exposure time, a Mechanism Controller (MC) for the control of the filter wheel and the shutter,
a UV sensitive camera, and an Electronics Unit (EU) which contains an embedded PC as instrument computer.

\subsubsection{Mechanisms}
The exposure time of the images acquired by SuFI is controlled by a mechanical shutter
\citep{Mueller2007}. A commercial two-blade shutter produced by the Japanese company Nikon
is used in combination with a cock mechanism developed at MPS. The two blades move time-delayed
to each other in the same direction and allow exposures of the SuFI camera with a precision better
than 0.1~ms which makes very short exposures possible.

The selection of the wavelength is done with the help of a filter wheel having six filter positions.
The change between two neighboring positions lasts about 1~s which possibly influences the maximal
reachable cadence of multiple-wavelength observing modes. The Planck curve\index{Planck function} decreases drastically
for short wavelengths so that unwanted leakage contributions from higher wavelength ranges can only
be avoided by the use of two filters. The two filters are mounted in two separate filter wheels
which are parallel to each other and rotate in opposite directions to minimize the torsional moment
induced into the SuFI structure. Only the highest wavelength requires only a single filter. The
filters are assigned to the filter wheel positions such that quick changes between photospheric
and chromospheric observations are possible \citep{Riethmueller2008a}. Because SuFI only observes
in five wavelength ranges, the sixth filter wheel position can be used for a lens to image the
entrance pupil which is beneficial during the integration of SuFI into the PFI structure.
Table~\ref{TabSufiFilters} lists all filters with their nominal central wavelength, the width of
the filter, the filter wheel position, and the typical exposure time.

   \begin{table}
   \caption{Assignment of the 6 filter wheel positions of SuFI.}
   \vskip3mm
   \label{TabSufiFilters}                                
   \centering                                            
   \begin{tabular}{l l l l l l}                          
   \hline                                                
   \noalign{\smallskip}
   Pos.         & $\lambda$    & FWHM       & \# Filters & Exp. Time  & Description                      \\
   \noalign{\smallskip}                                                                                                                 
                & (nm)         & (nm)       &            & (ms)       &                                  \\
   \hline                                                                                                                                  
   \noalign{\smallskip}                                                                                                                    
   \tt{O}       & \tt{388}     & \tt{~O.8}  & 2          & \tt{~~1OO} & CN band head, photosphere        \\
   \tt{1}       & \tt{396.8}   & \tt{~O.18} & 1          & \tt{~1OOO} & Ca\,{\sc ii}~H, chromosphere     \\
   \tt{2}       & \tt{3OO}     & \tt{~5}    & 2          & \tt{~~3OO} & continuum, photosphere           \\
   \tt{3}       & \tt{312}     & \tt{~1.2}  & 2          & \tt{~~2OO} & part of the OH band, photosphere \\
   \tt{4}       & \tt{214}     & \tt{1O}    & 2          & \tt{3OOOO} & upper photosphere                \\
   \tt{5}       & \tt{lens+ND} &            &            &            & pupil images during integration  \\
   \noalign{\smallskip}
   \hline                                                
   \noalign{\smallskip}
   \end{tabular}
   \end{table}

Shutter and filter wheel are controlled by the \Index{SuFI Mechanism Controller} which is commanded by the
\Index{SuFI Electronics Unit} via a serial RS422 interface \citep[see][]{Mueller2009}. Furthermore, the
SuFI Mechanism Controller switches the power of the camera and the filter wheel, controls the two
fans inside the pressurized camera electronics box, and controls the external trigger signal of the
camera. 

\subsubsection{Phase diversity prism}\label{PD_prism}
Wavefront deformations\index{wavefront deformation} can be determined by the method of Phase Diversity\index{phase diversity} (PD) in which a
target is simultaneously imaged twice with a well-known \Index{defocus} between the two images
\citep{Gonsalves1979,Paxman1992,Hirzberger2011}. The true object as well as the aberrations
can be retrieved from such observations. SuFI was designed as a PD imager in order to
have the possibility of correcting the observations for residual aberrations that can be
caused by thermoelastic deformations of the telescope and PFI structure. A PD image doubler
\citep{Grauf2007} is mounted directly in front of the camera head, so that each target is imaged
twice. The left half image is nominally focussed, the right half image is defocussed by 28~mm
which corresponds to one wave at 214~nm. This reduces indeed the effective field of view (which
is now rectangular) but allows an optimal reconstruction\index{image reconstruction} of the images within the post-flight data reduction
process.

\subsubsection{Camera}
\begin{figure*}
\centering
\includegraphics*[width=9cm]{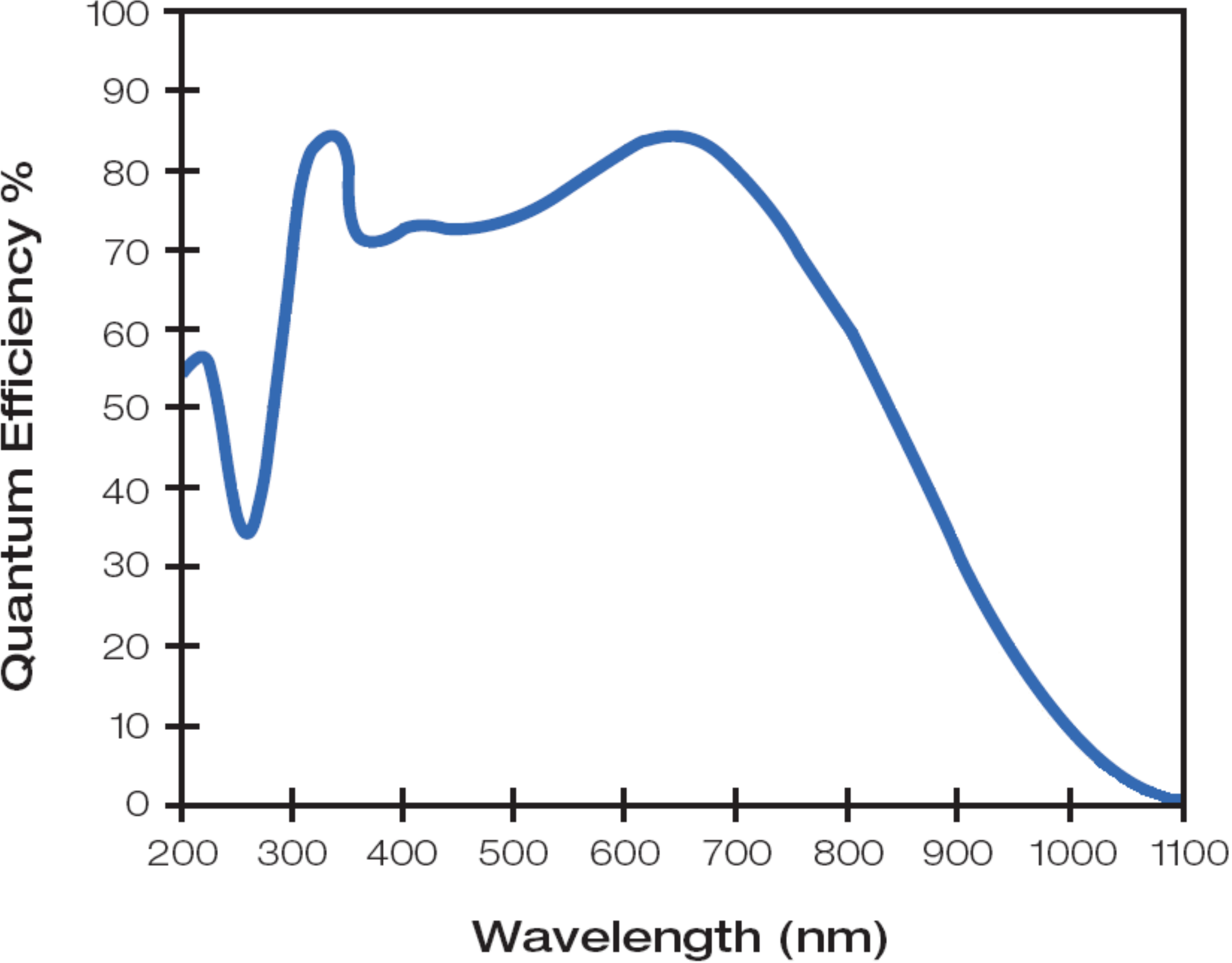}
\caption{Quantum efficiency\index{quantum efficiency} of the SITe back illuminated CCD S100AB-04 used in the \sunrise{} filter imager.
Adapted from the SITe S100A data sheet.}
\label{FigQuantumEfficiency}
\end{figure*}

The core of the \Index{SuFI} instrument is a BioXight BV20CCD camera of the company \Index{PixelVision} of Oregon, Inc.,
in which a Charged Coupled Device (CCD) sensor of the company Scientific Imaging Technologies, Inc. (SITe)
of the type S100AB-04 is used. This detector is a back illuminated $2\rm{K} \times 2\rm{K}$ CCD with
a pixel size of $12~\rm{\mu m} \times 12~\rm{\mu m}$, four read-out channels, 100\% \Index{filling factor},
and a dark current of only 1.82~electrons/pixel/sec at 240~K CCD temperature. The CCD sensor is cooled
by a \Index{Peltier element} whose hot side is in turn cooled by a liquid cooling system. The camera electronics
digitizes the analog CCD signals to 14~bit and the digital data are transferred to a PCI interface card
of type Lion2 inserted in the SuFI computer. Quantum efficiencies\index{quantum efficiency} of more than 50\%
(see Fig.~\ref{FigQuantumEfficiency}) and a maximal frame rate of 2.64~frames per second fulfill
the scientific requirements \citep{Schuessler2004} and make this CCD a nearly ideal sensor for SuFI.

The commercial camera had to be modified for utilization in SuFI. Since the \Index{Peltier element} can only
provide a temperature difference of at most 25~K, the element's hot side was connected to a radiator via
heat pipes and the camera head was separated from the camera electronics in order to allow two different
temperature levels for the two subsystems. The multitude of difficult to access hot spot components in
the camera electronics made a possible alteration of the electronics to vacuum suitability too risky,
so that a pressurized vessel was used to house the camera electronics. Two fans control the air circulation
inside the vessel which is connected to a second radiator. The two radiators emit the heat excess into
cold space. During the flight, the CCD was cooled by the \Index{Peltier element} to 268~K which led to the
optimal compromise between dark current, noise, and power consumption as determined by extensive
dark current measurements at different CCD temperatures \citep{Riethmueller2008b}.

Problems with the vacuum feedthroughs of the original fiber optic cables between camera electronics
and SuFI Electronics Unit led to an alteration to differential signaling via copper cables.
Finally, a SuFI power supply had to be developed at MPS for the 28~V primary power of the gondola.

\subsubsection{Electronics unit}
The \Index{SuFI Electronics Unit} controls all components of the SuFI instrument and connects the SuFI
camera with the ICU. Camera data are received by the EU, pre-processed, and transferred to the \Index{ICU}.
Additionally, commands are received from the ICU and executed, e.g. to move the SuFI mechanisms.

The hermetically sealed EU box contains an IBM compatible single board computer, the PCI camera interface card,
and a \Index{Sensor Interface Board} connected to the single board computer via a serial RS232 interface
\citep{Dackweiler2006}. Additionally, interfaces for external components are provided: a 100~MBit Ethernet
interface for connecting the ICU and a serial RS422 interface for connecting the SuFI MC.

The \Index{Sensor Interface Board} developed by the company DSI is identical to the board used by the ICU,
it controls the fans inside the pressurized EU box and it acquires several housekeeping values
\citep{vonderWall2005}. The single board computer is an AMPRO LittleBoard 800 \citep{AMPRO2005}
which possesses an Intel 1.4~GHz LV Pentium M~738 processor, 1~GB RAM, and a 4~GB compact flash disk.

\subsubsection{Flight software}
The main task of the \Index{SuFI software} is the execution of observing programs according to the observing
plan defined by the \sunrise{} team \citep{Schuessler2009}. Observing programs are sent from the ICU
to SuFI and require a flexible use of the instrument, i.e. all use cases mentioned in the following list
have to be covered:
\begin{itemize}
   \item SuFI shall acquire time series at fixed wavelengths in order to study highly dynamical solar
         surface phenomena at the maximum cadence.
   \item SuFI shall changes between a photospheric and a chromospheric wavelength as fast as possible.
   \item SuFI shall cycle through all five wavelengths in order to sample all available atmospheric heights.
   \item SuFI shall finish the observation after the acquisition of a given number of images.
   \item SuFI shall finish the observation if a given time is elapsed.
   \item SuFI shall observe a given selection of wavelengths for a given period of time, afterwards a cycle
         through all five wavelengths shall be done once and the entire procedure shall be then repeated.
\end{itemize}

The implementation of these use cases in software was carried out using the proven methods already
mentioned in the context of the ICU software (see section~\ref{SoftwareArchitecture}), i.e. the use of:
\begin{itemize}
   \item the \Index{TcpPipeLayer} for the communication between SuFI and ICU,
   \item the \Index{ACE} library for multi-threading, thread synchronization, error message handling, etc.,
   \item \Index{interface class}es,
   \item \Index{design pattern}s,
   \item \Index{memory checker}, \Index{call stack logger}, and \Index{profiler} for analyzing the run-time behavior of the software.
\end{itemize}

An observing program consists of a sequence of commands which are serially executed by the SuFI software.
In the normal case, the ICU sends such command sequences at a certain point in time (timestamp) according
to the \Index{SuFI timeline}\footnote{A detailed description of the ICU timeline philosophy is given in
\citet{Riethmueller2006e}.} which is stored on the ICU. It is also possible to send manual commands from
the ground to SuFI in order to manipulate or cancel running observing programs. For such manipulation
or cancelation commands, a second logical way of commanding has to exist because the first way is already
blocked by the execution of the observing program. Therefore, low- and high-priority commands are
introduced. Low-priority commands can be both, very simple commands or more complex commands needing
a long execution time, e.g. the acquisition of an image series. High-priority commands must be simple
and hence require a short execution time, e.g. the cancelation of an observing program ahead of time,
setting a flag, or modifying a parameter. For each command it has to be known a priori whether it is a simple
or a complex command.

\begin{figure*}
\centering
\includegraphics*[width=\textwidth]{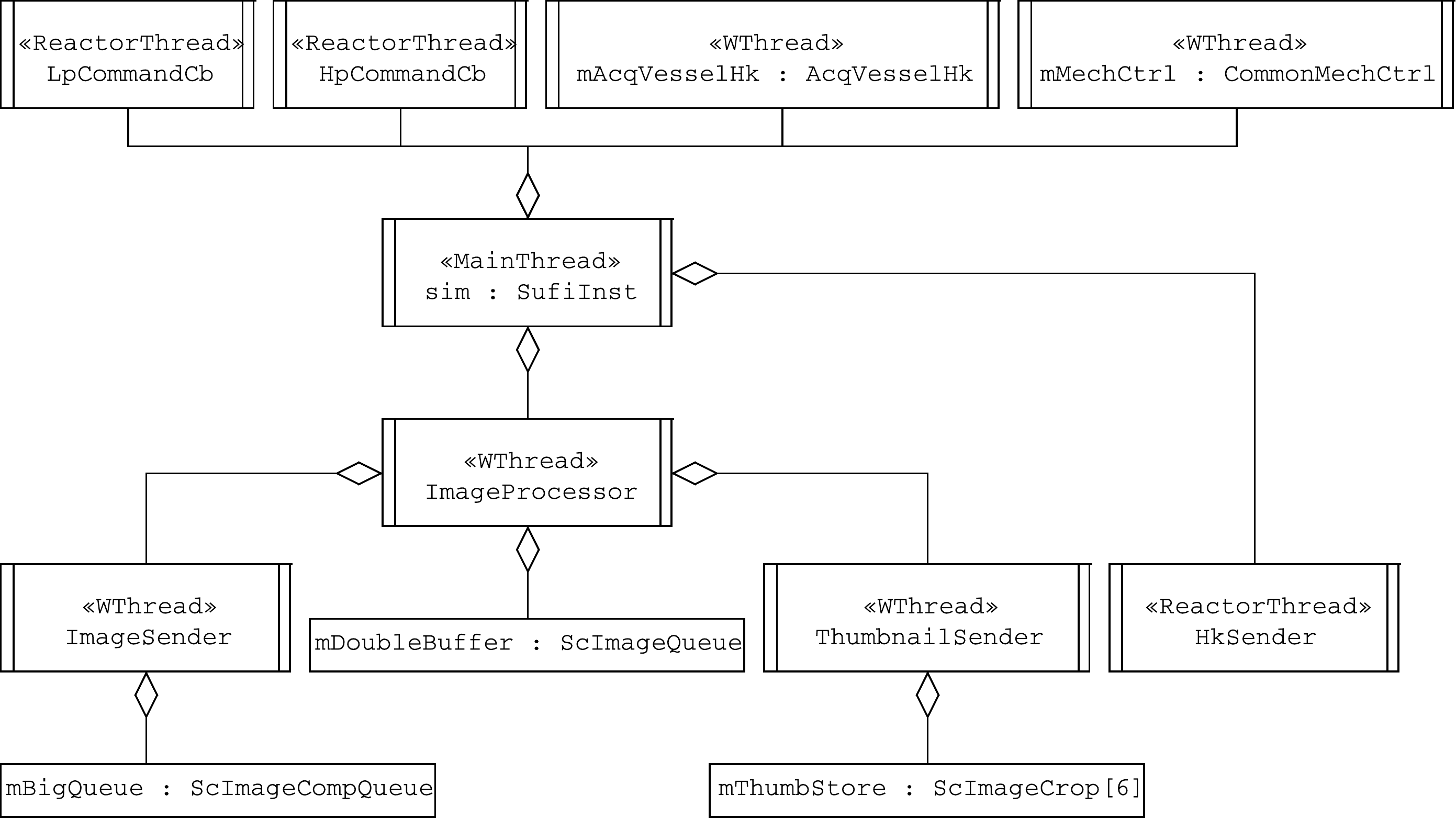}
\caption{Concurrency model of the SuFI software. See main text for an explanation.}
\label{FigThreadsOverview}
\end{figure*}

The \Index{concurrency model} of the SuFI software is displayed in Fig.~\ref{FigThreadsOverview}. The entry
point of the software is the construction of the application object of type SufiInst by the main thread
(central part of Fig.~\ref{FigThreadsOverview}). The main thread initializes the \Index{TcpPipeLayer} with
its pool of reactor threads. One of the reactor threads receives low-priority commands from the ICU
that are executed immediately in the case of a simple command or inserted into a command queue processed
serially by the main thread in the case of a complex command. A second reactor thread receives
high-priority commands from the ICU which are always executed immediately. A third reactor thread
(HkSender) works as alarm handler and transfers a current snapshot of the housekeeping table to
the ICU every 5~s. The ICU uses the regularly incoming Hk packets for a \Index{heartbeat monitor}ing of SuFI.
If the Hk stream becomes silent for longer periods, the ICU power cycles SuFI.

The main thread creates three further threads, the AcqVesselHk, the CommonMechCtrl, and the ImageProcessor
thread. The processing of housekeeping requests from the \Index{Sensor Interface Board} (e.g. pressure and
temperature inside the pressurized EU vessel) using synchronous function calls lasts about 9~s and is
hence implemented in a separate thread. This AcqVesselHk thread puts the housekeeping values into a
global Hk table which always contains the latest Hk values of every SuFI component, e.g. the current
filter position, which can be theoretically changed every second. The global Hk table decouples
Hk acquisition and Hk transfer done by the HkSender thread. The CommonMechCtrl thread is an active
base class which provides an interface to access the SuFI Mechanism Controller. There are two
implementations of this interface. The first one simulates the MC hardware and is only needed
during the software development phase for situations in which the real hardware is not available.
During the flight, the second implementation is used which accesses the real MC. Again, MC commands
are queued and housekeeping values of the MC are put into the global Hk table.

\begin{figure*}
\centering
\includegraphics*[width=\textwidth]{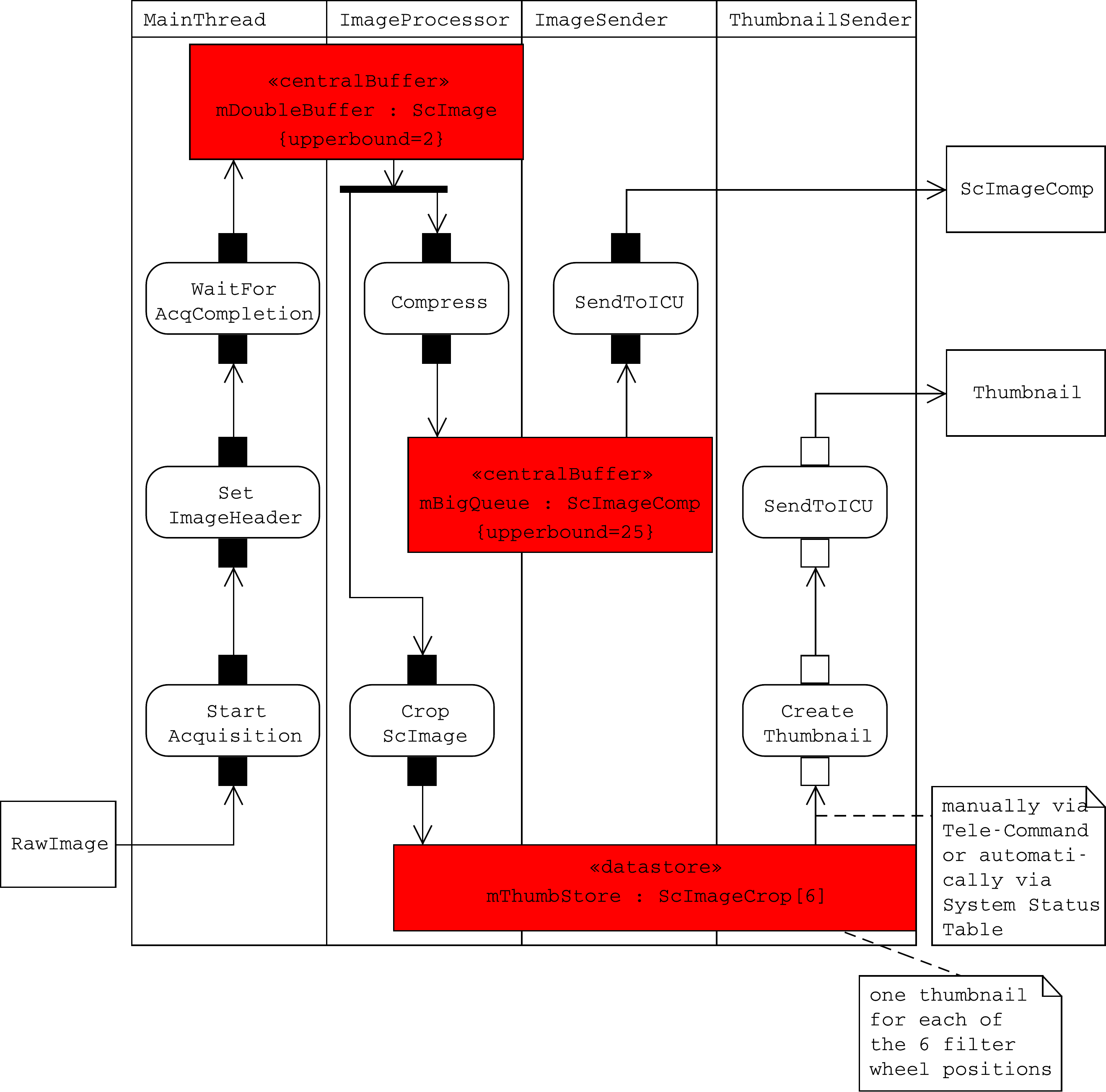}
\caption{Overview of the image data flow of SuFI and its distribution over 4 threads to reach a maximum cadence.}
\label{FigImageDataFlow}
\end{figure*}

The ImageProcessor thread allows the compression of image data and the creation of \Index{thumbnail}s not only
to be done in parallel to moving the filter wheel but also simultaneously with the transfer of data to the ICU.
For this purpose, the ImageProcessor thread creates the ImageSender thread and the ThumbnailSender thread.
The data flow of an image acquisition and its distribution over the various threads is shown in
Fig.~\ref{FigImageDataFlow}. The main thread initiates the read-out of the CCD into the free half of the
\Index{double buffer} for ScImage objects via the StartAcquisition action, where an ScImage object provides
memory for a raw image and an image header. Then the main thread determines the current filter wheel
position and exposure time from the MC in order to use these values together with other values from the
global Hk table to fill the image header of the current ScImage object. In parallel to the CCD read-out,
the filter wheel moves to the position of the next observation. If the CCD read-out is completed, the
current half of the double buffer is tagged as ready for further processing.

While one half of the double buffer is being filled with image data, the previous image contained in
the other half of the \Index{double buffer} is processed by the ImageProcessor thread simultaneously. Such
an ScImage object is compressed and inserted into the mBigQueue queue as ScImageComp object.
mBigQueue can contain at most 25 images and must provide such a large amount of memory because
sufficient buffer memory is needed for temporary interruptions of the data transfer to the ICU
(e.g. during the switch over from one four-chain of the \Index{DSS} to the next one). Whenever mBigQueue
contains an image, the ImageSender thread tries to transfer the image to the ICU. If the system
is overloaded, mBigQueue gets full and the oldest ScImageComp object is overwritten by a new one.
Additionally, the ImageProcessor thread creates thumbnails from the double buffered ScImage objects.
Because of the low telemetry rate, the amount of data required by a thumbnail needs to be considerably
reduced by cropping (it is sufficient to down link the focussed half image), binning, bit truncation,
and image compression. The parameters of the thumbnail creation are arbitrary within the valid ranges.
For each filter wheel position, an up-to-date thumbnail is stored in mThumbStore as an ScImageCrop object.
Thumbnails can be packetized and sent to the ICU by the ThumbnailSender thread, if the automatic
thumbnail sending mode is switched on and/or a thumbnail is requested manually and if telemetry
bandwidth is available for thumbnails. Whether this is the case is known from the System Status Table
\citep{Kolleck2009} which is broadcast from the ICU to the instruments every 3~s and contains
system information, e.g. the GPS position and altitude of the balloon, status of the aperture door,
the F2 mechanism, and the mirrors M2, M3, M4, or the pointing quality of the gondola.

\begin{figure*}
\centering
\includegraphics*[width=10cm]{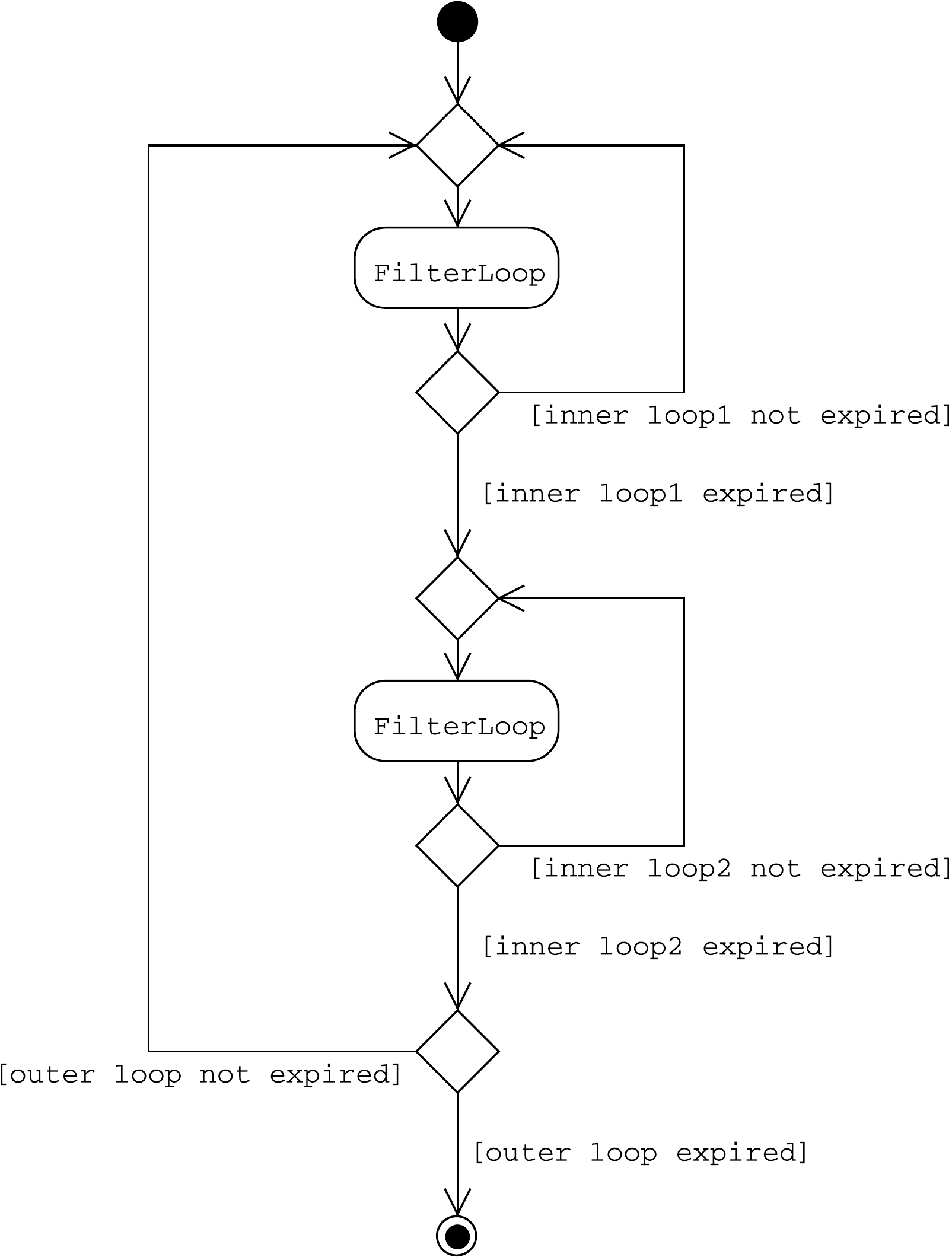}
\caption{Activity diagram visualizing an observing run of SuFI\index{SuFI observing run}.}
\label{FigObservingRun}
\end{figure*}

The acquisition of an image is repeated via the loop structure shown in Fig.~\ref{FigObservingRun}.
The serial processing of two inner loops within an outer loop guarantees that all the use cases
mentioned above are covered, in particular the last mentioned use case with the highest
requirements. The FilterLoop activity shown in Fig.~\ref{FigObservingRun} cycles through an
array of 20 filter wheel positions that are commanded one after the other. A negative filter wheel
position is thereby ignored, so that less than 20 positions can also be used in such a loop.
After each image acquisition, a possible delay leads to the user-defined cadence. The positioning
accuracy of the filter wheel is ensured to be high by not rotating the wheel always in one direction
but such that after one wavelength cycle the wheel is rotated back to the first filter position.
For a full five wavelengths cycle the filter positions can be commanded, e.g. in the order 0,1,2,3,4
(fast cycling, high return time) or in the order 0,2,4,3,1 (return time more uniformly
distributed over several steps), just to mention two possibilities. A 16~bit integer value determines
the completion of the FilterLoop activity. A positive value is interpreted as number of seconds, a
negative value as number of cycles.

During the flight, SuFI was controlled from ground with the help of a \Index{GSEOS} based \Index{EGSE} software, similar to
the ICU case. A second \Index{Qt} based SuFI EGSE software was used for displaying and analyzing \Index{thumbnail}s.

\subsubsection{Data reduction}
Some days after the landing of the gondola on Somerset Island on 2009 June 13, the two pressurized vessels
containing the data storages were safely recovered and transported to MPS. After a successful incoming
inspection and operation with the ICU qualification model, a backup of all data was done and the \Index{RAID}~5
format was decoded. The raw format of the images was converted into the \Index{Flexible Image Transport System} \citep[\Index{FITS},][]{Wells1981}
format commonly used in solar physics and was named the level~0 data set. Each image was saved in a separate
FITS file containing an extensive header which was defined similar to the header of the \hinode{} data.
The header consists of two parts. The first part is commonly defined for SuFI and IMaX and contains mainly
system-specific entries, e.g. observing time, pointing coordinates, flight altitude and position. The second
part contains instrument-specific entries, e.g. pressure and temperatures of the SuFI EU box, CCD temperature,
etc. Details can be found in the \Index{SuFI Data Manual} \citep{Feller2010}. 

Correcting the data for dark current and flat field as well as interpolating the bad pixels led to the level~1
data set. Some housekeeping values in the FITS headers had to be corrected by the delay time occurred in flight
by the transport of system-specific data with the help of the \Index{System Status Table} \citep{Kolleck2009}.
Finally, the two types of \Index{phase diversity} reconstruction\index{image reconstruction} led to the level~2 and level~3 data in which the
focussed and defocussed half images are brought together. Owing to a narrow transition region between the two
image halves as well as the slight offset of the position of this transition region from the image center,
the effective field of view of the SuFI level~2 and level~3 data is about 14\arcsec{}$\times$40\arcsec{}.
While for the level~2 PD reconstruction each image was partitioned into sub-apertures and the reconstruction
was done for each image and sub-aperture individually, the level~3 PD reconstruction was done with a single
wavefront retrieved by averaging over several sub-apertures and images of a time series. More details about
the PD reconstruction of SuFI data can be found in \citet{Hirzberger2011}.

A \Index{data mining} tool was developed for the navigation through the circa 150000 SuFI images. The tool allows
an easy selection of images according to freely definable image properties.

\subsection{Imaging magnetograph}

The {\bf I}maging {\bf Ma}gnetograph\index{magnetograph} e{\bf }Xperiment (\Index{IMaX}) is an imaging \Index{spectropolarimeter} which operates
in the strongly Zeeman sensitive Fe\,{\sc i} line at 5250.2~\AA{}. A spectral resolution of about 85~m\AA{}
is reached by the use of a tunable $\mathrm{LiNbO_3}$ etalon in double pass combined with a 1~\AA{} broad
prefilter. As needed for full \Index{Stokes vector} polarimetry, the polarimetric part of IMaX consists of two liquid
crystal \Index{retarder}s which are switched between four different polarization states at a frequency of 4~Hz. An
accumulation of images is possible in real-time in order to increase the signal-to-noise ratio. A polarizing
beam-splitter allows the use of two synchronized $\mathrm{1K \times 1K}$ CCD cameras in a \Index{dual-beam configuration}.
The simultaneous acquisition of the same field of view of 50\arcsec{}$\times$50\arcsec{}
in two oppositely linearly polarized states minimizes \Index{cross talk} between the \Index{Stokes parameter}s due to residual
\Index{pointing jitter}. Before and after an observing program, a phase diversity plate can be inserted in the
beam of one of the cameras. Lower wavefront aberrations can be determined by such PD measurements which
are then used in the post-facto data reduction for \Index{image reconstruction} purposes.

IMaX can be operated in various modes, where the number of accumulations ($N_A$) as well as the number
($N_{\lambda}$) and position of wavelength points can be modified. In addition to the full \Index{Stokes vector}
mode ($N_P=4$), a longitudinal mode ($N_P=2$) is also possible, which only records Stokes~$I$ and $V$
images. Depending on the scientific problem, the best compromise between cadence, signal-to-noise ratio,
and spectral sampling can be chosen. An overview of the IMaX modes used during the June 2009 flight is
given in table~\ref{TabImaxModes}. Detailed information on \Index{IMaX} can be found in \citet{MartinezPillet2011}.

   \begin{table}
   \caption{Observing modes of IMaX. Adapted from \citet{MartinezPillet2011}.}
   \vskip3mm
   \label{TabImaxModes}
   \centering
   \begin{tabular}{l r r r r r l}
   \hline
   \noalign{\smallskip}
   Observing mode & $N_P$ & $N_{\lambda}$ & $N_A$ & Duration & S/N  & Line samples             \\
   \noalign{\smallskip}
                  &       &               &       & (s)      &      & (m\AA{})                 \\
   \hline                                                                                                                    
   \noalign{\smallskip}
   V5-6           & 4     & 5             & 6     & 33       & 1000 & -80, -40, +40, +80, +227 \\
   V5-3           & 4     & 5             & 3     & 18       & 740  & -80, -40, +40, +80, +227 \\
   V3-6           & 4     & 3             & 6     & 20       & 1000 & -60, +60, +227           \\
   L3-2           & 2     & 3             & 2     & 8        & 1000 & -60, +60, +227           \\
   L12-2          & 2     & 12            & 2     & 31       & 1000 & -192.5...+192.5 each 35  \\
   \noalign{\smallskip}
   \hline                                                
   \noalign{\smallskip}
   \end{tabular}
   \end{table}

\begingroup
\hypersetup{linkcolor=white} 
\chapter[Brightness, distribution, and evolution of sunspot umbral dots]{Brightness, distribution, and evolution of sunspot umbral dots$^1$}\label{Ud1Chapter}\footnote{Published in Astronomy \& Astrophysics, 492, 233 (2008), see \citet{Riethmueller2008d}.}
\endgroup

\section{Introduction}

   The investigation of the complex fine structure of umbrae and penumbrae is crucial
   to understanding the subsurface energy transport in sunspots. The energy transport from the
   solar interior to the solar surface outside magnetic features is mainly determined by convection,
   visible as granulation in images of the quiet photosphere. The strong and nearly vertical umbral
   magnetic field suppresses normal overturning convection inside the umbra. However, it is believed
   that some form of residual magnetoconvection is responsible for much of the remaining energy
   transport and manifests itself in the form of fine structures, such as light bridges (LBs) or
   \Index{umbral dot}s (UDs). In the present paper we consider UDs, which contribute up to 37\,\%
   of the radiative umbral flux according to \citet{Adjabshirzadeh1983}.
   Different models have been proposed to explain the umbral dots. \citet{Choudhuri1986}
   postulated that UDs are thin columns of field-free hot gas between the cluster of small magnetic
   \Index{flux tube}s that form the subsurface structure of a sunspot according to \citet{Parker1979}.
   According to this model, a UD is formed when an upwelling brings hot material into the photosphere.
   An alternative model has been proposed by \citet{Weiss1990} who consider UDs to be spatially
   modulated oscillations in a strong magnetic field.

   A more recent, promising approach is presented by \citet{Schuessler2006}, who used numerical simulations
   of three-dimensional radiative magnetoconvection to improve the physical understanding of the umbral
   fine structure. The simulations exhibit the emergence of small-scale upflow plumes that start off
   like oscillatory convection columns below the solar surface but turn into narrow overturning cells
   driven by the strong radiative cooling around optical depth unity. Most of those UDs show a central
   \Index{dark lane}. The presence of dark lanes in penumbral and umbral fine structures has already
   been observed several times, cf. \citet{Scharmer2002,Langhans2007,Scharmer2007}. The verification of
   the predicted dark lanes in large UDs by \citet{Bharti2007} and \citet{Rimmele2008},
   as well as the verification of the predicted photospheric stratification of bright peripheral UDs
   by \citet{Riethmueller2008c}, support the Sch\"ussler \& V\"ogler model of UDs. There
   is now a need to learn more about this phenomenon, with a statistically robust analysis of UD
   properties and evolution being a promising means of achieving this aim.

   The most detailed analyses of UDs are more than 10 years old \citep{Sobotka1997a,Sobotka1997b}
   and are based on data observed with the 50-cm \Index{SVST} (\Index{Swedish Vacuum Solar Telescope}), cf. the
   recent reviews of umbral fine structures by \citet{Solanki2003,Thomas2004} and \citet{Sobotka2006}.
   The more recent papers of \citet{Tritschler2002,Hartkorn2003} and \citet{Sobotka2005} have
   concentrated on individual properties and lack, e.g., the determination of UD trajectories.
   Furthermore, the possibility of improving the spatial resolution with the help of modern \Index{image reconstruction}
   algorithms is only used by \citet{Tritschler2002}. The present paper aims to
   overcome these shortcomings, by employing data from the 1-m \Index{SST} (\Index{Swedish Solar Telescope}) equipped
   with an \Index{adaptive optics} system, by restoring the data employing \Index{MFBD} (multi-frame blind deconvolution\index{Multi-Frame Blind Deconvolution}),
   and determining the evolution of UD parameters whenever possible.

\section{Observations and data reduction}

   The data employed here were acquired on September 7, 2004 with the Swedish Solar Telescope
   at the Observatorio del Roque de los Muchachos on La Palma, Spain. Technical details of
   the SST are described by \citet{Scharmer2003a}. Wavefront aberrations caused by the telescope
   and by the turbulent atmosphere of the Earth were partially corrected by the \Index{adaptive optics} system,
   explained in \citet{Scharmer2003b}. The science camera was a Kodak Megaplus CCD with a pixel size
   of 9~$\mu$m and a \Index{plate scale} of 0.041$^{\prime\prime}$ (30~km on the Sun) per pixel. The camera was
   equipped with an interference filter at the wavelength of the 705.7~nm of the titanium
   oxide band head, the FWHM of this filter was 0.71~nm. The theoretical \Index{diffraction limit}
   of the telescope at the TiO wavelength is 0.18$^{\prime\prime}$ (130km). Due to the high sensitivity
   to umbral temperatures of TiO lines, the TiO band head is a good diagnostic wavelength range for imaging
   umbral features \citep{Berdyugina2003}. A wavelength in the red was chosen also in order to ensure a more
   homogeneous time series due to the more benign seeing at these wavelengths. Acquisition lasted from
   08:27~UT to 10:17~UT, i.e. a total of 110~min. The images were obtained in a frame selection mode that
   saved only the 8 best images of a 20-second-interval. The exposure time was 10~ms. The telescope pointed
   to the \Index{sunspot} of the active region NOAA~10667 at cos~$\theta$~=~0.95, i.e. relatively close to the solar
   disk center ($\theta$ is the heliocentric angle).

   The data were dark current and flat field corrected, reconstructed via the \Index{MFBD} technique
   \citep{Loefdahl2002}, derotated, destretched \citep{November1988}, and subsonic filtered
   with a cut-off phase velocity of 5~km\,s$^{-1}$ \citep{Title1989}. The obtained
   time series consists of 310 images with a spatial resolution in the range of
   $\sim$0.18$^{\prime\prime}$-0.25$^{\prime\prime}$, as we estimated from radially averaged power spectra.
   The field of view (FOV) is 37$^{\prime\prime}$~$\times$~59$^{\prime\prime}$ and contains the entire
   considered sunspot whose umbra is divided into two parts by a \Index{light bridge}. UDs in both
   parts of the umbra are analyzed in the next section.

\section{Data analysis}\label{Ud1DataAnalysis}

\begin{figure}
   \centering
   \includegraphics[width=0.5\linewidth-1mm]{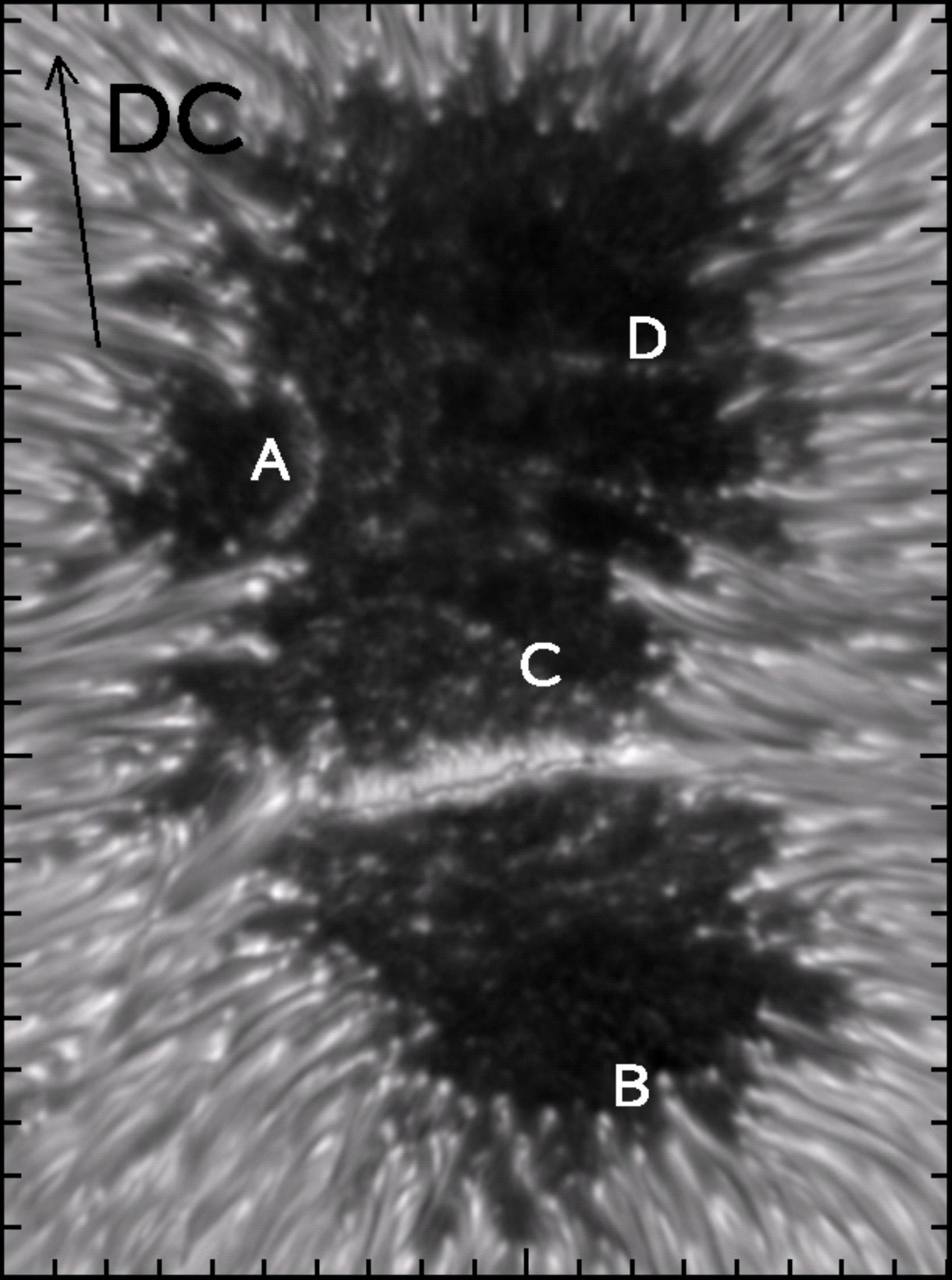}
   \includegraphics[width=0.5\linewidth-1mm]{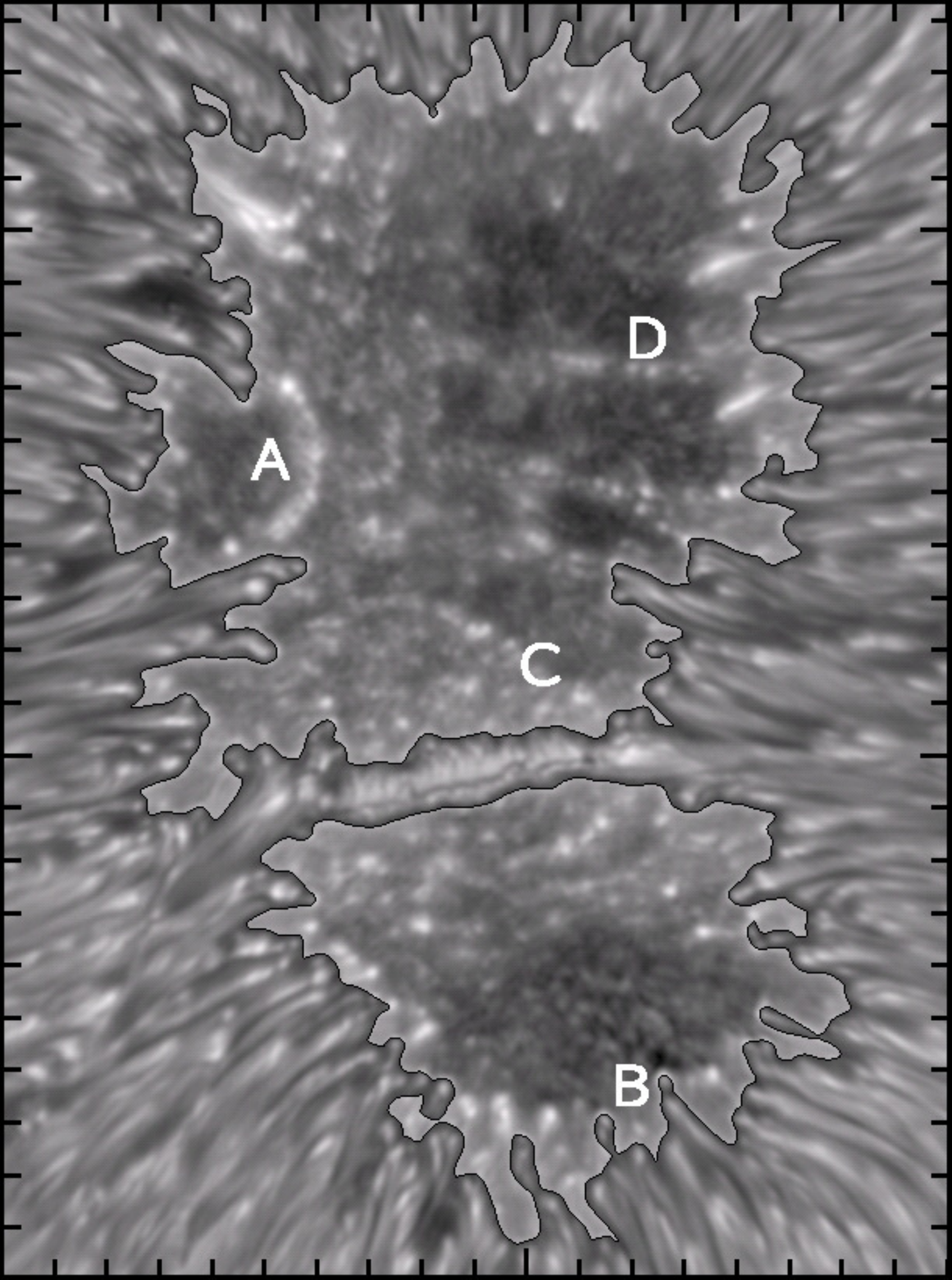}
   \caption{Frame taken at 09:53:20 UT: Best quality image plotted at normal contrast (left panel)
   and with increased umbral brightness (right panel). The black contour lines outline the umbra (see text
   for details). The direction to solar disk center (DC) is indicated by the arrow. Locations of
   umbral dots (UDs) that are discussed further in the text are marked by the letters A-D.
   Minor tick marks are given in Mm.}
   \label{FigBestImage}
   \end{figure}

   The detailed analysis of 310 images requires an automated algorithm for the identification of
   the thousands of umbral dots they contain. A specific algorithmic challenge is the fact
   that UDs as well as the local umbral background between them cover a broad range of intensities.
   At a normal contrast (left panel of Fig.~\ref{FigBestImage}) UDs are mainly visible near the
   penumbra. By displaying the square root of the umbral brightness instead of the
   brightness itself numerous UDs within the dark umbral background become visible as well
   (right panel).

   Our automated UD analysis starts with isolating the umbra, which is done by thresholding a lowpass
   filtered image (averaging a squared environment of 11~$\times$~11 pixels) at 35\,\% of the mean
   intensity of the quiet photosphere ($I_{ph}$). From the resulting set of contours we select only
   those longer than 3~Mm in order to avoid larger UDs from being connected to the umbral boundary.
   In the particular case of the studied penumbra the results are almost identical to identifying
   the two longest contours with the umbral boundary. The thus obtained umbral boundary is visible
   in the right panel of Fig.~\ref{FigBestImage} as the black contour line. This method for isolating
   the umbral boundary automatically ensures that a local brightening at the end of a penumbral fibril
   is only considered to be a UD if it is isolated from the penumbra in the sense that the intensity
   between the UD and the penumbral fibril falls below the applied threshold of 0.35~$I_{ph}$.

\begin{figure}
   \centering
   \includegraphics[width=\linewidth-1mm]{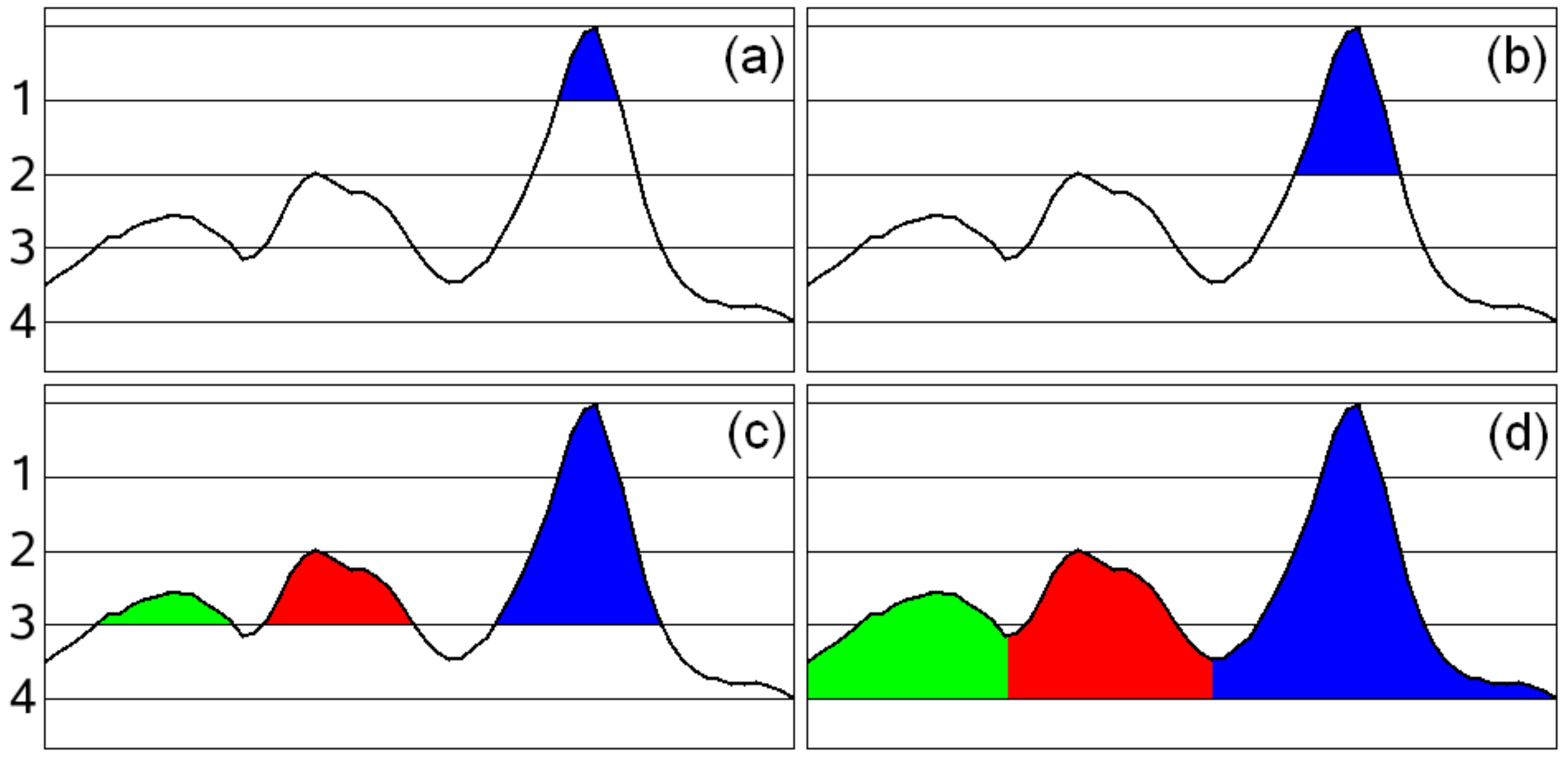}
   \caption{Illustration of \Index{multilevel tracking} algorithm with 4 levels applied to a typical intensity profile.}
   \label{FigMLT}
   \end{figure}

   In the next step the UDs in each of the 310 images (recorded at a cadence of 20.57~s) are
   detected. For this purpose several algorithms were tested, e.g. a method where, starting
   from the UD center, 8 equally distributed rays are followed until they reach the UD boundary which
   is defined as the position where the intensity drops below 50\,\% of the maximum intensity above the
   local umbral background. The resulting 8 boundary points lead to a polygon that is a good approximation
   of the UD boundary. The method does not work properly for UDs with a partly concave boundary and it
   cannot easily separate UDs that are close to each other. Finally, the \Index{multilevel tracking}
   algorithm of \citet{Bovelet2001}, which provided the best results in detecting the UD boundaries,
   was chosen. First the \Index{MLT} algorithm determines the global extrema of the umbral intensities and
   subdivides this range into equidistant levels. Bovelet and Wiehr used MLT to distinguish between
   granules and intergranular lanes of the quiet Sun and found that three MLT levels are sufficient
   for their purpose. Since umbral dots cover a broad range of intensities we have to use a noticeably
   higher number of levels. We normalized our best quality image to $I_{ph}$ and found an umbral
   intensity range from 0.36-0.96~$I_{ph}$ (Note: This range is only valid for the best quality image,
   other images may reach lower or higher umbral intensities). We found that 25 MLT levels is the
   optimal compromise between detecting as many UDs as possible and avoiding the misinterpretation
   of noise as UDs. Whereas small umbral dots, obtained with this choice of levels, have a
   typical contrast relative to the local background of about 0.05~$I_{ph}$, the noise level
   is about 0.005~$I_{ph}$ (see Fig.~\ref{FigMLT} for a typical intensity profile).
   Starting with the highest intensity level all pixels are found whose intensity exceeds this level.
   This leads to several bounded two-dimensional structures, that are tagged in a unique way, which
   is indicated by different colors in the one-dimensional illustration given in Fig.~\ref{FigMLT}.
   The obtained closed structures are extended pixel by pixel as long as the intensity is greater
   than the next lower level. After that the algorithm searches through the whole umbra again to find
   all pixels whose intensity is greater than the next lower level, which may lead to some newly
   detected closed structures. This procedure is repeated until the minimal intensity level is reached.
   At the end every umbral pixel belongs to exactly one closed structure. The mode of operation of the
   MLT algorithm is illustrated for one dimension in Fig.~\ref{FigMLT} for the case of 4 levels.

\begin{figure}
   \centering
   \includegraphics[width=\linewidth-1mm]{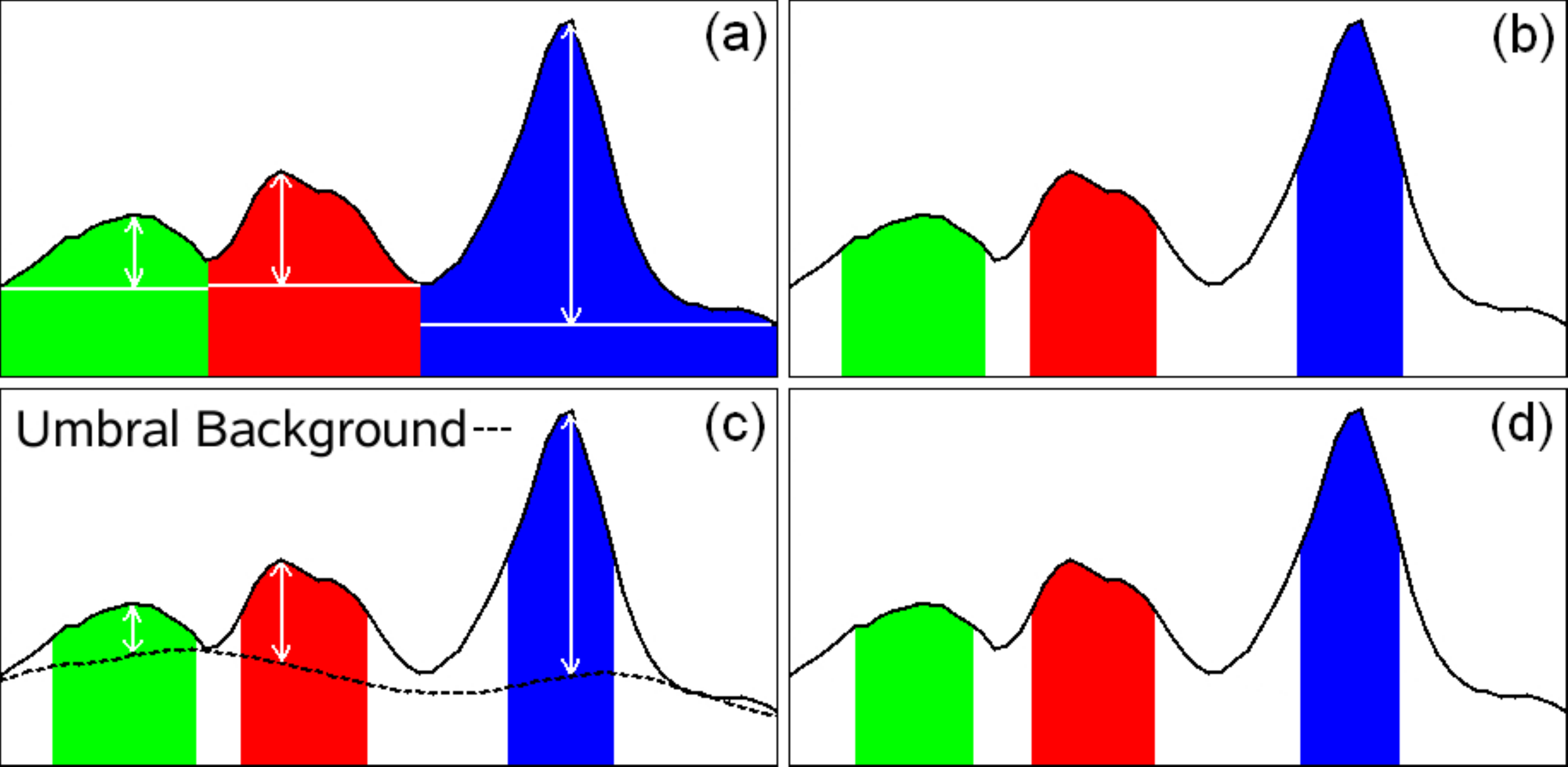}
   \caption{Determination of the boundary of umbral dots.}
   \label{FigCuttingPixels}
   \end{figure}

   The minimal ($I_{Min}$) and maximal ($I_{Peak}$) intensity of each closed structure is determined
   and all pixels that have an intensity lower than 50\,\% of this min-max range (white arrows in
   Fig.~\ref{FigCuttingPixels}a) are cut. This leads to a first estimate of the UD boundaries
   (see Fig.~\ref{FigCuttingPixels}b) that are used to determine the local umbral background intensities
   ($I_{bg}$), i.e. the intensities that would be observed in the absence of all UDs. We applied
   the method used by \citet{Sobotka2005} that approximates the local umbral background by a 2D surface
   fitted to the grid of local intensity minima, using the method of thin-plate splines
   \citep{Barrodale1993}. Since the local umbral background intensities are known now
   (dashed line in Fig.~\ref{FigCuttingPixels}c), we determine the exact UD boundaries by cutting
   all pixels lower than 50\,\% of the maximum intensity above the local umbral background (see white arrows in
   Fig.~\ref{FigCuttingPixels}c). Fig.~\ref{FigCuttingPixels}d illustrates that the resulting UD
   boundaries are similar to our first estimate in Fig.~\ref{FigCuttingPixels}b so that we don't need
   further iterations (this was the case with most identified UDs). Employing this procedure we found,
   on average, 323 UDs per image. Sometimes, our algorithm recognizes strong elongated bright structures
   as UDs that we would not consider as a UD by visual inspection of the images. Via a spot-check on
   selected images we estimate that the number of misidentifications is lower than 1\,\%. Due
   to the large number of detected UDs we are not able to remove the misidentifications by hand and
   accepted them as noise.

   Since a UD is an extended structure we determine the coordinates of the brightest pixel (peak intensity)
   and save them as the UD's position. This method is applicable because the noise was sufficiently reduced
   by our subsonic filter, as demonstrated by a typical intensity profile in Fig.~\ref{FigMLT}. We also
   determine the UD's diameter, defined as the diameter of a circle of area equal to that within the boundary
   of the UD. After the positions of all UDs of every image are known, the motions and trajectories of UDs
   are determined. The continuation of a trajectory in the image at the next (previous) time step is determined
   by finding the UD that is closest to the UD's current position. If no UD can be found within a 5~pixel
   neighborhood (theoretical \Index{diffraction limit}) of the current UD then the tracking stops. In a loop over all images every UD is tracked
   backward in time until its birth and forward in time until its death. The tracking must be
   tolerant to the occasional image with lower image quality in which the UD may not be correctly identified
   (specially the smaller and fainter ones). In practice we allow for a gap of up to two images. If a UD is
   present at nearly the same location on both sides of the gap, the tracking is continued. In this
   manner we found 12836 UD trajectories that are weakly lowpass filtered in space (averaging the
   positions of the same UD in 15 consecutive images) in order to further reduce seeing-induced noise.
   From now, we call such a smoothed trajectory simply $trajectory$. We note that 5949 of the
   12836 UDs are only identified in a single image. We decided not to ignore them, because these bright dots in
   the umbra are detected rather well, even if only for a very short time. We assigned a zero trajectory length
   and a lifetime of 20.57~s to these 5949 UDs.

\section{Results}

   \subsection{Qualitative results}\label{QualResults}

   A first impression of the temporal evolution of the smallest umbral structures is reached by making
   a movie of the reconstructed time series of images. Some interesting phenomena are found by the
   visual inspection of this movie and are explained briefly below.

   The \Index{sunspot} has two umbrae, a smaller and a roughly twice larger one separated by a \Index{light bridge} (LB).
   The LB contains a clearly visible \Index{dark lane} in agreement with the observations of, e.g.,
   \citet{Berger2003}. The lane is closer to the limbward edge of the LB. Possibly this is because the
   sunspot was observed at a heliocentric angle of $\theta~=~\mathrm{18^\circ}$, so that projection
   effects may cause the observed asymmetry \citep[e.g.][]{Lites2004}. However, the LB also displays
   another major asymmetry: the movie exhibits many UDs that are born within the LB and move into the
   larger umbra, i.e. towards the solar disk center (arrow in Fig.~\ref{FigBestImage}), while almost
   no UD leaves the LB into the smaller umbra, i.e. towards the solar limb (antiparallel to the arrow).

\begin{figure}
   \centering
   \includegraphics[width=\linewidth]{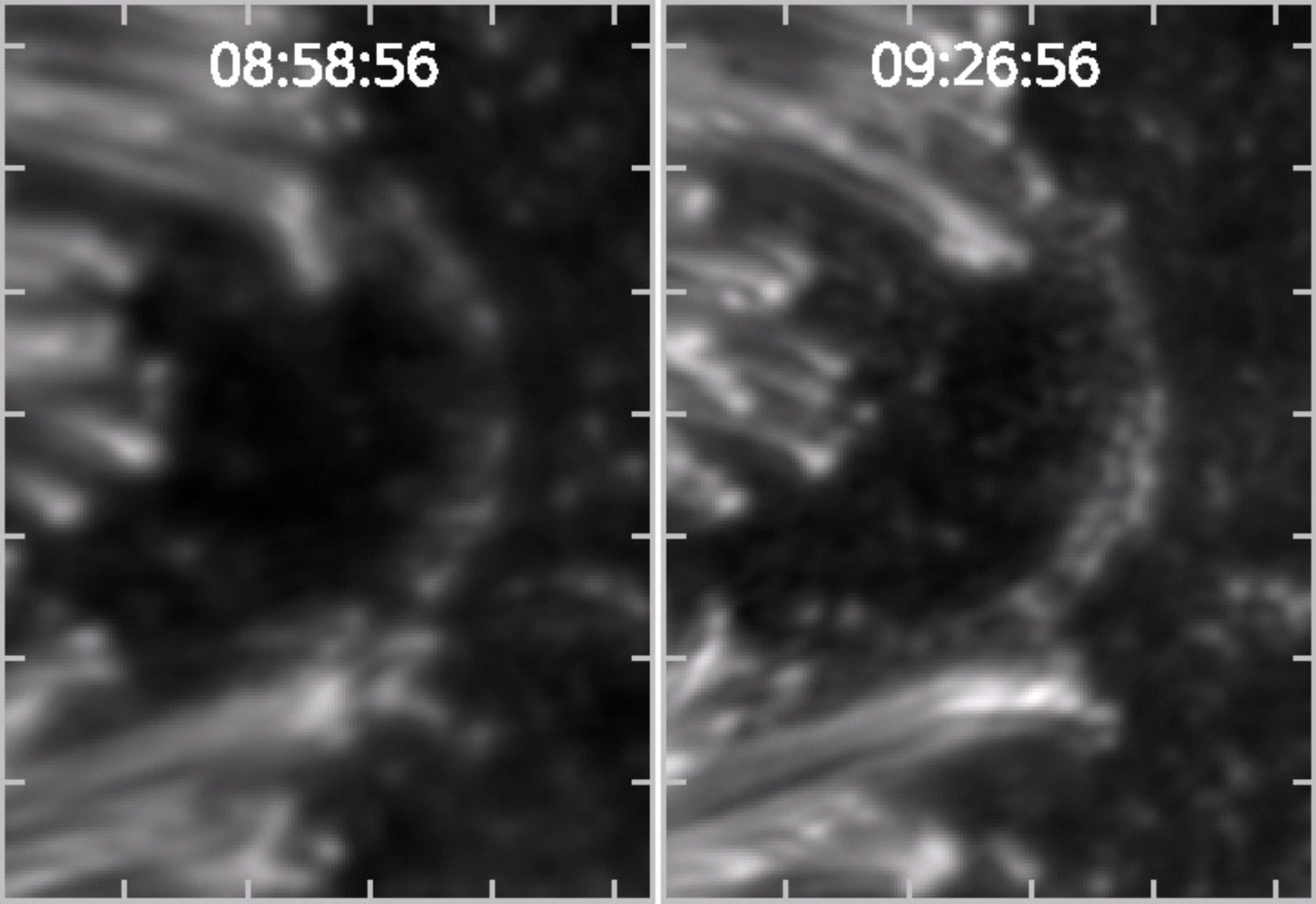}
   \caption{Most prominent chain of umbral dots in the studied umbra, located near label A in
   Fig.~\ref{FigBestImage} (FOV is 5.4~$\times$~7.4~Mm). The left panel shows the chain composed of
   several typical UDs along a curved line at 08:58:56 UT. The right panel shows the same chain
   28~min later. At this later time the lower half of the chain looks similar to a narrow light bridge.
   The typical \Index{dark lane} is clearly recognizable.}
   \label{FigUdChainVersions}
   \end{figure}

   The data clearly show that bright UDs often form chains. The most prominent chains often start
   from a penumbral filament and UDs are found to move along the chain until they dissolve. The
   horizontal motion of the UDs is preferentially along the chain and is directed from the ends
   of the chain to its center, where the UDs disappear. Fig.~\ref{FigUdChainVersions} displays the
   most prominent chain of UDs in the observed umbra. The length of this chain is about 3200~km and
   it is about 350~km wide. In the left panel, from 08:58:56 UT, the chain appears as a simple
   succession of UDs. One can see the same region in the right panel 28~minutes later. The appearance
   of the chain changed and now the lower part of the chain looks similar to a narrow light bridge.
   The typical \Index{dark lane} of a LB can be seen clearly, even if we degrade the image quality to the
   lower level of the left panel by convoluting the image with a point spread function of a circular
   pupil (not shown). This degradation was carried out to compensate for the higher spatial resolution
   of the later image. High spatial resolution is demonstrated by the presence of dark-cored penumbral
   filaments \citep[see][]{Langhans2007}, which are generally observed only at the highest resolution
   in the blue. The presence of the \Index{dark lane} is an indication that the chain eventually evolved into
   a fully developed LB some hours after the end of the recording \citep{Katsukawa2007}.

\begin{figure}
   \centering
   \includegraphics[width=\linewidth]{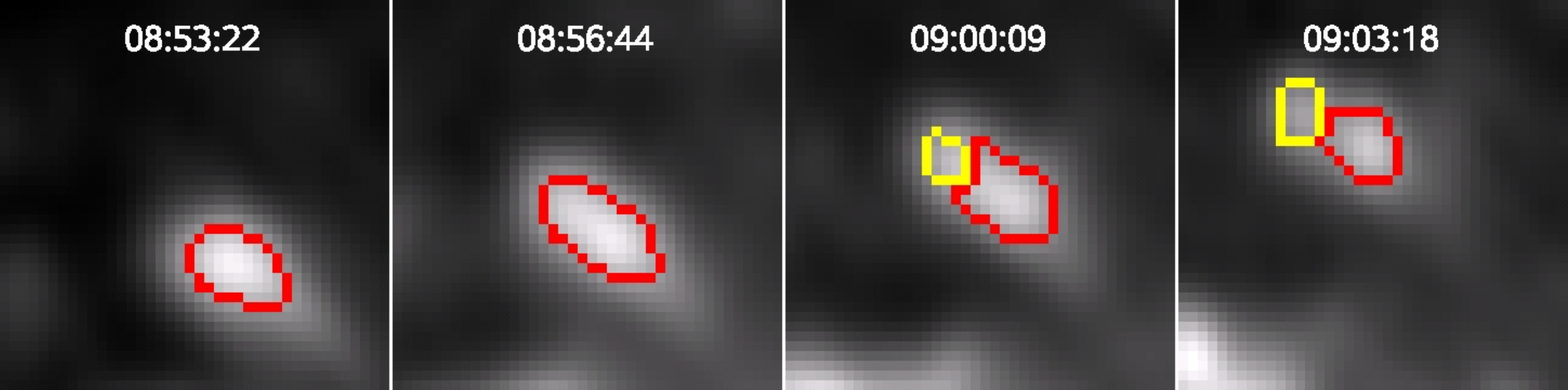}
   \caption{Splitting of an umbral dot near the position B in Fig.~\ref{FigBestImage}
   (FOV is 1.2~$\times$~1.2~Mm). The red and yellow lines are the UD boundaries as detected by the
   method explained in the main text.}
   \label{FigUdSplit}
   \end{figure}

\begin{figure}
   \centering
   \includegraphics[width=\linewidth]{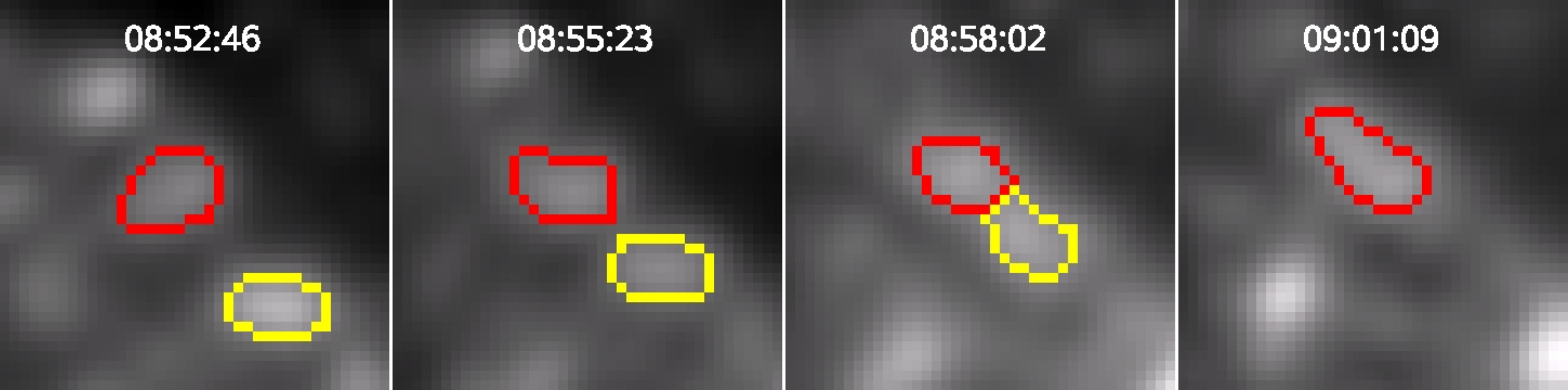}
   \caption{Merging of two umbral dots near position C in Fig.~\ref{FigBestImage}
   (FOV is 1.2~$\times$~1.2~Mm).}
   \label{FigUdMerge}
   \end{figure}

   In some exceptional cases we observe the splitting of a single UD into two parts that continue their
   life as independent UDs. Such a splitting is displayed in Fig.~\ref{FigUdSplit}. It concerns a UD
   located close to the \Index{penumbra}, near point B in Fig.~\ref{FigBestImage}. The
   opposite case, the merging of two UDs into a single one, can also be observed in a few rare cases:
   See Fig.~\ref{FigUdMerge} for an example. Nearly identical phenomena in the temporal evolution of
   penumbral grains were observed by \citet{Hirzberger2002}.

\begin{figure}
   \centering
   \includegraphics[width=\linewidth]{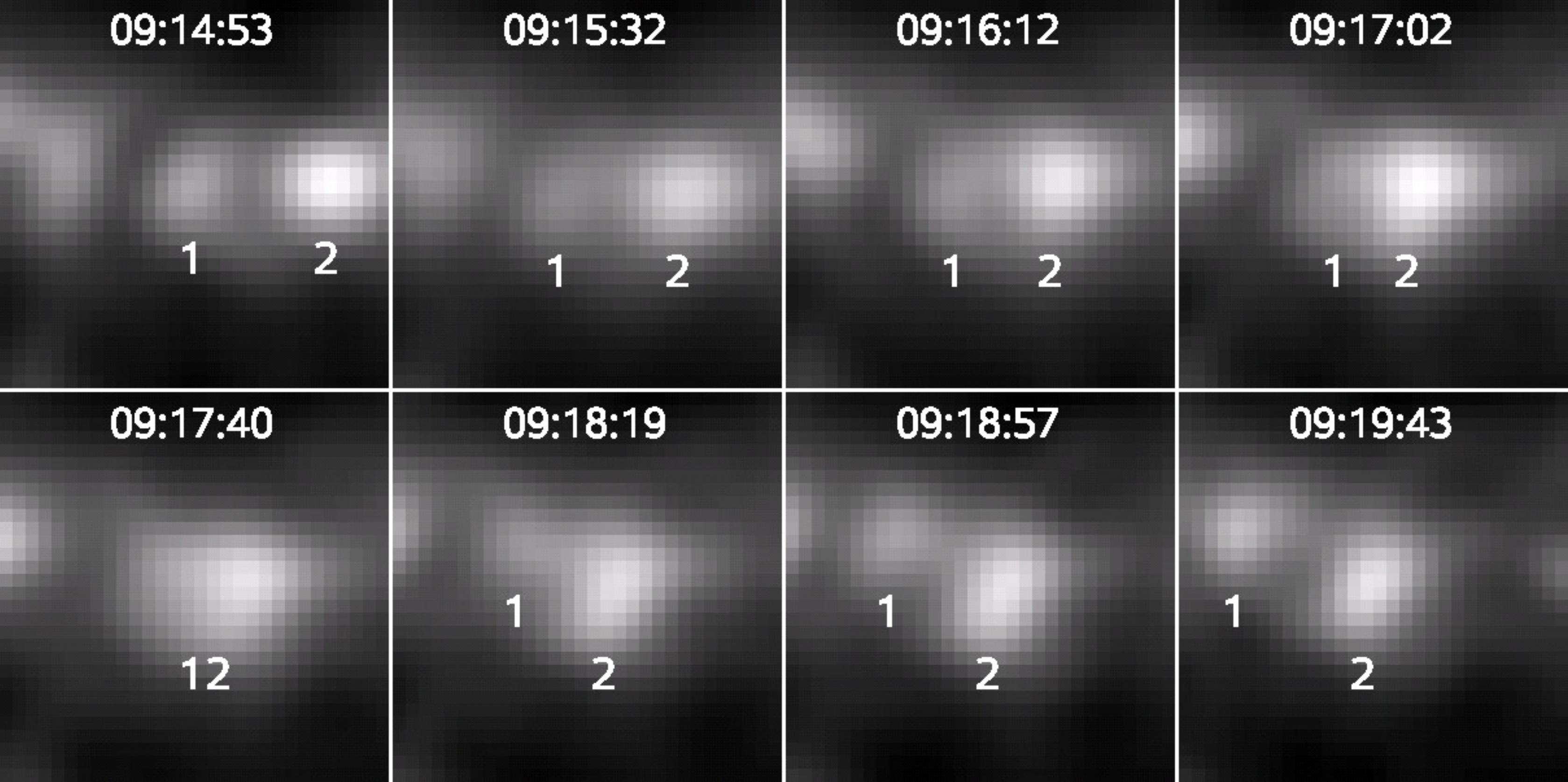}
   \caption{Elastic collision of two umbral dots. The sequence of events is seen near the position marked
   D in Fig.~\ref{FigBestImage} (FOV is 0.9~$\times$~0.9~Mm). UD~2 moves towards UD~1
   and merges with it, before separating from it again.}
   \label{FigUdNudge}
   \end{figure}

   An interesting sequence of events is illustrated in Fig.~\ref{FigUdNudge}. First the merging of two
   UDs can be observed followed immediately afterwards by the splitting of the resulting, unified UD
   into two new UDs. The sequence looks similar to an elastic impact in a Newton pendulum consisting of
   two balls: The nearly motionless UD~1 is hit by the moving UD~2. The impact of the two UDs brings
   the second UD to a standstill while the first UD starts to move roughly in the direction of the first UD.
   Note that this is simply an empirical description and we do not propose that this is what physically
   happens (or that the unified UD breaks up into the same parcels of gas which united to form it).

   If two UDs come close to each other then the visual impression of a single UD exhibiting a
   \Index{dark lane} can occur, see last panel of Fig.~\ref{FigUdNudge} for an example. None of the UDs in our data
   set seems to stay in such a state for a significant fraction of the UD lifetime. Since the dark lanes as
   seen in the simulations of \citet{Schuessler2006} are visible for most of the UD lifetime, we conclude
   that we find no clear evidence for such dark lanes. This difference to the results of \citet{Bharti2007}
   and \citet{Rimmele2008} may stem from the different wavelengths of the analyzed data. The wavelength can
   have a remarkable effect on the detected fine structure \citep[e.g.][]{Zakharov2008}.

   The properties of a UD change along its trajectory. In order to assign a property to the entire
   trajectory we either average over all points along the trajectory (which is expressed by
   introducing an upper index "Mean") or we determine the maximum value reached by that parameter
   over all trajectory points (which is expressed by the upper index "Max"), e.g., $D^{Mean}$ means
   the UD diameter averaged along the trajectory, while $I_{Peak}^{Max}$ is the maximum value of all
   peak intensities along a UD trajectory (where the peak intensity is the largest intensity within
   the UD boundary at a given point in time).

\begin{figure}
   \centering
   \includegraphics[width=0.5\linewidth-1mm]{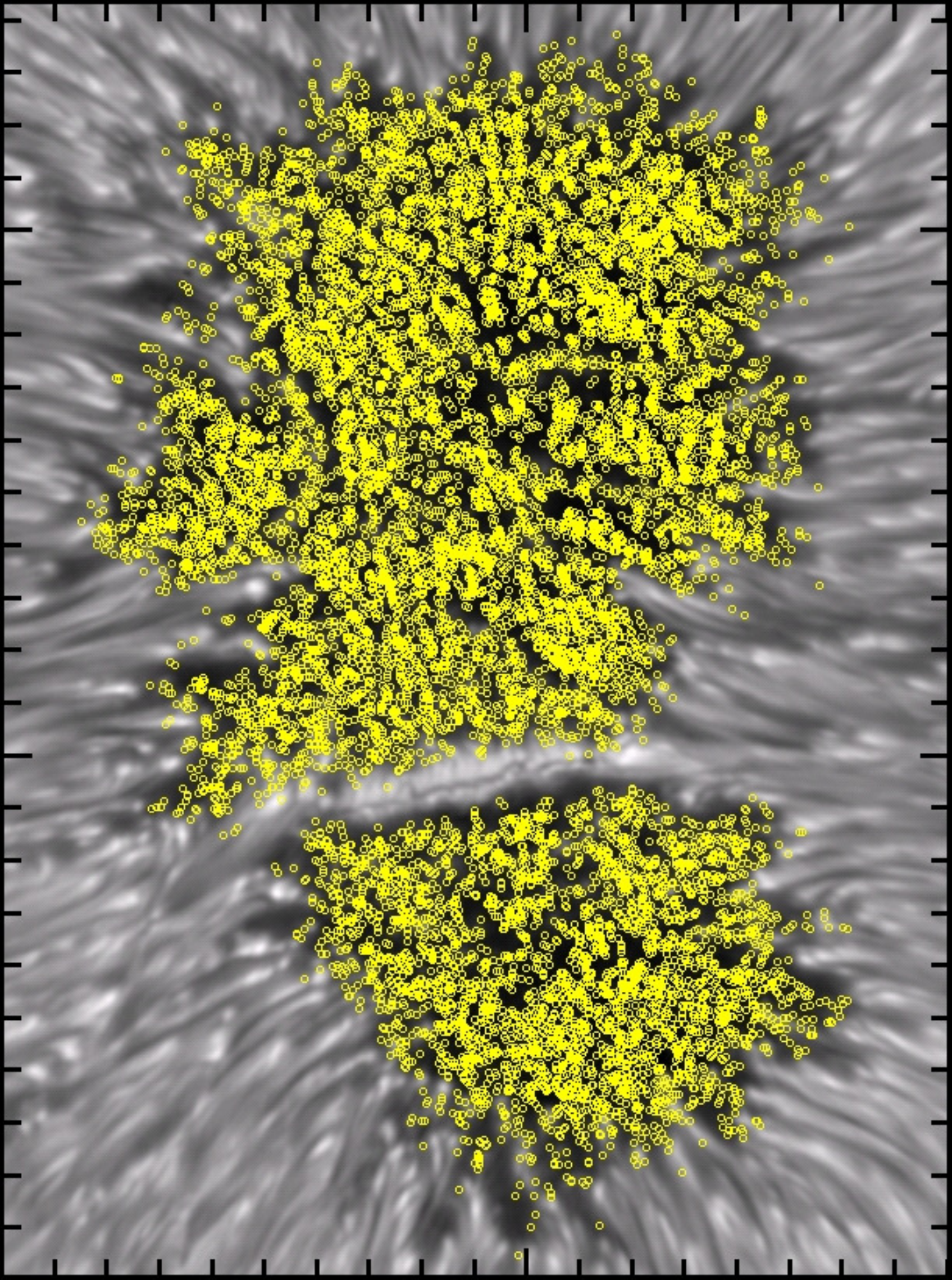}
   \includegraphics[width=0.5\linewidth-1mm]{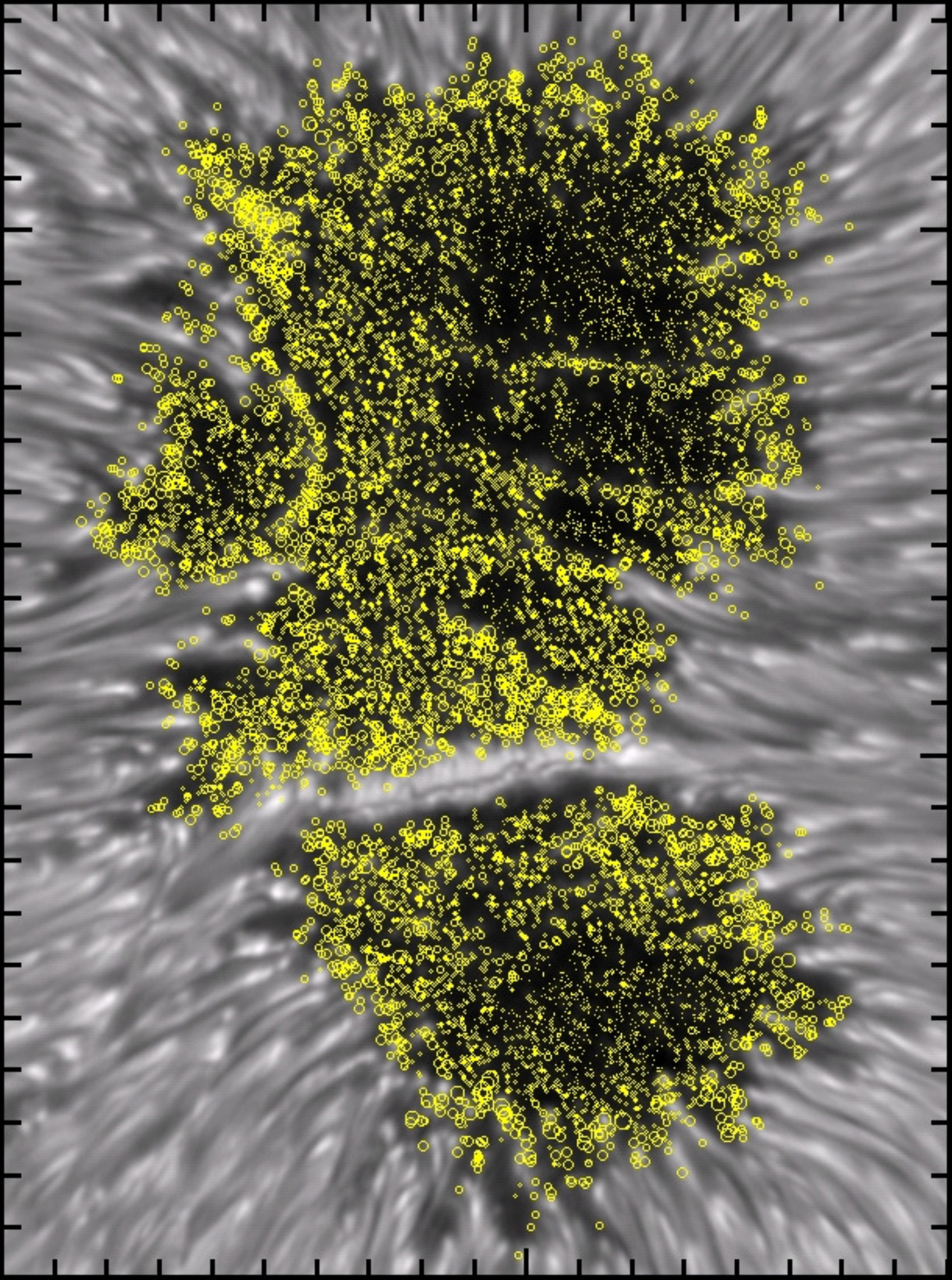}
   \caption{Spatial distribution of UD occurrence (left panel) and UD maximum brightness (right panel).
   Each UD is plotted only once along its entire trajectory, at the position of its maximum brightness.}
   \label{FigUdOverview}
   \end{figure}

   Fig.~\ref{FigUdOverview} shows the best image with all identified UDs marked by circles. The
   circles are centered on the positions of $I_{Peak}^{Max}$, i.e. the position of maximum intensity.
   In the left panel the circles have a constant radius, while their radii are proportional to $I_{Peak}^{Max}$
   in the right panel. As a result we get an impression of the spatial distribution of UD occurrence
   as well as of the spatial distribution of UD brightness. Obviously, there is hardly any part of
   the umbra which does not support umbral dots. Only very localized small voids are visible.
   The brightness distribution of UDs, however, is rather inhomogeneous, with clear concentrations
   of bright UDs and regions harboring mainly dark UDs (mainly in the upper right part of the
   upper umbra and in the lower part of the lower one). Since the UDs of a chain (like the chain close
   to label A in Fig.~\ref{FigBestImage} and shown in more detail in Fig.~\ref{FigUdChainVersions})
   appear to move along the chain and hardly in the direction perpendicular to it, there is often a narrow
   void directly beside the chain.

   \subsection{Quantitative properties}

\begin{figure}
   \centering
   \includegraphics[width=100mm]{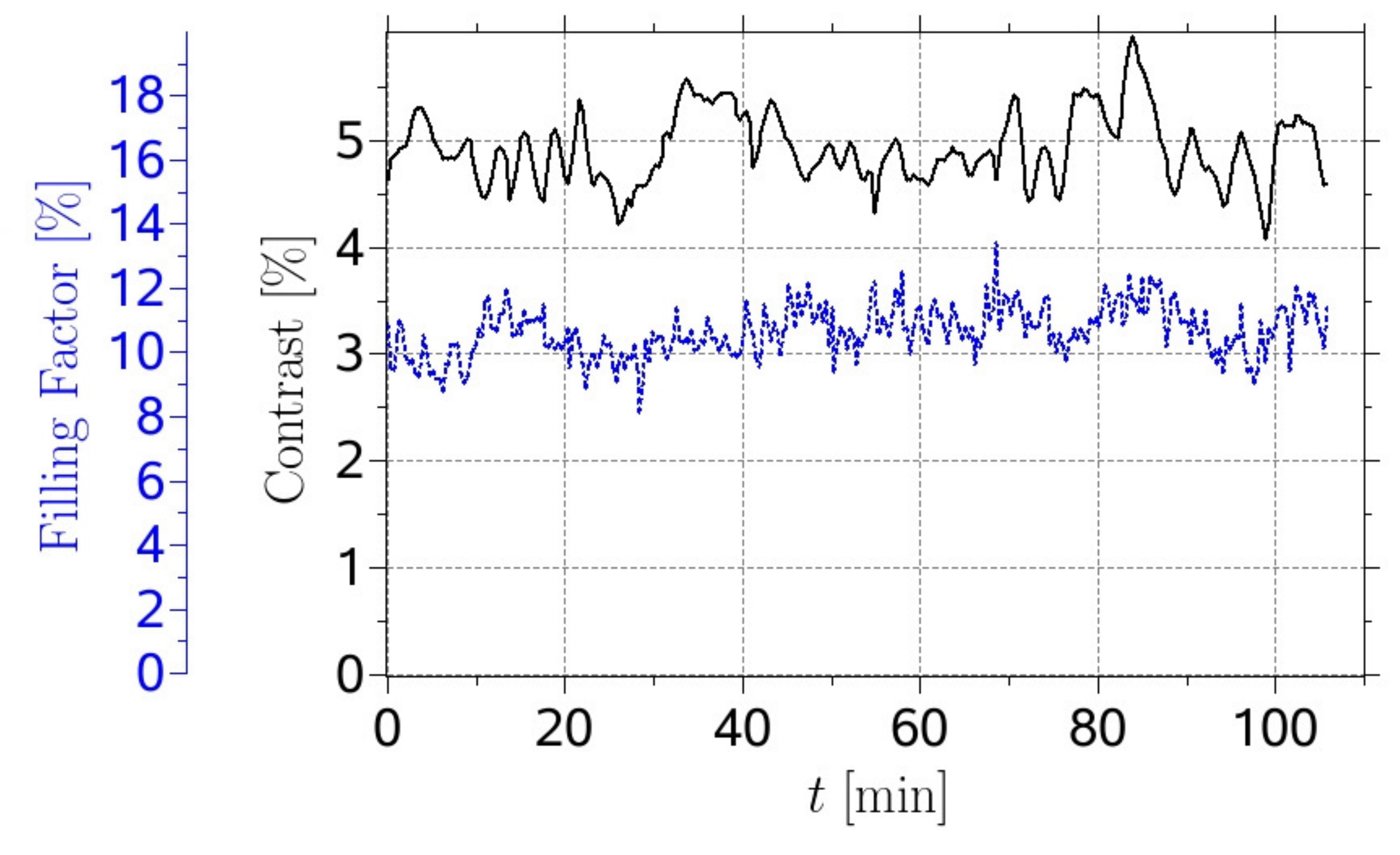}
   \caption{Dependence of UD \Index{filling factor} (dotted blue line) and granulation contrast
   (solid black line) on the interval of time, $t$, since the first image of the series was recorded.}
   \label{FigFillingFactor}
   \end{figure}

   The \Index{filling factor}, i.e. the sum of all UD areas relative to the total umbral area, is correlated
   to the image quality. Essentially, the filling factor is constant over the entire period of
   observations (see Fig.~\ref{FigFillingFactor}), which is important for the later determination of the
   time dependence of their properties. On average we determine a value of about 11\,\%. Fig.~\ref{FigFillingFactor}
   shows also the image contrast from an undisturbed granulation area outside the sunspot which demonstrates the
   high homogeneity of the image quality in our time series.

\begin{figure}
   \centering
   \includegraphics[width=85mm]{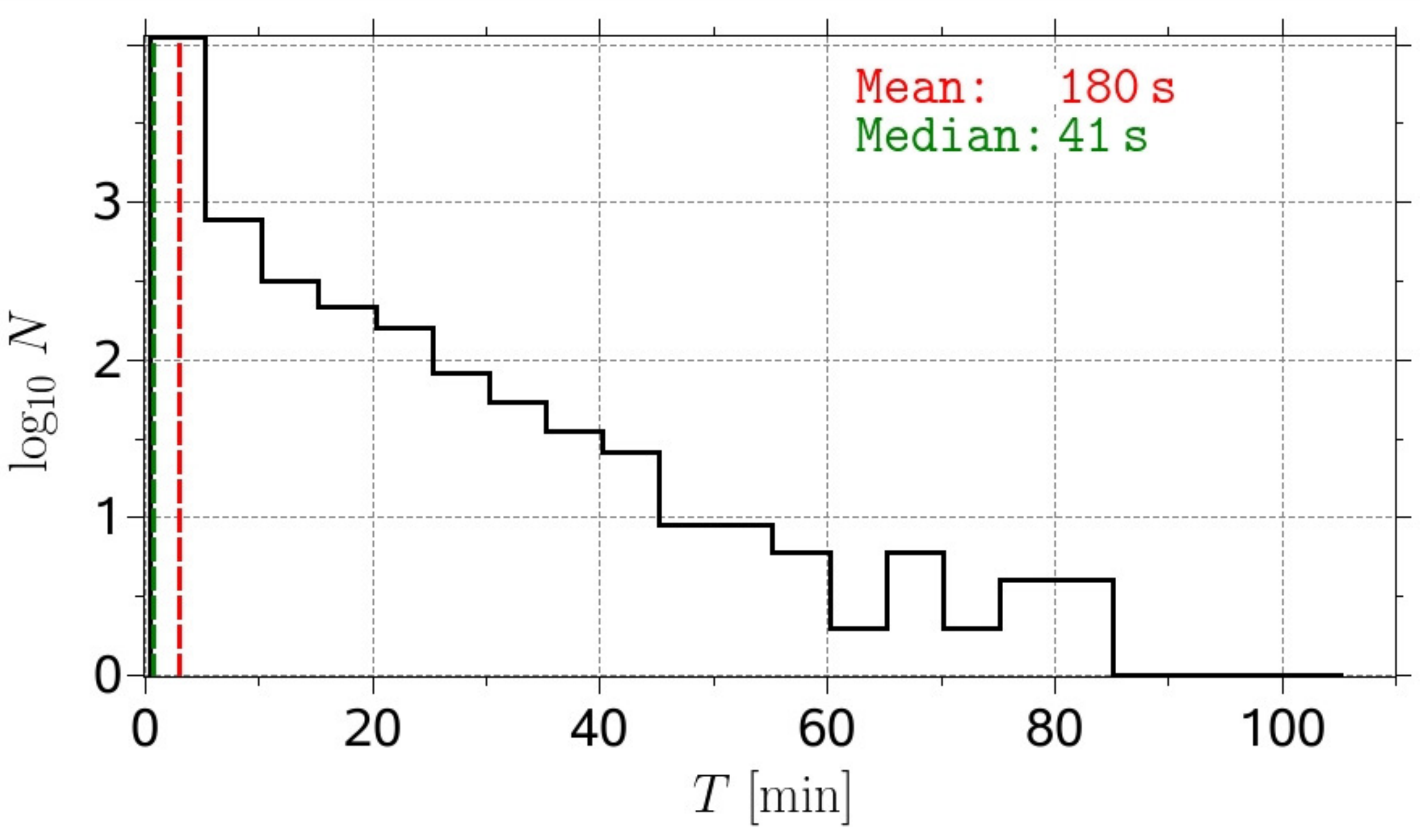}
   \caption{Histogram of lifetimes $T$ of all 12836 UD trajectories (solid line) and their mean and
   median value (dashed lines and text labels). The bin size is 300~s.}
   \label{FigHistLifetime}
   \end{figure}

   The histogram of the UD lifetimes is displayed in Fig.~\ref{FigHistLifetime} with a logarithmic $y$ axis.
   One can see that most of the UDs live for a short time. These short-lived UDs move over
   short distances which leads to physically nonsensical velocities due to the discretization of lifetime
   and distance. Consequently, when discussing trajectories
   and velocities of UDs we only consider the 2899 trajectories of UDs with lifetimes greater than 150~s.
   The histogram is nearly linear for lifetimes between 5 and 60 minutes, which, due to the logarithmic
   vertical scale, suggests an exponential distribution of lifetimes. The excess of UDs with short
   lifetimes may partly be due to seeing. Also given in Fig.~\ref{FigHistLifetime} is the mean lifetime
   (180~s) and the median lifetime (41~s). If we consider only the 2899 trajectories of UDs with lifetimes
   greater than 150~s then we find a mean lifetime of 630~s and a median of 390~s. Note that 281 UDs are
   already present in the first image and 344 UDs are still present in the last image, whereas
   only one UD survives the whole sequence. Ignoring those 625 UDs reduces the mean lifetime
   from 180~s to 152~s. The median lifetime as well as the shape of the histogram do not change, because
   the number of incomplete UD trajectories is small compared to the total number of UDs. Thus we decided
   to neglect this effect and consider all 12836 UDs in the following text.

\begin{figure*}
   \centering
   \includegraphics[width=\linewidth]{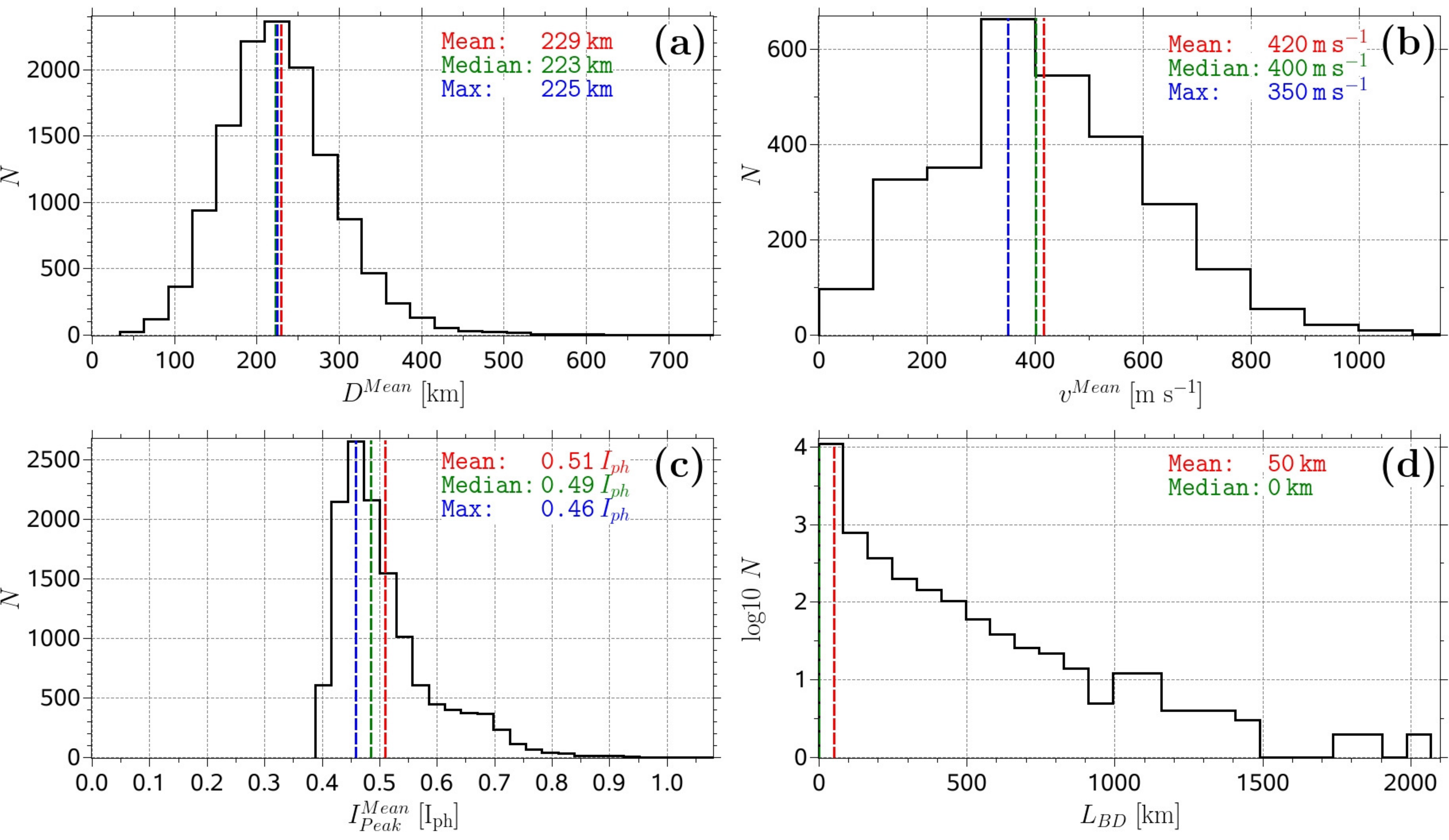}
   \caption{Histogram of mean diameters (a), mean horizontal velocities (b), mean peak intensities (c), and
   distances between birth and death position (d). (a), (c), and (d) are plotted for all 12836 UD trajectories
   and (b) for the 2899 trajectories of UDs that lived longer than 150~s. The location of the
   maximum, the mean, and the median of the distribution is indicated in each frame. The bin sizes
   are 30~km for (a), 100~m\,s$^{-1}$ for (b), 0.03~$I_{ph}$ for (c), and 90~km for (d).}
   \label{FigHistOverview}
   \end{figure*}

\begin{figure}
   \centering
   \includegraphics[width=100mm]{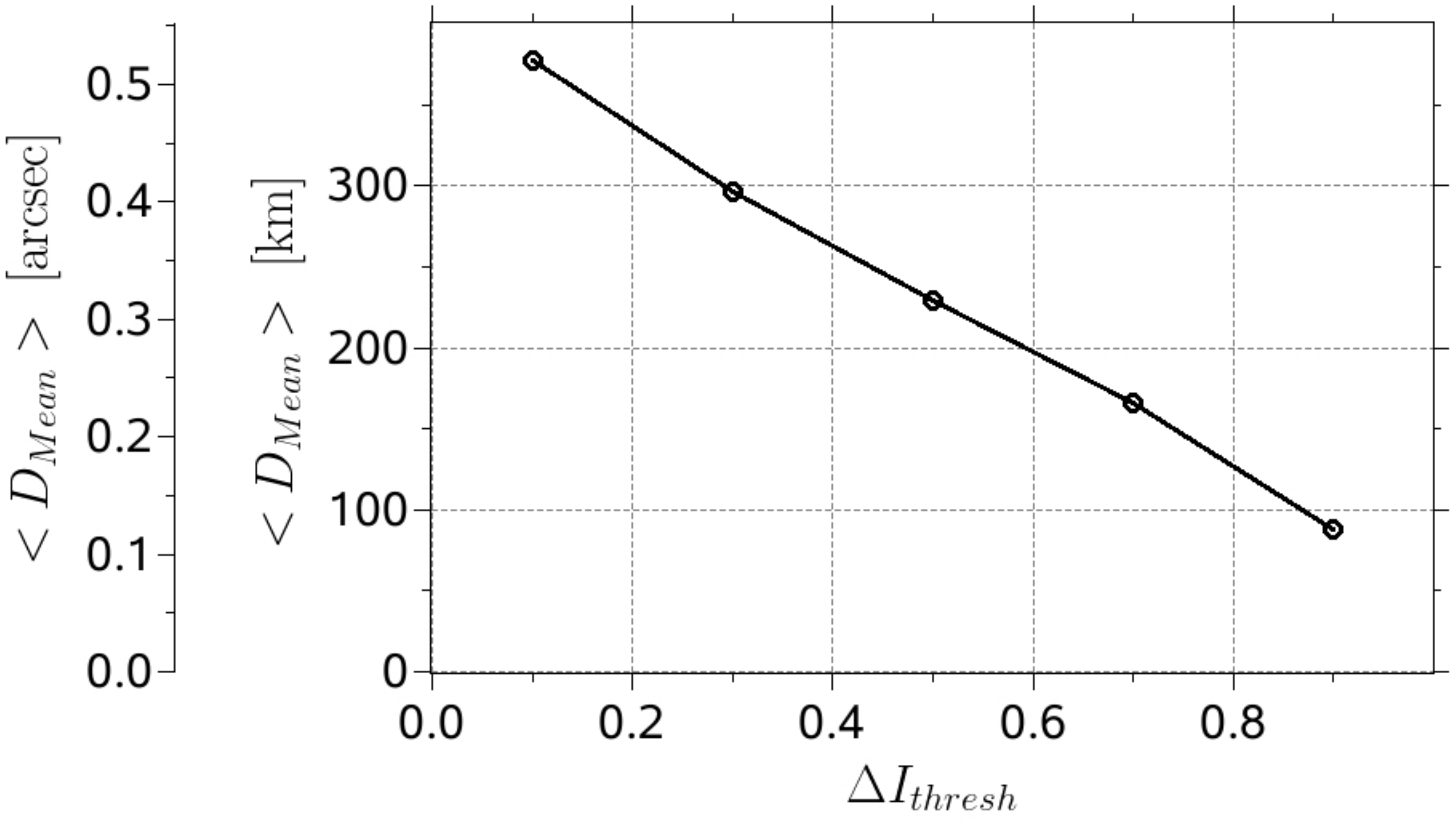}
   \caption{Mean UD diameter averaged over all trajectories as a function of the intensity threshold
   $\Delta I_{thresh}$ (defined in main text) that is used to determine the UD boundaries
   (see also Fig.~\ref{FigCuttingPixels} and its explanation).}
   \label{FigDiameterVsBoundThresh}
   \end{figure}

   The histogram of the mean UD diameters is plotted in Fig.~\ref{FigHistOverview}~a.
   The mean UD diameters vary between 50 and 750~km. The UDs have a mean diameter of around
   229~km and 95\,\% of the UDs are spatially resolved at our \Index{diffraction limit} of 130~km.
   \citet{Sobotka2005} found 175~km for the mean UD diameter. They determined
   the UD boundaries by finding all pixels with downward concavity, whereas we used all pixels whose
   intensity is greater than 50\,\% of the $I_{Peak}-I_{bg}$ range,
   $\Delta I_{thresh}=(I_{thresh}-I_{bg})/(I_{Peak}-I_{bg})=0.5$. As one can see in
   Fig.~\ref{FigDiameterVsBoundThresh} we would also find a mean UD diameter of 175~km if a
   threshold of about $\Delta I_{thresh}=0.67$ would be used. Nevertheless, we prefer to continue
   our analysis with a value of 0.5 in analogy to the FWHM definition. Clearly, the employed threshold
   influences the \Index{filling factor} as well, roughly quadratically. Remarkably, the histogram is nearly
   symmetric, which supports the conclusion that most UDs have been resolved.

   The histogram of the mean horizontal velocities, i.e. the quotient of trajectory length and lifetime,
   is plotted in Fig.~\ref{FigHistOverview}~b and exhibits a broad distribution from 0 to more than
   1~km\,s$^{-1}$ with a significant maximum at 350~m\,s$^{-1}$. The velocity distribution is slightly
   asymmetric with a small tail to higher velocities. Fig.~\ref{FigHistOverview}~c shows the histogram of
   the mean peak intensities, i.e. the mean of all peak intensities of the points along the trajectory.
   All intensities are normalized to the mean intensity of the quiet photosphere ($I_{ph}$). Just 3
   of the 12836 UDs reach a brightness greater than that of the quiet Sun, whereas most of the UDs
   are about half as bright as the quiet photosphere. The distribution is asymmetric, with a tail to
   higher intensities. Note that these brightnesses are strongly wavelength dependent and cannot
   be easily compared with values published in the literature \citep{Solanki2003}. Lastly, the
   histogram of the distances between the UD's birth and death position ($L_{BD}$) is given in
   Fig.~\ref{FigHistOverview}~d. With increasing distance the number of UDs decreases exponentially,
   so that only a few UDs travel over long distances in their life, most UDs do not leave the vicinity
   of their birth position. The maximum observed birth-death distance is 2~Mm, about 20\,\% of the upper umbra's
   diameter of roughly 10~Mm.

\begin{figure}
   \centering
   \includegraphics[width=\linewidth-1mm]{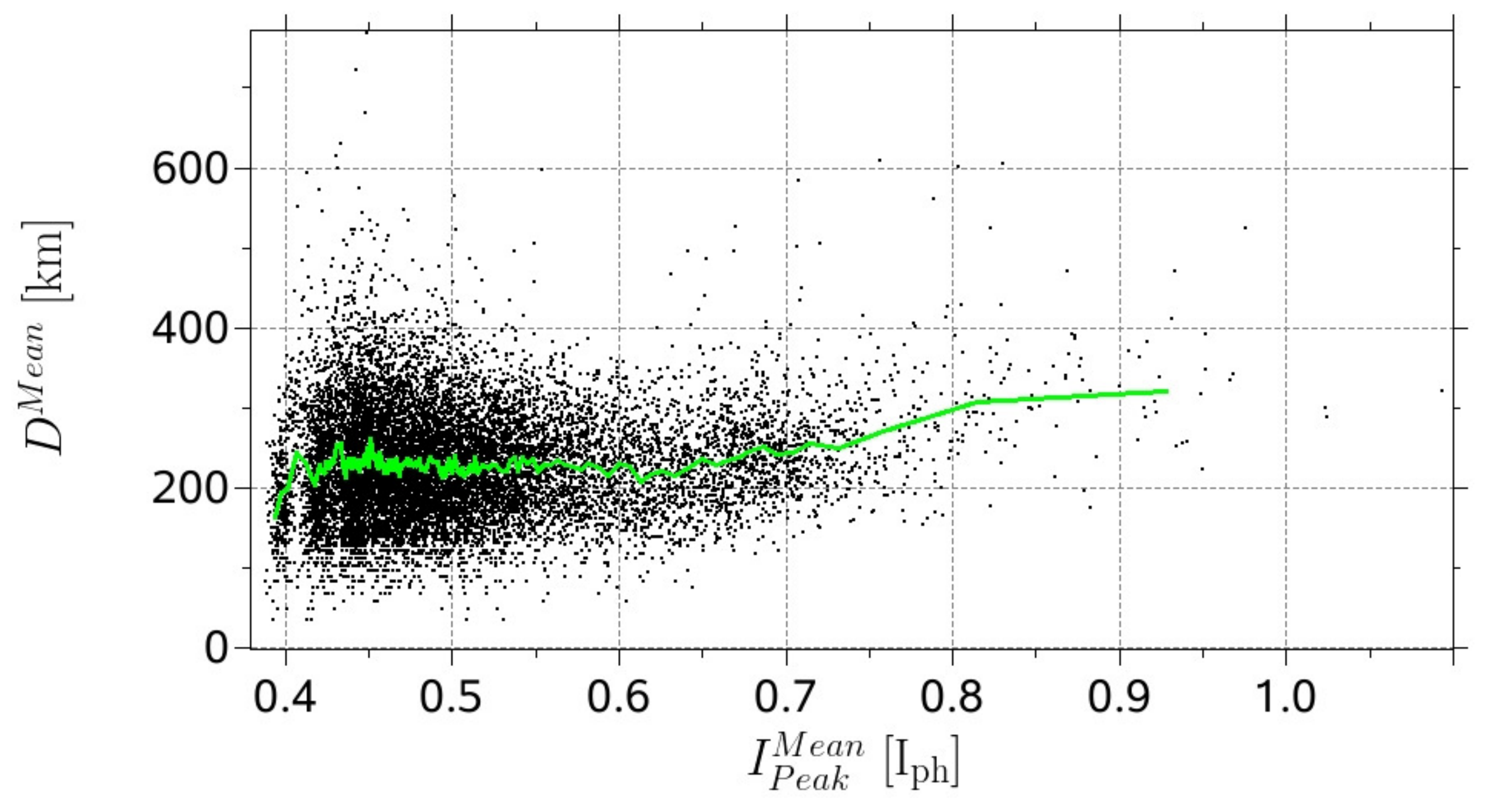}
   \caption{Scatterplot of mean UD diameter versus mean peak intensity. The solid green line connects binned values.}
   \label{FigScatterMeanDiameterVsMeanPeakIntensity}
   \end{figure}

   A UD size versus UD peak intensity scatterplot (Fig.~\ref{FigScatterMeanDiameterVsMeanPeakIntensity})
   reveals that there is a weak correlation between UD size and brightness. The brightest UDs are
   not the biggest ones and large UDs are not the brightest ones. The solid green line connects binned values
   (obtained by averaging 100 data points with similar intensities) and shows that bright UDs are on
   average a bit larger than small ones. The relation found by \citet{Tritschler2002} is qualitatively
   confirmed, although they only considered UD intensities in individual snapshots while we tracked the
   temporal development of the UDs over their lifetimes.

\begin{figure}
   \centering
   \includegraphics[width=\linewidth-1mm]{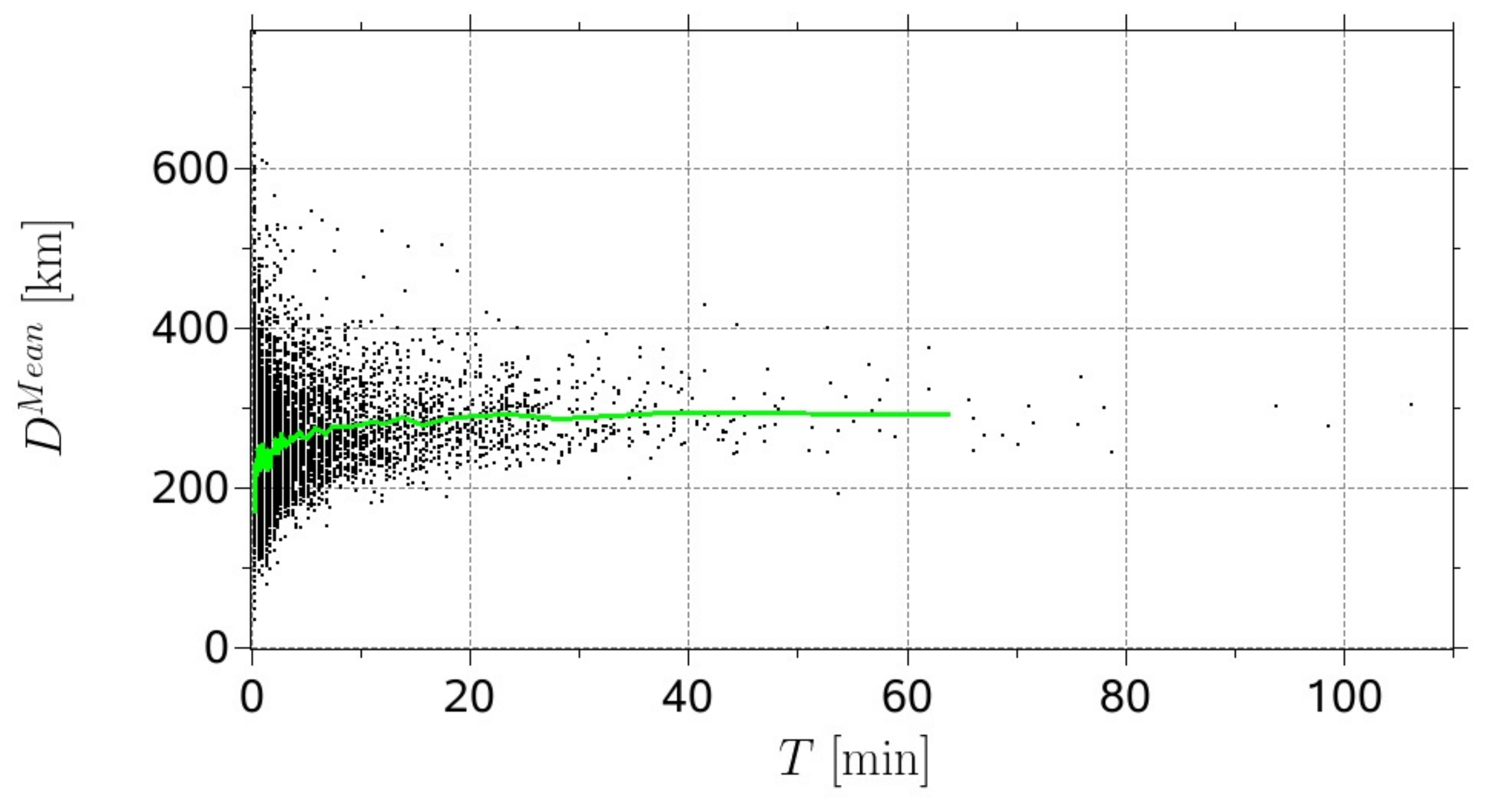}
   \caption{Scatterplot of mean UD size versus lifetime. The solid green line connects binned values.}
   \label{FigScatterMeanDiameterVsLifetime}
   \end{figure}

   The relation between a UD's mean size (i.e. the size averaged over the lifetime) and its lifetime is
   plotted in Fig.~\ref{FigScatterMeanDiameterVsLifetime} (the green curve is obtained after binning over
   100 data points). The binned values show an increase in $D^{Mean}$ with increasing lifetimes for
   short-lived UDs. For the longer lived ones size and lifetime do not correlate. The UD sizes scatter
   more for short lifetimes. All long-lived UDs are of intermediate size of around 290~km. The large,
   short-lived UDs are all present in the first image of the time series, so that their lifetime would
   actually be larger if we had started our observation earlier.

   In the literature we often find a separation into two UD classes, e.g. \citet{GrossmannDoerth1986}
   find a difference between peripheral and central UDs, i.e. between UDs that are born
   close to the umbra-penumbra boundary and UDs that are born deep in the umbra, whereas
   \citet{Hartkorn2003} and \citet{Sobotka1997b} distinguish bright and dark UDs. We use
   different properties to find reasonable distinctions between types of UDs. E.g. from now on
   a given UD is called a peripheral UD (PUD) if the UD's birth position is closer than 400~km
   to the umbral boundary, otherwise it is termed a central UD (CUD). (The selected
   threshold comes from the histogram of the distances between the UD's birth position and the
   umbral boundary (not shown) which shows a maximum at around 400~km.) Alternatively, if the
   distance traveled between birth and death position is larger than 750~km then we call it
   a mobile UD, otherwise a stationary UD. (We plotted the trajectories of all UDs
   whose $L_{BD}$ was greater than a threshold, which was determined by starting at a
   small value and increasing it step by step. We stopped at 750~km which is the
   smallest $L_{BD}$ at which no trajectories occurred anymore in the central part of the umbra.)
   The aim here is not to separate UDs into distinct
   classes by a single property, e.g. a histogram of $L_{BD}$ (Fig.~\ref{FigHistOverview}~d)
   does not show two peaks, even if restricted to long-lived UDs. However, as we shall see below
   the most mobile UDs are formed near the penumbra, while the least mobile ones are mainly
   formed deep in the umbra. Such a distinction may help to guide theory towards a better
   understanding of the origin and evolution of UDs with different properties and at different
   locations.

   \begin{figure*}
   \centering
   \includegraphics[width=0.5\linewidth-1mm]{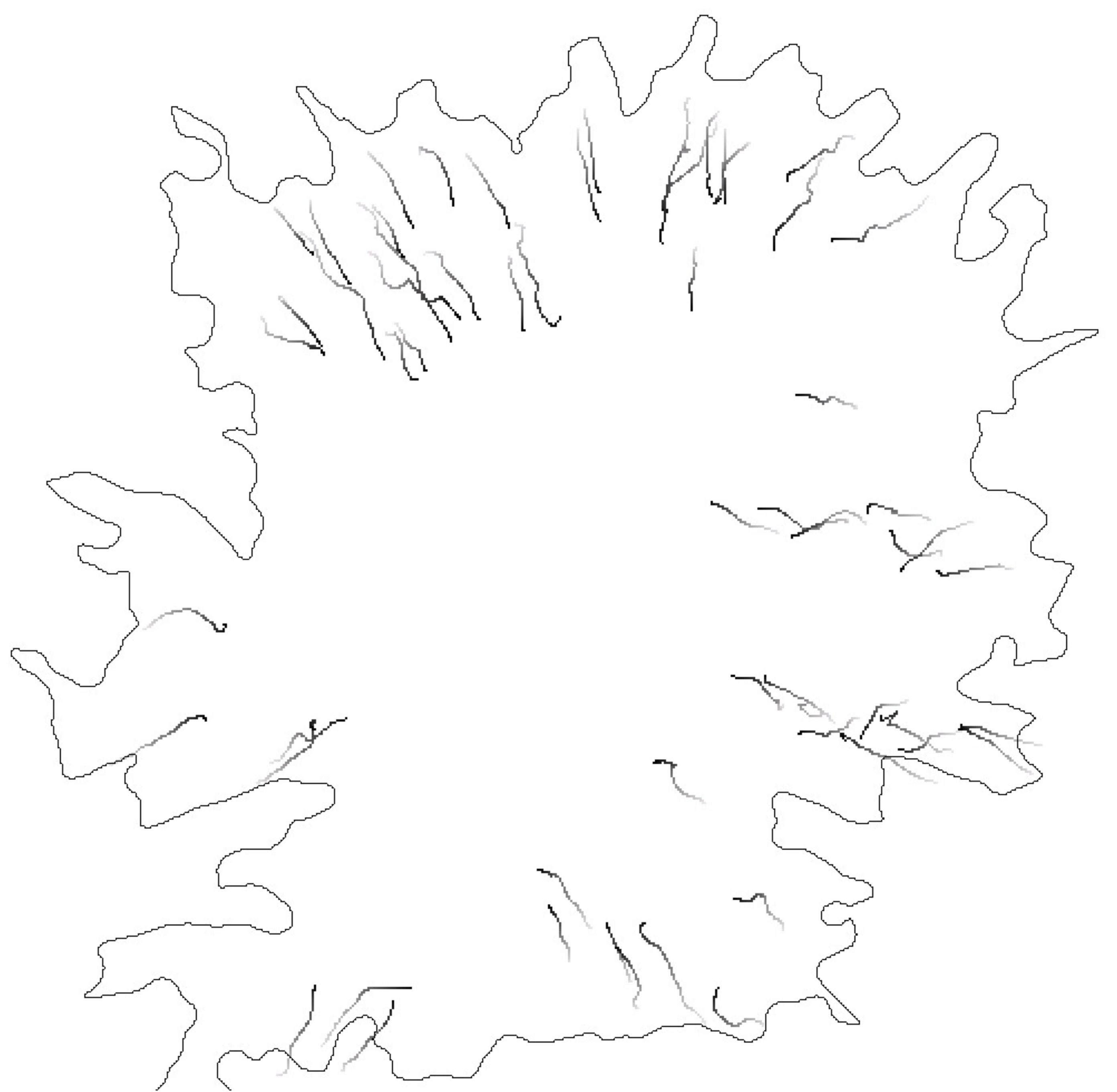}
   \includegraphics[width=0.5\linewidth-1mm]{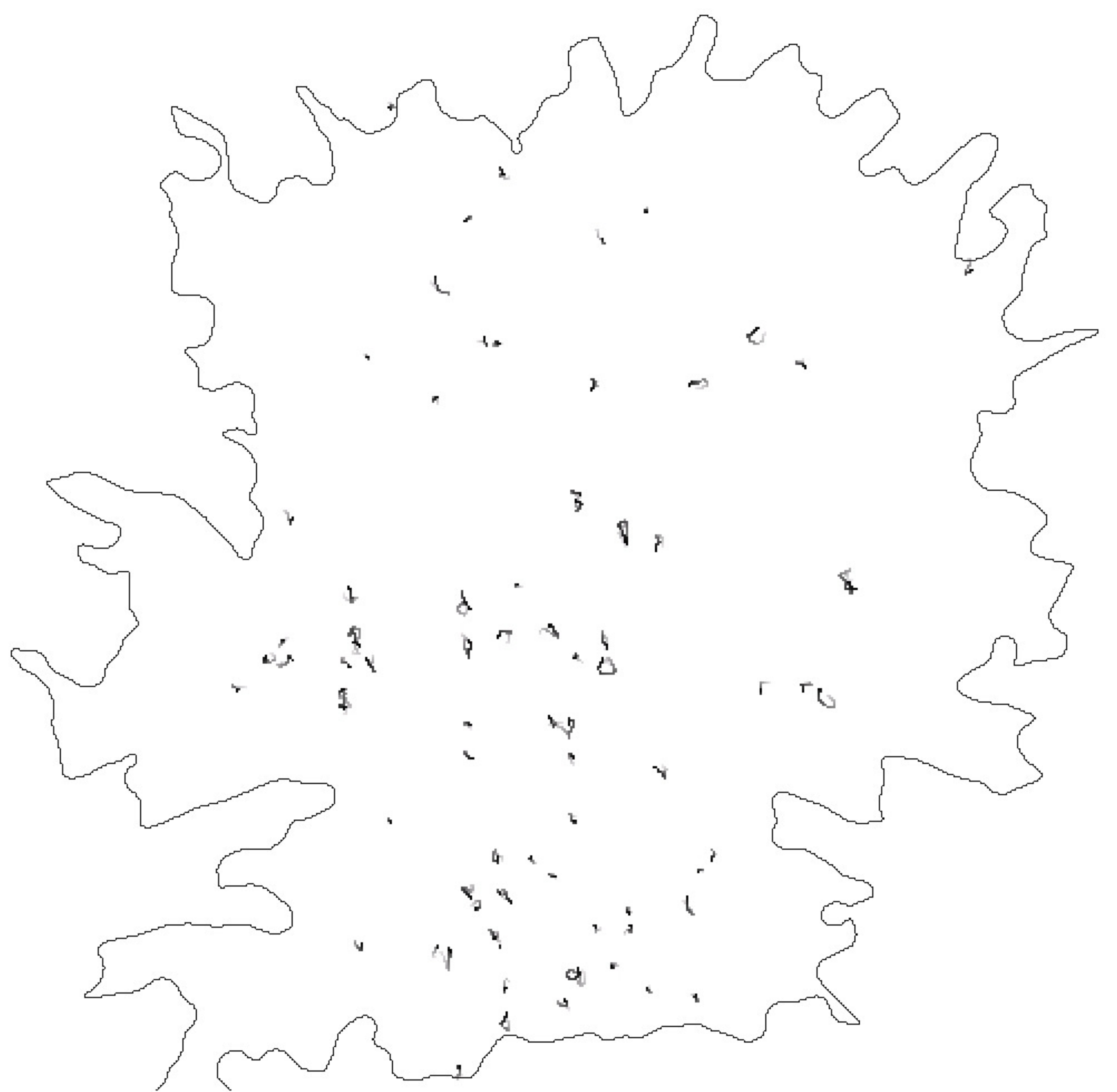}
   \caption{UD trajectories in the upper umbra. The bright ends of the trajectories denote the positions
   of the UD's birth and the dark ends show the position at death. The black contour line corresponds to the
   umbral boundary as detected in the best quality image. The left panel shows all UDs whose distance between
   the birth and the death position was greater than 750~km, the right panel shows all UDs with a birth-death
   distance smaller than 150~km whose lifetime was greater than 800~s.}
   \label{FigUdTrajectoriesBirthDeathDist}
   \end{figure*}

   The strong concentration of bright UDs near the umbral boundary (or along proto-light bridges)
   is already clear from Fig.~\ref{FigUdOverview} (right panel). Fig.~\ref{FigUdTrajectoriesBirthDeathDist}
   shows a separation by birth-death distance. The left panel displays the longest UD trajectories, i.e.
   only the mobile UDs are shown. These UDs are also relatively long-lived. (Smallest lifetime of this UD
   class, that contains 85 UDs, is 15~min.) The right panel shows trajectories of long-lived UDs with a
   small birth-death distance. A clear separation by the birth position is readily identifiable. Almost
   all trajectories with a large birth-death distance start close to the umbra-penumbra boundary
   and these UDs move nearly radially into the umbra. A visual inspection of the movie of the reconstructed
   time series of images shows that many of these UDs are former penumbral grains that broke away from the
   penumbra. In contrast, many of the UDs with a small birth-death distance are born in
   the umbral interior and move along a closed loop or jitter around their birth position. We cannot say
   if this jitter has a physical background or if it is caused by residual seeing-induced noise. As mentioned
   in section~\ref{Ud1DataAnalysis} we determined the umbral boundary individually for each image but we show only that
   corresponding to the best quality image as black contour line in Fig.~\ref{FigUdTrajectoriesBirthDeathDist},
   \ref{FigUdTrajectoriesDiameterAndPeakIntensity}, and \ref{FigLightbridge}. Consequently some trajectories
   (or parts of them) are outside the black contour line, although they are always inside the umbra at the time
   of their occurrence (see, e.g., the bottom-left corner of the left plot of
   Fig.~\ref{FigUdTrajectoriesBirthDeathDist}).

\begin{figure}
   \centering
   \includegraphics[width=100mm]{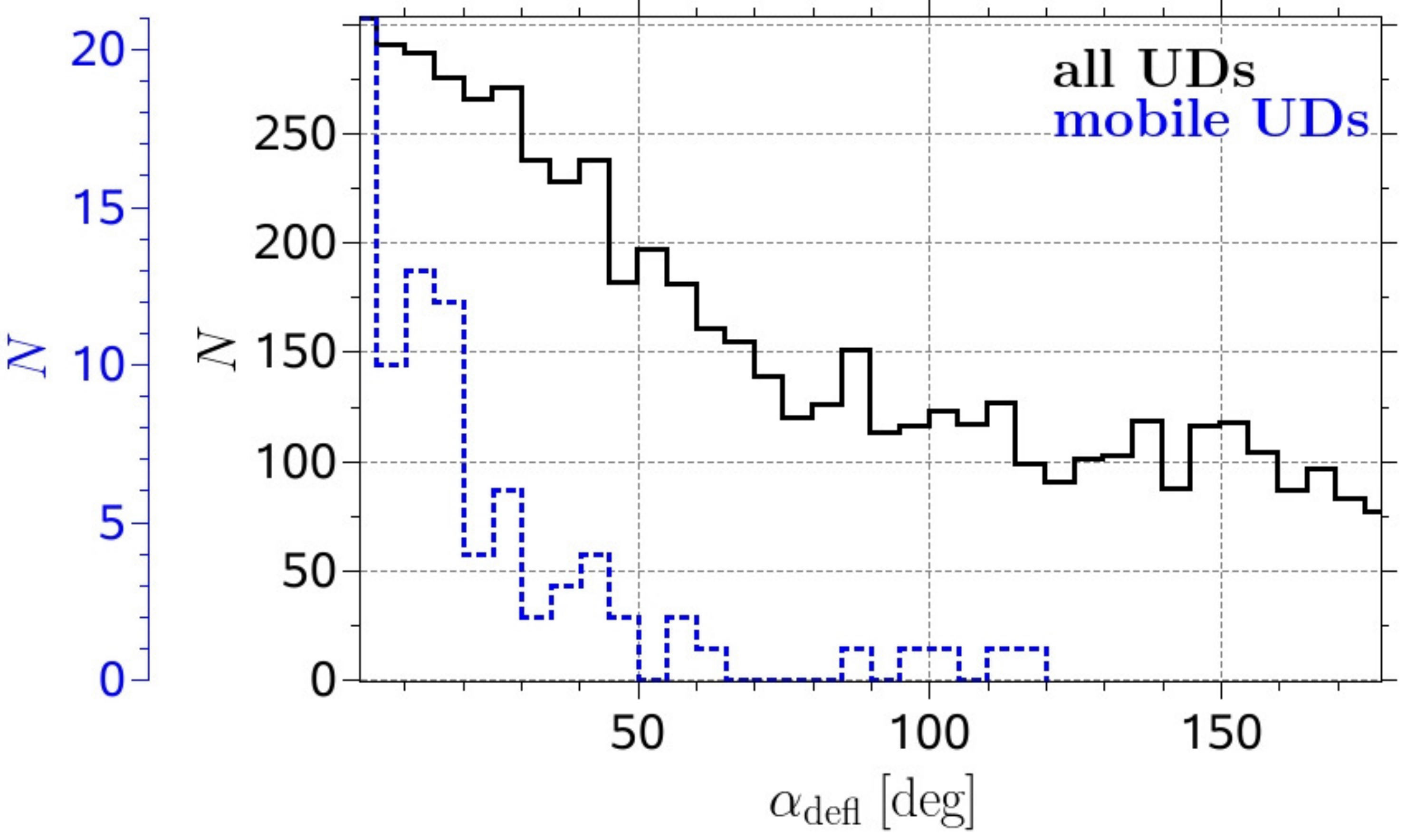}
   \caption{Histogram of radial deflection angles $\alpha_{\rm{defl}}$ of all UD trajectories (solid black line)
   and of the mobile UD trajectories (dotted blue line).}
   \label{FigHistRadialDeflection}
   \end{figure}

\begin{figure*}
   \centering
   \includegraphics[width=0.5\linewidth-1mm]{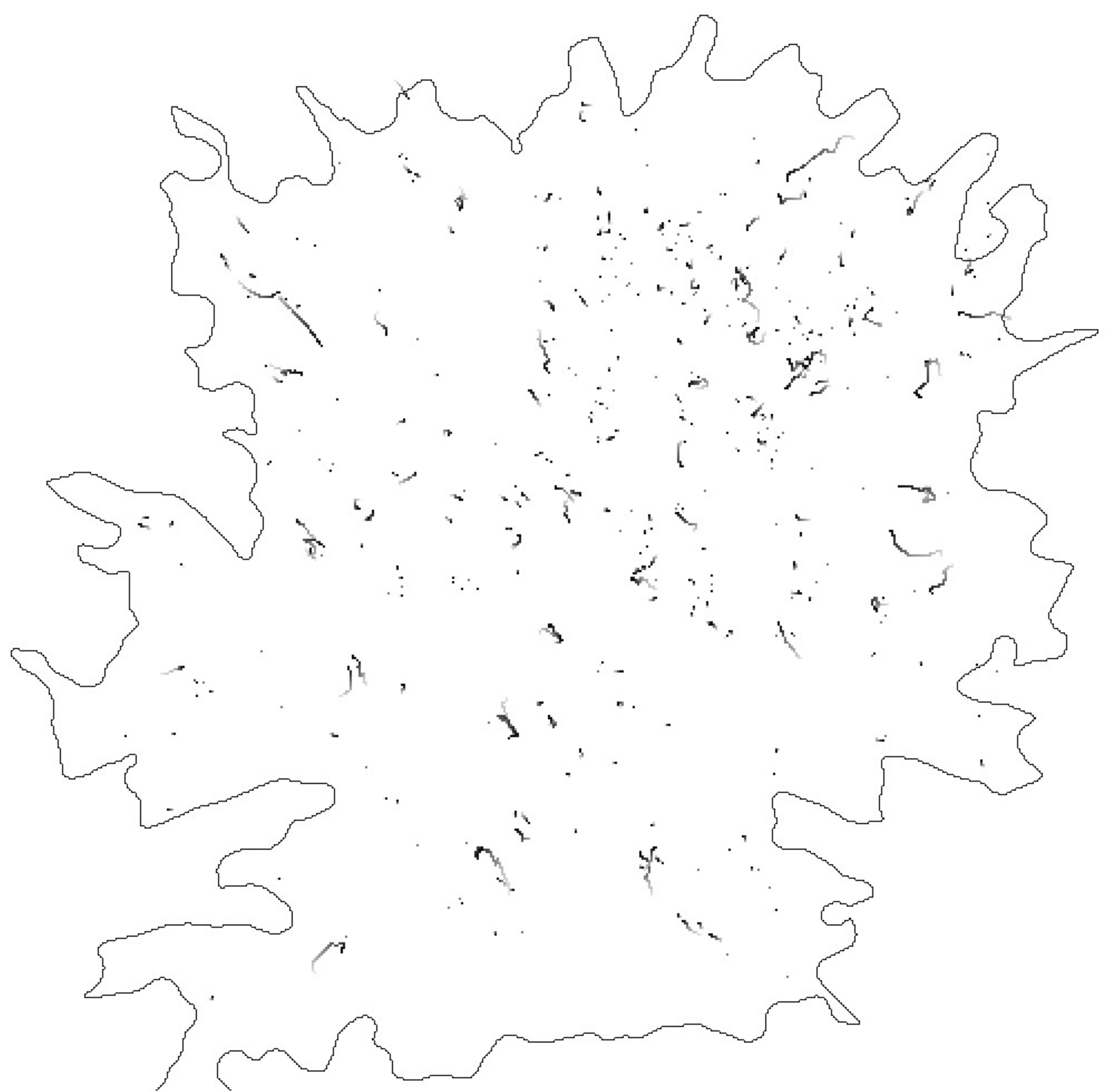}
   \includegraphics[width=0.5\linewidth-1mm]{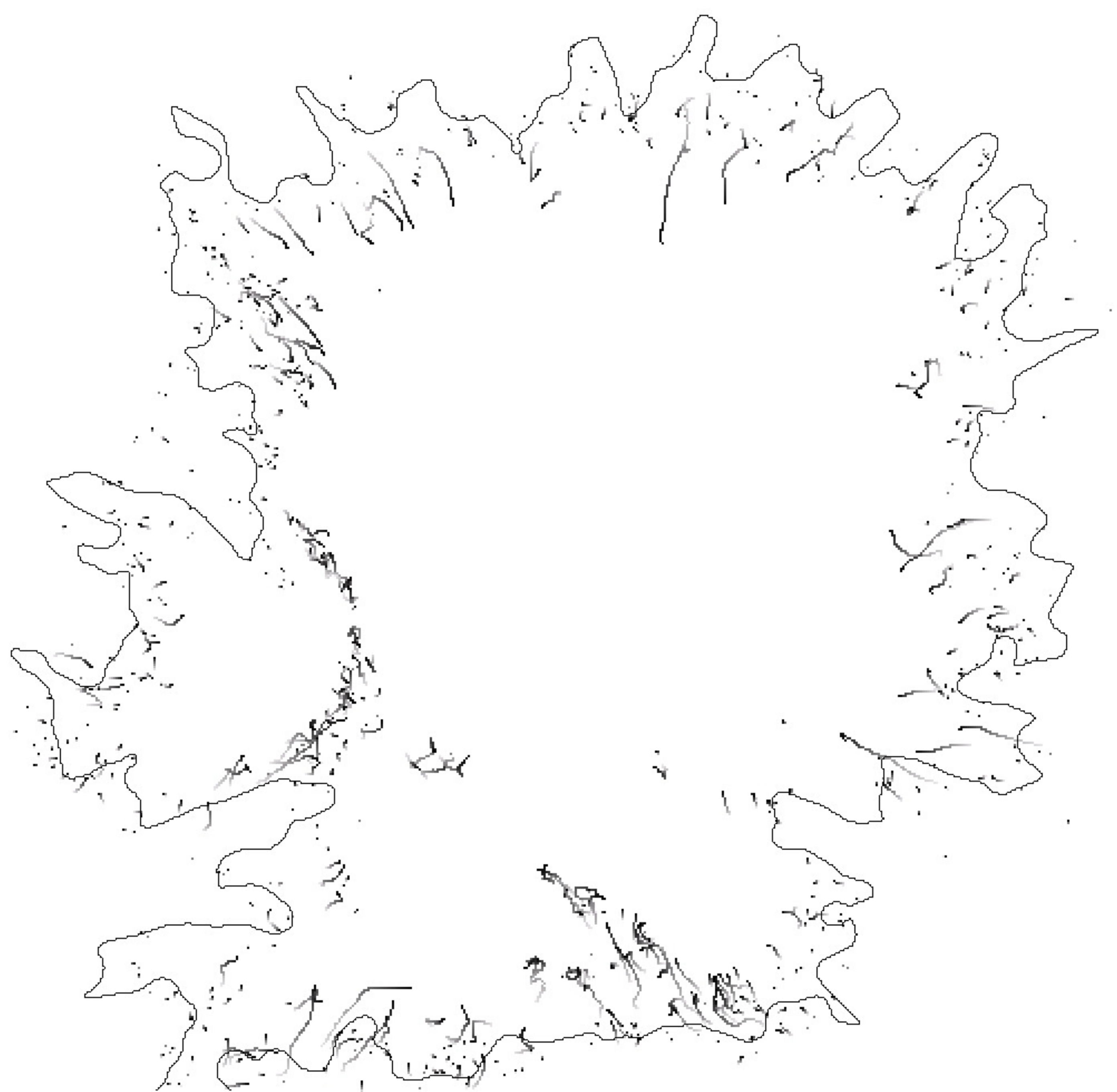}
   \caption{UD trajectories in the upper umbra. The left panel shows all UDs with mean diameter
   greater than 350~km and the right panel shows all UDs with a mean peak intensity greater than 0.65~$I_{ph}$.}
   \label{FigUdTrajectoriesDiameterAndPeakIntensity}
   \end{figure*}

\begin{figure}
   \centering
   \includegraphics[width=\linewidth-1mm]{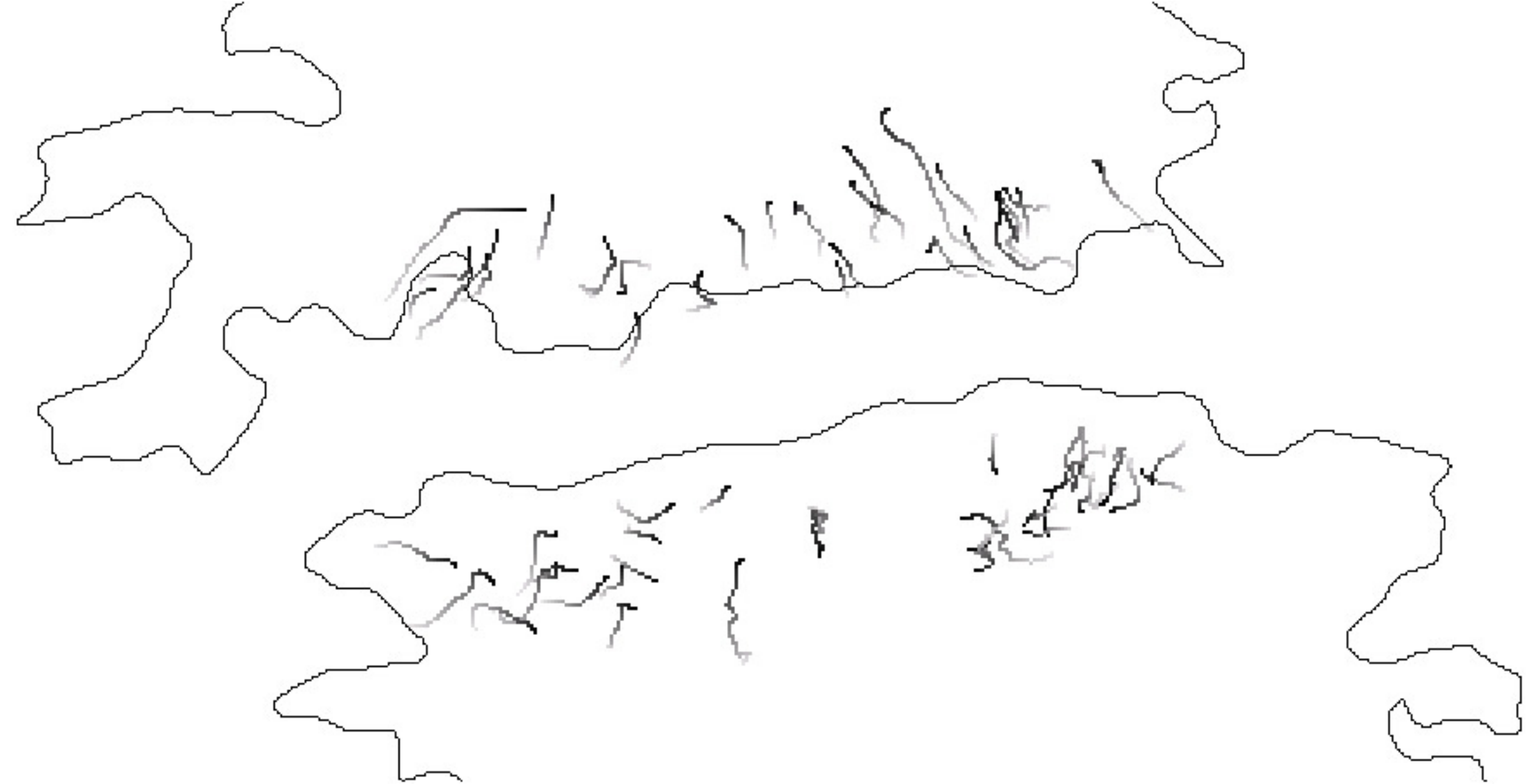}
   \caption{UD trajectories near the light bridge with birth-death distance greater than 300~km.}
   \label{FigLightbridge}
   \end{figure}

   Let us now consider more quantitatively the fact that the mobile UDs prefer to move radially towards the
   umbral center. Fig.~\ref{FigHistRadialDeflection} displays a histogram of the UD's deflection angles
   $\alpha_{\rm{defl}}$ that is defined as angle between the line connecting the umbral center and the UD's
   birth position and the line connecting the UD's birth and death position. Radially directed inward flow
   will lead to $\alpha_{\rm{defl}}~=~\mathrm{0^\circ}$ and an outward flow to
   $\alpha_{\rm{defl}}~=~\mathrm{180^\circ}$. Obviously, this definition makes sense only for the 5689 UD
   trajectories that have different birth and death positions. The solid black line in
   Fig.~\ref{FigHistRadialDeflection} shows the histogram for all these UDs and exhibits a clear tendency
   for a radially directed inward motion (42\,\% of the UDs are found to have a deflection angle lower than
   $\mathrm{45^\circ}$). This tendency is much more significant if we only consider mobile UDs, see the
   dotted blue line (88\,\% of the mobile UDs are found to have a deflection angle lower than
   $\mathrm{45^\circ}$). However, histograms calculated for central UDs and for peripheral UDs (not shown)
   lead, in principle, to the same shape as for all UDs, i.e. UDs born close to the penumbra do not show a
   significantly higher tendency of radially inward directed motion than the central UDs.

   The trajectories of large UDs are drawn in Fig.~\ref{FigUdTrajectoriesDiameterAndPeakIntensity} (left panel).
   They are born throughout the umbra, with a tendency to cluster in the darker part of the umbra. The vicinity
   of the light bridge is avoided. This UD class contains long trajectories as well as short ones. The
   trajectories of the brightest UDs can be seen in the right panel of
   Fig.~\ref{FigUdTrajectoriesDiameterAndPeakIntensity}. These UDs all emerge close to the penumbra, the light
   bridge, or the prominent UD chain (label A in Fig.~\ref{FigBestImage}).

   Many UDs with a preferred direction of motion arise near the \Index{light bridge}. Most of the UDs that emerge
   on the disk center side of the light bridge (i.e. into the large umbra) move away from the light bridge
   while many of the limbside UDs (i.e. those formed in the small umbra) move towards the light bridge
   (see Fig.~\ref{FigLightbridge}). A high density of UDs is formed by splitting off the light bridge,
   but all in one direction, in which the LB is corrugated and unsharp (disk center, large umbra side),
   while on its other straight and sharp side nearly no UDs leave the LB. 

   \begin{sidewaystable*}
   \caption{Characteristic values of UD parameters for different UD classes defined in the main text.}
   \vskip3mm
   \label{CharValues}                                    
   \centering                                            
   \begin{tabular}{l l l l l l l l l l}                  
   \hline                                                
   \noalign{\smallskip}
   Class               & Condition       & $N$        & $D^{Mean}$           & $v^{Mean}$            & $I_{Peak}^{Mean}$       & $(I_{Peak}/I_{bg})^{Mean}$ & $T$                    & $L_{BD}$               & $L_{Traj}$             \\
   \noalign{\smallskip}                                                                                                                                                                                                               
                       &                 &            & [km]                 & [m\,s$^{-1}$]         & [$I_{ph}$]              &                            & [s]                    & [km]                   & [km]                   \\
   \noalign{\smallskip}                                                                                                                                                                                                                  
   \hline                                                                                                                                                                                                                                
   \noalign{\smallskip}                                                                                                                                                                                                                  
      all UDs          &                 & \tt{12836} & \tt{229}$\pm$\tt{68} &                       & \tt{O.51}$\pm$\tt{O.O9} & \tt{1.1O}$\pm$\tt{O.O8}    & \tt{~18O}$\pm$\tt{39O} & \tt{~~5O}$\pm$\tt{13O} & \tt{~~7O}$\pm$\tt{2OO} \\
      all UDs          & $T>$\tt{15O}\,s & \tt{~2899} & \tt{272}$\pm$\tt{53} & \tt{42O}$\pm$\tt{19O} & \tt{O.55}$\pm$\tt{O.1O} & \tt{1.17}$\pm$\tt{O.1O}    & \tt{~63O}$\pm$\tt{63O} & \tt{~19O}$\pm$\tt{22O} & \tt{~29O}$\pm$\tt{35O} \\
      peripheral UDs   & $T>$\tt{15O}\,s & \tt{~~621} & \tt{252}$\pm$\tt{44} & \tt{45O}$\pm$\tt{21O} & \tt{O.65}$\pm$\tt{O.O7} & \tt{1.23}$\pm$\tt{O.11}    & \tt{~56O}$\pm$\tt{55O} & \tt{~21O}$\pm$\tt{25O} & \tt{~29O}$\pm$\tt{35O} \\
      central UDs      & $T>$\tt{15O}\,s & \tt{~2278} & \tt{278}$\pm$\tt{53} & \tt{41O}$\pm$\tt{19O} & \tt{O.53}$\pm$\tt{O.O9} & \tt{1.15}$\pm$\tt{O.O9}    & \tt{~65O}$\pm$\tt{65O} & \tt{~18O}$\pm$\tt{21O} & \tt{~3OO}$\pm$\tt{35O} \\
      mobile UDs       & $T>$\tt{15O}\,s & \tt{~~~85} & \tt{287}$\pm$\tt{33} & \tt{68O}$\pm$\tt{14O} & \tt{O.64}$\pm$\tt{O.O8} & \tt{1.29}$\pm$\tt{O.O9}    & \tt{227O}$\pm$\tt{83O} & \tt{1O8O}$\pm$\tt{3OO} & \tt{147O}$\pm$\tt{41O} \\
      stationary UDs   & $T>$\tt{15O}\,s & \tt{~2814} & \tt{272}$\pm$\tt{53} & \tt{41O}$\pm$\tt{19O} & \tt{O.55}$\pm$\tt{O.1O} & \tt{1.16}$\pm$\tt{O.1O}    & \tt{~58O}$\pm$\tt{55O} & \tt{~16O}$\pm$\tt{15O} & \tt{~26O}$\pm$\tt{27O} \\
      chain A UDs      & $T>$\tt{15O}\,s & \tt{~~~48} & \tt{3O1}$\pm$\tt{45} & \tt{48O}$\pm$\tt{19O} & \tt{O.68}$\pm$\tt{O.O6} & \tt{1.29}$\pm$\tt{O.O8}    & \tt{~96O}$\pm$\tt{71O} & \tt{~24O}$\pm$\tt{23O} & \tt{~48O}$\pm$\tt{4OO} \\ 
      light bridge UDs & $T>$\tt{15O}\,s & \tt{~~217} & \tt{254}$\pm$\tt{41} & \tt{43O}$\pm$\tt{19O} & \tt{O.64}$\pm$\tt{O.O8} & \tt{1.23}$\pm$\tt{O.11}    & \tt{~8OO}$\pm$\tt{8OO} & \tt{~22O}$\pm$\tt{24O} & \tt{~37O}$\pm$\tt{43O} \\ 
   \noalign{\smallskip}
   \hline                                                
   \end{tabular}
   \end{sidewaystable*}

   The mean diameter $D^{Mean}$, mean horizontal velocity $v^{Mean}$, mean peak intensity $I_{Peak}^{Mean}$,
   mean intensity contrast $(I_{Peak}/I_{bg})^{Mean}$, lifetime $T$, birth-death distance $L_{BD}$, trajectory
   length $L_{Traj}$ (see definitions of these quantities in the previous text), and the number of UDs $N$ of
   the types or classes mentioned earlier in this section are summarized in Table~\ref{CharValues}. We averaged
   over all trajectories of a UD class. The standard deviation $\sigma$ is given after each average value.
   As mentioned earlier, we consider only UDs that lived longer than 150~s in all cases in which we calculate
   the mean velocity; only the first line includes all UDs. The last two rows consider only UDs that are
   close to chain A or the light bridge, respectively. According to Table~\ref{CharValues} the difference
   between peripheral UDs and central UDs is not so large (using the simple categorization described above).
   The largest relative difference is in the brightness. For all other parameters the difference is less
   than $1\sigma$. In contrast to that, the difference between mobile UDs and stationary UDs is more
   significant. A relatively large difference of more than $1\sigma$ is found for the mean horizontal
   velocity, for the mean intensity contrast, for the lifetime, and, again for the brightness. On average
   mobile UDs are brighter, they move faster, and they live longer than stationary UDs but they have similar
   sizes. 

   Sometimes one can observe complete chains of successive UDs that are very close to each other (see label
   A in Fig.~\ref{FigBestImage} and the left panel of Fig.~\ref{FigUdChainVersions}). According to
   Table~\ref{CharValues} these UDs are relatively bright and they move along the chain from the endpoints
   of the chain towards its center, where they disappear. Their mean peak intensity is 0.68~$I_{ph}$ which
   is significantly higher than 0.51~$I_{ph}$, the average of all UDs, but is comparable to that of the
   peripheral and mobile UDs. The mean diameter as well as the mean velocity of the UDs within the chain
   is slightly above average. Table~\ref{CharValues} also reveals that UDs that are born close to the
   light bridge show on average a significantly higher brightness and contrast than the mean UD but all
   other properties do not show remarkable differences.

\begin{figure}
   \centering
   \includegraphics[width=\linewidth-1mm]{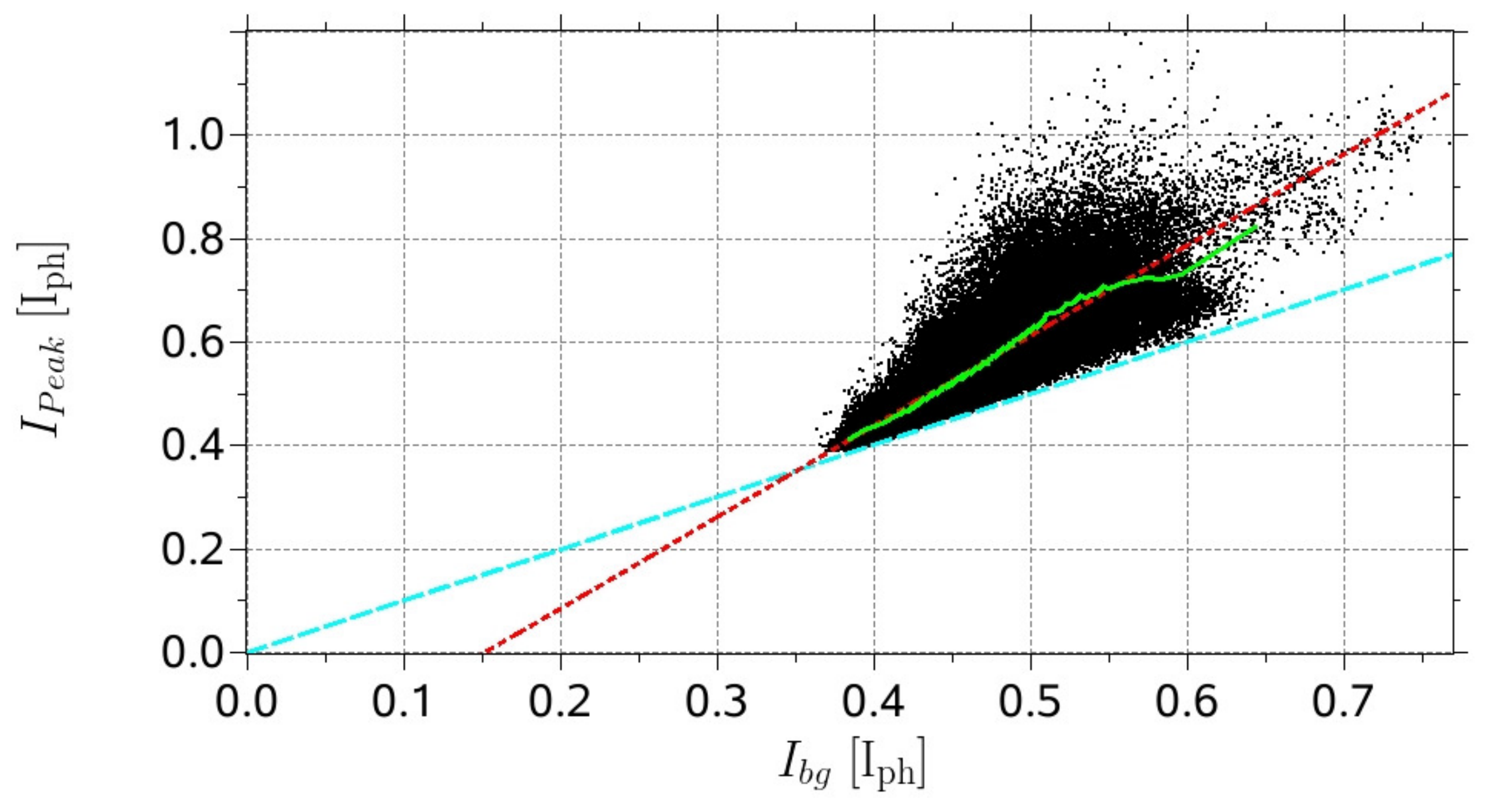}
   \caption{UD peak intensity versus umbral background intensity. The solid green line connects binned values.
   The dashed cyan line displays the theoretical, lower limit of the peak intensities. A linear fit to the data
   results in the dotted red line.}
   \label{FigIPeakVsIbg}
   \end{figure}

   The umbral background is brightest near the penumbra and gets darker towards the center of the umbra.
   Fig.~\ref{FigIPeakVsIbg} shows that the UD peak intensity correlates with the umbral background
   intensity, with the mean ratio $I_{Peak}/I_{bg}$~=~1.2~$\pm$~0.1 (intensity contrast). \citet{Sobotka2005}
   found an intensity contrast of 1.8 for a wavelength of 451~nm and 1.6 for the wavelength 602~nm. Obviously,
   the intensity contrast not only depends on the wavelengths but also on the spatial resolution
   that can be slightly different even if we compare data from the same telescope. Additionally,
   the intensity contrast can also be affected by the subsonic filter and the \Index{MFBD} image restoration we applied.
   Importantly, however,
   both binning the points in Fig.~\ref{FigIPeakVsIbg} (solid green curve) and a linear regression (dotted red curve)
   indicate that the contrast increases nearly linearly with $I_{bg}$. A scatter plot of the UD intensity
   contrast versus the shortest distance between the UD birth position and the penumbra (not shown) reveals
   that this UD contrast is not constant over the umbra. The closer the UD is born to the \Index{penumbra} the higher
   its intensity contrast, although the contrast does not drop as rapidly from the umbral boundary as the
   UD brightness does, so that partly, the dependence on distance is due to the dependence on $I_{bg}$
   (Fig.~\ref{FigIPeakVsIbg}). We also found that on average the long-lived UDs have a higher contrast than
   the short-lived ones.

   In a further step we subdivided the umbra into several boxes and determined the probability that a UD
   is born in such a box. The map obtained in this manner shows a uniform distribution of the UD birth
   probability (not plotted). Only for very small box sizes do the dark umbral nuclei (see left panel of
   Fig.~\ref{FigUdOverview}) become visible as locations of reduced UD production.

\begin{figure*}
   \centering
   \includegraphics[width=\linewidth]{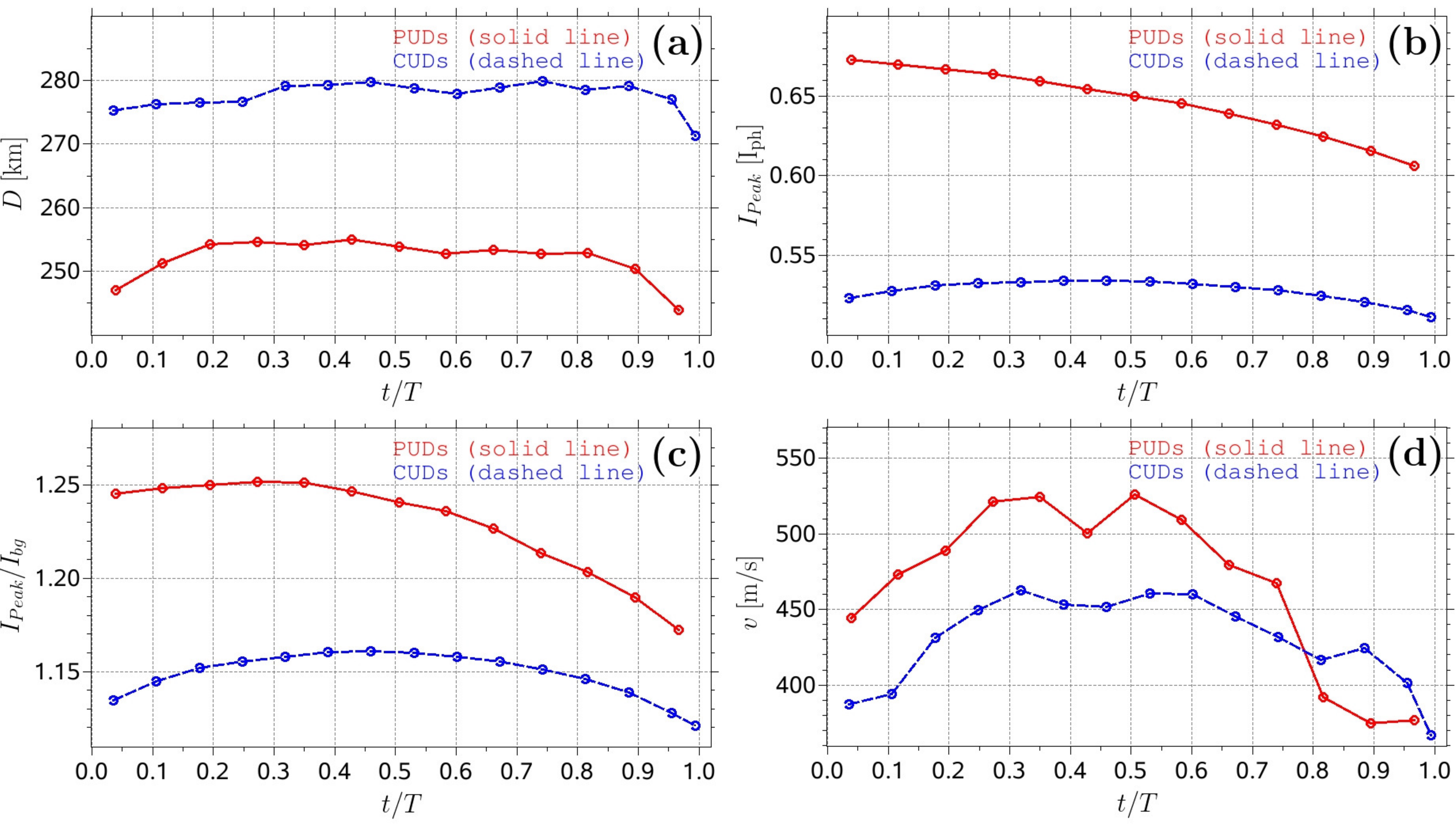}
   \caption{Temporal evolution of the UD diameter (a), UD peak intensity (b), intensity contrast (c), and
   velocity (d) separated for peripheral UDs (solid red line) and central UDs (dashed blue line).
   The UD lifetime is normalized to unity.}
   \label{FigTempEvolution}
   \end{figure*}

   Finally, we are interested in the temporal evolution of the UD properties over their lifetimes. To this
   end we normalize all UD lifetimes to unity and average the temporal evolution of the diameters, peak
   intensities, intensity contrasts ($I_{Peak}/I_{bg}$), and horizontal velocities of the 2899 trajectories
   with $T~>~150~s$. In order to weight all trajectories equally, i.e. independently of their lifetime, we
   up-sample all trajectories to 310 points of time via interpolation. (Our time series contains 310 images,
   so that no trajectory can consist of more than 310 points.) Then we averaged the UD parameters with the
   help of our binning method, which is applied separately for the 621 PUDs and the 2278 CUDs.
   Each bin contains 15000 points for PUDs and 50000 points for CUDs. The results are plotted in
   Fig.~\ref{FigTempEvolution} and show that the mean PUD is smaller, brighter and moves faster than the
   mean CUD (as could already be deduced from Table~\ref{CharValues}). More importantly, there are distinct
   differences in their mean evolution. Whereas both types of UDs share the property that their diameters
   evolve rather moderately over time (the increase after birth and the decrease before death are less than
   10\,\% of the maximum diameter (see panel a), the evolution of their brightness (panel b) and in
   particular of their contrast (panel c) differ considerably. While the mean CUD displays an initial
   gentle brightening followed by an equally gentle darkening, the mean PUD darkens continuously. The small
   magnitude of the change in brightness may be due to the fact that we have averaged over UDs with very
   different absolute intensity. More information may be gleaned from the contrast, i.e. the peak intensity
   divided by the local umbral background intensity, plotted in panel (c). The mean PUD initially remains
   almost constant, exhibiting a slight maximum at around 1/3 of the mean lifetime before dropping
   rapidly over the remaining portion of its life. The contrast of the CUDs displays a much more symmetric
   evolution. The birth velocity of the mean PUD is nearly 50~m\,s$^{-1}$ higher than for the mean CUD,
   while the velocity at death of the two UD types is similar, see panel (d). Both velocity curves show
   an initial increase, followed by a decrease. As in the case of the contrast the velocity profile is
   much more symmetric for the CUDs.

\section{Discussion and conclusions}

   We have analyzed a time series of images of a mature sunspot close to solar disk center.
   Due to the excellent image quality we were able to resolve thousands of UDs. Exhaustive UD analyses
   can be found in earlier papers \citep{Sobotka1997a,Sobotka1997b,Hartkorn2003,Sobotka2005,Sobotka2006},
   but the present article is the first detailed UD study of a long time series of reconstructed images
   with the consistent high resolution of a 1-meter telescope.

   Trajectories, lifetimes, diameters, horizontal velocities, peak intensities, and distances between
   birth and death locations were determined by tracking single UDs over the time series. These
   characteristic values were used to look for reasonable separations into UD classes. In the following
   we summarize the obtained results and compare them with other investigations in the literature:

   \begin{enumerate}
      \item There is hardly any part of the umbra which does not support UDs, but the UD brightnesses
      depend strongly on the location within the umbra, which confirms the previous observation of
      \citet{Sobotka1997b}.
      \item The histogram of lifetimes shows an exponential distribution, i.e. a UD does not have a
      typical lifetime. More than 3/4 of all studied UDs lived less than 150~s and their motion was
      negligible. The exponential distribution is in qualitative agreement with the results obtained by
      \citet{Sobotka1997a,Sobotka1999}. Quantitatively, \citet{Sobotka1997a} obtains a median lifetime
      of 6~min for an umbra of about 6~Mm diameter and a median of 12~min for a 4~Mm pore
      \citep{Sobotka1999}, whereas we find a median value of 0.7~min for a roughly 10~Mm umbra. Note
      that these median values depend strongly on algorithmic constraints as well as on the cadence of
      the time series. For example, the method used by \citet{Sobotka1997a} cannot lead to lifetimes
      shorter than 1.5~min. If we only consider UDs with lifetimes greater than 1.5~min our median
      increases to 4.1~min. Irrespectively of which of these two values we use, our results are
      consistent with the conclusion of \citet{Sobotka1999} that UDs are more stable in a weak magnetic
      field if we assume a direct correlation between umbral diameter and magnetic field strength
      \citep[cf.][]{Kopp1992}. Alternatively, due to the strong dependence of umbral brightness on
      umbral size \citep{Mathew2007} the lifetime may be influenced mainly by the radiative flux or
      umbral temperature. The scatterplot of the mean umbral background intensities versus the UD
      lifetimes (not shown) is consistent with both possible explanations mentioned above: UDs live longer
      in brighter parts of the umbra. Due to the non-linear, monotonically decreasing relation between
      magnetic field strength and background intensity as observed by \citet{Kopp1992} and confirmed
      by \citet{MartinezPillet1993,SolankiEtAl1993}, this implies that UDs live longer in regions of weak field.
      \item The histogram of mean diameters exhibits a maximum at 225~km (0.31$^{\prime\prime}$) and
      descends from there towards the \Index{diffraction limit}, so that we expect the majority of UDs to have
      been spatially resolved. This seems not to be the case in many of the earlier papers because
      there a monotonic decrease was obtained towards higher diameters
      \citep[see][]{Sobotka1997a,Sobotka1999}. \citet{Sobotka2005} also analyzed data obtained with
      the 1-meter SST. These data lead to a histogram that is qualitatively similar to ours. The mean
      diameter of 175~km (0.24$^{\prime\prime}$), as well as the average \Index{filling factor} of 9\,\% is,
      however, noticeable smaller. This difference can be explained by the use of a different method
      to determine the UD boundary. An increase of our brightness threshold to determine the UD
      boundary leads to smaller diameters and to lower filling factors. \citet{Hamedivafa2008} used
      an improved method of image segmentation and also found a mean diameter of 230~km and a similar
      shaped histogram.
      \item The mean horizontal velocity of those of our UDs that live longer than 150~s is 420~m\,s$^{-1}$
      which is significantly higher than the 210~m\,s$^{-1}$ reported by \citet{Molowny1994} and higher than
      the 320~m\,s$^{-1}$ found by \citet{Sobotka1999}. In both studies the mean horizontal velocity
      was calculated by means of least-squares linear fits of the x and y coordinates of all trajectory
      points, which leads to an underestimation of velocity in case of curved trajectories. Our
      histogram of horizontal velocities shows a maximum at 350~m\,s$^{-1}$, whereas some UDs can reach
      velocities above 1~km\,s$^{-1}$. This is in qualitative agreement with the histograms of
      \citet{Kitai1986}, \citet{Molowny1994}, and \citet{Hamedivafa2008} but disagrees with the results of
      \citet{Sobotka1997b,Sobotka1999} whose histograms do not exhibit a maximum; they peak at zero
      velocity and show a monotonic decrease toward higher velocities of up to 1~km\,s$^{-1}$.
      The majority of our UDs moves irregularly around the birth position. However, there are some
      mobile UDs that travel over long distances within their lifetime. Almost all mobile UDs emerge
      close to the umbral border, i.e. near the penumbra, they are brighter than the average and their
      horizontal motion is preferentially directed towards the center of the umbra. The mean velocity
      of our mobile UDs is 680~m\,s$^{-1}$, which is in good agreement with the recent observation
      of \citet{Katsukawa2007} who found a mean velocity of 700~m\,s$^{-1}$.
      \item The relation between mean UD size and lifetime is non-linear. On average, the size of UDs
      increases with lifetime, which was also found by \citet{Sobotka1997a}, but in contrast to their
      work we find a narrow size distribution of around 290~km for long-lived UDs.
      \item UDs that were born close to the penumbra show a significantly higher contrast than
      the UDs of the umbral interior. The mean UD intensity contrast $I_{Peak}/I_{bg}$ is 1.2 which is
      smaller than the value of 1.6 reported by \citet{Sobotka2005}. This may partly be due to the
      longer wavelength of our observation. Additionally, our statistical ensemble contains many more UDs.
      In particular we took many UDs with low contrast into account, made possible by the multilevel
      tracking technique \citep{Bovelet2001} we employed to identify UDs. Consequently, we believe that
      the lower UD contrast we find is not due to a lower resolution or higher \Index{stray light}, but rather
      to differences in identification of UDs and in particular the difference in wavelength. We stress
      that the UD contrast itself depends on the background intensity; the higher the intensity the
      stronger the contrast. Also it cannot be ruled out that there could be systematically
      different contrasts between different sunspots due to intrinsicly different physical properties of
      the spots.
      \item Whereas the temporal variation of the UD diameter is qualitatively similar for UDs formed
      close to the penumbra (PUDs) and those formed in the body of the umbra (CUDs), their intensity
      contrast and horizontal velocity display contrasting evolutions. The mean PUD shows a continuous
      darkening which is in agreement with the results of \citet{Kitai2007} for a single typical PUD.
      The typical CUD of \citet{Kitai2007} is found to increase in brightness linearly and then to darken
      linearly with time, whereas our results for the mean CUD show a non-linear increase in brightness
      until nearby half of the lifetime followed by an again non-linear decrease. The clear difference
      between PUDs and CUDs in the behavior of their contrast and mean velocity may be a result of the
      different origin of the two types of features \citep{Kitai2007}. We confirm from visual inspection
      of a subset of PUDs that PUDs are formed when penumbral grains cross the umbral boundary.
   \end{enumerate}

   A comparison of the results with the simulations of \citet{Schuessler2006} shows a better agreement
   with CUDs, than PUDs. For example, the simulated UDs display a gradual increase in contrast
   followed by a gradual decrease, just as CUDs. They also display little proper motion. This qualitative
   agreement further strengthens the interpretation of UDs as localized columns of overturning convection
   proposed by \citet{Schuessler2006}. Hinode data had earlier suggested the presence of \Index{dark lane}s in
   large UDs \citep{Bharti2007} and revealed a decrease in the magnetic field strength with depth,
   as well as an upflow associated with a temperature enhancement \citep{Riethmueller2008c}, in good
   qualitative agreement with the simulations. A detailed analysis of the simulations similar to the one
   carried out here would allow a more quantitative comparison.

   The PUDs have significantly different evolution histories than the simulated features. They start
   at a higher speed \citep{Kitai1986} and in particular display their maximum brightness right after the
   beginning of their life \citep[cf.][]{Kitai2007}. This, combined with the fact that they are born very close
   to the penumbra, or actually by breaking away from the penumbra \citep{Thomas2004}, and move radially
   towards the umbral center \citep{Kitai1986} supports that these are two distinct types of UDs based on
   their origin and evolution, although their physical structure is relatively similar \citep{Riethmueller2008c}.

\begingroup
\hypersetup{linkcolor=white} 
\chapter[Stratification of sunspot umbral dots]{Stratifications of sunspot umbral dots from inversion of Stokes profiles recorded by \hinode{}$^1$}\label{Ud2Chapter}\footnote{Published in The Astrophysical Journal Letters, 678, 157 (2008), see \citet{Riethmueller2008c}.}
\endgroup

\section{Introduction}

   The energy transport immediately below the solar surface is mainly determined by convective
   processes that are visible as granulation patterns in white-light images of the quiet photosphere.
   This convection is suppressed inside sunspot umbrae due to the strong vertical magnetic field, but some
   form of magnetoconvection \citep{Weiss2002} is needed to explain the observed umbral brightnesses.
   Umbral fine structure such as light bridges or \Index{umbral dot}s (UDs), dotlike bright features inside umbrae,
   may well be manifestations of magnetoconvection. Different models have been proposed to explain UDs,
   e.g., columns of field-free hot gas in between a bundle of thin magnetic flux ropes
   \citep{Parker1979,Choudhuri1986}, or spatially modulated oscillations in a strong magnetic field
   \citep{Weiss1990}. Recent numerical simulations of three-dimensional radiative magnetoconvection
   \citep{Schuessler2006} reveal convective plumes that penetrate through the solar surface and look very
   much like UDs. Although recent broadband images may have spatially resolved UDs
   \citep{Sobotka2005,Riethmueller2008d}, spectropolarimetry is needed to learn more about their physical nature.
   Previous spectroscopic observations led to heterogeneous results. \citet{Kneer1973} found that
   UDs exhibit upflows of 3~km~s$^{-1}$ and a 50\% weaker magnetic field compared to the nearby umbra,
   whereas \citet{Lites1991} and \citet{Tritschler1997} reported
   little field weakening. Finally, \citet{SocasNavarro2004} observed a weakening
   of 500~G and upflows of a few 100~m~s$^{-1}$. More details can be found in the reviews of umbral fine
   structure by \citet{Solanki2003} and \citet{Sobotka2006}. One reason for the
   difference in results has been the influence of scattered light and variable seeing, which affect
   the different analyzed data sets to varying degrees. It therefore seems worthwhile to invert
   Stokes profiles obtained by the \Index{spectropolarimeter} (\Index{SP}) on the \hinode{} spacecraft. The usefulness
   of \hinode{} data for the study of UDs was demonstrated by \citet{Bharti2007}, who found
   that large UDs show \Index{dark lane}s whose existence had been predicted by \citet{Schuessler2006}.

\section{Observations and data reduction}

   The data employed here were acquired by the spectropolarimeter \citep{Lites2001}
   of the Solar Optical Telescope \citep[SOT,][]{Suematsu2008} onboard \hinode{}. They are
   composed of full Stokes spectra in the Fe~I line pair around 6302~{\AA} and the nearby continuum
   of a sunspot of NOAA AR~10933 recorded from 12:43 to 12:59~UT on 2007 January 5
   using the 0.16$^{\prime\prime}$x164$^{\prime\prime}$ slit. At this time the sunspot was located at
   a heliocentric angle of 4$^\circ$, i.e. very close to disk center. The observations covered the
   spectral range from 6300.89 to 6303.26~{\AA}, with a sampling of 21~{m\AA}~pixel$^{-1}$. The SP
   was operated in its normal map mode, i.e. both the sampling along the slit and the slit-scan
   sampling were 0.16$^{\prime\prime}$, so that the spatial resolution should be close to the
   \Index{diffraction limit} of $1.22~\lambda/D = 0.32^{\prime\prime}$. The integration time per slit position
   was 4.8~s which reduced the noise level to $10^{-3}~I_c$.

\begin{figure}
   \centering
   \includegraphics[width=110mm]{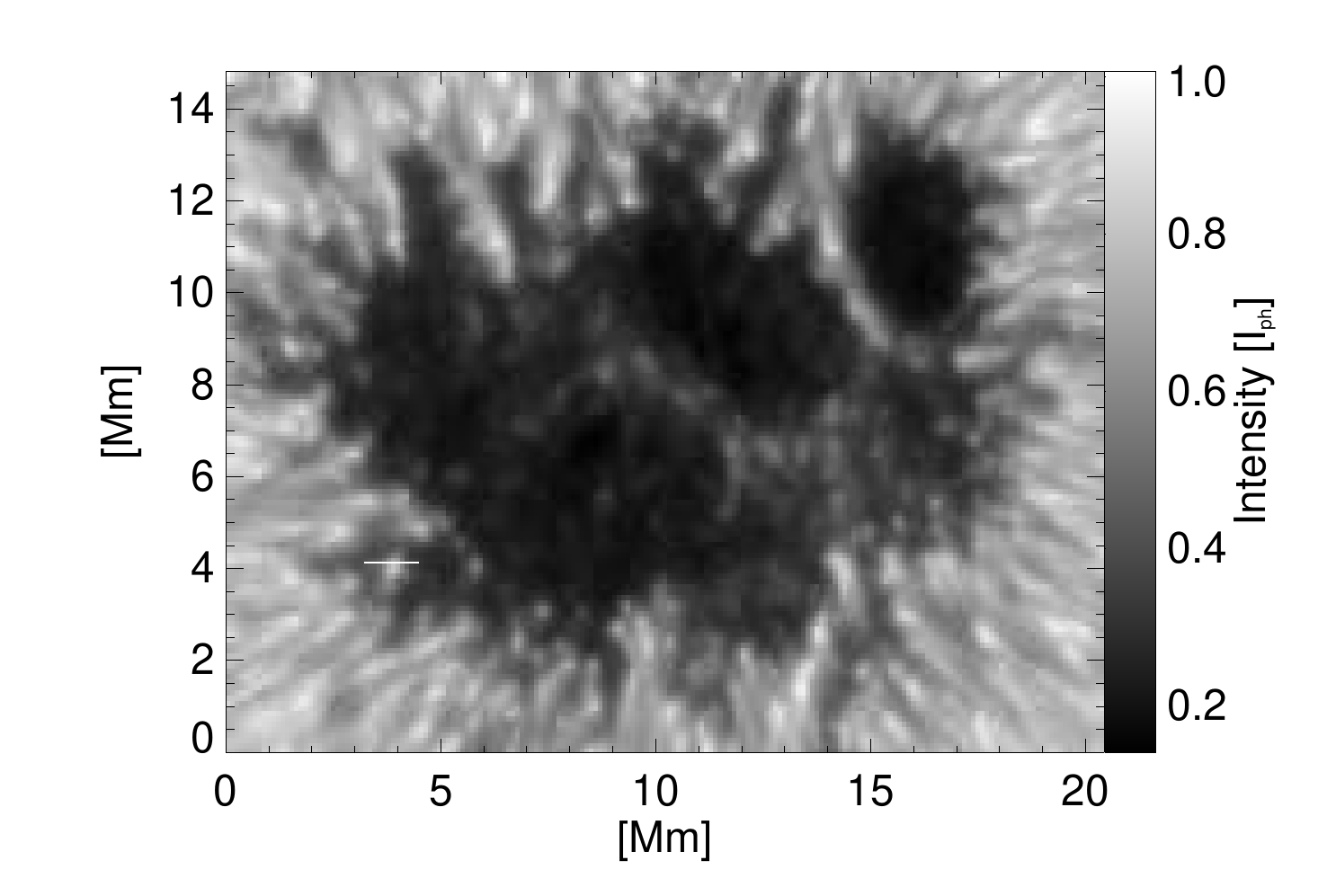}
   \caption{Continuum intensity map of the sunspot NOAA~10933 as observed by the \hinode{} SOT/SP on 2007
   January 5. Heliocentric angle is $\theta$~=~4$^\circ$. Intensities are normalized to the intensity
   level of the quiet photosphere $I_{ph}$. The white line at (4,4)~Mm marks the cut through an umbral
   dot (UD) that is discussed in greater detail.}
   \label{FigUmbra}
   \end{figure}

   The data were corrected for dark current, flat field, and \Index{instrumental polarization} with the help of the
   SolarSoft package\footnote{See \href{http://www.lmsal.com/solarsoft}{http://www.lmsal.com/solarsoft}}.
   A continuum intensity image (put together from the slit scan) of the chosen \Index{umbra} is shown in
   Fig.~\ref{FigUmbra}. Due to the large slit length we are always able to find a sufficiently extensive
   region of quiet Sun that is used to normalize intensities.

\section{Data analysis}

   To obtain atmospheric stratifications of temperature ($T$), magnetic field strength ($B$), and line-of-sight
   velocity ($v_{LOS}$) we use the inversion code \Index{SPINOR} described by \citet{Frutiger2000b}.
   This code incorporates the \Index{STOPRO} routines \citep{Solanki1987}, which compute synthetic Stokes
   profiles of one or more lines upon input of their atomic data and one or more model atmospheres.
   Local \Index{thermodynamic equilibrium} conditions are assumed and the Unno-Rachkovsky\index{Unno-Rachkovsky equations} \Index{radiative transfer equation}s
   are solved. The inversions use an optical depth scale as the appropriate coordinate
   for radiative transfer problems. For reasons of comparability we use the optical depth at 500~nm
   ($\tau_{500}$). Starting with an initial guess model, the synthetic profiles were iteratively
   fitted to observed data using \Index{response function}s (RFs) and the \Index{merit function} $\chi^2$
   \citep{RuizCobo1992,Frutiger2000a} is minimized. With the help of the RFs we find that the
   Fe~I line pair at 6302~{\AA} is mainly formed within the $\log(\tau_{500})$ interval [$-3,0$], which
   corresponds to a height range of about 400~km under \Index{hydrostatic equilibrium} conditions in the umbra.
   The free parameters are defined at the four nodes $-3$, $-2$, $-1$, and $0$ of the $\log(\tau_{500})$ grid.
   The atmospheric stratification is then interpolated using splines onto a 10 times finer $\log(\tau_{500})$
   grid.

   The first step of our analysis is the wavelength calibration required to determine line-of-sight
   (LOS) velocities. For every slit position we average the Stokes $I$ profiles of all locations
   along the slit whose total polarization $P = \int(Q^2 + U^2 + V^2)^{1/2}d\lambda$ is negligible, since
   those locations are assumed to represent the quiet Sun. This mean $I$ profile is used to fit \Index{Voigt profile}s
   to the two Fe~I lines from which the line center wavelengths are determined. The convective \Index{blueshift}
   of 140~m~s$^{-1}$ \citep[see][]{MartinezPillet1997,Dravins1981} is then removed.

   The next step is to find an appropriate model atmosphere. Since we are interested in the atmospheric
   stratification of temperature, magnetic field strength, and LOS velocity within a UD, these
   three atmospheric parameters are assumed to be height dependent, whereas field inclination and azimuth angle,
   \Index{micro-turbulence}, and \Index{macro-turbulence} are assumed to be height independent. We experimented intensively
   with adding a second model component to represent the \Index{stray light}, but the inversion results did not
   improve significantly, confirming the almost negligible stray light in the \Index{SP}. Therefore, in the
   interests of a robust inversion, we forbore from adding a stray light component, thus reducing the number
   of free parameters.

   Lastly, we have to find initial guesses for all free parameters. We use an initial temperature stratification
   according to the umbral core model L of \citet{Maltby1986} and assume a vertical magnetic
   field of 2000~G and zero LOS velocity at all heights. Initial guesses for \Index{micro-turbulence} and \Index{macro-turbulence}
   are 0.1 and 2~km~s$^{-1}$, respectively. Other initial guesses gave very similar results, except for
   a limited number of outliers. For these, repeating the inversion with an initial guess close to the final
   result of one of the neighboring pixels returned values consistent with those obtained for the other pixels.

\section{Inversion results}

\begin{figure}
   \centering
   \includegraphics[width=\linewidth]{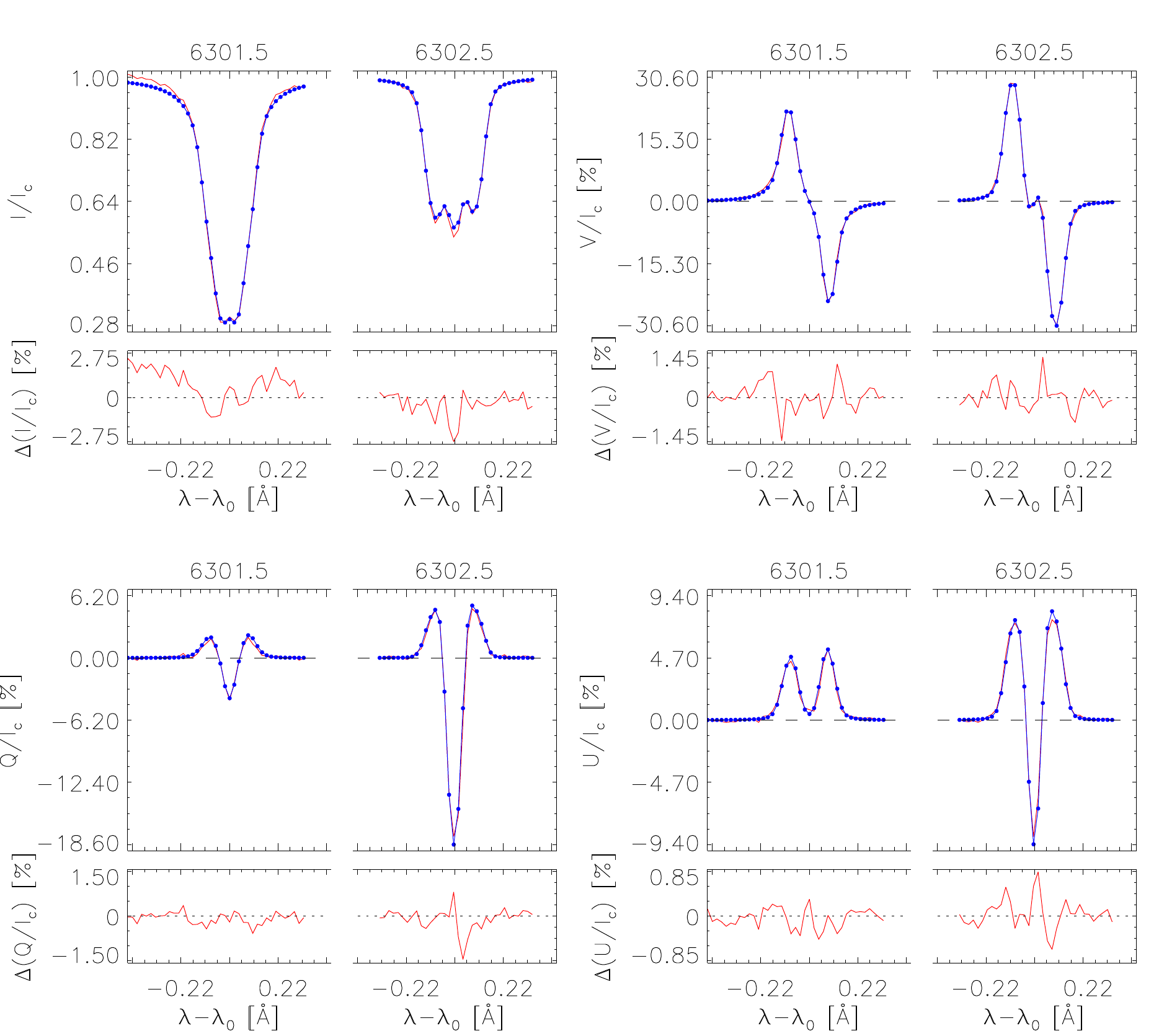}
   \caption{Stokes $I$, $V$, $Q$ and $U$ profiles from the center of the UD marked in Fig.~\ref{FigUmbra}.
   Red lines are the measured, blue lines the best-fit profiles, i.e. the inversion result. The bottom parts
   of each panel show the difference between the two on an expanded scale.}
   \label{FigProfileUdBright}
   \end{figure}

\begin{figure}
   \centering
   \includegraphics[width=\linewidth]{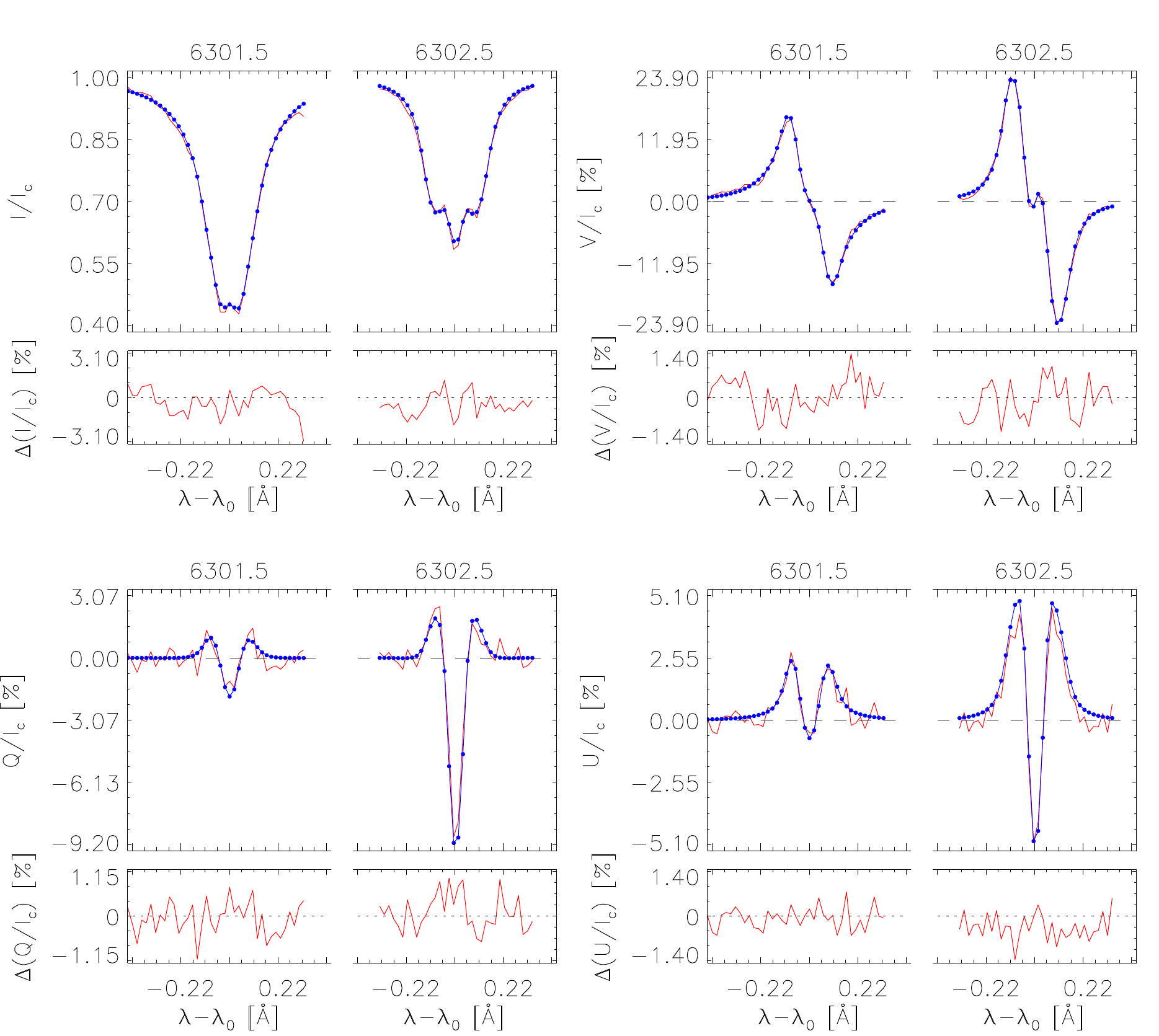}
   \caption{The same as Fig.~\ref{FigProfileUdBright}, but for Stokes $I$, $V$, $Q$ and $U$ profiles of
   the diffuse background near the UD.}
   \label{FigProfileUdDark}
   \end{figure}

   We analyzed a total of 51 UDs, which were identified by applying the multilevel tracking (MLT) algorithm
   \citep{Bovelet2001,Riethmueller2008d}. For each UD
   the location of its core was identified, a cut was made through it, reaching to the neighboring diffuse
   background (DB), and the profiles from all the pixels along this cut were inverted. We first discuss the
   results for the UD marked in Fig.~\ref{FigUmbra}, chosen because of its brightness, which leads
   to particularly small error bars. A comparison of the measured profiles with the best-fit profiles resulted
   from the inversion can be seen in Fig.~\ref{FigProfileUdBright} for the UD and in Fig.~\ref{FigProfileUdDark}
   for the DB selected as the location of lowest continuum intensity in a 1.4~$\times$~1.4~Mm$^2$ environment
   of the UD center. Due to the low signal in the dark background the measured DB profiles are much noisier
   than the UD center's profiles, but in general, the Stokes spectra can be fitted remarkably well.

\begin{figure}
   \centering
   \includegraphics[width=120mm]{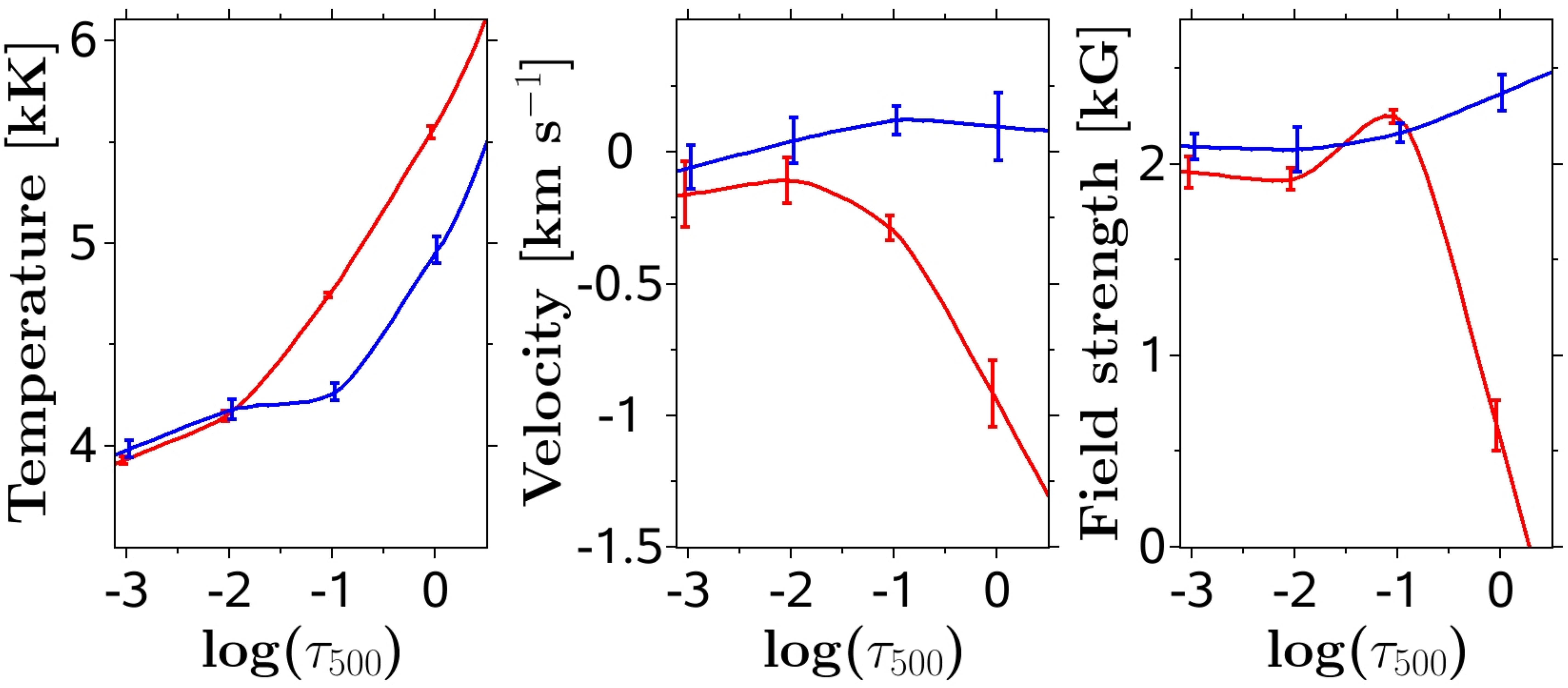}
   \caption{Atmospheric stratification obtained from the Stokes profiles at the location of the UD's center
   (red lines) and from the Stokes profiles of the diffuse background near the UD (blue lines).
   The formal errors of the inversion at the used optical depth nodes are indicated by bars. Negative
   LOS velocity values indicate upflows.}
   \label{FigSingleAtm}
   \end{figure}

   The stratification of the retrieved atmospheric parameters $T$, $v_{LOS}$, and $B$ in the center of the UD
   and in the DB are plotted in Fig.~\ref{FigSingleAtm}. In the upper photosphere ($-3 \le \log(\tau_{500})
   \le -2$) the error bars overlap; i.e. we find little significant difference between UD and DB. In the
   deeper photosphere, however, the inversions return strongly different stratifications. Thus, the UD
   temperature is higher than the DB temperature, consistent with the intensity enhancement of the UD in
   the continuum map. The LOS velocity (which is identical to the vertical velocity due to the small
   heliocentric angle) exhibits strong upflows in the UD center, whereas the DB is nearly at rest.
   The magnetic field strength is roughly 2~kG for the heights $-3 \le \log(\tau_{500}) \le -1$. Below
   $\log(\tau_{500}) = -1$ the field strength of the UD decreases strongly with depth, whereas the field
   strength of the DB increases moderately.

\begin{figure}
   \centering
   \includegraphics[width=120mm]{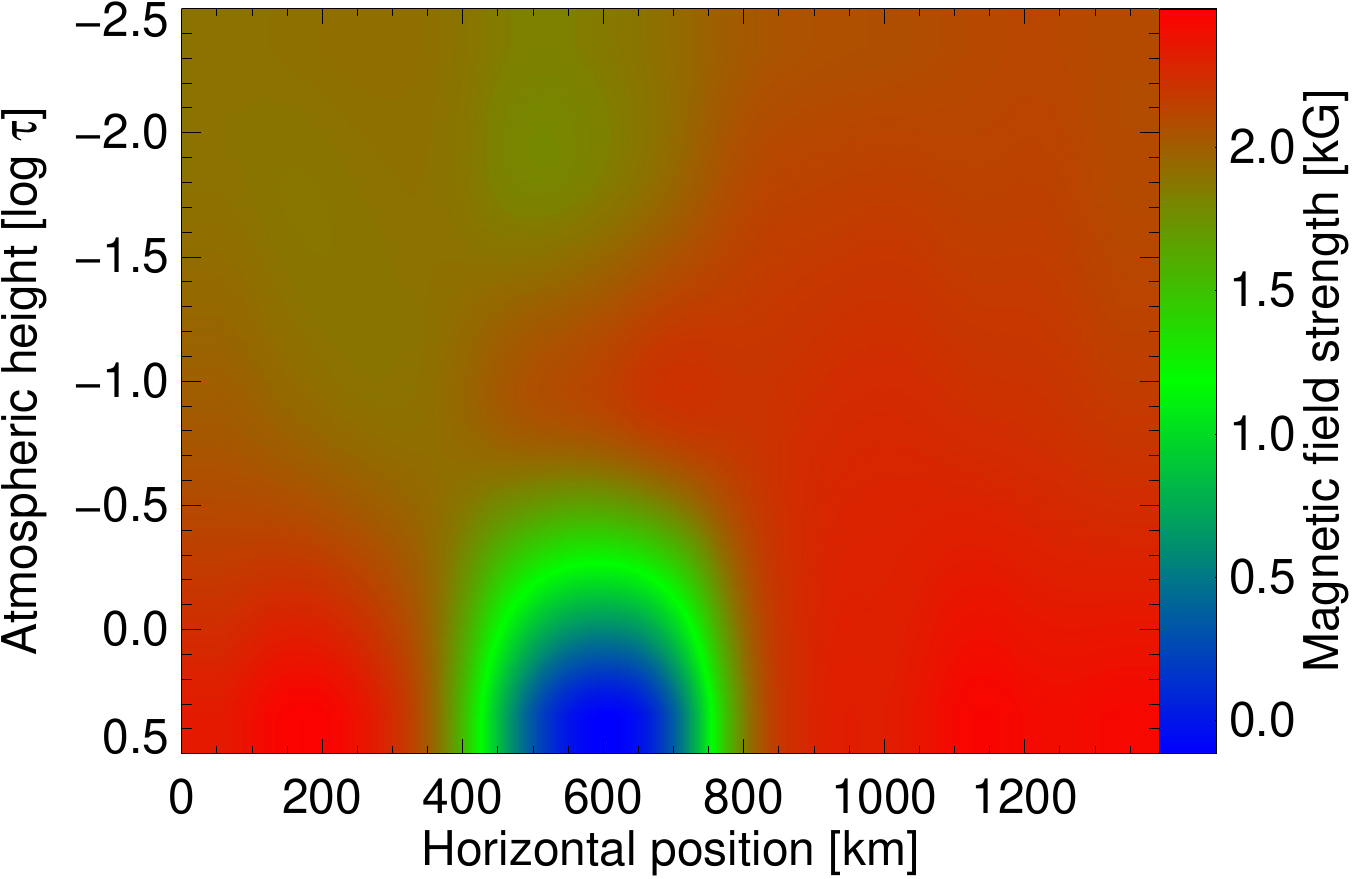}\\
   \includegraphics[width=120mm]{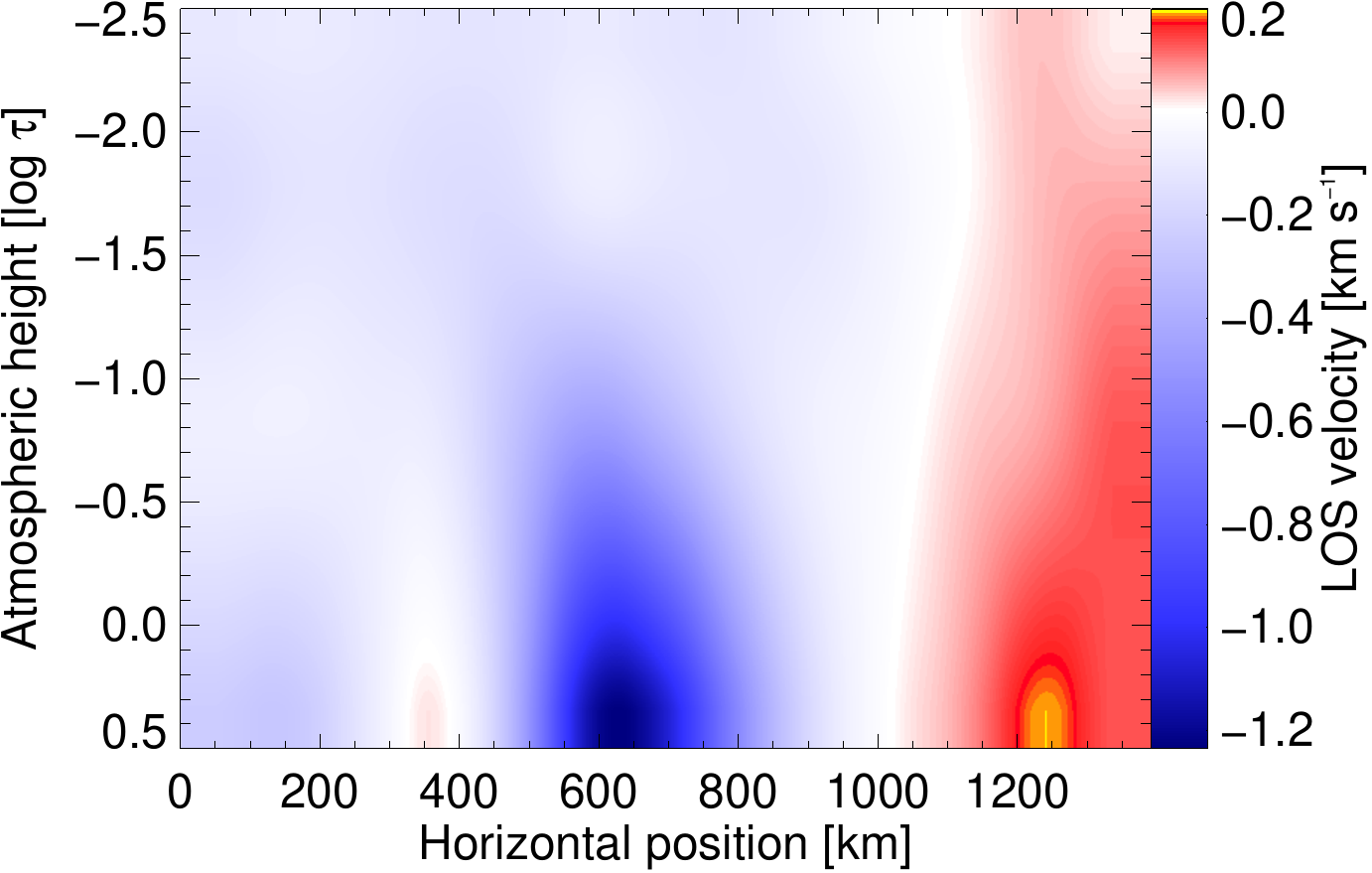}
   \caption{Vertical cut through the UD marked in Fig.~\ref{FigUmbra} in the direction indicated by the
   white line. Colors of the top panel indicate magnetic field strength. The bottom panel shows LOS velocity.
   Negative velocities are upflows.}
   \label{FigVerticalCut}
   \end{figure}

   The vertical cuts of magnetic field strength and LOS velocity through 13 pixels lying along the white
   line in Fig.~\ref{FigUmbra} are shown in Fig.~\ref{FigVerticalCut}. Jumps from one pixel to the next
   were smoothed through interpolation. There is clear evidence for a localized decrease in UD field
   strength in the low photosphere, co-located with an upflow that extends higher into the atmosphere and
   a weak downflow on at least one side. The magnetic fields are 4$^\circ$ more inclined in the UD than
   they are in the DB around the UD. Fig.~\ref{FigVerticalCut} looks remarkably like Fig.~2 of
   \citet{Schuessler2006}, in spite of the fact that Fig.~\ref{FigVerticalCut}
   is plotted on an optical depth scale in the vertical direction and is thus distorted by an unknown
   amount relative to a corresponding figure on a geometrical scale.

\begin{figure}
   \centering
   \includegraphics[width=120mm]{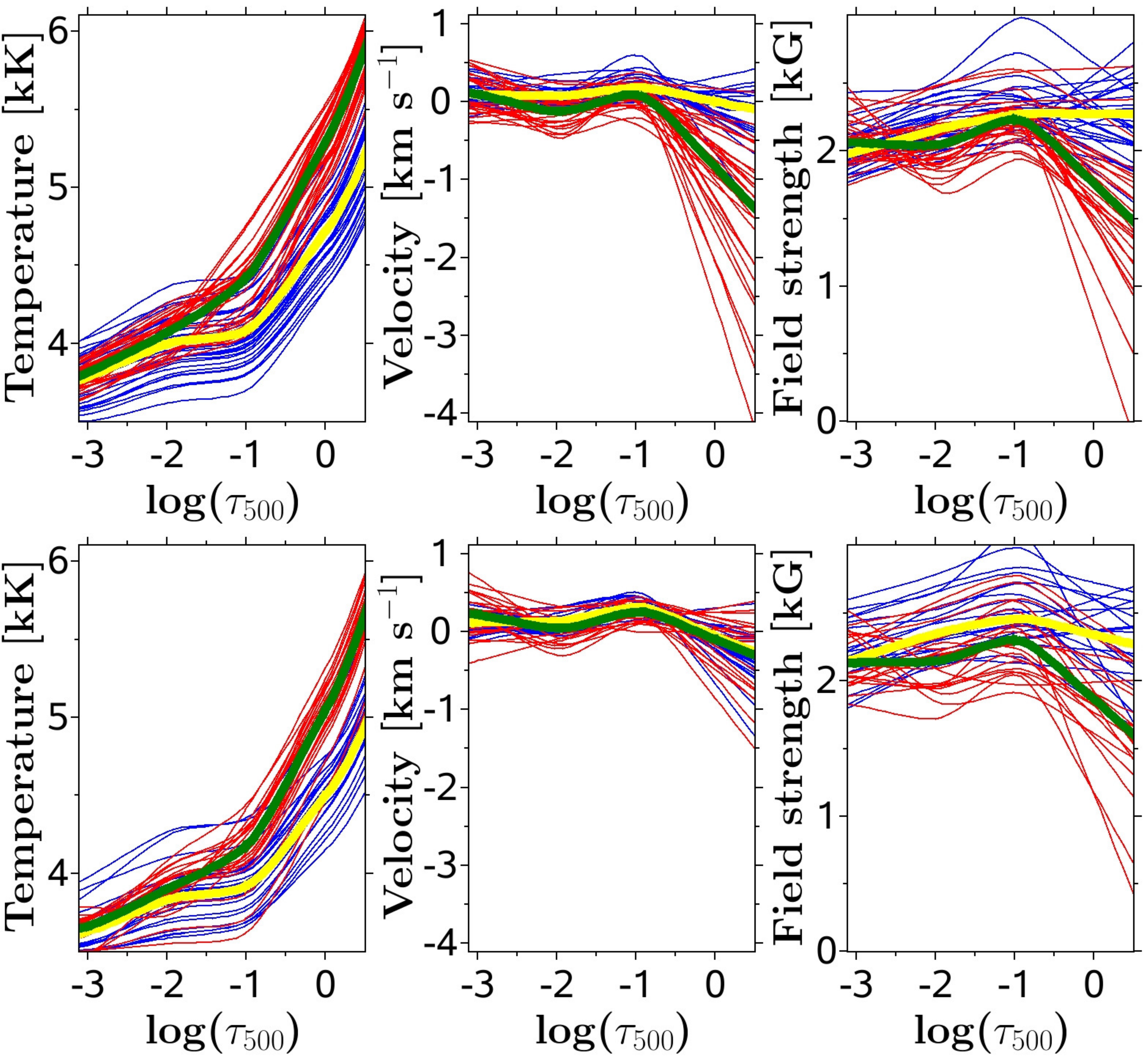}
   \caption{Atmospheric stratifications of peripheral umbral dots (top 3 panels) and central umbral dots
   (bottom 3 panels). The red lines show the stratification at the location of the UD's center and the
   blue lines correspond to the nearby diffuse background. The green line is the weighted average of all
   red lines and the yellow line is the weighted average of all blue lines, where we used the reciprocal
   error bars as weighting factors.}
   \label{FigMultipleAtms}
   \end{figure}

   Next we discuss all 51 analyzed UDs. In the literature we often find a separation into two UD regimes.
   For example, \citet{GrossmannDoerth1986} differentiate between peripheral
   UDs (PUDs) and central UDs (CUDs), i.e. between UDs that are born close to the umbra-penumbra boundary
   and UDs that are born deep in the umbra. We follow this distinction and plot the obtained stratifications
   of the 30 PUDs (distance to umbra-penumbra boundary less than 2000~km) in the top panels of
   Fig.~\ref{FigMultipleAtms}, while the remaining 21 CUDs are represented in the bottom panels of
   Fig.~\ref{FigMultipleAtms}. The results largely mirror those obtained for the UD discussed above. In the
   upper atmosphere UDs center and DB do not differ in their mean values of $T$, $v_{LOS}$, and $B$.
   On average, the CUDs are about 150~K cooler than the PUDs in the upper atmosphere, just as the DB
   around the CUDs is cooler than the DB around the PUDs. At $\log(\tau_{500}) = 0$ we find that PUDs are
   570~K hotter than the local DB and CUDs are 550~K hotter than the DB in their vicinity. The magnetic
   field strength at $\log(\tau_{500}) = 0$ is weakened by about 510~G for PUDs and 480~G for CUDs, whereas
   only PUDs exhibit significant upflows of about 800 m~s$^{-1}$. The mean LOS velocity shows no difference between
   CUD centers and DB. In order to make sure that an upflow is not being missed due to the lower S/N ratio
   of the CUD Stokes profiles, we have also averaged the Stokes profiles of all the CUDs. An inversion
   of there averaged Stokes profiles gave a result that agrees with the averaged stratifications ($green~line$)
   in the bottom panels of Fig.~\ref{FigMultipleAtms} within the error bars. This suggests that any upflow
   velocity in CUDs is mostly restricted to layers below the surface or is too concentrated or too weak to be
   detected by the inversions. Finally, we find that the magnetic field of the PUDs is on average
   4$^\circ$ more horizontal than for their DB. We see no inclination difference for CUDs.

\section{Discussion}

   We identified 30 peripheral and 21 central umbral dots in \hinode{} spectropolarimetric data of a sunspot
   within 4$^\circ$ of disk center. With the help of Stokes profile inversions of the Fe~I lines at
   630~nm we determined the stratifications of temperature, magnetic field strength, and LOS velocity.
   The present work differs from that of \citet{SocasNavarro2004} in the superior quality
   of the employed data with twice the spatial resolution and practically no scattered light. This allows
   a detailed determination of the atmospheric stratification. The higher spatial resolution of the
   \hinode{} SP data also allows us to, for the first time, reconstruct both the horizontal and the vertical
   structure of UDs. We also extended the analysis to a more numerous statistical ensemble of 51 UDs.

   Vertical cuts through UDs provide a remarkable confirmation of the results of MHD simulations of
   \citet{Schuessler2006}: both show that UDs differ from their surroundings
   mainly in the lowest visible layers, where the temperature is enhanced and the magnetic field is
   weakened. We found a temperature enhancement of 550~K and a magnetic field reduction of about 500~G
   (at optical depth unity). In addition, PUDs display upflow velocities of 800~m~s$^{-1}$ on average,
   again in good agreement with the simulations. There are also some differences between our results
   and those of \citet{Schuessler2006}. Thus, according to our inversions the
   magnetic field strength of the DB is somewhat depth dependent. This was not the case for the MHD
   simulations due to the used periodic boundary conditions. Furthermore, although some of the UDs
   display a weak downflow bounding the strong central upflow (see Fig.~\ref{FigVerticalCut}),
   these are neither as narrow nor as strong as the downflows at the ends of \Index{dark lane}s as reported
   by \citet{Schuessler2006}, probably due to the limited spatial resolution of our data. We may also
   be missing some of the narrow downflows by considering only single cuts across individual UDs.

   \citet{SocasNavarro2004} reported 10$^\circ$ more inclined magnetic fields in PUDs.
   This result is qualitatively confirmed by our work; we find an inclination increase of 4$^\circ$ for PUDs
   but no increase for CUDs, which can be assumed as a further hint that the main part of the CUD structure
   is below the surface. These results can be interpreted in terms of the strong DB fields expanding with
   height and closing over the UD, as proposed by \citet{SocasNavarro2004}.

\begingroup
\hypersetup{linkcolor=white} 
\chapter[Vertical flows and mass flux balance of sunspot umbral dots]{Vertical flows and mass flux balance of sunspot umbral dots$^1$}\label{Ud3Chapter}\footnote{Submitted to Astronomy \& Astrophysics Letters, see \citet{Riethmueller2013}.}
\endgroup

\section{Introduction}

   Umbral dots (UDs) are small brightness enhancements in sunspot umbrae
   or pores and were first detected by \citet{Chevalier1916a}. The strong
   vertical magnetic field in umbrae suppresses the energy transport by convection
   \citep{Biermann1941}, but some form of remaining heat transport is needed to
   explain the observed umbral brightness \citep{Adjabshirzadeh1983}. Magnetoconvection
   in umbral fine structure, such as UDs and light bridges, is thought to be the main
   contributor to the energy transport in the umbra \citep{Weiss2002}, see
   reviews by \citet{Solanki2003,Sobotka2006,Borrero2011}.

   Progress in the physical understanding of \Index{umbral dot}s was made by the
   numerical simulations of three-dimensional radiative magnetoconvection
   \citep{Schuessler2006,Bharti2010}. Most of the simulated UDs have a horizontally
   elongated shape and show a central \Index{dark lane} in their bolometric intensity images.
   In the deepest photospheric layers, the inner parts of UDs exhibit magnetic
   field weakenings and upflow velocities. The simulated UDs are surrounded by downflows that
   are often concentrated in narrow downflow channels at the endpoints of the dark
   lanes \citep{Schuessler2006}. Higher up in the photosphere, the UDs in the simulations
   do not differ significantly from the diffuse background.

   Considerable efforts on the observational side were made to test these theoretical
   predictions. Dark lanes inside UDs were found in the observations of \citet{Bharti2007}
   with the 50~cm \hinode{} telescope and by \citet{Rimmele2008}, who observed with the
   76~cm \Index{Dunn Solar Telescope}. However, \citet{Louis2012} analyzed straylight corrected
   \hinode{}/BFI data and did not find \Index{dark lane}s in their observed UDs, which leaves
   room for doubt whether the observed phenomena are really identical with the synthetic ones.
   The UDs described by \citet{Bharti2007} differ from those reported by \citet{Schuessler2006}
   in that the area of the observed features is an order of magnitude larger, possibly they
   are the remains of a decayed \Index{light bridge}.

   More important than the \Index{dark lane}s are the flows, since they are central to the convective
   nature of the UDs. \citet{Riethmueller2008c}, using inversions of \hinode{}/SP data,
   discovered upflows in the deep layers of peripheral UDs (PUDs) but not in central UDs (CUDs),
   while downflows were not detected. Later, \citet{Ortiz2010} studied a small pore recorded
   with the \Index{CRISP} instrument of the 1~m Swedish Solar Telescope and found irregular
   and diffuse downflows in the range 500-1000~m/s for a small set of 5 UDs. In contrast,
   in their recent study, \citet{Watanabe2012} analyzed a larger set of 339 UDs, also
   observed with \Index{CRISP}, and found significant UD upflows, but no systematic downflow
   signals. Thus, the existence of downflows in or around UDs remains uncertain, so that
   the fate of the material flowing up in UDs is unclear. The depth-dependent inversions
   of full Stokes profiles done by \citet{SocasNavarro2004} and later at higher resolution
   by \citet{Riethmueller2008c} revealed a temperature enhancement and a field weakening
   for the UDs compared to the nearby umbral background, with both being strongest in the
   deepest observed layers.

   Since the observational picture is inhomogeneous, there is a need for a further UD study
   with high spatial and spectral resolution being of utmost importance. In this work, the improved
   Stokes inversion method of \citet{vanNoort2012} is applied to \hinode{}/SP data
   \citep[see][]{vanNoort2013}. This so-called 2D inversion\index{inversiontwodim@2D inversion} method allows the depth-dependent
   structure to be obtained basically as it would be in the absence of the telescope's
   \Index{point spread function} (\Index{PSF}).

\section{Observation, data reduction, and analysis}\label{SecObs}

\begin{figure}
   \centering
   \includegraphics[width=120mm]{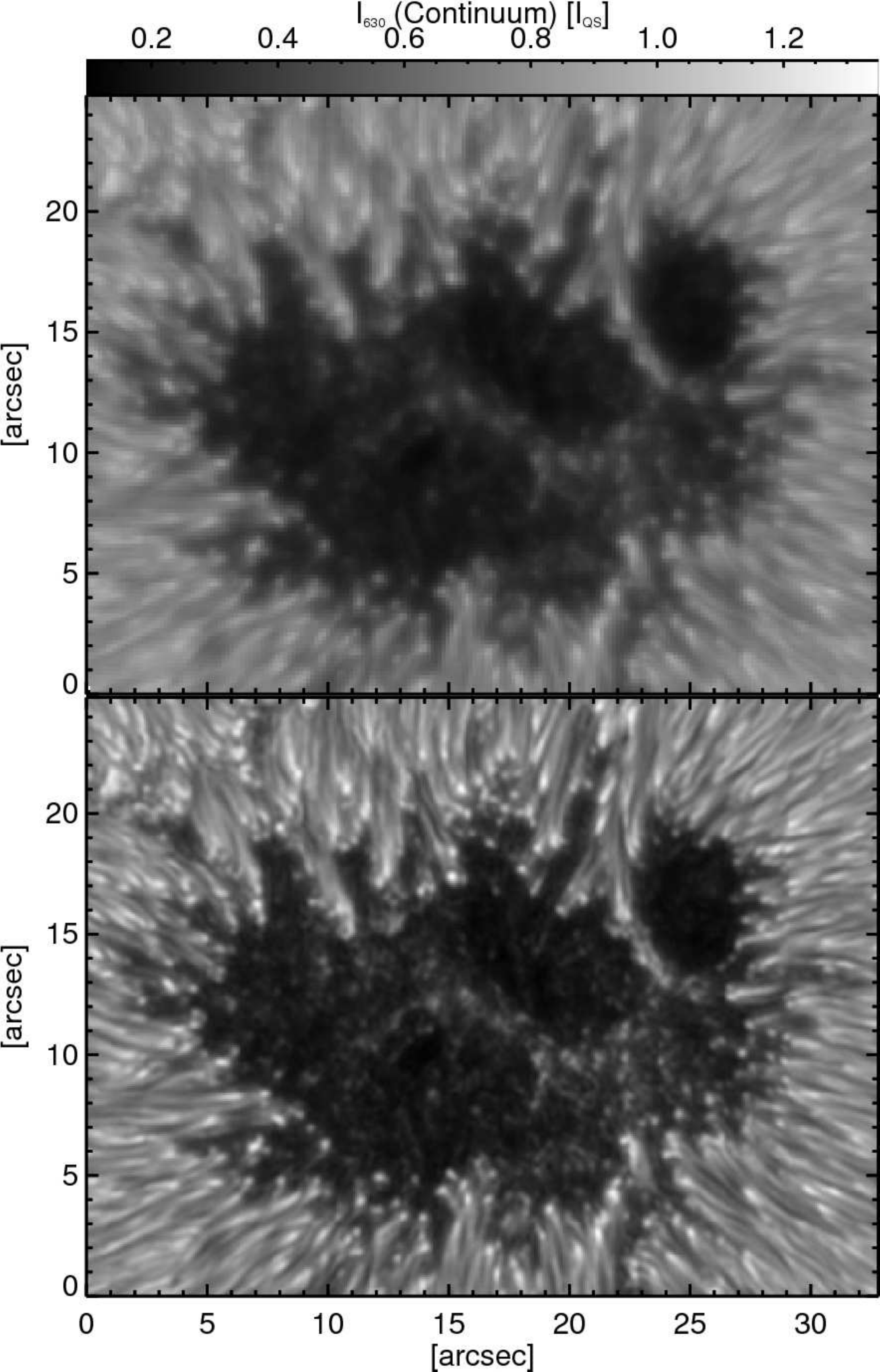}
   \caption{Stokes~$I$ continuum intensity of the \hinode{}/SP map of a \Index{sunspot} \Index{umbra} of NOAA AR~10933 at
   original resolution (top panel). The Stokes~$I$ continuum resulting from the 2D inversion\index{inversiontwodim@2D inversion} is shown in
   the bottom panel. The intensity is normalized to the mean quiet-Sun intensity $I_{\rm{QS}}$.}
   \label{Fig1}
   \end{figure}

   The data we analyzed in this study were recorded from 12:43 to 13:00~UT on 2007 January 5
   with the spectropolarimeter \citep[SP,][]{Lites2001} of the Solar Optical Telescope
   \citep[SOT,][]{Tsuneta2008} on the \hinode{} spacecraft \citep{Kosugi2007}.
   SP was operated in its normal map mode, i.e. the integration time per slit position
   was 4.8~s, resulting in a noise level of $10^{-3}$ (in units of the continuum intensity).
   The sampling along the slit, the slit width as well as the scanning step size were 0\carcsec{}16,
   the spectral sampling in the considered range from 6300.89 to 6303.26~{\AA} was 21~{m\AA}~pixel$^{-1}$.
   The center of the observed umbra was located very close to the disk center, at a heliocentric angle
   of 2.6$^\circ$. The full Stokes profiles were corrected for dark current as well as flat field effects
   and calibrated with the $sp\_prep$ routine of the SolarSoft package. Part of the calibrated Stokes~$I$
   continuum intensity map obtained by \hinode{} SP is shown in the upper panel of Fig.~\ref{Fig1}. The original FOV
   is much larger and contains quiet-Sun regions that are used for the intensity normalization.

   Under the assumption of local \Index{thermodynamic equilibrium}, the Stokes profiles of the Fe\,{\sc i}
   6301.5~{\AA} and 6302.5~{\AA} lines were inverted by applying the version of the \Index{SPINOR} inversion
   code \citep{Frutiger2000a,Frutiger2000b} extended by \citet{vanNoort2012}. In this version of the code,
   the instrumental effects responsible for the spectral and spatial degradation of the observational data
   are taken into account, so that the inverted parameters correspond to spatially deconvolved
   values (but without the added noise that deconvolution generally introduces). The observational data are
   spatially upsampled by a factor of two, so that the input and output data of the \Index{SPINOR} inversion have
   a sampling of 0\carcsec{}08 per pixel \citep[for details, see][]{vanNoort2013}. The spatial PSF
   used by the code is taken from \citet{Danilovic2008} and considers the 0.5~m clear aperture of the SOT,
   the primary mirror's central obscuration, the three spiders, and a \Index{defocus} of 0.1~waves. Height dependent
   temperature, LOS velocity, magnetic field strength, field inclination, field azimuth, and \Index{micro-turbulence}
   are determined at three $\log{\tau_{500}}$ nodes: $-2.5$, $-0.9$, and $0$. More details of the inversion
   of this spot are provided by \citet{vanNoort2013} and \citet{Tiwari2013}.

   A continuum map obtained from the best-fit Stokes~$I$ profiles of the 2D inversion\index{inversiontwodim@2D inversion} result can be seen
   in the bottom panel of Fig.~\ref{Fig1}. Since the deconvolution of the data with the theoretical spatial
   \Index{PSF} is now indirectly part of the inversion process, the contrast is significantly enhanced and
   umbral dots can be identified much more clearly than in the original data. Hence, the continuum map
   in the bottom panel of Fig.~\ref{Fig1} is used for a manual detection of the location of the most
   prominent 81 UDs which we divided into 25 central UDs (CUDs) and 56 peripheral UDs (PUDs) depending
   on their distance to the umbra-penumbra boundary. Once the locations of the UDs' centers are known,
   the UD boundaries are determined from the continuum map by a \Index{multilevel tracking} (\Index{MLT}) algorithm
   \citep[see][]{Bovelet2001}, using 25 equidistant intensity levels. The resulting contiguous MLT
   structures are then cut at 50\% of the local min-max intensity range, which is taken as the UD boundary.
   A detailed description of the use of the MLT algorithm for isolating UDs is given by
   \citet{Riethmueller2008d}. 

   The knowledge of the UD boundaries allowed us to average UD properties over all pixels within the
   UD boundary. Stratifications of temperature, LOS velocity, and field strength of the UDs were then determined
   as such averages. The same quantities were also determined for the UDs' diffuse background (DB), defined
   as the average over all pixels in a 400~km wide ring around the UD boundary. UD and DB quantities were
   retrieved for optical depths between $\log{\tau_{500}}=-2.5$ and 0 in steps of $0.5$.

   The LOS velocity maps of the inversion result show a clear $p$-mode\index{pmode@$p$-mode} pattern with a spatial wavelength of
   about 10\arcsec{} that has to be removed to avoid any $p$-mode influence on our results. The usually employed
   technique of \Index{Fourier filter}ing in three-dimensional $k\omega$-space cannot be applied in our case because
   only a single map of the observational data was available for inversion. We therefore removed the
   $p$-modes in the LOS velocity maps at all used $\log{\tau_{500}}$ nodes by applying a highpass filter
   (implemented as the difference between the original velocity map and its running boxcar, $21\times{}21$ pixels,
   filtered counterpart). Since our results depend on a careful zero velocity determination, we re-calibrated
   the velocities even if the highpass filter already roughly removed the velocity offset. To achieve this
   we assume that the dark core of the umbra is at rest. The darker part of the umbra is identified by
   thresholding the lowpass filtered continuum image ($11\times{}11$ pixels) at 50\% of the intensity range.
   We further excluded a circle of 600~km radius around each of the 81 identified UDs and subtracted the
   mean velocity of the remaining dark umbral pixels from the velocity maps at each optical depth. This
   procedure was found to be robust in the sense that changing the threshold for identifying the darkest part
   of the umbra by $\pm10\%$, or increasing the radius of the exclusion zone around the UDs by 200~km did not
   influence our results.

\section{Results}

   Even at the significantly improved image quality provided by the inversion, we were not able
   to find \Index{dark lane}s in the central parts of the UDs as reported by \citet{Schuessler2006} in
   MHD simulations and by \citet{Bharti2007} in other deconvolved \hinode{} images.

\begin{figure}
   \centering
   \includegraphics[width=\linewidth]{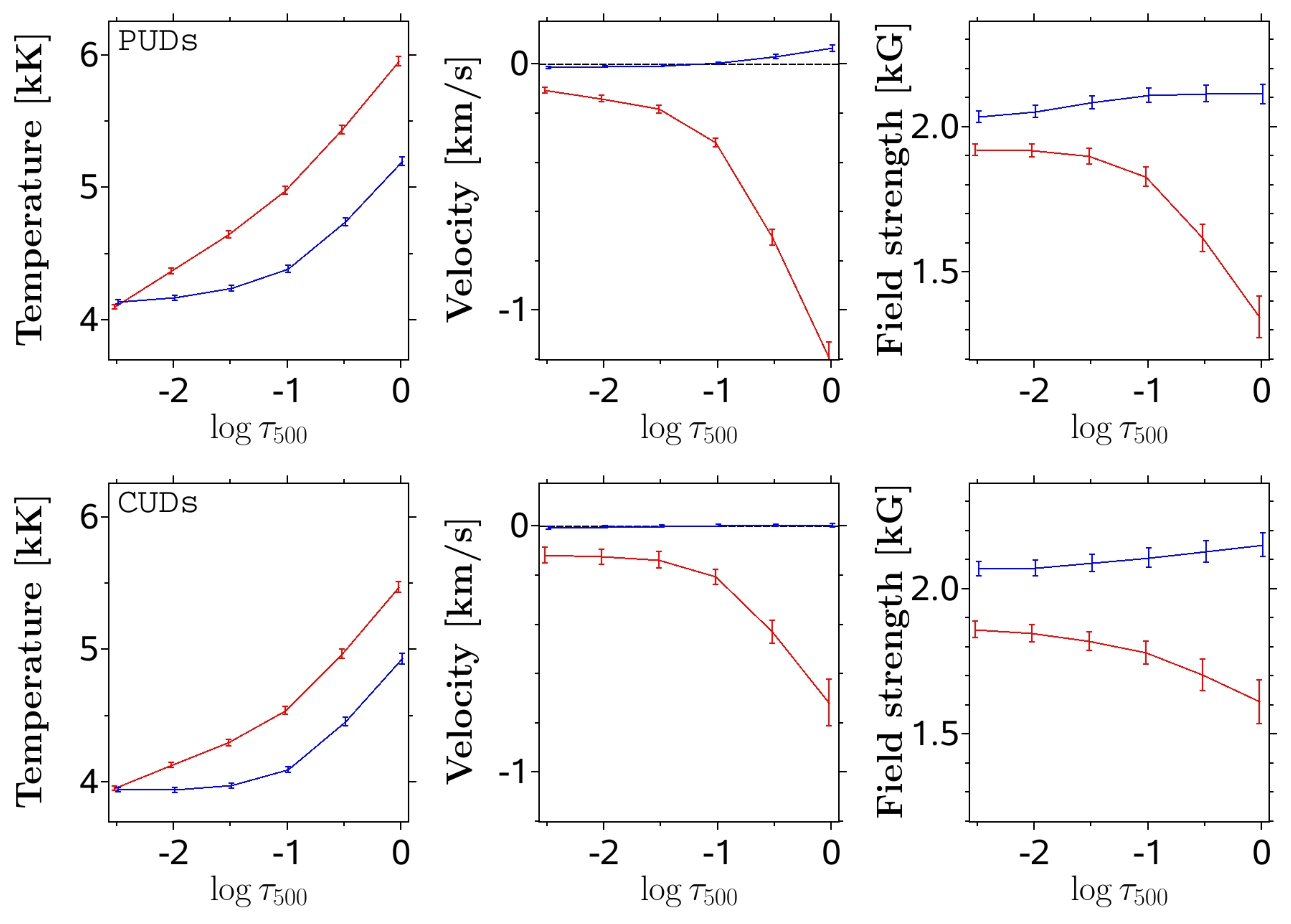}
   \caption{Optical depth dependence of temperature, LOS velocity, and magnetic field strength averaged
   over 56 peripheral umbral dots (top 3 panels) and 25 central umbral dots (bottom 3 panels). The error
   bars denote standard deviations of the mean ($\sigma/\sqrt{N}$). The red lines exhibit the
   stratifications of the mean UD while the blue lines correspond to the mean nearby diffuse background
   (see main text for details). Negative LOS velocity values indicate upflows.}
   \label{Fig2}
   \end{figure}

   The stratifications of temperature, velocity, and field strength, averaged separately over all PUDs
   and CUDs, are displayed in Fig.~\ref{Fig2}. While in the upper photosphere ($\log{\tau_{500}}=-2.5$)
   the considered properties hardly differ between the mean UD and DB, they deviate significantly from
   each other in the deep photosphere ($\log{\tau_{500}}=0$). Compared to their DB we find at optical
   depth unity a temperature enhancement and a field weakening of 760~K and 770~G, respectively, for the
   mean PUD, while for CUD the values are 540~K and 540~G. The mean UD magnetic field weakens with depth
   and the weakening in PUDs is more enhanced than the one in CUDs. The field strength of the mean DB
   increases with depth, as expected.

\begin{figure}
   \centering
   \includegraphics[width=\linewidth]{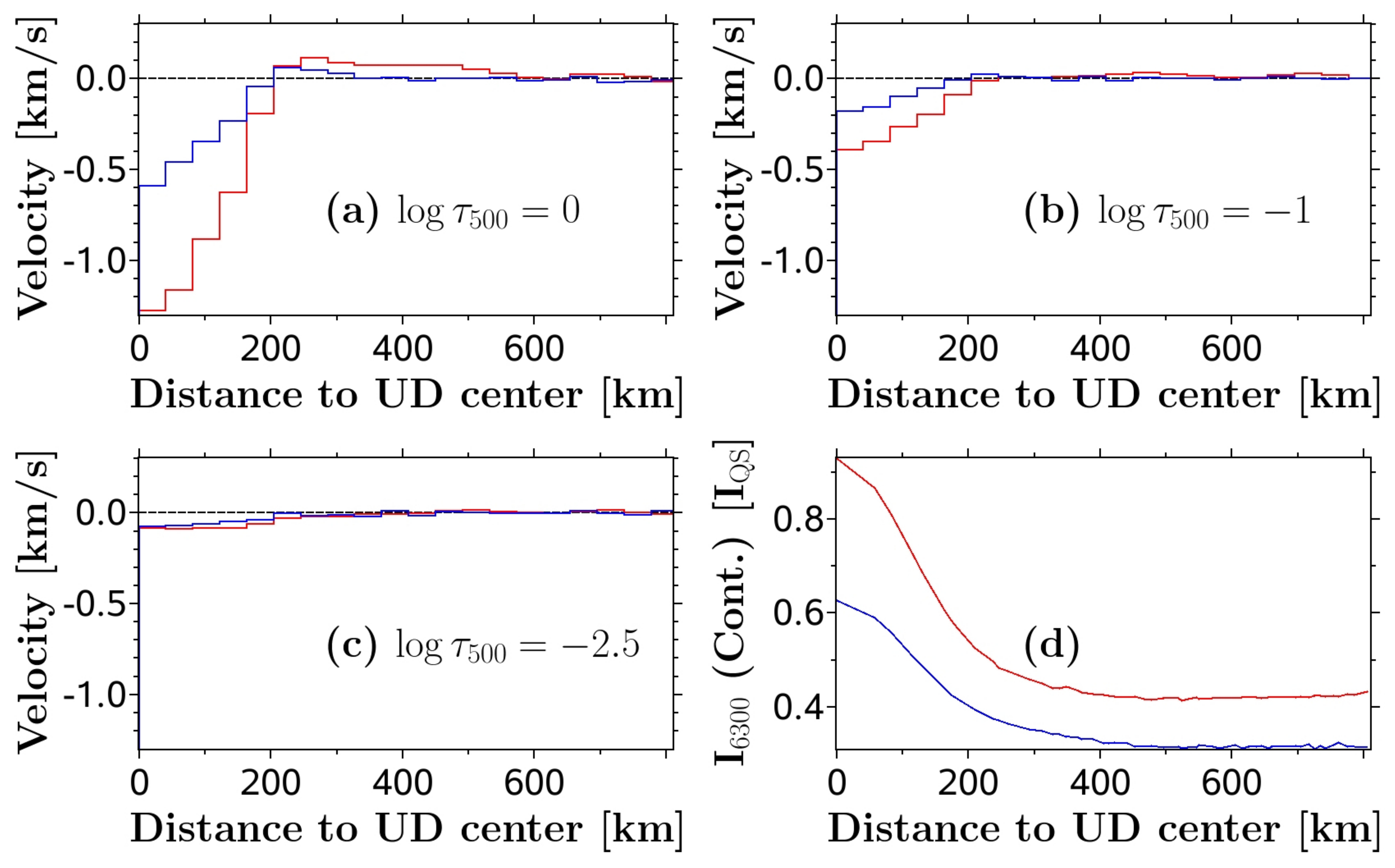}
   \caption{Panel~(a)-(c): LOS velocity at constant optical depth as function of the distance to the UD center averaged
   over all azimuthal angles of the peripheral (red lines) and central (blue lines) UDs. The optical depths
   are given as text labels. Panel~(d): Mean continuum intensity profile of the two UD classes.}
   \label{Fig3}
   \end{figure}

   The LOS velocity (which is virtually identical to the vertical velocity component due to the small
   heliocentric angle) of DB and UD is almost zero in the upper photosphere. Strong upflows of
   $-1200$~m/s and $-720$~m/s are found at optical depth unity for the mean PUD and CUD, while a weak
   but significant downflow of 64~m/s is found for the mean DB of the peripheral UDs only. These values
   should be compared with the uncertainty in the velocity averaged over the DB of PUDs,
   $\sigma/\sqrt{N}=14$~m/s ($\sigma$ - standard deviation, $N$ - number of UDs). The DB of the central UDs
   is on average at rest within the error bars. Further insight into the up- and downflows in and around
   UDs is given by Fig.~\ref{Fig3}. We averaged the velocities of all pixels at roughly the same distance
   to the UD center and plotted such mean velocities as a function of the distance to the UD's center,
   again separately for PUDs and CUDs. Note that this method is independent of any determination of the
   UD boundary. Panel~(a) of Fig.~\ref{Fig3} shows the velocities at optical depth unity. Between $0-200$~km
   distance from the UD's center, we find upflows. Then, from 200 to 600~km we see downflows (between 200
   and 350~km for CUDs), while for distances larger than 600~km the velocity is almost zero.  The downflows
   in the lower photosphere peak at a distance of roughly 220~km from the UD's center and have values of
   110~m/s and 60~m/s for the mean PUD and CUD, respectively. They are thus minute compared to the maximum
   upflows in the UDs of $-1280$~m/s and $-600$~m/s. Still, the upflows as well as the downflows are on
   average stronger for the PUDs than for the CUDs. Both up- and downflows increase rapidly with depth
   (compare the panels (a)-(c)). The upflows within the UDs are much weaker at $\log{\tau_{500}}=-1$
   (panel~(b)) and the downflows cannot be seen anymore. At $\log{\tau_{500}}=-2.5$ (panel (c)), there
   is almost no velocity signal at all. The mean intensity profiles are plotted in panel~(d) and revealed
   a half-width-half-maximum radius of 160~km for the PUDs and 140~km for the CUDs.

   We next calculated \Index{mass flux}es as the sum over all pixels within a 600~km vicinity of the UD center since
   for larger distances the velocity is negligible. The required densities are provided by the \Index{SPINOR} code
   under the assumption of \Index{hydrostatic equilibrium}. Table~\ref{MassFluxes} lists the upward and downward
   mass fluxes per UD, $M_u$ and $M_d$, for various optical depths and separated for the two UD classes.
   The uncertainties in Table~\ref{MassFluxes} are the standard deviations of the averages over all UDs
   of a given class ($\sigma$). The last two columns give the mass flux ratios. The mass flux increases
   strongly with depth due to the density and velocity increase. Even if the downflows are minute compared
   to the upflows, they cover a much larger area so that the upward and downward mass flux are roughly
   balanced within the uncertainties.
   \begin{table*}
   \caption{Upward and downward mass fluxes per umbral dot (computed within 600~km radii) and their ratios at various optical depths.}
   \vskip3mm
   \label{MassFluxes}                                
   \centering                                        
   \begin{tabular}{c c c c c c c}                    
   \hline                                            
   \noalign{\smallskip}
   $\log{\tau_{500}}$ & $M_u$                     & $M_u$                     & $M_d$                     & $M_d$                     & $M_u/M_d$          & $M_u/M_d$ \\
                      & PUD                       & CUD                       & PUD                       & CUD                       & PUD                & CUD       \\
                      & [\tt{1O}$^9\frac{kg}{s}$] & [\tt{1O}$^9\frac{kg}{s}$] & [\tt{1O}$^9\frac{kg}{s}$] & [\tt{1O}$^9\frac{kg}{s}$] & []                 & []        \\
   \noalign{\smallskip}                                                                                                                                      
   \hline                                                                                                                                                    
   \noalign{\smallskip}                                                                                                                                      
   \tt{~O~~}          & \tt{~89$\pm$22~}          & \tt{ 75$\pm$13~}          & \tt{1O5$\pm$26~}          & \tt{~69$\pm$1O~}          & \tt{O.85$\pm$O.42} & \tt{1.O9$\pm$O.35} \\
   \tt{-1~~}          & \tt{~17$\pm$4~~}          & \tt{ 22$\pm$4~~}          & \tt{~18$\pm$4~~}          & \tt{~21$\pm$4~~}          & \tt{O.98$\pm$O.45} & \tt{1.O2$\pm$O.36} \\
   \tt{-2~~}          & \tt{4.6$\pm$1.3}          & \tt{6.O$\pm$1.4}          & \tt{4.1$\pm$1.2}          & \tt{5.5$\pm$1.4}          & \tt{1.12$\pm$O.67} & \tt{1.1O$\pm$O.52} \\
   \tt{-2.5}          & \tt{2.5$\pm$O.8}          & \tt{3.3$\pm$O.9}          & \tt{2.2$\pm$O.7}          & \tt{2.9$\pm$O.8}          & \tt{1.17$\pm$O.72} & \tt{1.14$\pm$O.62} \\
   \hline                                            
   \end{tabular}
   \end{table*}

\section{Discussion and Conclusions}

   We have used a new \Index{inversion} technique to retrieve the atmospheric parameters of 81 UDs
   in a sunspot umbra from data recorded with the \Index{spectropolarimeter} onboard \hinode{}.
   In agreement with earlier studies \citep{SocasNavarro2004,Riethmueller2008c}, we find
   that in the deep photosphere the temperature is enhanced and the magnetic field is weakened
   in the UDs compared to their umbral surroundings. Table~\ref{LitComp} compares the main
   UD properties retrieved from the conventional and the improved inversion technique. For
   a direct comparison with \citet{Riethmueller2008c}, who reported peak values rather than
   spatial averages, we also give the peak values obtained from the new inversion in
   Table~\ref{LitComp}. In fact, nearly all values listed in Table~\ref{LitComp} are higher
   for the new inversion method, which emphasizes the considerably improved data quality reached by
   the implicit removal of the telescope's spatial \Index{PSF}.

   \begin{table}
   \vskip3mm
   \caption{Comparison of UD properties at the continuum formation height between \citet{Riethmueller2008c} and this study.}
   \vskip3mm
   \label{LitComp}                                   
   \centering                                        
   \begin{tabular}{c c c c c c c}                    
   \hline                                            
   \noalign{\smallskip}
   UD class                            & PUD        & PUD       & PUD       & CUD        & CUD       & CUD      \\
   study                               & \tt{2OO8}a & this      & this      & \tt{2OO8}a & this      & this     \\
   value type                          & peak       & peak      & avg       & peak       & peak      & avg      \\
   \hline
   \noalign{\smallskip}
   $T_{\rm{UD}}-T_{\rm{DB}}\,$[\tt{K}] & \tt{57O}   & \tt{112O} & \tt{~76O} & \tt{55O}   & \tt{~79O} & \tt{54O} \\
   $B_{\rm{DB}}-B_{\rm{UD}}\,$[\tt{G}] & \tt{51O}   & \tt{1O2O} & \tt{~77O} & \tt{48O}   & \tt{~72O} & \tt{54O} \\
   $v_{\rm{up}}\,$[\tt{m/s}]           & \tt{8OO}   & \tt{224O} & \tt{128O} & \tt{-~~}   & \tt{126O} & \tt{6OO} \\
   $v_{\rm{down}}\,$[\tt{m/s}]         & \tt{-~~}   & \tt{-~~~} & \tt{~11O} & \tt{-~~}   & \tt{-~~~} & \tt{~6O} \\
   \hline                                            
   \end{tabular}
   \end{table}

   The 2D inversion\index{inversiontwodim@2D inversion} results revealed clear upflow signals for both UD types, while \citet{Riethmueller2008c}
   could only find them for the PUDs. On average, UDs show upflows up to a radial distance of 200~km from
   their centers. In general, these upflows are stronger for PUDs than for CUDs. Between 200 and 600~km
   from a UD's center, we find small but significant downflows, whereas there is no relevant velocity signal
   further away. The velocity signal decreases rapidly with atmospheric height.

   Previous observational studies detected upflows, but could not detect downflows associated
   with UDs \citep[e.g.][]{SocasNavarro2004,Riethmueller2008c,Watanabe2012}, or at least not systematically
   \citep{Ortiz2010}. This raised the question where all the upflowing plasma ends up. The first systematic
   detection of downflows around UDs in this paper gives us the possibility of calculating upward and
   downward mass fluxes. Our finding of rather well-balanced mass fluxes depends on a careful velocity calibration,
   described in Section~\ref{SecObs}. If all umbral pixels had been used for the zero velocity determination,
   the zero velocity could possibly be blue-shifted due to the UD upflows, thus giving rise to artificial
   downflows. This effect is ruled out since our velocity re-calibration ignores the UDs as well as their
   600~km surroundings and uses the darkest parts of the umbra only.

   We further believe that the downflows seen in the top left panel of Fig.~\ref{Fig3} are real and not a result
   of ringing effects. Such effects could be caused by the nearly axisymmetrical shape of the spatial PSF used
   in our inversion and would have affected all quantities. However, plots of temperature and field strength versus
   the UD center distance (not shown) do not exhibit any signs of ringing. According to \citet{Schuessler2006}, the
   downflows are concentrated in narrow channels preferentially at the endpoints of the central \Index{dark lane}s
   of the UDs. Our spatial resolution is insufficient to detect the dark lanes nor the narrow downflow
   channels. The relatively small downflow signals become significantly larger than the noise only after
   the azimuthal averaging around UDs.

   The picture introduced by \citet{Schuessler2006} of UDs as a natural consequence
   of magnetoconvection in the strong vertical magnetic field of an umbra is confirmed by our study.
   In deep layers the rising hot plasma pushes the field to the side, weakening the field there. Around the
   continuum formation height the rising gas cools by radiative losses, turns over, and flows down around
   the UDs. Such evidence for overturning comes from mass balance between up- and downflows and from the fact
   that the central upflows are associated with hot material, whereas the peripheral downflows are cool.
   Furthermore, the rapid decrease of upward mass flux with height is also typical for overturning, overshooting
   convection. The fact that PUDs are associated with larger flow velocities, greater temperature enhancements,
   and stronger field reduction suggests a more vigorous convection than in the central umbra.

   We suggest observations at a spatial resolution higher than available here, e.g. with the reflight of
   the 1~m \sunrise{} telescope \citep{Solanki2010} or with the 1.5~m \solarC{} telescope, to determine
   if the UDs show the predicted central \Index{dark lane}s with narrow downflow channels at their endpoints.

\begingroup
\hypersetup{linkcolor=white} 
\chapter[Bright points in the quiet Sun as observed by \sunrise{}]{Bright points in the quiet Sun as observed in the visible and near-UV by the balloon-borne observatory \sunrise{}$^1$}\label{Bp1Chapter}\footnote{Published in The Astrophysical Journal Letters, 723, 169 (2010), see \citet{Riethmueller2010}.}
\endgroup

\section{Introduction}

   Bright points\index{bright point} (BPs) are small-scale brightness enhancements located in the
   darker \Index{intergranular lane}s. \citet{Dunn1973} and \citet{Mehltretter1974} were the first to describe
   these BPs in filter images taken in the far line wings of H$\alpha$ and Ca\,{\sc ii}~K,
   respectively. Common models consider BPs as radiative signatures of
   magnetic elements, which are often described by nearly vertical slender \Index{flux tube}s or sheets\index{flux sheet}
   \citep{Deinzer1984,SolankiEtAl1993}. The increased \Index{magnetic pressure} within the flux tube leads to
   its evacuation, and the lateral inflow of heat through the walls of the flux tube makes it hot
   and bright \citep{Spruit1976}. Consequently, BPs are often used as tracers of magnetic elements.

   The contrast of BPs relative to the average quiet-Sun intensity is of interest since it provides
   insight into the structure and thermodynamics of magnetic elements. Furthermore,
   the excess brightness of magnetic elements is an important contributor to variations of the total
   solar irradiance \citep{Solanki2002}. In the visible, the contrast of BPs is particularly high
   in wavelength regions that are dominated by absorption bands of temperature-sensitive molecules
   such as CH and CN \citep{Muller1984,Berger1995,Zakharov2005,Berger2007,Utz2009}. We expect that
   their contrast is also large in the \Index{ultraviolet} (\Index{UV}), but they have never been studied at
   wavelengths shorter than 388~nm.

   In this work, we extend the study of BPs, in particular their contrasts, to the UV spectral range
   down to 214~nm, using seeing-free images gathered by the balloon-borne 1-m aperture \sunrise{}
   telescope. This is particularly important owing to the finding that irradiance changes below
   400~nm produce over 60\% of the variation of the \Index{total solar irradiance} over the solar cycle
   \citep{Krivova2006}. The spectral and total irradiance variations are caused by the variation
   of magnetic flux at the solar surface, in particular in the form of small-scale magnetic elements
   \citep{Krivova2003,Krivova2006,Wenzler2006}.

\section{Observations, data reduction, and analysis}

   The data employed here were acquired during the 2009 June stratospheric flight of \sunrise{}. Technical
   details of the telescope are described by \citet{Barthol2011}. Image stabilization, feature tracking,
   and correction of low-order wavefront aberrations were achieved by the gondola's pointing system in
   conjunction with a six-element Shack-Hartmann correlating wavefront sensor\index{Shack-Hartmann wavefront sensor} \citep{Berkefeld2011}.

   Observations in the near-ultraviolet spectral domain between 214~nm and 397~nm were acquired with the
   Sunrise Filter Imager \citep[\Index{SuFI},][]{Gandorfer2011}. Simultaneously, the full Stokes \Index{magnetograph}
   \Index{IMaX} \citep{MartinezPillet2011} scanned the Fe\,{\sc i} line at 525.02~nm (\Index{Land\'e factor} $g=3$), 
   hence providing kinematic and magnetic information. An overview of the collected data and a description
   of some of the observed phenomena is given by \citet{Solanki2010}.

   We use three time series recorded from 00:36 to 00:59~UT, 01:31 to 02:00~UT, and 14:22 to 15:00~UT
   on 2009 June 9. At these times, the telescope pointed to quiet regions close to the disk center.
   The SuFI instrument recorded filtergrams centered at 397~nm (Ca\,{\sc ii}~H), 388~nm (CN), 312~nm, 300~nm,
   and 214~nm, while IMaX was operated in its standard vector spectropolarimeter mode, i.e., full Stokes
   observations at five wavelength points with six accumulations. The five wavelength points were set to
   $\lambda-\lambda_0=-80,-40,+40,+80,$ and $+227$~m\AA~relative to the center of the line. A summary
   of the filter widths (FWHM) and exposure times can be found in Table~\ref{ExpTimes}. The effective exposure
   time of an IMaX continuum image was 6~s (6 accumulations $\times$ 4 modulation states $\times$ 250~ms), while
   \Index{Dopplergram}s need five wavelength points; i.e., their effective exposure time was 30~s
   \citep[see][for more details]{MartinezPillet2011}.

   \begin{table}
   \caption{Exposure times of the used time series.}
   \vskip3mm
   \label{ExpTimes}                                      
   \centering                                            
   \begin{tabular}{l l l l l}                            
   \hline                                                
   \noalign{\smallskip}
   Central     & FWHM         & 00:36$-$00:59 & 01:31$-$02:00 & 14:22$-$15:00 \\
   Wavelength  & of Filter    & Exp. Time     & Exp. Time     & Exp. Time     \\
   \noalign{\smallskip}                                                                                                                           
   (nm)        & (nm)         & (ms)          & (ms)          & (ms)          \\
   \hline                                                                                                                                            
   \noalign{\smallskip}                                                                                                                              
   \tt{214}    & \tt{1O}      & \tt{...}      & \tt{...}      & \tt{3OOOO}    \\
   \tt{3OO}    & \tt{~5}      & \tt{5OO}      & \tt{5OO}      & \tt{~~25O}    \\
   \tt{312}    & \tt{~1.2}    & \tt{15O}      & \tt{15O}      & \tt{~~3OO}    \\
   \tt{388}    & \tt{~O.8}    & \tt{~8O}      & \tt{~8O}      & \tt{~~15O}    \\
   \tt{396.8}  & \tt{~O.18}   & \tt{96O}      & \tt{96O}      & \tt{~~9OO}    \\
   \tt{525.O2} & \tt{~O.OO85} & \tt{25O}      & \tt{25O}      & \tt{~~25O}    \\
   \noalign{\smallskip}
   \hline                                                
   \noalign{\smallskip}
   \end{tabular}
   \end{table}

   At a typical flight altitude of around 36~km, the 214~nm wavelength range was still strongly attenuated by
   the residual atmosphere of the Earth, so that an exposure time of 30~s was needed, even at the highest Sun
   elevations achieved during the flight. The 214~nm data were only acquired during the third time series.
   The cadence of the IMaX data was always 33~s, while it was 12~s for the SuFI data of the first
   two time series (i.e., all recorded SuFI wavelengths within 12~s) and it was 39~s for the last time series
   that included the 214~nm channel.

   The data were corrected for dark current and flat field. Additionally, the \Index{instrumental polarization}
   was removed from the IMaX data with the help of Mueller matrices\index{M\"uller matrix} determined by pre-flight
   \Index{polarimetric calibration}s. Finally, the in-flight phase-diversity\index{phase diversity} measurements (permanently for SuFI and intermittently
   for IMaX) were used to correct the images for low-order wavefront aberrations \citep{Hirzberger2011,
   MartinezPillet2011}. The phase-diversity reconstruction\index{image reconstruction} of the SuFI images was done using averaged
   wavefront errors \citep[level-3 data, see][]{Hirzberger2010}. All intensity images were normalized
   to the intensity level of the mean quiet Sun, $I_{\rm{QS}}$, which was defined as the average of the whole
   image. The images were then re-sampled to the common \Index{plate scale} of 0\carcsec{}0207 pixel$^{-1}$
   (original plate scale of SuFI's 300~nm images) via bilinear interpolation, and the common field of view
   (FOV) of 13\arcsec{}$\times$38\arcsec{} was determined. Residual noise was removed
   by applying a running boxcar filter of 3$\times$3 pixels. Its width corresponds to about half of the
   spatial resolution of about 0\carcsec{}15 reached by the considered observations (determined
   from radially averaged power spectra). As a proxy of the longitudinal magnetic field, the \Index{circular polarization degree}
   $\langle p_{\rm{circ}} \rangle$ averaged over the four points in the line ($-80,-40,+40,+80$~m\AA),
   \[ \langle p_{\rm{circ}} \rangle = \frac{1}{4} \sum_{i=1}^{4}{a_i\frac{V_i}{I_i}}~, \]
   was calculated with a reversed sign of the two red wavelength points ($\bfit{a}=[1,1,-1,-1]$) to avoid
   cancelation. The line-of-sight (LOS) component of the velocity vector was obtained by a Gaussian
   fit to the Stokes~$I$ profiles. The velocity maps were corrected for the wavelength shift over the
   FOV caused by the IMaX etalon, and a convective \Index{blueshift} of 200~m~s$^{-1}$ was removed
   \citep[see][]{MartinezPillet2011}.

   Only one set of IMaX and SuFI observations every five minutes was analyzed in order to allow BPs
   to evolve between images. In total, we analyzed 19 image sets distributed over the three time series.
   BPs were manually identified in the CN images at 388~nm, in order to be consistent with earlier work that
   concentrated on visible wavelengths \citep[e.g.][]{Zakharov2005,Berger2007}. At each of the other wavelengths,
   we then determined the local brightness maximum in a 11$\times$11 pixel (i.e., 0\carcsec{}22$\times$0\carcsec{}22)
   environment of the detected BP position at 388~nm. This method takes into account that
   the various wavelengths represent different atmospheric layers (most obvious for the Ca\,{\sc ii}~H images)
   so that inclined magnetic features may appear at slightly different horizontal positions at different wavelengths.
   In total, we detected 398 BPs, of which 211 were from the third time series including the 214~nm band.

\section{Results}

\begin{figure*}
   \centering
   \includegraphics*[width=\textwidth]{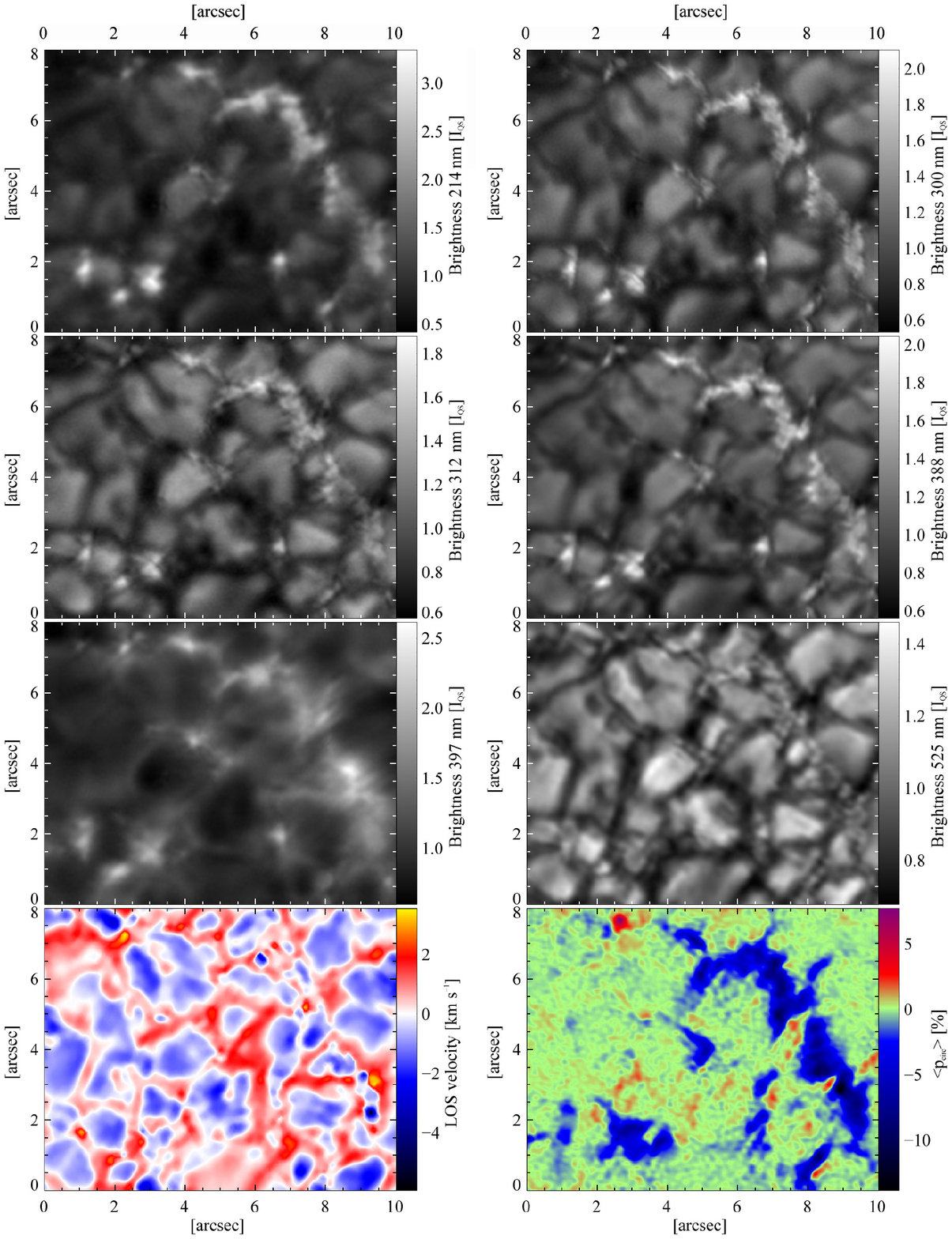}
   \caption{Intensity maps of the five wavelengths observed by SuFI (two first rows and the left panel of the third one)
   and of the continuum sample by IMaX (the right panel of the third row), all normalized to the corresponding mean
   intensity level of the quiet Sun, $I_{\rm{QS}}$. The LOS velocity obtained from a Gaussian fit (positive
   velocities correspond to downflows) and the mean \Index{circular polarization degree} (see the main text for definition)
   are shown in the bottom panels. }
   \label{FigFiltergrams}
   \end{figure*}

   Fig.~\ref{FigFiltergrams} shows BPs at 14:27:08 UT in a region of
   10\arcsec{}$\times$8\arcsec{} (subregion of the
   13\arcsec{}$\times$38\arcsec{} full FOV). The six upper panels display the five SuFI filtergrams and the
   IMaX Stokes~$I$ continuum image ($+227$~m\AA). Most of the bright features are visible in each
   filtergram. However, the contrast of these features at 525~nm is significantly lower than
   in the five shorter wavelengths. Some features show a prominent brightness enhancement in the Ca\,{\sc ii}~H
   image but are of moderate brightness at the other wavelengths, e.g., at position
   (8\carcsec{}6, 3\carcsec{}8). Most remarkable is the high contrast of the BPs at 214~nm,
   which strongly exceeds granulation brightness variations. It is this high contrast that lets the
   granulation appear relatively dark in the first panel of Figure 1, since the gray scales of all panels
   are adapted to the full min/max range.
   The bottom left panel shows the LOS component of the velocity. While the three BPs at the positions
   (1\carcsec{}0, 1\carcsec{}7), (1\carcsec{}8, 0\carcsec{}8), and 
   (2\carcsec{}6, 1\carcsec{}3) show strong downflows of up to 2.5~km~s$^{-1}$, most other BPs
   exhibit only moderate velocities. The bottom right panel displays the mean \Index{circular polarization degree}.
   Most of the BPs in the plotted region are associated with negative polarity. Nonetheless, several bipolar
   regions can be seen, for instance the two neighboring BPs at (2\carcsec{}3, 7\carcsec{}3)
   and at (2\carcsec{}6, 7\carcsec{}7), respectively.

\begin{figure*}
   \centering
   \includegraphics*[width=\textwidth]{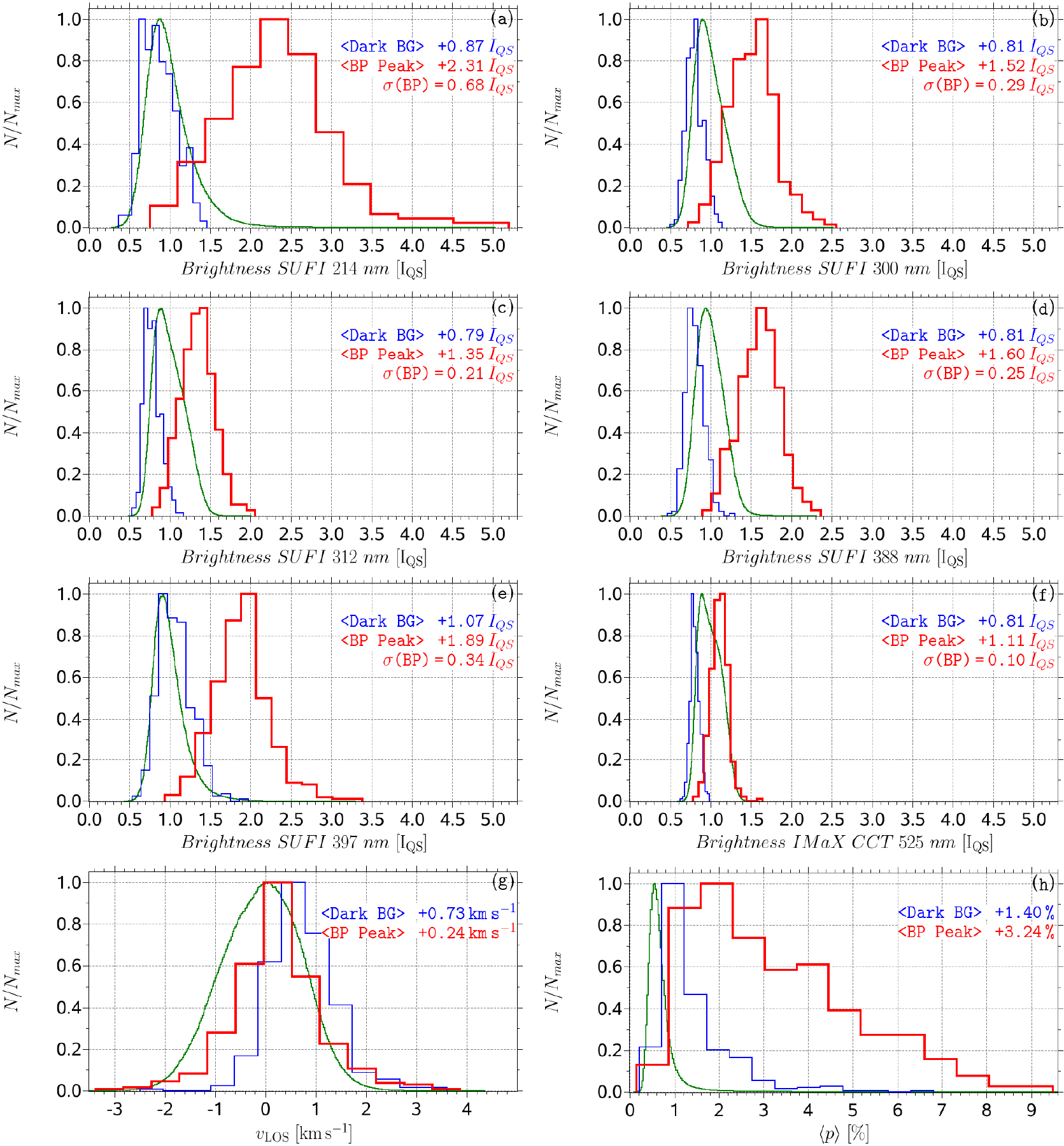}
   \caption{Brightness histograms of the six wavelengths observed by \sunrise{} together with histograms of
   the LOS velocity and the averaged polarization degree. Red lines correspond to histograms of
   the BPs' brightest pixel, blue lines to the darkest pixel in a 0\carcsec{}3 vicinity of the bright point,
   and the green lines denote histograms of all pixels in all frames. The mean values of the red and blue histograms
   as well as the standard deviations of the red brightness histograms are indicated as the text labels.}
   \label{FigHistograms}
   \end{figure*}

   Histograms of the BP peak intensity relative to the mean quiet-Sun intensity are displayed in
   Fig.~\ref{FigHistograms} as red lines in panels (a)$-$(f). The blue lines correspond to the brightness
   histograms of the darkest pixel whose distance to a BP's brightest pixel is less than 0\carcsec{}3
   (typical width of an \Index{intergranular lane}). The text labels denote the
   mean values. The red histograms are largely symmetric, although there is a tendency for a tail toward
   higher contrast values, in particular at 214~nm and 397~nm. As already indicated in
   Fig.~\ref{FigFiltergrams}, the largest average contrast is shown by BPs in the 214~nm image. The
   highest peak brightness of 5.0~$I_{\rm{QS}}$ is also reached at this wavelength. At first sight it might
   be surprising that some BPs are less bright than some of the pixels of the blue histograms. A closer
   look reveals, however, that all BPs are indeed bright relative to the pixels in their immediate
   surroundings. For the Ca\,{\sc ii}~H line, the mean intensity of the darkest pixels
   in the BPs' vicinity is, with 1.07~$I_{\rm{QS}}$, higher than for the other wavelengths, because the Ca
   structures are generally more diffuse and larger than our limit of 0\carcsec{}3.
   The green curves give histograms of the normalized intensity
   for all pixels in all frames, practically representing the intensity
   distribution of quiet-Sun granulation. A comparison of the maximum position
   of the red and green histograms clearly shows that the BPs are much more
   conspicuous in the UV than in the visible spectral range.
   The quiet-Sun histograms are more extended toward higher intensities than
   the histograms for the BP background, since the latter largely represents
   intergranular lanes. Exceptions are the histograms for Ca\,{\sc ii}~H, which
   shows reverse granulation and those for 214~nm, where the comparable width
   indicates structures intermediate between reverse and normal granulation\index{reverse granulation}
   \citep[see][]{Solanki2010}.

   Panel~(g) of Fig.~\ref{FigHistograms} shows the LOS velocity
   histograms for the brightest BP pixels (red), the darkest pixels of their
   vicinity (blue), and for all pixels (green). On average, the LOS velocity of
   the darkest pixels shows a downflow of 730~m~s$^{-1}$, while the BPs themselves
   are associated with a clearly weaker average downflow of 240~m~s$^{-1}$. However,
   the broad wings of the distribution contain BPs with significant upflows of up
   to $-3.1$~km~s$^{-1}$ or downflows of up to 3.6~km~s$^{-1}$. 7.5\% of the
   analyzed BPs have upflows with $v_{\rm{LOS}}<-1$~km~s$^{-1}$ and 15\% of the
   BPs show downflows with $v_{\rm{LOS}}>1$~km~s$^{-1}$. The
   red histogram (BP velocities) is located between the green histogram (dominated
   by the large number of pixels showing upflows) and the blue histogram (mainly
   intergranular pixels showing downflows). Our velocities are
   obtained from a Gaussian fit to the Stokes~$I$ profiles. The Stokes~$V$
   zero-crossings may possibly show different velocities if a BP is not
   spatially resolved.

   In contrast to the last panel of Fig.~\ref{FigFiltergrams}, where the signed
   net circular polarization was plotted, panel (h) of Fig.~\ref{FigHistograms}
   displays the histogram of the unsigned \Index{total polarization degree} $\langle p
   \rangle$ averaged over the four wavelength points in the Fe\,{\sc i} line,
   \[ \langle p \rangle = \frac{1}{4} \sum_{i=1}^{4}{\sqrt{\left( \frac{Q_i}{I_i}
   \right)^2+\left( \frac{U_i}{I_i} \right)^2+\left( \frac{V_i}{I_i} \right)^2}}~. \]
   This histogram is highly asymmetric. The polarization degree at the positions
   of the peak brightness reaches values of up to 9.1\%, the mean value is 3.24\%.
   Such large values of the polarization in 525.02~nm suggest that at least some
   of the magnetic features have been resolved \citep[see][for more details]{Lagg2010}.
   However, 4.3\% of the BPs are associated with $\langle p \rangle <1\%$.
   Although the polarization degree in the darkest surrounding pixels is weaker,
   almost all of them still show a significant polarization degree of more than
   $0.3\%$ (three times the noise level) with the strongest value being 6.6\%.
   Polarization usually displays larger structures than the BPs \citep[see the
   bottom right panel of Fig.~\ref{FigFiltergrams} and ][]{Title1996},
   which can also be concluded from the fact that the blue
   histogram shows larger polarization values than the green histogram for all
   pixels.

   The mean values and their standard deviations for all red and blue histograms in Fig.~\ref{FigHistograms} are
   summarized in Table~\ref{HistValues}. Columns labeled with BP refer to histograms of the BP peak value,
   while DB denotes the histograms of the dark background. The standard deviation of the BP brightness is
   $\approx0.5\left(\left\langle I/I_{\rm{QS}}\right\rangle-1\right)$ in all the SuFI channels, except Ca\,{\sc ii}~H.
   This relationship can be used to estimate the brightness histogram width for other wavelengths.
   The rightmost column of the table shows the ratio of the BP contrast (mean BP brightness value minus one)
   to the rms intensity contrast \citep[taken from][and calculated over the whole FOV of the level-3
   data]{Hirzberger2010}. This contrast is a measure of how strongly BPs stand out relative to the
   surrounding granulation (or reversed granulation\index{reverse granulation} and waves in the case of 397~nm). According to
   Table~\ref{HistValues}, images at 214~nm show the BPs most clearly.

   \begin{table}
   \begin{minipage}{\linewidth}
   \renewcommand{\footnoterule}{}
   \caption{Bright point histogram properties.\protect\footnote{Mean values and standard deviations of the red histograms referring to the bright points (labeled as BP) and blue histograms of the dark background (labeled as DB) of Fig.~\ref{FigHistograms}. The rightmost column compares the BP contrasts with the rms granulation contrast (see the main text for details).}}
   \vskip3mm
   \label{HistValues}                                    
   \centering                                            
   \begin{tabular}{l l l l l l}                          
   \hline                                                
   \noalign{\smallskip}
   Quantity                                    & Mean       & $\sigma$   & Mean       & $\sigma$   & Contrast  \\
                                               & BP         & BP         & DB         & DB         & Ratio     \\
   \hline                                                                                     
   \noalign{\smallskip}                                                                                                                                    
   \tt{$\left( I / I_{\rm{QS}} \right)_{214}$} & \tt{2.31}  & \tt{O.68}  & \tt{O.87}  & \tt{O.21}  & \tt{4.7}  \\
   \tt{$\left( I / I_{\rm{QS}} \right)_{3OO}$} & \tt{1.52}  & \tt{O.29}  & \tt{O.81}  & \tt{O.11}  & \tt{2.4}  \\
   \tt{$\left( I / I_{\rm{QS}} \right)_{312}$} & \tt{1.35}  & \tt{O.21}  & \tt{O.79}  & \tt{O.1O}  & \tt{1.8}  \\
   \tt{$\left( I / I_{\rm{QS}} \right)_{388}$} & \tt{1.6O}  & \tt{O.25}  & \tt{O.81}  & \tt{O.11}  & \tt{3.3}  \\
   \tt{$\left( I / I_{\rm{QS}} \right)_{397}$} & \tt{1.89}  & \tt{O.34}  & \tt{1.O7}  & \tt{O.2O}  & \tt{4.O}  \\
   \tt{$\left( I / I_{\rm{QS}} \right)_{525}$} & \tt{1.11}  & \tt{O.1O}  & \tt{O.81}  & \tt{O.O6}  & \tt{O.8}  \\
   \tt{$v_{\rm{LOS}}\,(\rm{m\,s^{-1}})$}       & \tt{24O}   & \tt{91O}   & \tt{73O}   & \tt{65O}   &           \\
   \tt{$\langle p \rangle\,(\rm{\%})$}         & \tt{3.24}  & \tt{1.82}  & \tt{1.4O}  & \tt{O.84}  &           \\
   \noalign{\smallskip}
   \hline                                                
   \noalign{\smallskip}
   \end{tabular}
   \end{minipage}
   \end{table}


\section{Summary and discussion}

   We identified 398 BPs in simultaneously observed photometric and polarimetric \sunrise{}
   images of a quiet-Sun region close to the disk center. Our data include three wavelengths in the near-UV
   in which the Sun was never observed before at high spatial resolution. We determined the peak brightness
   and the brightness of the dark background of every detected BP at each observed wavelength. The BPs' peak
   intensity reaches up to 5.0 times the mean quiet-Sun intensity $I_{\rm{QS}}$ at 214~nm. The mean peak intensity
   at that wavelength is 2.31~$I_{\rm{QS}}$. The 214~nm wavelength also displays the largest ratio of BP contrast
   to the rms of the intensity over the whole FOV (see the rightmost column of Table~\ref{HistValues}). This ratio
   is a measure of how prominent BPs are in an image at a particular wavelength. These values indicate that
   they are even more prominent at 214~nm than in the core of Ca\,{\sc ii}~H (for the 0.18~nm wide filter
   employed by SuFI).

   The value of the mean peak brightness at 388~nm (1.60~$I_{\rm{QS}}$) agrees exactly with the CN peak
   brightness obtained by \citet{Zakharov2005} from data recorded with the 1~m Swedish Solar Telescope
   (SST). Examples of BP brightness values (close to the disk center) given in the literature for the
   frequently observed $G$-band at 430~nm (CH molecule) are 1.45~$I_{\rm{QS}}$ \citep{Zakharov2005} and
   1.2~$I_{\rm{QS}}$ \citep{Berger2007}. Both studies analyzed SST data. \citet{Utz2009} used $G$-band data from
   the 50~cm Solar Optical Telescope onboard \hinode{} and found 1.3~$I_{\rm{QS}}$. These values are comparable
   with what we find in the near-UV, with the exception of 214~nm, which displays distinctly higher values.
   Note that the BP contrasts given in this study are not corrected for instrumental scattered light\index{stray light}.
   After this correction, the BP contrasts will most likely increase \citep[see, e.g.,][]{Feller2013,Wedemeyer2008, Mathew2009}.

   From the spectropolarimetric data we derived the LOS component of the velocity vector. The
   largest BP velocities reach $-3.1$~km~s$^{-1}$ in upflows and 3.6~km~s$^{-1}$ in downflows. The mean
   value is 240~m~s$^{-1}$, in good agreement with the average velocity of 260~m~s$^{-1}$ published
   by \citet{Beck2007} and reasonably consistent with the absence of Stokes~$V$ zero-crossing shifts
   found by \citet{Solanki1986} and \citet{MartinezPillet1997} in data with much lower spatial
   resolution. In contrast to this, \citet{GrossmannDoerth1996} reported a stronger mean downflow of
   800~m~s$^{-1}$ as derived from the zero-crossing of their Stokes~$V$ profiles and \citet{Sigwarth1999}
   found a velocity range of $\pm$5~km~s$^{-1}$ and a mean velocity of 500~m~s$^{-1}$. The nature of the
   BPs displaying strong up- or downflows in Stokes~$I$ will be investigated in a subsequent study.

   The polarization degree, averaged over the four points within the Fe\,{\sc i} line, is also calculated
   from the IMaX data and shows values up to 9.1\%. The mean BP polarization degree is 3.24\%, which
   is clearly above the mean signal of 1.40\% for the dark vicinity. Intriguingly, about 4\% of the
   BPs are associated with relatively weak \Index{magnetic flux} ($\langle p \rangle <1\%$). This raises interesting
   questions, since enhanced temperatures and hence brightness in magnetic elements (\Index{flux tube}s) is
   caused by evacuation, which in turn is proportional to $B^{2}$. Therefore, we would expect that BPs show
   strong Stokes~$V$ signals. Significantly inclined magnetic fields cannot explain the observed weak
   Stokes~$V$ signals, because we also measured weak Stokes~$Q$ and $U$ signals. Additionally, strongly
   evacuated \Index{flux tube}s are strongly buoyant and hence should be nearly vertical \citep{Schuessler1986}.
   \citet{Lagg2010} find a strong Stokes~$I$ line weakening for kilo-Gauss network patches due to
   temperature enhancements in the flux tubes. Such an absorption weakening can also lead to a weak
   Stokes~$V$ signal. Although the network patches analyzed by \citet{Lagg2010} are spatially resolved,
   we find many BPs that exhibit complex Stokes profiles, which is a clear indication that not all of
   the BPs are spatially resolved. Insufficient resolution can also contribute to weak Stokes~$V$ signals.

   In summary, we find high intensity contrasts of BPs in the near-UV range (including the first
   measurements below 388~nm), with values up to 5~$I_{\rm{QS}}$ at 214~nm. The simultaneous spectropolarimetric
   measurements confirm the close association of BPs with magnetic flux concentrations in intergranular
   downflow lanes. However, the majority of the BPs exhibit only weak downflows.

   The reasonably high cadence of \sunrise{} data (between 4~s and 39~s, depending on the number
   of observed wavelengths and their exposure times) and the high measured contrasts of
   BPs make detailed future studies of the dynamical properties of BPs very promising
   \citep[e.g.,][]{Jafarzadeh2013}. Of considerable additional benefit for such studies will
   be the possibility to compare the dynamics at different layers in the solar atmosphere,
   which are covered by the combination of SuFI and IMaX wavelength bands.

\begingroup
\hypersetup{linkcolor=white} 
\chapter[Comparison of bright points between observations and MHD simulations]{Comparison of solar photospheric bright points between \sunrise{} observations and MHD simulations$^1$}\label{Bp2Chapter}\footnote{To be submitted, see \citet{Riethmueller2013b}.}
\endgroup

\section{Introduction}
   Magnetic fields in the network and in active region plage are often concentrated into strong kilo-Gauss
   field elements \citep{Stenflo1973,Solanki2006,Ishikawa2007}. At high spatial resolution such elements
   appear as \Index{bright point}s \citep[BPs,][]{Berger2001} owing to the inflow of radiation from their walls into
   their evacuated interiors \citep{Spruit1976,Deinzer1984}. Many features of magnetic elements are known
   \citep[see][for an overview]{Solanki1993} and various aspects of the underlying model of magnetic flux
   tubes have been tested, but such tests have generally suffered from the fact that the data were not
   able to spatially resolve magnetic elements.

   This situation has changed with the availability of data recorded by the \sunrise{} balloon-borne
   observatory, which have allowed magnetic elements even in the internetwork quiet Sun to be spatially
   resolved \citep{Lagg2010}. Hence, it is now a good opportunity to revisit BPs and to compare their
   observational properties with predictions from state-of-the-art radiation MHD simulations.

   One motivation to study BPs is their contribution to variations of the total solar irradiance (TSI).
   Around the maximum of the solar activity cycle, the reduction of solar irradiance owing to dark
   sunspots and pores is overcompensated by an increased brightness of the BPs \citep{Froehlich2011}.
   As a result, TSI, i.e. the irradiance integrated over all wavelengths, is on average higher during
   the solar maximum than during minimum \citep{Willson1988}. The TSI variations are only weak over the
   solar cycle, but because 60\,\% \citep{Krivova2006} or even more \citep{Harder2009} of the variations
   in TSI are produced at wavelengths shorter than 4000\,\AA{}, the variations in the ultraviolet (UV)
   can be much more relevant. From recent stratospheric observations we know that the BP contrasts are
   particularly high in the UV \citep{Riethmueller2010}, i.e. the radiative properties of BPs possibly
   play an important role in influencing the Earth's climate. A variation of the UV irradiance changes
   the chemistry of the stratosphere, which can propagate into the troposphere and finally influence
   the climate \citep{London1994,Larkin2000,Haigh2010,Gray2010,Ermolli2012}.

   In addition, there are reasons intrinsic to solar physics why BPs are of interest. Firstly, BPs
   are the most easily visible signatures of strong-field magnetic elements and hence have been widely
   observed \citep[e.g.][]{Muller1984,Berger1995,Berger2001,Utz2009}. Such magnetic elements carry
   much of the magnetic energy, even if they harbor only a small fraction of the magnetic flux.
   Secondly, the flux tubes guide magneto-hydrodynamic waves, which could contribute to coronal heating
   \citep{Roberts1983,Choudhuri1993} and may be related to the waves found to run along spicules
   \citep{DePontieu2007}. Additionally, flux-tube motions can lead to field-line braiding and the
   build up of energy, which may be released through nanoflares \citep{Parker1988}.

   The first science flight of the \sunrise{} observatory revealed the very high contrasts of
   BPs in the UV \citep{Riethmueller2010}. Here we follow up on this work by carrying out a more in-depth
   analysis of a quiet-Sun region as observed by \sunrise{} and by comparing the data with numerical
   simulations of three-dimensional radiative magnetoconvection. Hence we further explore the interplay
   between observation and simulation which has been so fruitful in the past. The MHD simulation data were
   degraded with known instrumental effects that were present during the \sunrise{} observations so that
   they can be compared directly with the observational data. We extend existing studies by considering
   many more observational quantities (intensity at multiple wavelengths, line-of-sight (LOS) velocity,
   spectral line width, and polarization degree) when comparing observational data and MHD simulations,
   which allowed us to test the realism of the MHD simulations far more stringently than by simply comparing
   intensities. We carried out such comparisons in two steps, first for all pixels in the images and later
   restricted to just the pixels identified as lying within BPs. After we satisfied ourselves that
   the simulations give a reasonable representation of the observations, we used the original,
   undegraded simulations to learn more about the BPs and their underlying magnetic features.

\section{Observations, simulations, and degradation}
\subsection{Observations}
   The data we used in this study were acquired with the balloon-borne 1-m aperture \sunrise{}
   telescope that flew from Kiruna in northern Sweden to Somerset Island in northern Canada
   in June 2009 \citep{Barthol2011,Solanki2010}. At the flight altitude of roughly 35\,km,
   99\,\% of the air mass was below the observatory, so that the disturbing influence of the
   Earth's atmosphere (seeing) was minimized. A fast tip-tilt mirror, which was controlled by a
   correlating wavefront sensor, reduced the residual pointing jitter and the onboard adaptive
   optics system corrected the images for low order wavefront aberrations \citep{Berkefeld2011}.
   Two instruments were operated simultaneously: the Sunrise Filter Imager
   \citep[SuFI;][]{Gandorfer2011} and the Imaging Magnetograph eXperiment
   \citep[IMaX;][]{MartinezPillet2011}.

   During the herein considered time series, recorded from 23:00 to 24:00 UT on 2009 June 10,
   the telescope pointed to a quiet-Sun region close to the center of the solar disk ($\mu=1.0$).
   SuFI observed the Sun at the wavelengths 2995\,\AA{} (33\,\AA{} FWHM, 325\,ms exposure time,
   mainly atomic spectral lines), 3118\,\AA{} (8.5\,\AA{} FWHM, 300\,ms exposure time, part of the
   OH band), 3877\,\AA{} (5.6\,\AA{} FWHM, 65\,ms exposure time, CN band), and 3973\,\AA{}
   (1.8\,\AA{} FWHM, 750\,ms exposure time, Ca\,{\sc ii}~H line). Dark current and flat field
   corrections were applied to the SuFI data. The images were phase-diversity (PD) reconstructed
   using the wavefront errors retrieved from the in-flight PD measurements via a PD prism in front
   of the camera. The reconstructed data are referred to as level-2 data, see
   \citet{Hirzberger2010,Hirzberger2011}. Here we concentrated on the 3118\,\AA{} and the
   3877\,\AA{} bands. The spectra in these bands, taken from the NSO spectral atlas of
   \citet{Kurucz1984} are plotted in the upper two panels of Fig.~\ref{FigSynthSpectra} (black lines)
   along with the filter profiles (green lines).


   IMaX scanned the Fe\,{\sc i} line at 5250.2\,\AA{} (Land\'e factor $g=3$) in its L12-2 mode,
   i.e. only Stokes~$I$ and $V$ were measured, at twelve scan positions with two accumulations.
   The twelve scan positions were set to $\lambda-\lambda_0=-192.5,...,+192.5$\,m\AA{} relative
   to the center of the average quiet-Sun profile of the line, in steps of 35\,m\AA. The effective
   spectral resolution of IMaX was 85\,m\AA{} (full-width-half-maximum (FWHM) value). The data were
   corrected for dark current and flat field and interference fringes were removed with a manually
   designed Fourier filter. The IMaX data were then reconstructed with the help of the phase diversity
   technique. Additionally, the instrumental polarization and the residual cross-talk with intensity
   was removed. Stray light was not removed from either data set. All intensity images were divided
   by the mean quiet-Sun value, $I_{\rm{QS}}$, that was defined as the average of the image.

   Finally, the Stokes~$I$ profiles were fitted with a Gaussian function to retrieve the spectral line
   parameters: LOS velocity and line width. The increased reliability of the retrieved parameters for
   more scanned line positions was the main driver for using data with twelve scan positions instead
   of five as in the so-called V5-6 mode of IMaX, which was employed in our former study
   \citep{Riethmueller2010}. Since we were studying BPs, thought to be associated with strong-field,
   relatively vertical magnetic features, the Stokes~$Q$ and $U$ profiles were considered to be less
   important for the present work. This assumption is supported by the analysis of \citet{Jafarzadeh2013},
   who found that BPs extend nearly vertically in height. Fig.~\ref{FigSynthSpectra} shows relevant parts
   of the solar spectrum taken from an NSO spectral atlas \citep{Kurucz1984}. The spectrum centered on the
   Fe\,{\sc i} line at 5250.2\,\AA{} is plotted in the bottom panel. The simulated profile of the
   5250.2\,\AA{} line in the presence of an upflow of 5\,km~s$^{-1}$ is overplotted as a dotted red line
   in order to demonstrate that even strong up- (or downflows) can be safely identified with the L12-2 mode
   of IMaX (see the blue arrows marking the wavelengths sampled by IMaX in L12-2 mode), but can be missed
   or missidentified with only five scan positions (see red arrows). The image quality of the \sunrise{}
   data was strongly dependent on the gondola's varying pointing stability. L12-2 data were recorded over
   only a relatively short period of time during the \sunrise{} flight, when the pointing stability was not
   particularly good. Therefore the analyzed L12-2 data have a somewhat worse image quality compared to
   the better spatially resolved V5-6 data analyzed by \citet{Riethmueller2010}. The LOS velocities were
   corrected for the wavelength shift over the FOV caused by the IMaX etalon \citep[see][]{MartinezPillet2011}.
   In this work, negative LOS velocities correspond to upflows.

   We selected the nine data sets acquired at 23:05:08, 23:09:20, 23:20:22, 23:26:09, 23:31:56, 23:36:39,
   23:42:26, 23:47:09, and 23:53:28 UT for an in-depth study from the one hour time series. The selection
   was done so that the time interval between two consecutive sets was on average five minutes so as to
   give the BPs some time to evolve between two analyzed data sets. For each data set we checked that the
   pointing stability of the gondola and hence the image quality was as good as any among the L12-2 data,
   although it was found to be somewhat lower than of the best V5-6 data.

\subsection{Simulations}
   The three-dimensional non-ideal compressible radiation MHD simulations considered here were
   calculated with the MURaM code which solves a system of equations consisting of the continuity
   equation, the momentum equation, the energy equation, the induction equation, and the equation
   of state \citep{Voegler2005b}. The radiative energy exchange rate of the energy equation is determined
   by a non-gray radiative transfer module under the assumption of local thermal equilibrium (LTE).
   The equation of state takes into account effects of partial ionization because they influence the
   efficiency of the convective energy transport. Periodic boundary conditions were used in the
   horizontal directions. A free in- and outflow of matter was allowed at the bottom boundary
   of the computational box under the constraint of total mass conservation, while the top boundary
   was closed (i.e. zero vertical velocity). A statistically relaxed purely hydrodynamical simulation
   was used as initial condition. From tests with different magnetic fluxes, we estimated the
   mean unsigned vertical magnetic flux density of our quiet-Sun observations to correspond roughly
   to a simulation with a starting value of 30\,G and hence a unipolar homogeneous vertical magnetic
   field of $B_{\rm z} = 30$\,G was introduced into the hydrodynamical simulation (see
   section~\ref{CompAllPixels} for a more precise estimate of the mean flux). The simulation was
   run for a further 3 hours of solar time to reach and stay for a sufficiently long time in a
   statistically stationary state. 30 equidistant snapshots covering 141\,min of solar time were then
   used for this study. The data cubes cover 6\,Mm in both horizontal directions with a cell
   size of 10.42\,km (0\carcsec{}014). In the vertical direction they extend 1.4\,Mm at 14\,km cell size.
   On average, unit optical depth for the continuum at 5000\,\AA{} is reached about 500\,km below
   the upper boundary. To evaluate the dependence of our MHD results on the mean magnetic flux, we also
   calculated 10 snapshots each taken from simulation runs with an initial mean unsigned vertical flux
   density of 0\,G (purely hydrodynamical run), 50\,G, and 200\,G, while all other parameters were
   kept identical to the 30\,G run.

   The output of the MURaM code consisted of data cubes of the density, velocity (x, y, z component),
   total energy density, magnetic field (x, y, z component) as well as gas pressure and temperature.
   For a direct comparison with the observations, these quantities had to be converted into
   Stokes profiles. This was done by a forward calculation with the SPINOR inversion code
   \citep{Frutiger2000a,Frutiger2000b,Berdyugina2003} that uses the STOPRO routines \citep{Solanki1987}
   to compute synthetic Stokes spectra for atomic and molecular spectral lines assuming LTE
   and solving the Unno-Rachkovsky radiative transfer equations \citep{Rachkovsky1962}.
   All spectral line syntheses in this paper were carried out for the center of the solar disk ($\mu=1.0$).

\begin{landscape}
   \begin{figure*}
   \centering
   \includegraphics[height=120mm]{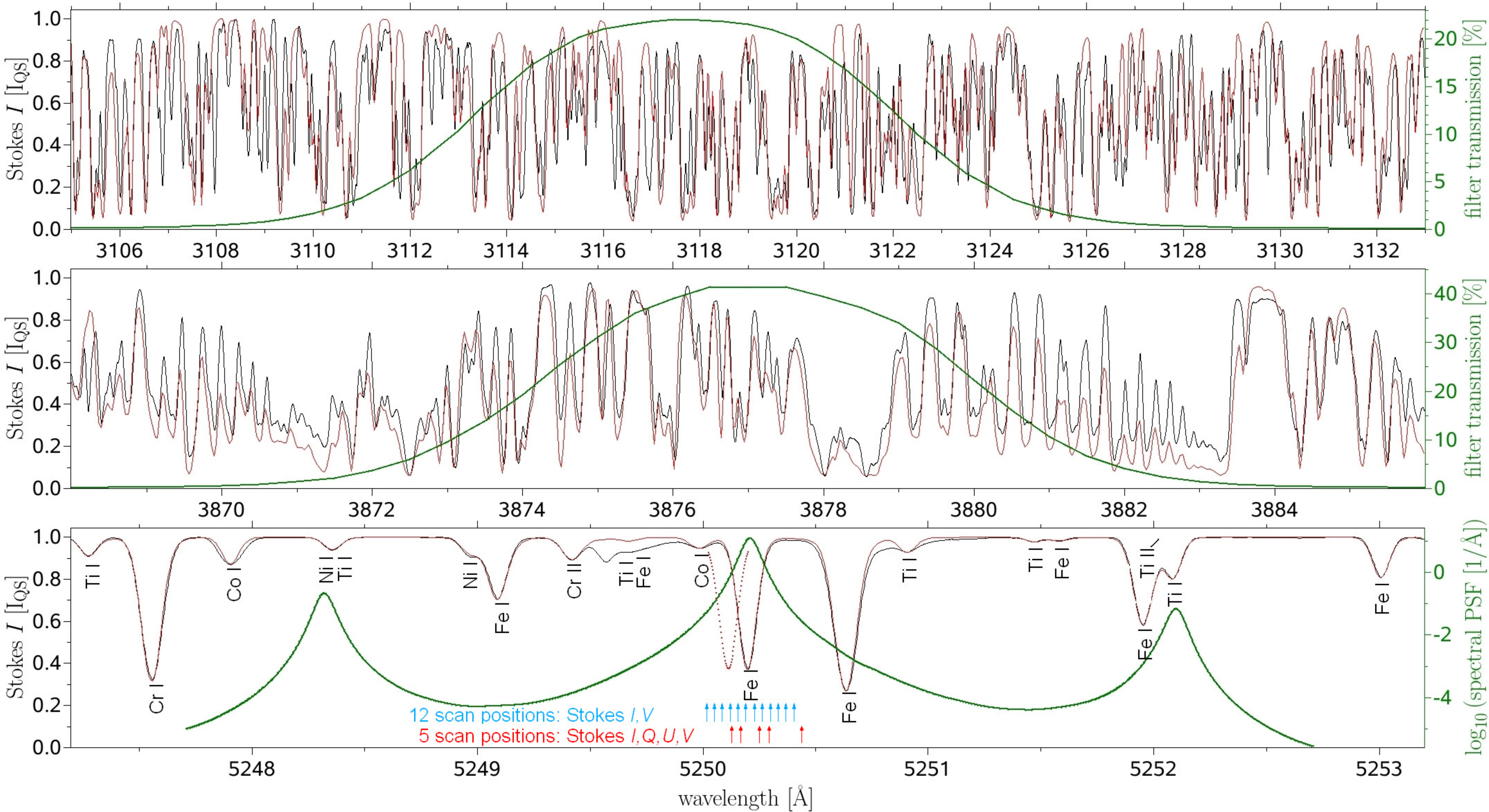}
   \caption{Excerpts from an NSO spectral atlas \citep[black lines;][]{Kurucz1984}, spatially averaged synthetic
   spectra (red lines), and instrumental filter profiles (green lines referring to the scale on the right side
   of the figure) for the part of the OH band around 3118\,\AA{} (top panel), the CN band at 3877\,\AA{}
   (middle panel), and the Fe\,{\sc i} line at 5250.2\,\AA{} (bottom panel). The dotted red line in the bottom
   panel simulates the Doppler shifted profile of the Fe\,{\sc i} line produced by an upflow
   of 5\,km~s$^{-1}$. The red arrows indicate the five scan positions of the IMaX V5-6 mode, while the twelve
   blue arrows mark them for the L12-2 mode.}
   \label{FigSynthSpectra}
   \end{figure*}
   \end{landscape}

   For SuFI's spectral range at 3118\,\AA{} we synthesized 538 spectral lines (spectral
   sampling was 50~Samples/\AA{}), for the 3877\,\AA{} range 354 spectral lines (33~Samples/\AA{}),
   and finally, for the IMaX instrument we synthesized 20 spectral lines (128~Samples/\AA{}), i.e.
   the Fe\,{\sc i} line at 5250.2\,\AA{} itself and 19 of its neighboring lines which could possibly
   contribute to the synthesized Stokes signals owing to the secondary peaks of IMaX's spectral point spread
   function (PSF). For the three considered wavelength ranges, the spatially averaged synthetic spectra
   (calculated for a snapshot with 30\,G average vertical field) are shown in Fig.~\ref{FigSynthSpectra}
   (red lines) and compared with the average observed spectra of quiet Sun taken from the NSO atlas
   \citep{Kurucz1984}. The synthetic spectra were convolved with a Gaussian of
   FWHM~=~1.93\,km~s$^{-1}\sqrt{2/\ln{2}}~\lambda/c$ to fit the spectral resolution of the NSO atlas.
   The spectral PSF of IMaX (green line in the bottom panel) includes the IMaX prefilter and is plotted
   on a logarithmic scale for a better visibility of the secondary peaks at 5248.32\,\AA{} and 5252.10\,\AA{}.

\subsection{Synthetic instrumental effects}
   One of the most sensitive parts of this work was the introduction of synthetic instrumental effects,
   i.e. the degradation of the synthetic data to the same spatial and spectral resolution, to the same
   stray light contamination and noise level as the observed data. The various degradation steps significantly
   influenced the values of the parameters we compared between simulation and observation. We found that only
   if all the relevant effects of the \sunrise{} instrumentation are known to relatively high precision,
   is the comparison between synthetic and observed data meaningful. All the used degradation steps are
   explained in the following. They were applied in the same order as described below.

\subsubsection*{Spectral resolution and sampling}\label{SpecResol}
   The synthetic Stokes profiles output from the \Index{SPINOR} code had perfect spectral resolution and
   high spectral sampling which had to be reduced to the values of the \sunrise{} instruments. The
   transmission profiles of the SuFI filters as well as the spectral PSF of IMaX (including the pre-filter)
   were measured in the laboratory before the launch of \sunrise{}. In the case of SuFI,
   the transmission profiles of the two SuFI filters (3118\,\AA{} and 3877\,\AA{}) were re-sampled to
   the wavelength grid points of the synthetic intensity profiles. Then the filter transmission profile were
   multiplied by the intensity profile point by point and the products were summed up. Such a scalar product
   gave the intensity of a pixel of a synthetic SuFI image. In the case of IMaX, the synthetic Stokes
   profiles were convolved with the spectral PSF of IMaX. Finally, Stokes images at the twelve scan
   positions of IMaX's L12-2 mode were retrieved.

\subsubsection*{Spatial resolution and residual \Index{pointing jitter}}\label{SpatResol}
   The spatial resolution of the \sunrise{} data was limited by the 1-m aperture, the considered wavelength,
   and the quality of the gondola's pointing. The theoretical \Index{diffraction limit} was not fully reached
   during the first science flight of \sunrise{} owing to the residual pointing jitter. From azimuthally
   averaged power spectra of the SuFI and IMaX intensity images, a precise determination of the spatial
   resolution was not possible, but allowed a rough estimate between 0\carcsec{}20 and 0\carcsec{}24
   for the data analyzed here, which are not as highly resolved as data used for earlier publications.
   In the ideal case of a perfect knowledge of the \sunrise{} PSF and no pointing jitter, a spatial
   degradation of the synthesized data would not be needed because we compared with reconstructed
   \sunrise{} data. Theoretically, the deconvolution with the PSF reconstructs the original rms
   contrasts. In the non-ideal case of \sunrise{}, the residual pointing jitter during the observation
   was taken into account by convolving all synthetic Stokes images with a two-dimensional Gaussian
   of FWHM~=~0\carcsec{}23. This FWHM value led to the best match between the rms contrasts of the
   observed and synthesized IMaX continuum images as we found after some tests with different FWHM values.

\subsubsection*{Stray light}
   In the case of ground-based solar observations, the atmospheric \Index{stray light} is a significant part
   of the total amount of stray light. Owing to the flight altitude of approximately 35\,km on average,
   we expect that atmospheric stray light is negligible, so that we have to deal with instrumental stray
   light alone. By observing the solar limb, intensity profiles of the limb could be recorded for the
   considered wavelengths. These solar limb profiles were then compared with the intrinsic limb profiles
   of the Sun taken from the literature \citep{Dunn1968} which allowed the stray light MTFs to be
   calculated. The stray light MTFs were then multiplied with the synthetic Stokes images in Fourier space.
   Details about the determination of the stray light affecting \sunrise{} data are given by \citet{Feller2013}.

\subsubsection*{Noise}
   The noise level of IMaX's Stokes $V$ images was determined at the continuum wavelength (+192.5\,m\AA{}
   offset from the core of the line), which is generally free of $V$ signals. The signals found by
   \citet{Borrero2010} are sufficiently rare not to influence the noise determination significantly.
   The histograms of the nine PD reconstructed Stokes $V$ continuum images we considered here showed a
   clear Gaussian shape with a standard deviation of $3.3 \times 10^{-3} I_{\rm{QS}}$ ($I_{\rm{QS}}$
   is the mean continuum intensity of the quiet Sun).

   The retrieval of the Stokes $I$ noise level was more difficult because here the standard deviation is
   not entirely determined by the noise, but also by the granulation pattern. For that reason, we determined
   standard deviations for small regions within granules, assuming the signal to be nearly constant
   for such small regions. We found similar values as for Stokes $V$, which confirmed our assumption of small
   intrinsic variations within the small patches considered. Consequently, we added a Gaussian noise with
   $\sigma=3.3 \times 10^{-3} I_{\rm{QS}}$ to all synthetic Stokes images.

\subsubsection*{Plate scale}
   The cell size of the simulation data was 10.42\,km and had to be adapted to the pixel size of the IMaX
   observation of 40.10\,km or to that of the SuFI observation of 15.24\,km, respectively. Various tests showed
   that the adopted \Index{plate scale} hardly influenced the parameters we considered and hence did not affect our
   results (mainly because it is significantly smaller than the width of the spatial PSF). For the sake of
   simplicity, we therefore skipped this degradation step in the following study.

\subsubsection*{Retrieval of line parameters}\label{ParamRetrieval}
   In the last step, we fitted a Gaussian to the twelve points of the Stokes $I$ profile of the 5250.2\,\AA{}
   line for each pixel to determine the quantities LOS velocity and line width (as FWHM value). The
   Stokes~$V$ profiles were used for calculations of the \Index{circular polarization degree}, defined as
   \begin{equation}\label{Eq_Mean_CPi}
   \langle p_{\rm{circ}} \rangle = \frac{1}{12} \sum_{i=1}^{12}{\left| \frac{V_i}{I_i} \right|}~,
   \end{equation}
   where the averaging was done over the twelve scan positions of the 5250.2\,\AA{} line. For SuFI data,
   only the intensity in each filter was available.

\section{Results}

\begin{landscape}
   \begin{figure*}
   \centering
   \includegraphics[height=120mm]{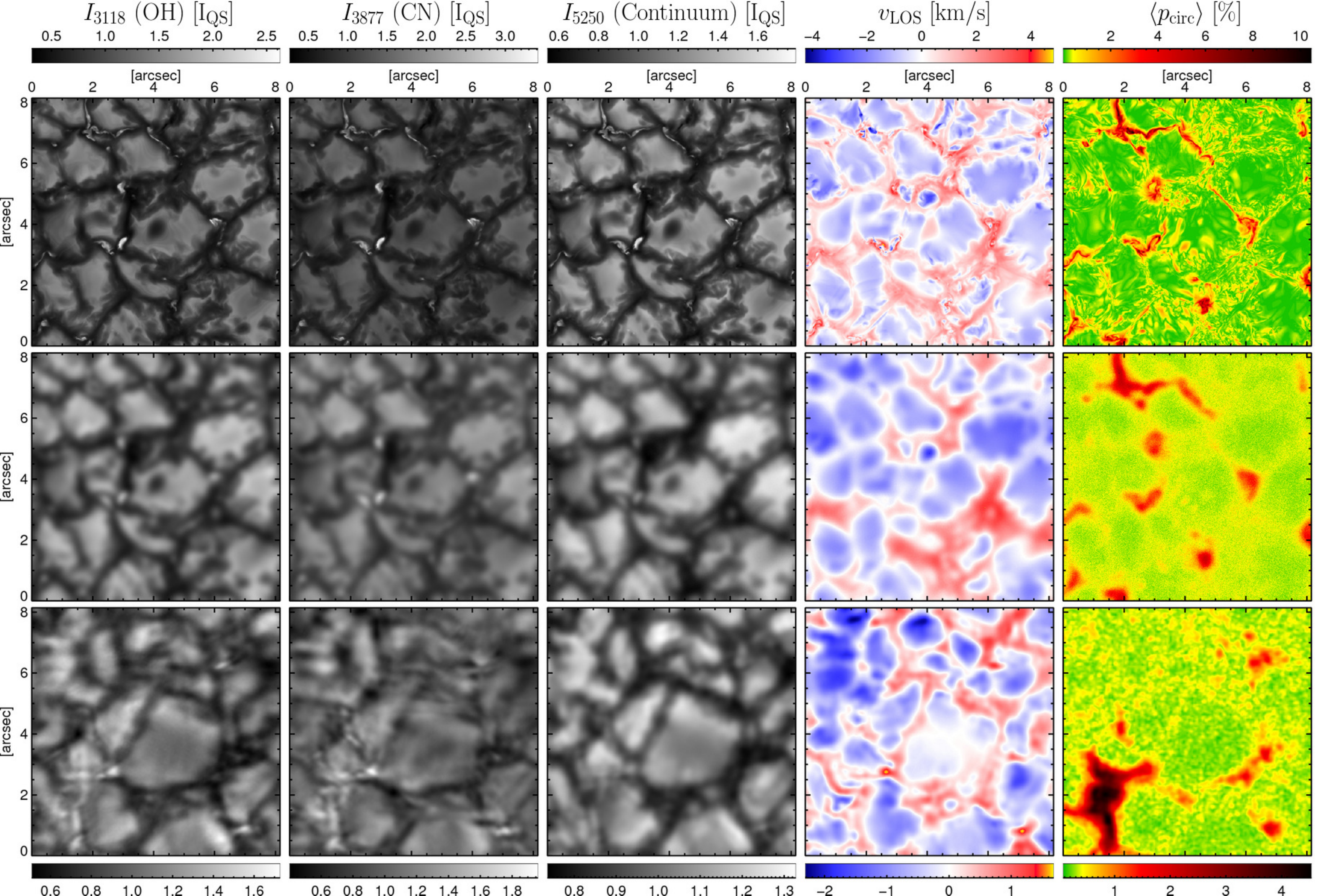}
   \caption{Intensity maps in the wavelength bands 3118\,\AA{} and 3877\,\AA{}, as well as for the continuum at
   5250.4\,\AA{} (three first columns), all normalized to the corresponding mean quiet-Sun intensity
   level, $I_{\rm{QS}}$. The LOS velocity (positive velocities correspond to downflows) and the circular
   polarization degree (see main text for definition) are shown in the fourth and fifth columns, respectively.
   The color bars at the top refer to the undegraded MHD data shown in the first row, while the lower color
   bars apply to the degraded MHD data (second row) and to the data obtained from the \sunrise{}
   observatory (third row).}
   \label{FigImgOverview}
   \end{figure*}
   \end{landscape}

   Fig.~\ref{FigImgOverview} contrasts the MHD data with the observational data. The two topmost rows of
   panels depict a snapshot of the original and the degraded MHD data, respectively. The bottom panels
   exhibit an 8.1\arcsec{}$\times$8.1\arcsec{} quiet-Sun region (a subregion of the
   13\arcsec{}$\times$37\arcsec{} common FOV of SuFI and IMaX) as observed with \sunrise{} at 23:05:08 UT.
   Several bright granules, separated by darker intergranular lanes, can be seen in the displayed
   quiet-Sun regions. The undegraded MHD data show many small bright features within the \Index{dark lane}s.
   Only the largest of these features can be identified as BPs in the degraded MHD data, the smaller ones
   are smeared out by the degradation. The BP contrasts exceed the granulation contrasts in the observed
   OH and CN image, but not for the observed 5250\,\AA{} image, which confirms the results of
   \citet{Riethmueller2010}. Some of the observational BP flux concentrations (see e.g. the one at
   position (1.5\arcsec{},2.0\arcsec{})) have collected more flux and are bigger than the largest
   BPs in the degraded simulations. The degraded data look fairly similar to the observations (excepting
   the largest and brightest BPs). In particular, the granulation contrasts in the degraded simulated
   granulation is very similar to that in the observational data at all three wavelengths. This gives
   us the confidence to proceed with a more detailed analysis with the help of histograms of various
   quantities.

\subsection{Simulations versus observations: all pixels}\label{CompAllPixels}
\begin{figure}
   \centering
   \includegraphics[width=100mm]{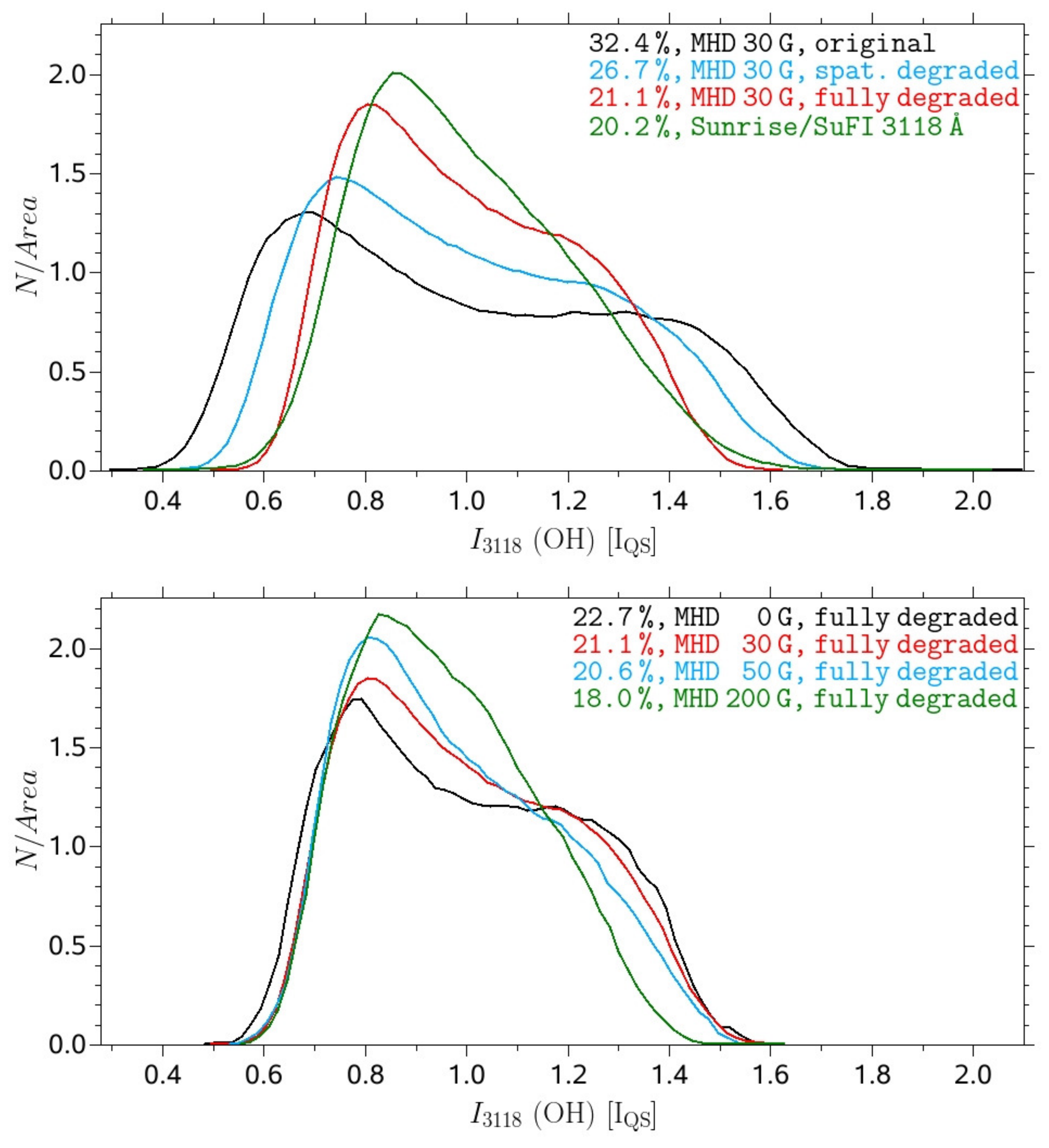}
   \caption{Intensity histograms of all pixels for the OH band data around 3118\,\AA{}. Top panel: The
   black line corresponds to the original 30\,G MHD simulation, the blue line to the spatially degraded data,
   and the red line to the fully degraded simulation data. The green line displays the \sunrise{}
   observations. Bottom panel: Influence of the MHD simulations' mean flux density on the fully
   degraded OH intensity histogram. The black line shows a purely hydrodynamical simulation, i.e. without
   any magnetic field. The mean unsigned vertical flux density was 30\,G for the histogram colored in red,
   50\,G for the blue line, and 200\,G for the green line. RMS contrasts are indicated in the text labels.}
   \label{FigOHHist}
   \end{figure}

\begin{figure}
   \centering
   \includegraphics[width=100mm]{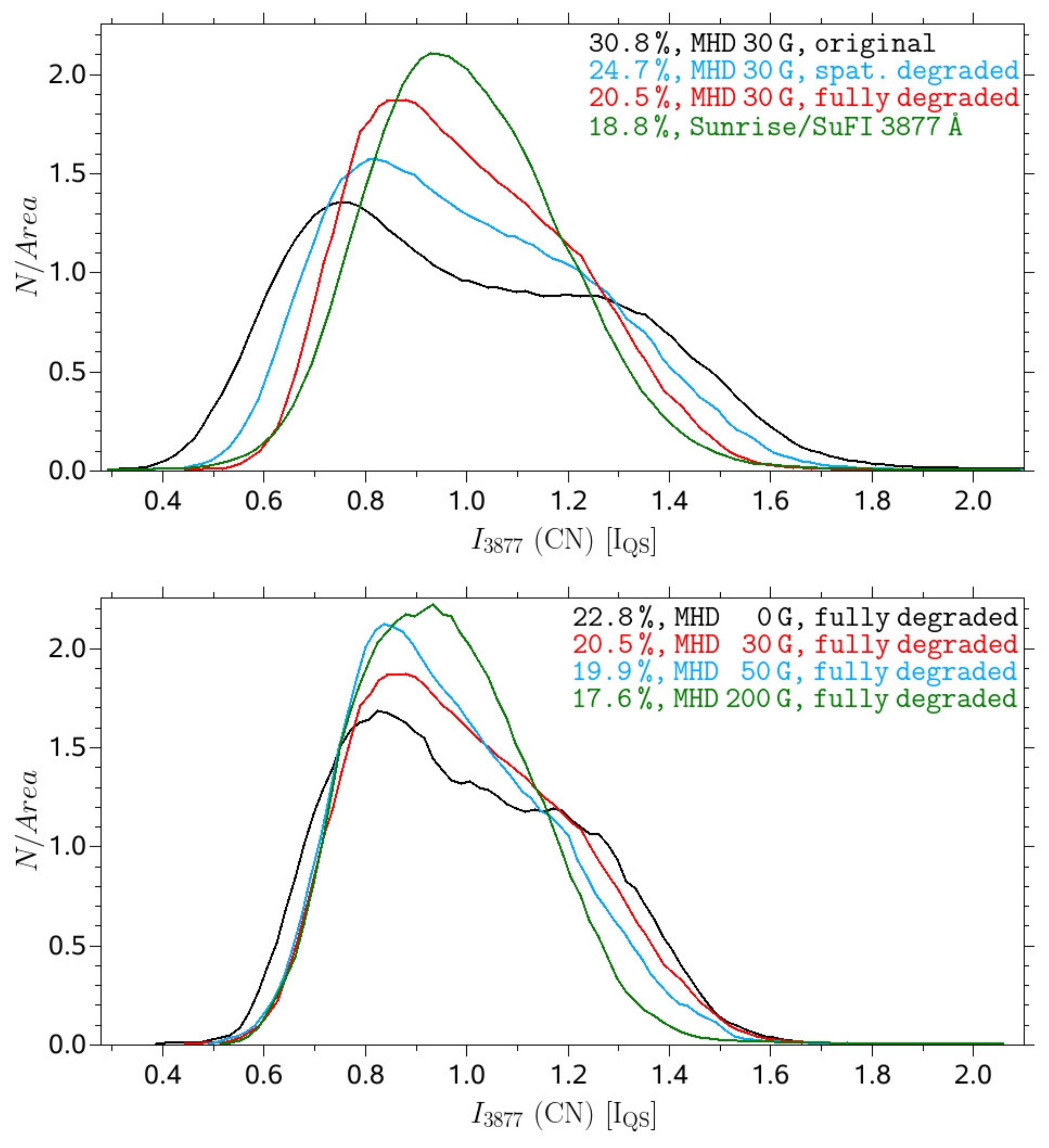}
   \caption{The same as Fig.~\ref{FigOHHist}, but for the CN band data around 3877\,\AA{}.}
   \label{FigCNHist}
   \end{figure}

\begin{figure}
   \centering
   \includegraphics[width=100mm]{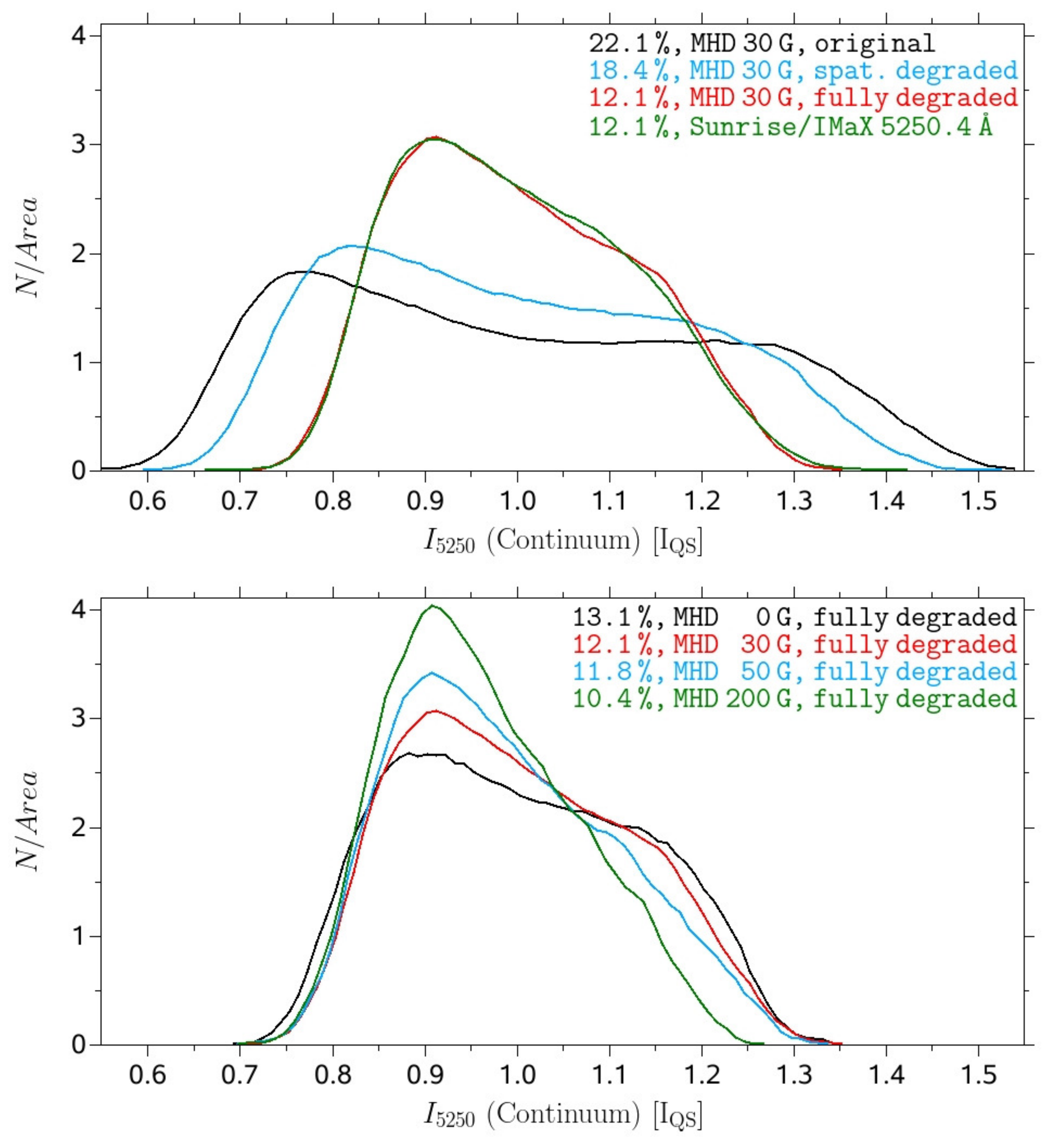}
   \caption{The same as Fig.~\ref{FigOHHist}, but for the continuum at 5250.4\,\AA{}.}
   \label{FigCCTHist}
   \end{figure}

\begin{figure}
   \centering
   \includegraphics[width=100mm]{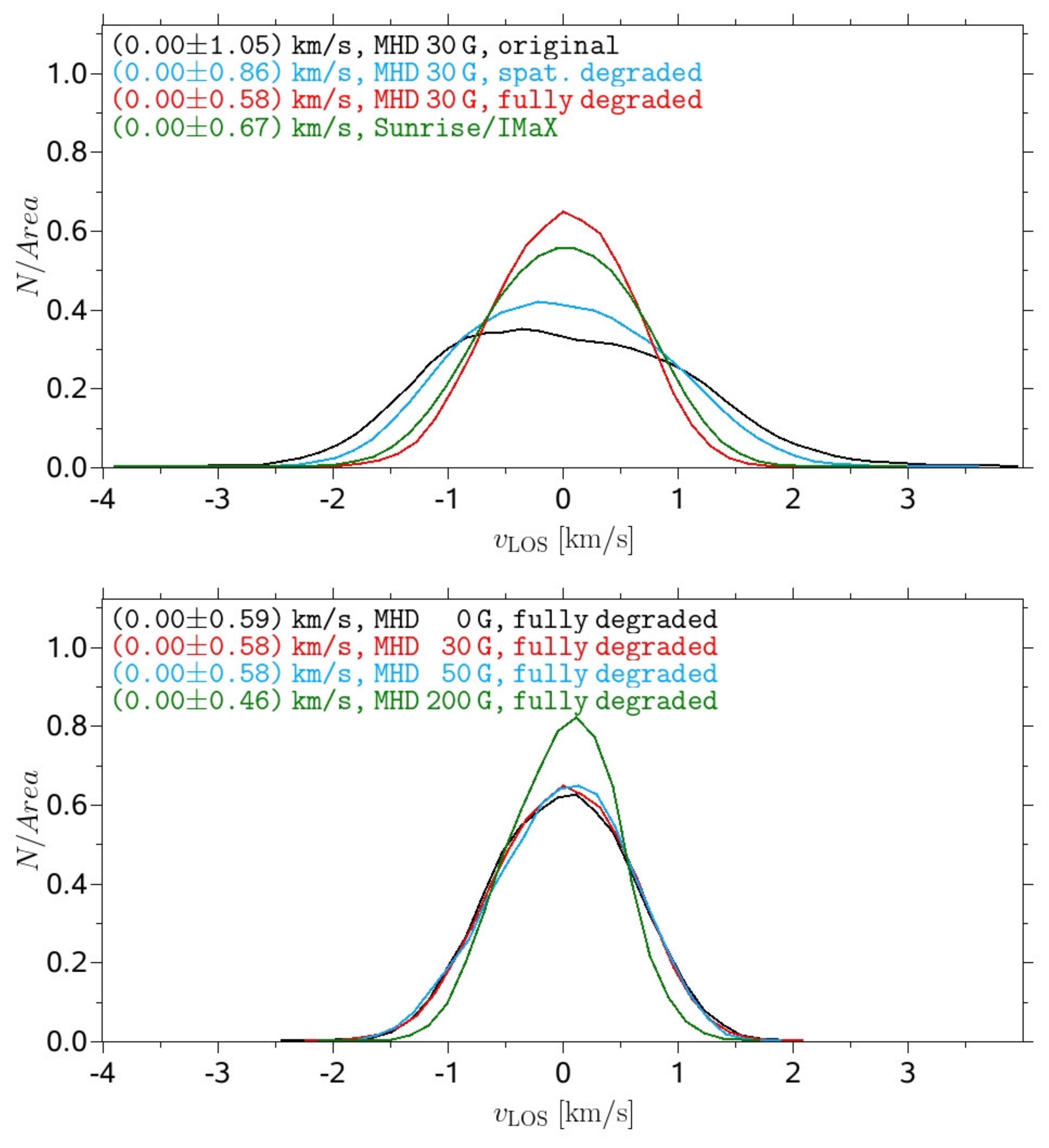}
   \caption{Same as Fig.~\ref{FigOHHist}, but for the LOS velocity as retrieved from a Gaussian fit to
   the Stokes~$I$ profile of Fe\,{\sc i} 5250.2\,\AA{}. Mean values and their standard deviations are
   indicated in the text labels. Negative velocities are upflows.}
   \label{FigLMINHist}
   \end{figure}

\begin{figure}
   \centering
   \includegraphics[width=100mm]{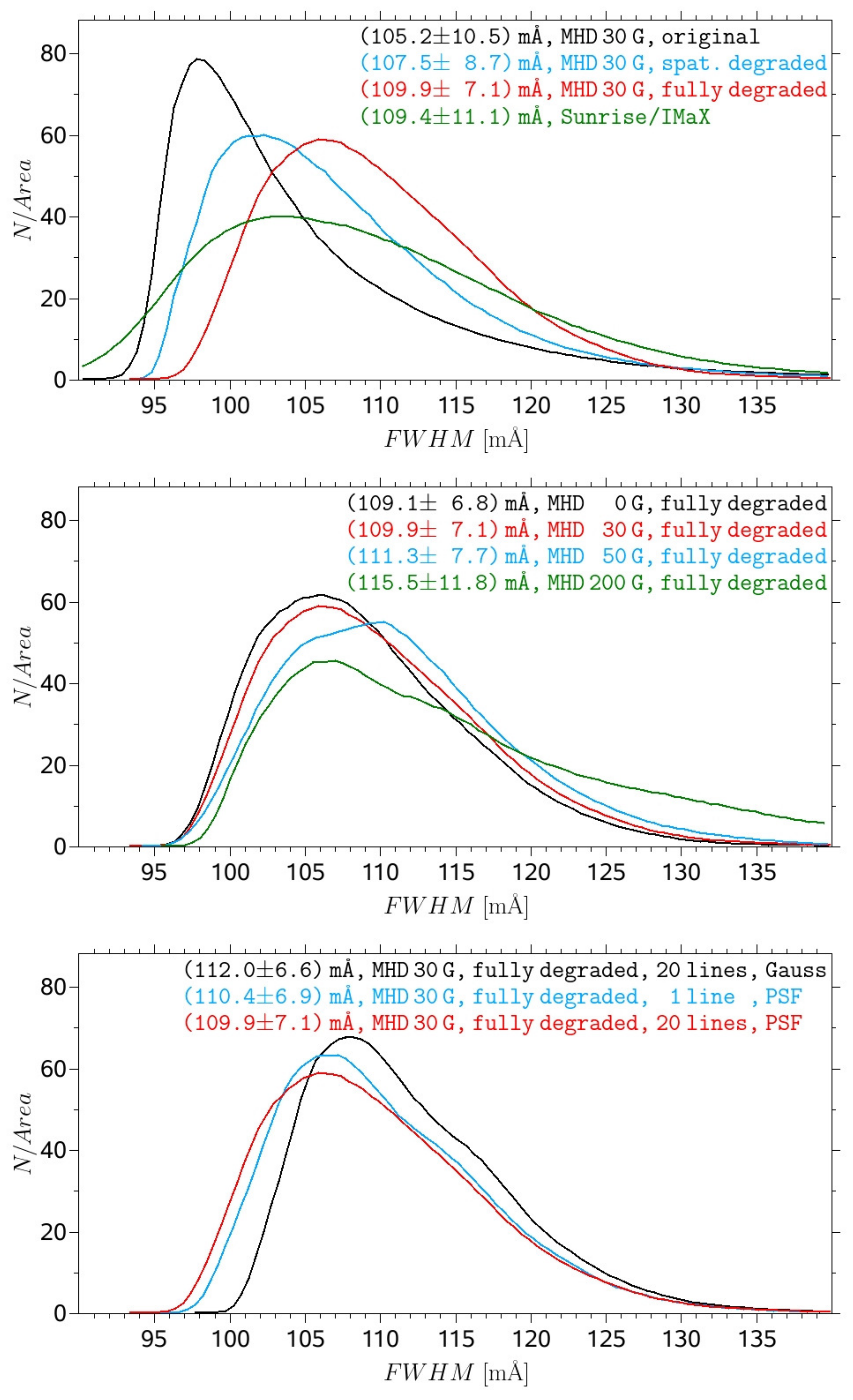}
   \caption{Top and middle panel: Same as Fig.~\ref{FigOHHist}, but for the spectral line width (FWHM)
   as retrieved from a Gaussian fit to the Stokes~$I$ profiles of Fe\,{\sc i} 5250.2\,\AA{}. Bottom panel:
   Influence of different approximations of the spectral synthesis and degradation on the histograms of
   the spectral line width of the degraded 30\,G MHD data. The spectra were fully degraded in all three
   plotted cases. The black line corresponds to the histogram obtained when the filter transmission profile
   was approximated by an 85\,m\AA{} Gaussian instead of the measured spectral PSF. The blue line corresponds
   to a spectral synthesis of only the 5250.2\,\AA{} line, i.e. the 19 neighboring lines were not
   synthesized. For comparison reasons, the red line of the top panel is plotted again and corresponds
   to the 20-lines synthesis and spectral degradation with the measured spectral PSF.}
   \label{FigFWHMHist}
   \end{figure}

\begin{figure}
   \centering
   \includegraphics[width=100mm]{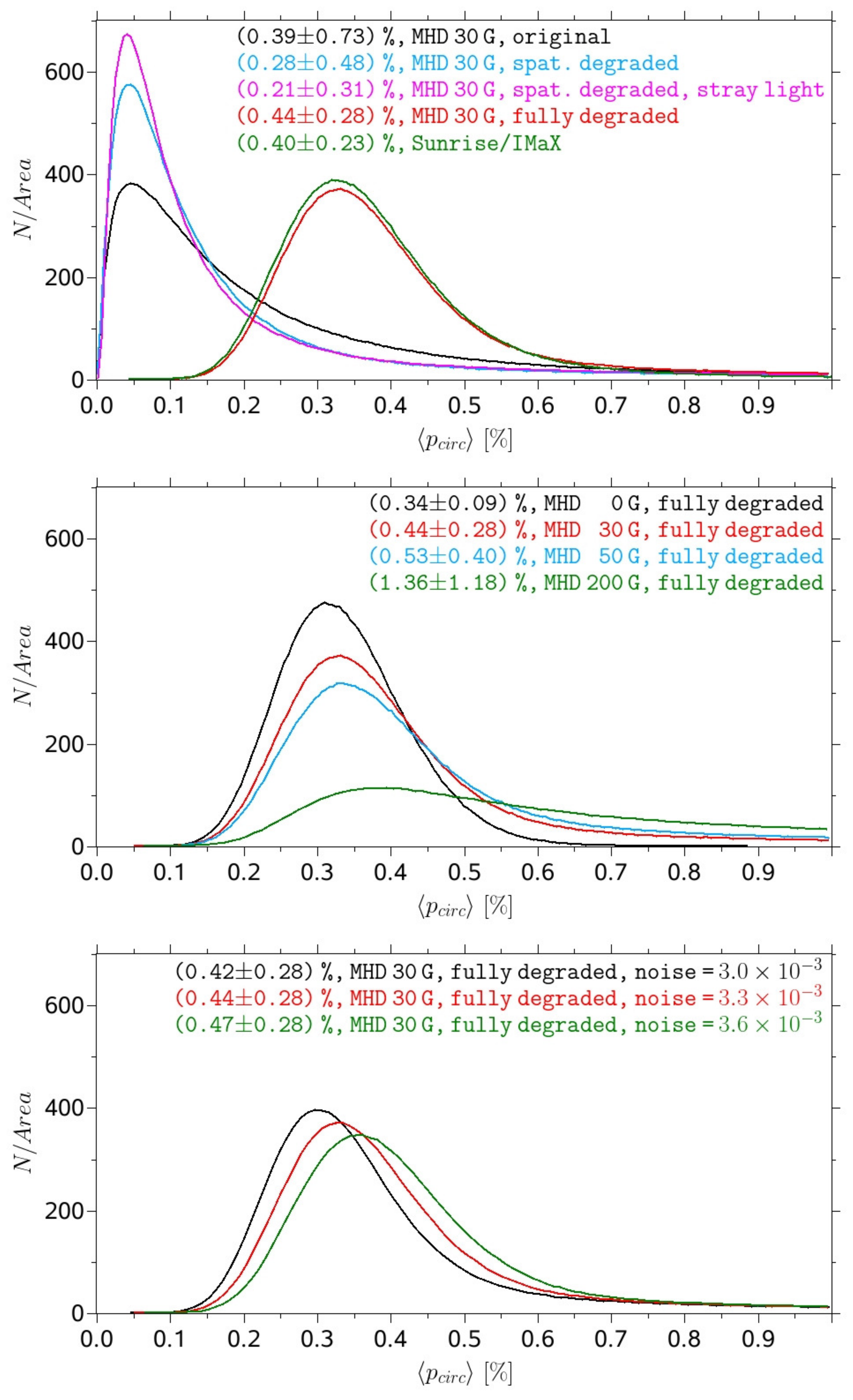}
   \caption{Histograms of the \Index{circular polarization degree} $\langle p_{\rm{circ}} \rangle$ as retrieved
   from the Stokes~$I$ and $V$ profiles of Fe\,{\sc i} 5250.2\,\AA{} (see Eq.~(\ref{Eq_Mean_CPi})
   for definition). Top panel: The black line corresponds to the original 30\,G MHD simulation, the blue
   line to the spatially degraded data, the magenta line represents the histogram after stray light
   contamination has been added on top. Finally, the red line corresponds to the fully degraded 30\,G
   simulation. The green line displays the \sunrise{}/IMaX observations. Middle panel: Influence of the
   MHD simulations' mean flux density on the fully degraded polarization histogram. The black line shows
   a purely hydrodynamical simulation, i.e. without any magnetic field. The mean unsigned vertical flux
   density was 30\,G for the histogram colored in red, 50\,G for the blue line, and 200\,G for the green
   line. The noise level was always $3.30 \times 10^{-3}~I_{\rm{QS}}$. Bottom panel: Influence of the
   noise level. The flux density was always 30\,G. The noise levels in units of the mean quiet-Sun
   intensity are printed as text labels.}
   \label{FigMean_CPHist}
   \end{figure}

   All pixels of all frames contributed to the histograms plotted in Figs.~\ref{FigOHHist}~to~\ref{FigMean_CPHist}.
   In order to ease comparisons between the relatively similarly shaped histograms, the integral over the
   histograms was always normalized to one.

   Fig.~\ref{FigOHHist} exhibits histograms of the normalized intensities of the OH data at 3118\,\AA{}.
   The top panel reveals the influence of the various degradation steps on the histogram of the 30\,G MHD data.
   The histogram of the original MHD data is drawn in black. "Original" means that the Stokes~$I$ spectra
   are multiplied with SuFI's filter transmission profile but no other degradation steps have been applied.
   The blue line represents spectrally and spatially degraded data, while the red line displays the
   fully degraded data, i.e. after spectral, spatial, stray light, and noise degradation. A comparison
   of the black with the blue line shows the influence of the spatial degradation. The noise hardly changes
   the histograms in Figs.~\ref{FigOHHist}~to~\ref{FigFWHMHist} and hence the difference between the blue
   and the red line is mainly due to stray light.

   The fully degraded MHD data are then compared with the SuFI observations colored in green. The
   degradation of the simulated data reduces the rms contrast in the OH band from 32.4\,\% down to 21.1\,\%
   which is 0.9\,\% higher than the observational contrast of 20.2\,\%. Though all histograms show a
   certain amount of asymmetry, the histogram of undegraded simulated data indicates a superposition of two
   populations: the first one consists of intergranular pixels with low intensities, the second one
   contains bright pixels from the granules or from bright points. This superposition is still somewhat
   visible after degradation. The observational histogram does not reveal such a clear superposition of
   two populations.

   The bottom panel of Fig.~\ref{FigOHHist} displays OH intensity histograms of fully degraded MHD data
   for different mean unsigned vertical flux densities. The $I_{\rm{QS}}$ for the intensity normalization
   in the considered spectral band is determined from the 30\,G data (closest to our observations). The
   rms contrast is highest in the field-free case, 22.7\,\%, decreasing to 18\,\% in the 200\,G simulation
   (typical mean flux density of a plage region). This reflects the fact that the convection is inhibited
   by magnetic fields \citep{Biermann1941}. Without any magnetic field, there are no BPs and hence
   the fraction of the area covered by dark intergranular lanes is relatively large. With increasing
   magnetic flux the number density of BPs increases (see discussion of Fig.~\ref{FigBPMean_CPHist})
   which reduces the area fraction of the darker regions in the intergranular lanes and the rms contrast
   is reduced as well, but we think that this effect plays only a minor role. Note that the black line
   of the bottom panel of Fig.~\ref{FigOHHist} has not only more dark pixels than the others but it has
   also more bright pixels. The superposition of two populations is most pronounced in the field-free
   data but it is not visible in the 200\,G data.

   The intensity histograms of the CN band (3877\,\AA{}) are displayed in Fig.~\ref{FigCNHist}. Compared
   with the OH histograms, the superposition of two populations is somewhat less pronounced for the
   simulation data. The rms contrast of the simulated 30\,G data is reduced from 30.8\,\% to 20.5\,\%
   after degradation, which is 1.7\,\% higher than the 18.8\,\% contrast of the SuFI data. The main
   difference between the observation and the degraded MHD data is the stronger asymmetry of the synthetic
   histogram. The rms contrast reduces from 22.8\,\% to 17.6\,\% if a 200\,G magnetic field is present.

   Intensity histograms of the IMaX continuum wavelength $5250.2\,\rm{\AA}+192.5\,\rm{m\AA}=5250.4\,\rm{\AA}$
   are plotted in Fig.~\ref{FigCCTHist}. In Figs.~\ref{FigCCTHist}~to~\ref{FigMean_CPHist}, the term
   "original" MHD data (black lines in the top panels) implies that the Stokes~$I$ spectra are convolved with
   IMaX's spectral PSF, but are not otherwise degraded. The full degradation of the 30\,G MHD data leads
   to a decrease in the rms contrast from 22.1\,\% to 12.1\,\%, which is exactly the observational contrast.
   A very good match between the degraded MHD data and the observational data is not only found for the rms
   contrast (which is not a big surprise; see section~\ref{SpatResol}), but also for the shape of the
   histograms. The superposition of two populations can be seen for the undegraded data, but it is not so
   clear for the degraded simulations. The fully degraded 200\,G MHD data have a 2.7\,\% lower contrast
   than the corresponding field-free data.

   Fig.~\ref{FigLMINHist} exhibits histograms of the LOS velocity as determined from the Gaussian fit of
   the twelve scan positions of the Stokes~$I$ profiles of the 5250.2\,\AA{} line. Since an absolute
   wavelength calibration of the \sunrise{}/IMaX data has not been done, we decided to force the
   observational LOS velocities to have the same mean value as the degraded 30\,G MHD data. The standard
   deviation of the 30\,G simulation is reduced by the degradation from 1050\,m~s$^{-1}$ to 580\,m~s$^{-1}$,
   which is close to the standard deviation of the observed value of 670\,m~s$^{-1}$. The histograms
   are only weakly asymmetric. Small mean flux densities (0-50\,G) led to almost identical velocity
   histograms. Increasing the average vertical field to 200\,G significantly impedes the convection,
   reducing the standard deviation of the velocities to 460\,m~s$^{-1}$.

   Histograms of the 5250.2\,\AA{} line widths are displayed in Fig.~\ref{FigFWHMHist}. The degradation
   of the 30\,G simulation data (top panel) causes a shift of the position of the histogram's maximum
   towards larger line widths coupled with an increase in the width of the histogram, bringing it closer
   to the histogram of the observed values. Thus the histograms of the degraded simulations and the
   observations display a reasonable match for the mean values, but a significant mismatch of their widths,
   i.e. the mean as well as the most common line widths of the simulated profiles are close to the
   observed values, but the scatter of the line widths is clearly larger for the observational data.
   The middle panel of Fig.~\ref{FigFWHMHist} shows how the MHD line width histogram depends on the mean
   flux density. A larger magnetic flux increases the mean value of line widths as well as the standard
   deviation mainly owing to an increased number of larger line width values (partly due to enhanced
   Zeeman splitting). The number of small line width values is hardly influenced by the mean flux
   density and hence none of the considered fluxes matches the observational histogram well.

   In the bottom panel of Fig.~\ref{FigFWHMHist} we demonstrate the influence of the secondary peaks of IMaX's
   spectral PSF on the degraded 30\,G MHD line width histogram. In particular the approximation of the spectral
   PSF of IMaX by a Gaussian function led to a significantly increased discrepancy between observation and
   simulation. The best match was reached by doing a full 20-lines synthesis and using the measured spectral
   PSF for the degradation. Note, that convolving with a 85\,m\AA{} Gaussian, or restricting the line
   synthesis to just Fe\,{\sc i} 5250.2\,\AA{} left the histograms of all other quantities considered
   in this study practically unchanged.

   Histograms of the \Index{circular polarization degree} $\langle p_{\rm{circ}} \rangle$ as defined in
   Eq.~(\ref{Eq_Mean_CPi}) are given in Fig.~\ref{FigMean_CPHist}. As for the line widths, we found
   a strong asymmetry in all histograms. This time the match between the degraded 30\,G simulation
   and the observation was remarkably good. The spatial degradation as well as the \Index{stray light}
   contamination made the histogram more narrow and left the maximum position nearly unchanged.
   The influence of the noise was dominant as can be seen by comparing the magenta line (spectral +
   spatial + stray light degradation) and the red line (spectral + spatial + stray light + noise
   degradation). The noise broadens the histogram and shifts the maximum position towards higher
   polarization values. Note that in contrast to $\langle p_{\rm{circ}} \rangle$ the histograms
   of all other parameters considered so far were hardly affected by noise.

   The middle panel of Fig.~\ref{FigMean_CPHist} displays the dependence of the fully degraded MHD
   histogram on the mean vertical flux density. The maximum position and the width of the histograms
   increase with mean flux density. The 0\,G simulation (which had no magnetic field at all) has no
   intrinsic polarization, so that the black line is entirely due to noise that was introduced as part
   of the degradation. This pure noise histogram differs clearly from the observational histogram,
   so that from such variations of the mean MHD flux we estimated that the \sunrise{} data correspond to
   an average LOS field of circa 25\,G. (Even the 30\,G simulations shown in the top panel of
   Fig.~\ref{FigMean_CPHist} exhibit a slightly too large mean flux density compared to our observation.)
   We note, however, that all simulations considered here started with an initially homogeneous, vertical
   and unipolar field. A comparison with simulations with a different initial condition (or a different
   Reynolds number) could result in a different estimate of the magnetic field in the observed region
   \citep[see, e.g.,][]{Danilovic2010,Pietarila2010}. Also how exactly the \Index{circular polarization degree}
   is normalized (e.g. division by $I_{\rm{i}}$ as in Eq.~(\ref{Eq_Mean_CPi}), division by $I_{\rm{12}}$
   (local continuum), or division by $\langle I_{\rm{12}} \rangle = I_{\rm{QS}}$, i.e. the mean continuum,
   see Eq.~(\ref{Eq_Mean_CPc})) can have a slight influence on the estimated average field strength.
   Additionally, a possible \Index{cross talk} from Stokes~$Q$ or $U$ into Stokes~$V$, which can not be corrected
   due to the lack of Stokes~$Q$ and $U$ signals in the IMaX L12 data, could influence our estimate of the
   mean MHD flux.

   The dependence of the degraded 30\,G MHD polarization histogram on the noise level, which was added to
   all Stokes images, is shown in the bottom panel of Fig.~\ref{FigMean_CPHist}. Three noise levels between
   $3.00 \times 10^{-3} I_{\rm{QS}}$ and $3.60 \times 10^{-3} I_{\rm{QS}}$ (see text labels) are plotted
   in different colors. Again, the position of the maximum and the width of the histograms increase with
   noise levels. Nevertheless, noise level and mean MHD flux density can both be fitted to the
   observational histogram unambiguously since small variations of the mean flux density mainly changes the
   amplitude and the width of the histogram, while a small variation of the noise level mainly shifts the
   maximum position. The best-fit noise level was determined to be $\sigma_{\rm{fit}} = 3.30
   \times 10^{-3} I_{\rm{QS}}$, which is somewhat lower than the standard deviation of the observed
   Stokes $V$ continuum signals, $\sigma_{\rm{all}} = 3.77 \times 10^{-3} I_{\rm{QS}}$, because the
   assumption of a signal-free Stokes $V$ continuum is not entirely true. Such a signal was found by
   \citet{Borrero2010}, but was restricted to 0.005\,\% of all their spatial pixels. Since we do not
   know how strongly the area covered by these signals increases with decreasing threshold, we use the
   $\sigma_{\rm{fit}}$ value as an approximation of the true noise level.

\subsection{Simulations versus observations: bright points}\label{CompBPs}
   We now compare BP properties between \sunrise{} observations and the degraded 30\,G simulations.
   From the \Index{flux tube} paradigm one would expect kilo-Gauss fields for the BPs and hence strong polarization
   signals, but \citet[][see their Fig.~2, panel h]{Riethmueller2010} find that the majority of the BPs is
   only weakly polarized. We use the BP detection method applied by \citet{Riethmueller2010} for a direct
   comparison with their results.

   First of all, we need an additional step in the pre-processing of the observational data because
   IMaX and SuFI differed significantly in their \Index{plate scale}s and also we had a slight difference
   in the plate scales of the various SuFI wavelengths. Therefore we re-sampled all data to the
   common plate scale of 0\carcsec{}0207 Pixel$^{-1}$ (original plate scale of the SuFI 2995\,\AA{}
   images) via bilinear interpolation. With the help of cross correlation functions the two
   instruments' common field of view (FOV) was identified. Finally, the larger IMaX FOV was cropped
   to the smaller 13\arcsec{}$\times$37\arcsec{} SuFI FOV.

   We then manually identified the BPs' peak intensity in the CN images because of their good visibility
   in that molecular band \citep{Schuessler2003}. The local intensity maximum in an 11$\times$11~pixel
   (i.e., 0\carcsec{}22$\times$0\carcsec{}22) patch surrounding each BP detected at 3877\,\AA{}
   was determined for each of the other wavelengths. This method takes into account that the various
   wavelengths are formed in different atmospheric layers so that inclined features may appear at
   slightly different horizontal positions at different wavelengths. Also, the dark background (DB)
   close to the BPs was retrieved. It was defined as the darkest pixel within 0\carcsec{}3 of each BP.
   We extended the simple and manual BP detection method of \citet{Riethmueller2010}
   by the determination of the BP boundaries with the help of a \Index{multilevel tracking} (\Index{MLT}) algorithm
   \citep[see][]{Bovelet2001}. The MLT algorithm determined the intensity range of the CN images
   and subdivided this into 25 equidistant levels. Starting with the highest intensity level all
   pixels were found whose intensity exceeds this level. This led to several contiguous two-dimensional
   structures, which were tagged with a unique number. The obtained structures were extended
   pixel by pixel as long as the intensity was greater than the next lower level. Then the algorithm
   searched through the whole image again to find all pixels whose intensity was greater than the
   next lower level, which often led to newly detected contiguous structures. This procedure was
   repeated until the minimal intensity level was reached. At the end, every pixel belonged to exactly
   one contiguous structure. Finally, all pixels that had an intensity lower than 50\,\% of the
   local min-max range were rejected, which led to the boundary of the BPs. A more detailed description
   of MLT, including some illustrative figures, was given by \citet{Riethmueller2008d} where the algorithm
   was applied to the detection of umbral dots. The BP boundary detection via MLT was applied to the SuFI
   intensity images, to the IMaX continuum intensity images, and also to the maps of the $\langle
   p_{\rm{circ}} \rangle$ (see section~\ref{ParamRetrieval}). We identified 121 BPs in the nine observational
   data sets. This number is limited by few L12 IMaX data sets available. The corresponding BP number density
   was 0.05 BPs per $\rm{Mm}^2$, which is 1.7 times higher than the value found by \citet{Jafarzadeh2013}
   in the Ca\,{\sc ii}~H channel of \sunrise{}/SuFI. The difference likely stems from
   the restriction to small BPs by \citet{Jafarzadeh2013}.

   We applied the same manual detection method to the degraded CN images of the 30 simulation data sets
   with 30\,G field strength. Here we found 277 BPs (0.26 BPs per $\rm{Mm}^2$). We also detected BPs in
   the 30 undegraded CN images. Owing to the much higher spatial resolution of the undegraded data, we found
   a lot more BPs there, in total 898 (0.83 BPs per $\rm{Mm}^2$), although many of them were relatively small.
   (Histograms for the BP diameter are given below, see Fig.~\ref{FigBPDiameterHist}.) In case of the
   synthetic data, the boundaries of our manually detected BPs were obtained by applying the MLT algorithm
   to the CN, OH, and 5250.4\,\AA{} intensity images, to the maps of the \Index{circular polarization degree},
   and since we also discuss BP properties which were a direct output of the MHD calculations such as the
   magnetic field strength, we additionally applied the MLT algorithm to the field strength maps at
   constant optical depth $\log(\tau)=0$ and $\log(\tau)=-2$. The determination of various boundaries
   for the same set of BPs takes into account that the magnetic features change in size with height and
   do not always overlap one to one with their brightness enhancements.

\begin{figure}
   \centering
   \includegraphics[width=100mm]{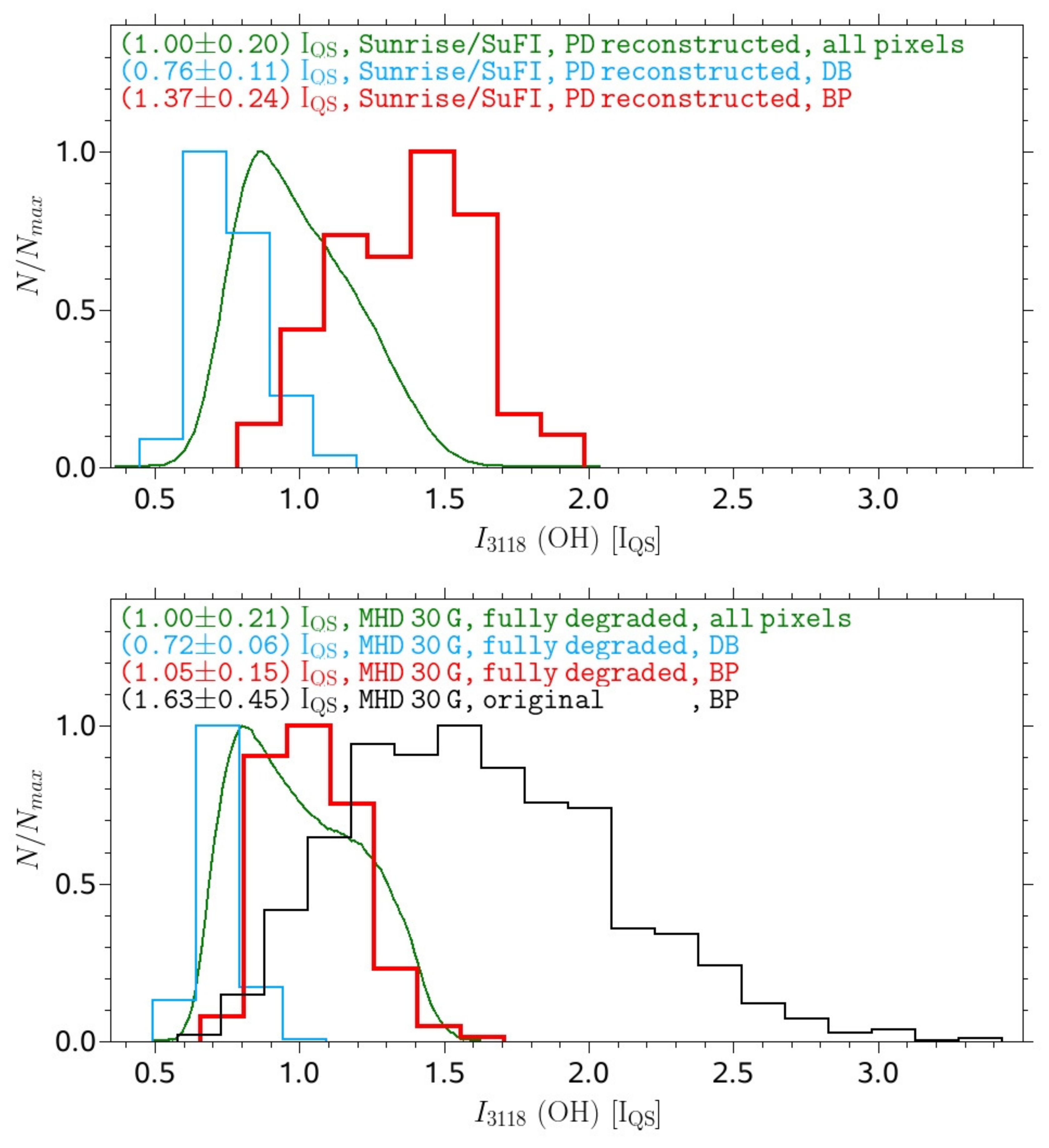}
   \caption{Histograms of the bright point (BP) peak intensity in the 3118\,\AA{} OH band (red lines),
   the intensity histograms of the BPs' dark background (blue lines), and the histograms of all pixels
   (green lines). Mean values and standard deviations are given in the text labels. The top panel
   shows the histograms obtained from the observational data recorded by the SuFI instrument, the bottom
   panel displays the same for the degraded 30\,G MHD simulations. The BP histogram of the undegraded MHD
   data is indicated in the bottom panel by the black line.}
   \label{FigBPOHHist}
   \end{figure}

\begin{figure}
   \centering
   \includegraphics[width=100mm]{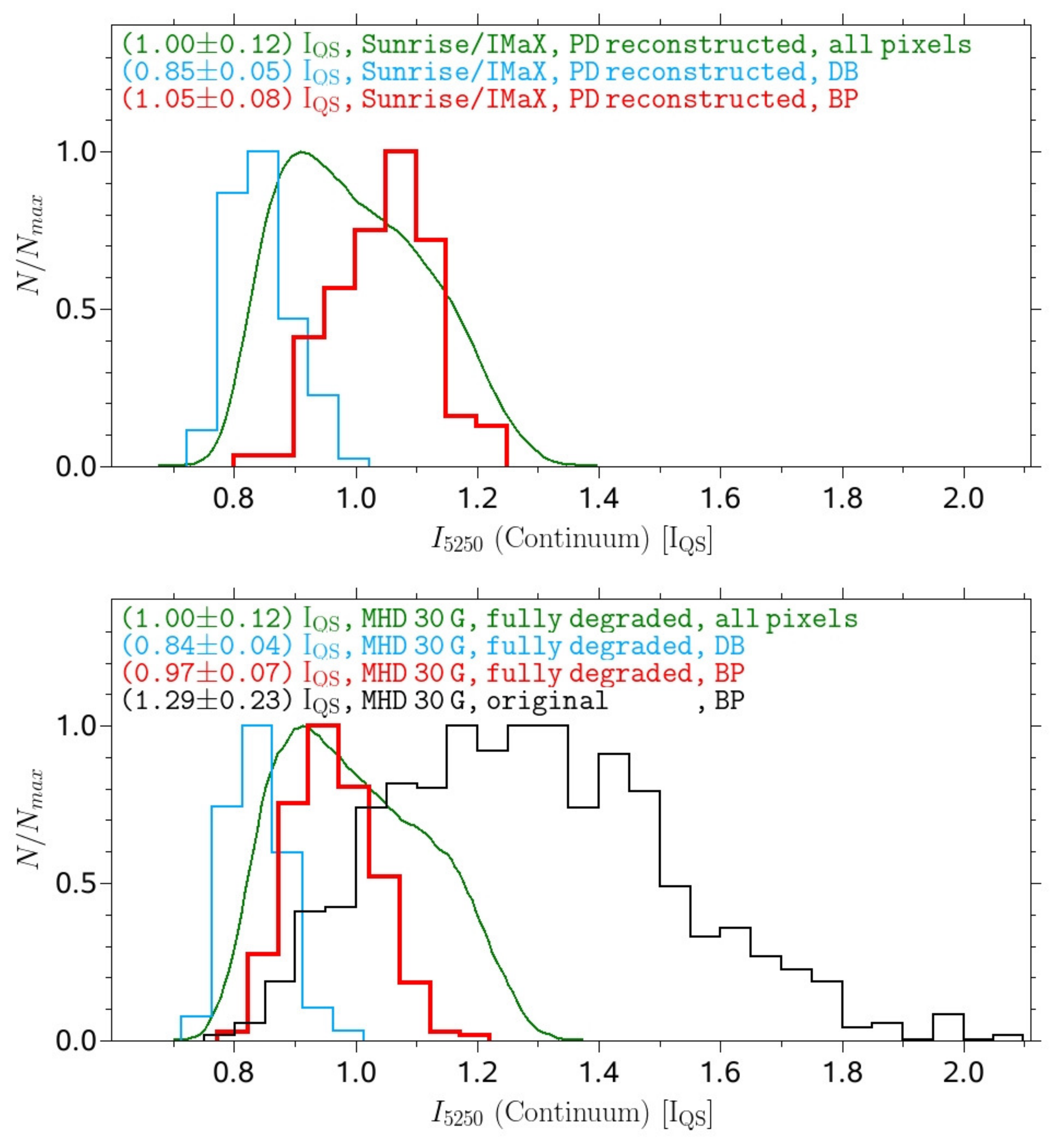}
   \caption{Same as Fig.~\ref{FigBPOHHist}, but for the peak intensity at the 5250.4\,\AA{} continuum.}
   \label{FigBPCCTHist}
   \end{figure}

\begin{figure}
   \centering
   \includegraphics[width=100mm]{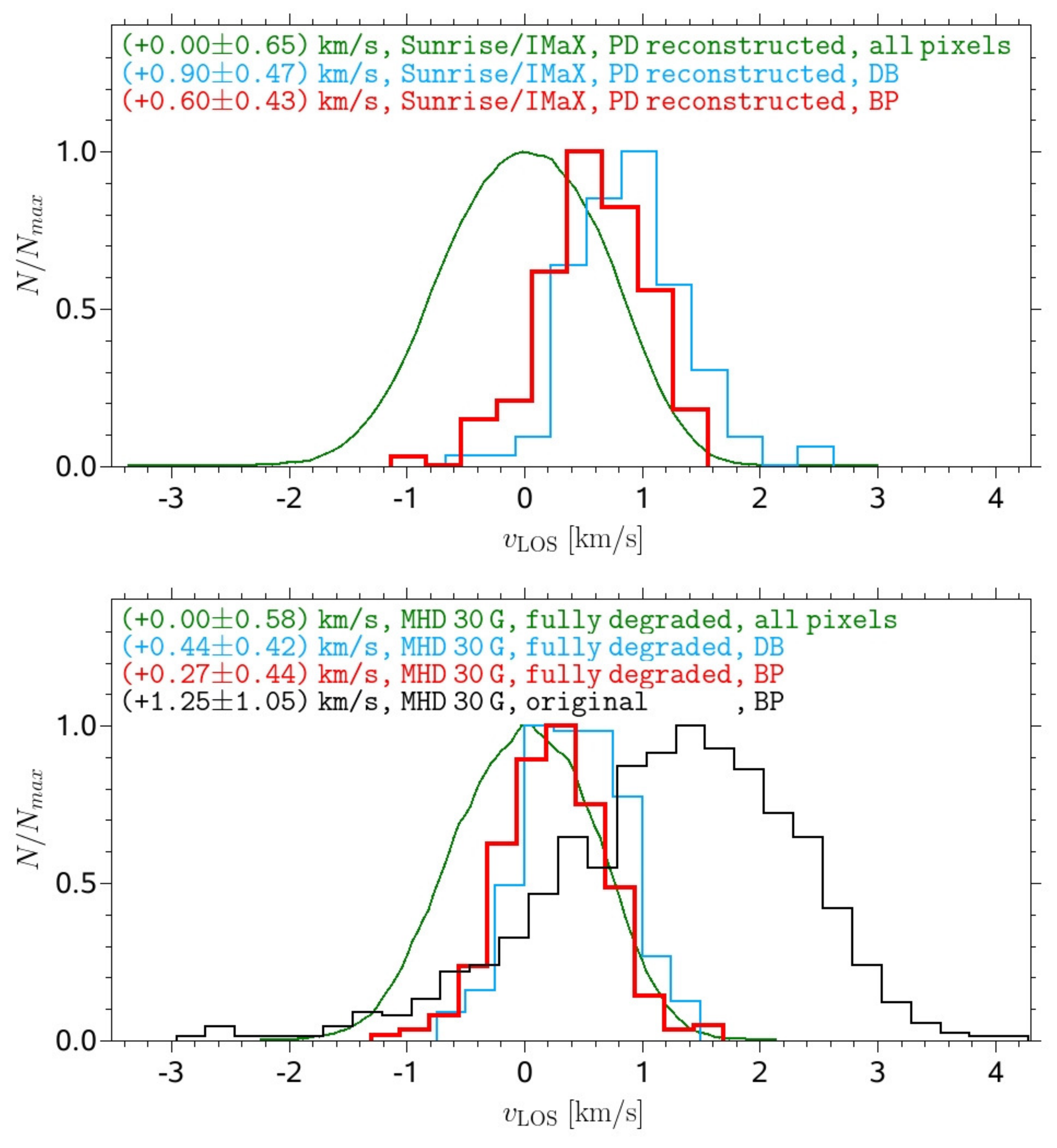}
   \caption{Same as Fig.~\ref{FigBPOHHist}, but for the LOS velocity.}
   \label{FigBPLMINHist}
   \end{figure}

\begin{figure}
   \centering
   \includegraphics[width=100mm]{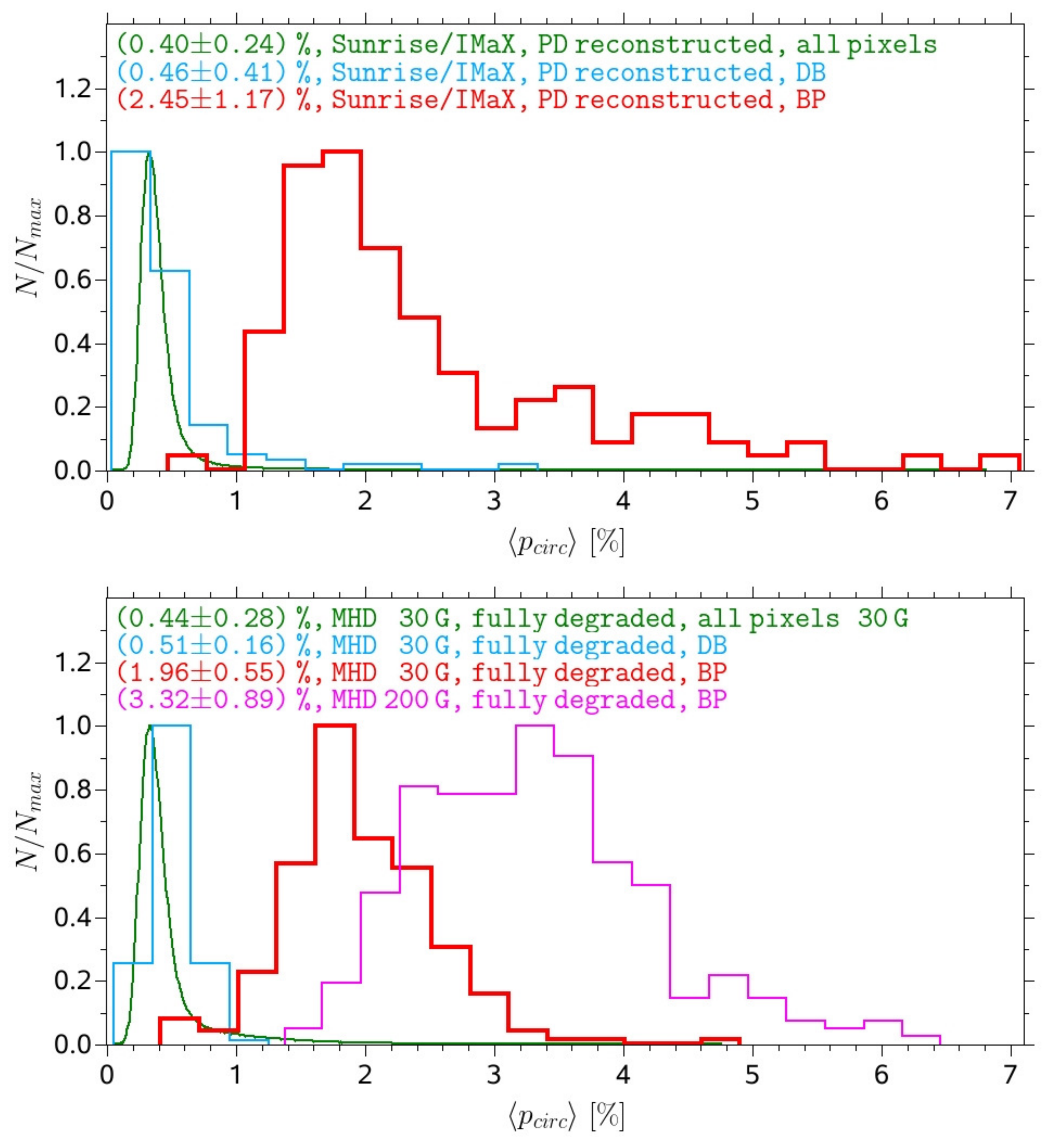}
   \caption{Same as Fig.~\ref{FigBPOHHist} for the spatial peak value of the \Index{circular polarization degree}.
   Additionally, the magenta line shows the histogram of the BP polarization for the degraded 200\,G
   MHD simulations.}
   \label{FigBPMean_CPHist}
   \end{figure}

   Histograms of various quantities in BPs are displayed as red lines in
   Figs.~\ref{FigBPOHHist}~to~\ref{FigBPMean_CPHist}. For comparison, the blue lines represent histograms
   of the same parameters in the dark background (DB, defined as the darkest pixel within 0\carcsec{}3
   of each BP's peak position). Likewise for comparison, we plot in green the corresponding
   histograms over all pixels in the images. These histograms have already been plotted in
   Figs.~\ref{FigOHHist}, \ref{FigCCTHist}, \ref{FigLMINHist}, and \ref{FigMean_CPHist}.\footnote{There
   are slight differences between the histograms over all pixels plotted in Figs.~\ref{FigOHHist},
   \ref{FigCCTHist}, \ref{FigLMINHist}, and \ref{FigMean_CPHist} on the one hand, and those in
   Figs.~\ref{FigBPOHHist}, \ref{FigBPCCTHist}, \ref{FigBPLMINHist}, and \ref{FigBPMean_CPHist} on the
   other hand. This is because the former are obtained from data of the original \Index{plate scale} and FOV,
   but the latter are produced from the re-sampled data of the common FOV.} The histograms of
   Figs.~\ref{FigBPOHHist}~to~\ref{FigBPMean_CPHist} are normalized to their maximum which turns out
   to be more favorable for comparison between histograms of significantly different shapes than
   normalization by their integrals. Figures~\ref{FigBPOHHist}~to~\ref{FigBPMean_CPHist} compare
   the set of 121 BPs detected from the \sunrise{} observations (upper panels) with the set of
   277 BPs detected from the degraded 30\,G MHD simulations (bottom panels). Additionally, the bottom
   panels in Figs.~\ref{FigBPOHHist}~to~\ref{FigBPLMINHist} show the histogram of the 898 BPs which
   we detected in the undegraded MHD data (black lines), so that the influence of the degradation is
   visible.

   Histograms of the BPs' peak intensity (intensity of the brightest pixel of a BP) in the OH band at
   3118\,\AA{} are drawn in Fig.~\ref{FigBPOHHist}. As expected, the highest BP intensities coincide
   with the highest intensities found in these images. The observed BPs exhibit a peak intensity
   range of $(0.86-1.97)~I_{\rm{QS}}$ (with a mean value of $1.37~I_{\rm{QS}}$), which is considerably
   higher than the range of $(0.73-1.58)~I_{\rm{QS}}$ with a mean value of $1.05~I_{\rm{QS}}$ obtained
   from the degraded simulations. Hence the observed BPs are brighter. The observed DB histogram is
   also broader than the simulated one.

   Fig.~\ref{FigBPCCTHist} shows histograms of the BPs' peak intensity in the continuum at 5250.4\,\AA{}.
   Again, the range covered by the observational BP histogram, $(0.82-1.24)~I_{\rm{QS}}$, is larger than
   that covered by the simulational BP histogram, $(0.80-1.18)~I_{\rm{QS}}$, although the discrepancy is
   not so marked. In addition, we found an almost perfect agreement between observation and simulation
   for the histograms of all pixels and a rather good match for the DB histograms. Also in contrast to
   the OH band, the highest intensities in the 5250.4\,\AA{} continuum images do not belong to BPs but
   to the brightest parts of granules.

   As mentioned above, the LOS velocity was retrieved from a Gaussian fit to Fe\,{\sc i} 5250.2\,\AA{}
   Stokes~$I$. The LOS velocities of all pixels belonging to a particular BP in the 5250.4\,\AA{}
   continuum are averaged and this averaged velocity is then assigned to that BP. The red lines of
   Fig.~\ref{FigBPLMINHist} exhibit histograms of such spatially averaged BP velocities. The standard
   deviations of the observational histograms agree with the degraded simulations, but the observed
   BPs and their DB display larger downflows. Clearly, the BPs' LOS velocity lies on the downflow side
   of the distribution for the whole map. The velocities of the observational BPs range between
   $-980$\,m~s$^{-1}$ and 1510\,m~s$^{-1}$, with an average downflow of 600\,m~s$^{-1}$. The downflow
   of the DB is on average 300\,m~s$^{-1}$ stronger. The BPs of the simulations show LOS velocities
   ranging from $-1180$\,m~s$^{-1}$ to 1620\,m~s$^{-1}$ with a mean downflow of 270\,m~s$^{-1}$. Here,
   the downflow of the mean DB is only 170\,m~s$^{-1}$ stronger than for the BPs. The original, undegraded
   simulations exhibit much larger downflows.

   Histograms of the BPs' \Index{circular polarization degree} (peak polarization within the BP boundary
   as determined from the polarization maps) can be seen in Fig.~\ref{FigBPMean_CPHist}. Firstly, a
   comparison of the upper panel of Fig.~\ref{FigBPMean_CPHist} (IMaX observations with 12 scan positions)
   with Fig.~2, panel~(h) of \citet{Riethmueller2010} (IMaX observations with 5 scan positions) shows that
   even for a significantly improved wavelength sampling, we find that most BPs are only weakly polarized.
   The strongest BP polarization degrees observed in the V5-6 mode reached 9.1\,\%, while they reached
   only 6.8\,\% in the L12-2 mode. This difference is caused by the fact that we averaged over all twelve
   scan steps in the L12-2 mode and hence over more scan positions close to the continuum (i.e. low
   polarization signals) than in the V5-6 mode. By interpolating Stokes~$V$ to the wavelength positions
   of the V5-6 mode we obtained similar polarization degrees as found in the V5-6 data by
   \citet{Riethmueller2010}. Also, the noise levels of the two data sets were comparable. We also
   calculated the histogram of the \Index{circular polarization degree} of the set of V5-6 BPs analyzed by
   \citet{Riethmueller2010} (not shown) and compared it with their histogram of the \Index{total polarization degree}
   shown in Fig.~2, panel~(h) of \citet{Riethmueller2010}. From the similarity of the two histograms
   we concluded that the Stokes~$Q$ and $U$ signals are negligible for a statistical BP study.

   A comparison of the observed and degraded 30\,G simulation BP $\langle p_{\rm{circ}} \rangle$ histograms
   revealed a good agreement for most of the BPs, but a population of BPs showing strong $\langle
   p_{\rm{circ}} \rangle$ is found only in the observations. We can rule out a possible over-reconstruction
   of the IMaX data as the cause of these large $\langle p_{\rm{circ}} \rangle$ because the rms contrasts
   matched rather well. Since we also applied the BP boundary detection via MLT to the circular
   polarization maps, we were able to determine the effective diameter of the polarized features, defined as
   the diameter of a circle of area equal to that within the $\langle p_{\rm{circ}} \rangle$ boundary of
   the BP. We found that 6 of the 121 observed BPs had $\langle p_{\rm{circ}} \rangle > 4.8\,\%$
   (strongest BP polarization of the 30\,G simulation) and their diameter was on average 399\,km, while the
   other 115 BPs had a mean diameter of only 299\,km. The long tail towards stronger polarization degrees,
   which we found in the observational BP histogram, is caused by large and strongly polarized BPs which
   were not present in the 30\,G simulations. For comparison reasons, we also plotted the histogram of
   $\langle p_{\rm{circ}} \rangle$ of the 285 BPs that we detected in the 10 degraded snapshots of our
   200\,G simulations (magenta line of the bottom panel of Fig.~\ref{FigBPMean_CPHist}). There the mean
   polarization degree was 3.32\,\%, the strongest value being 6.4\,\%. The number density was 0.79 BPs
   per $\rm{Mm}^2$.

\subsection{Why is the majority of bright points weakly polarized?}\label{ExplWeakBPs}
   The bottom panel of Fig.~\ref{FigBPMean_CPHist} exhibits many synthetic BPs which are only weakly
   polarized. The same was noted by \citet{Riethmueller2010}. This weak polarization could be due to
   the fact that most BPs have intrinsically weak fields, contrary to the standard \Index{flux tube} theory
   \citep{Spruit1976}, or they are very highly inclined, nearly horizontal, also contrary to theory
   for strong fields \citep{Schuessler1986}. Alternatively, they could be spatially unresolved at the
   spatial resolution reached by \sunrise{}, or the weak Stokes~$V$ could also be caused by thermal
   weakening of Fe\,{\sc i} 5250.2\,\AA{} in BPs. We searched for the cause by analyzing the simulation
   data.

   BPs are often modeled by nearly vertical slender flux tubes. In the flux tube model, only magnetic
   field strengths in the kilo-Gauss range can explain the brightnesses that are observed in BPs. The field
   increases the magnetic pressure which leads to an evacuation inside the tube and hence a depressed
   optical depth unity surface. The lateral inflow of heat through the walls of the flux tube makes it
   hot and bright.

   To determine what polarization signals can be expected for kilo-Gauss fields, we synthesized Stokes
   profiles for a standard atmosphere, the HSRASP \citep{Chapman1979}, assigned a zero velocity and a
   height independent field strength of 1\,kG. The synthetic profiles were then convolved with the
   spectral PSF of IMaX and the Stokes~$V$ values at the twelve scan positions of the IMaX L12-2 mode
   were used to calculate $\langle p_{\rm{circ}} \rangle$ according to Eq.~(\ref{Eq_Mean_CPi}).
   A value of 10.71\,\% was obtained. This value is significantly higher than the mean value of
   $\langle p_{\rm{circ}} \rangle$ of 1.96\,\% obtained from the fully degraded data.

\begin{figure}
   \centering
   \includegraphics[width=100mm]{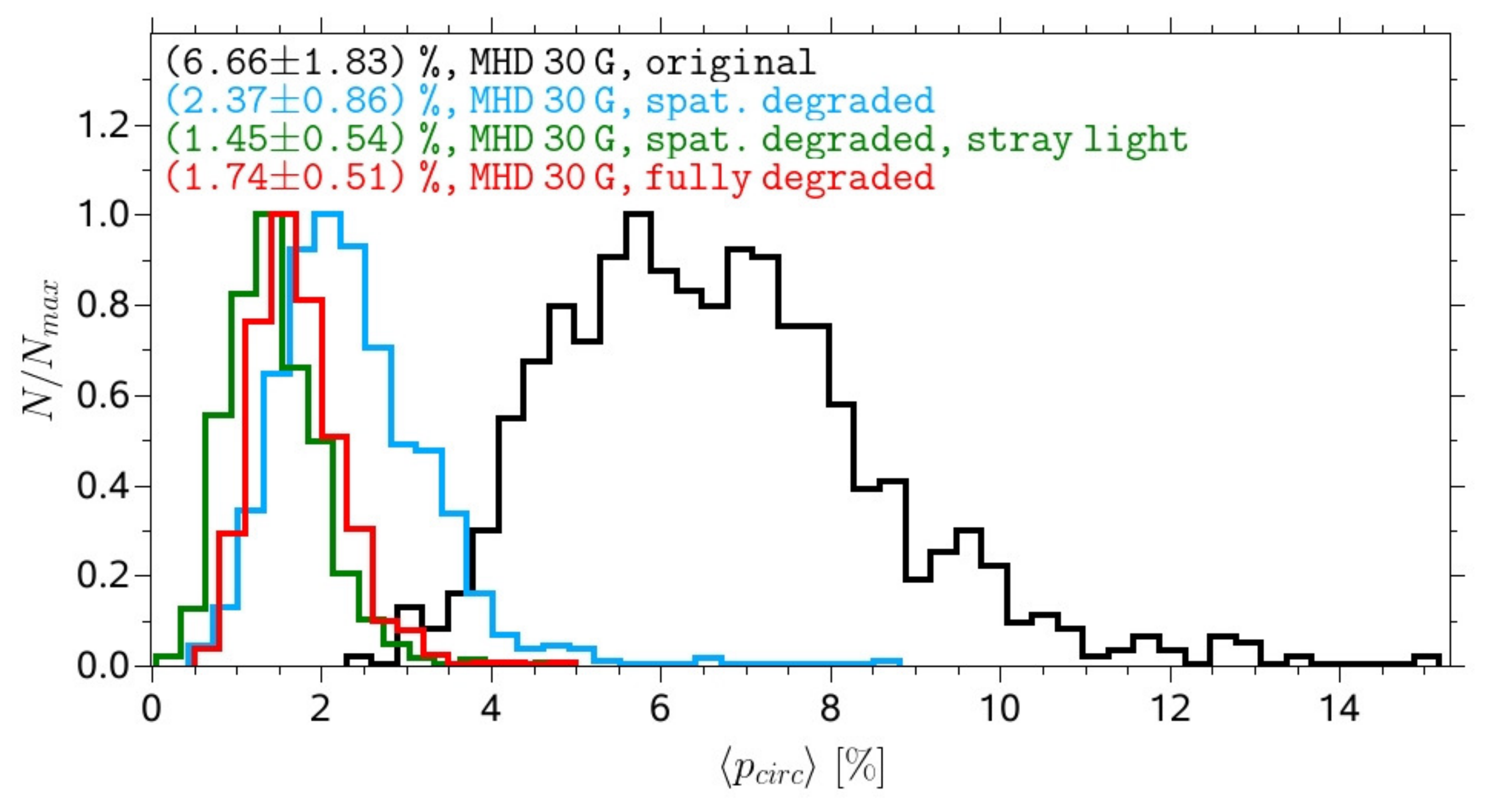}
   \caption{Influence of the various degradation steps on the histogram of the BP circular polarization
   degree. The degradation was applied to the 30\,G MHD simulations. The black line corresponds
   to the data that were only spectrally degraded. An additional spatial degradation is shown in blue. The
   green line contains also the stray light contamination and the fully degraded BPs are colored in red,
   i.e. including noise.}
   \label{FigBPMean_CP2Hist}
   \end{figure}

   To analyze the influence of the several degradation steps on the mean BP polarization, we used
   the set of 898 BPs that we detected in the undegraded CN images for the original\footnote{Again,
   "original" means only spectrally degraded, because this step was part of our synthesis. No
   further degradation steps had been applied at this stage.} MHD data and determined their peak
   $\langle p_{\rm{circ}} \rangle$ values (see black line in Fig.~\ref{FigBPMean_CP2Hist}), for the data
   that were spectrally and spatially degraded (blue line), for the additionally stray light contaminated
   images (magenta line), and the fully degraded data (red line of Fig.~\ref{FigBPMean_CP2Hist}).
   The spatial degradation reduced the mean BP polarization from 6.66\,\% down to 2.37\,\%. The stray light
   led to a further reduction down to 1.45\,\% and the noise increased the mean value to 1.74\,\%.
   In contrast to the histograms of all pixels (top panel of Fig.~\ref{FigMean_CPHist}), here the noise
   was not the main contributor but the spatial degradation, because the BP $\langle p_{\rm{circ}} \rangle$
   values were much higher than the noise level.

   Note that the red line in the bottom panel of Fig.~\ref{FigBPMean_CPHist} and the red line
   of Fig.~\ref{FigBPMean_CP2Hist} are not identical. Both histograms were retrieved from the
   same fully degraded $\langle p_{\rm{circ}} \rangle$ maps, but the lower panel of
   Fig.~\ref{FigBPMean_CPHist} was calculated for the 277 BPs that were detected in the degraded
   CN images while Fig.~\ref{FigBPMean_CP2Hist} displays the 898 BPs detected from the undegraded
   CN images, which contained many small BPs and hence led to a smaller mean value of 1.74\,\%
   compared to 1.96\,\% for the 277 BPs detected in the degraded CN images.

\begin{figure*}
   \centering
   \includegraphics[width=\textwidth]{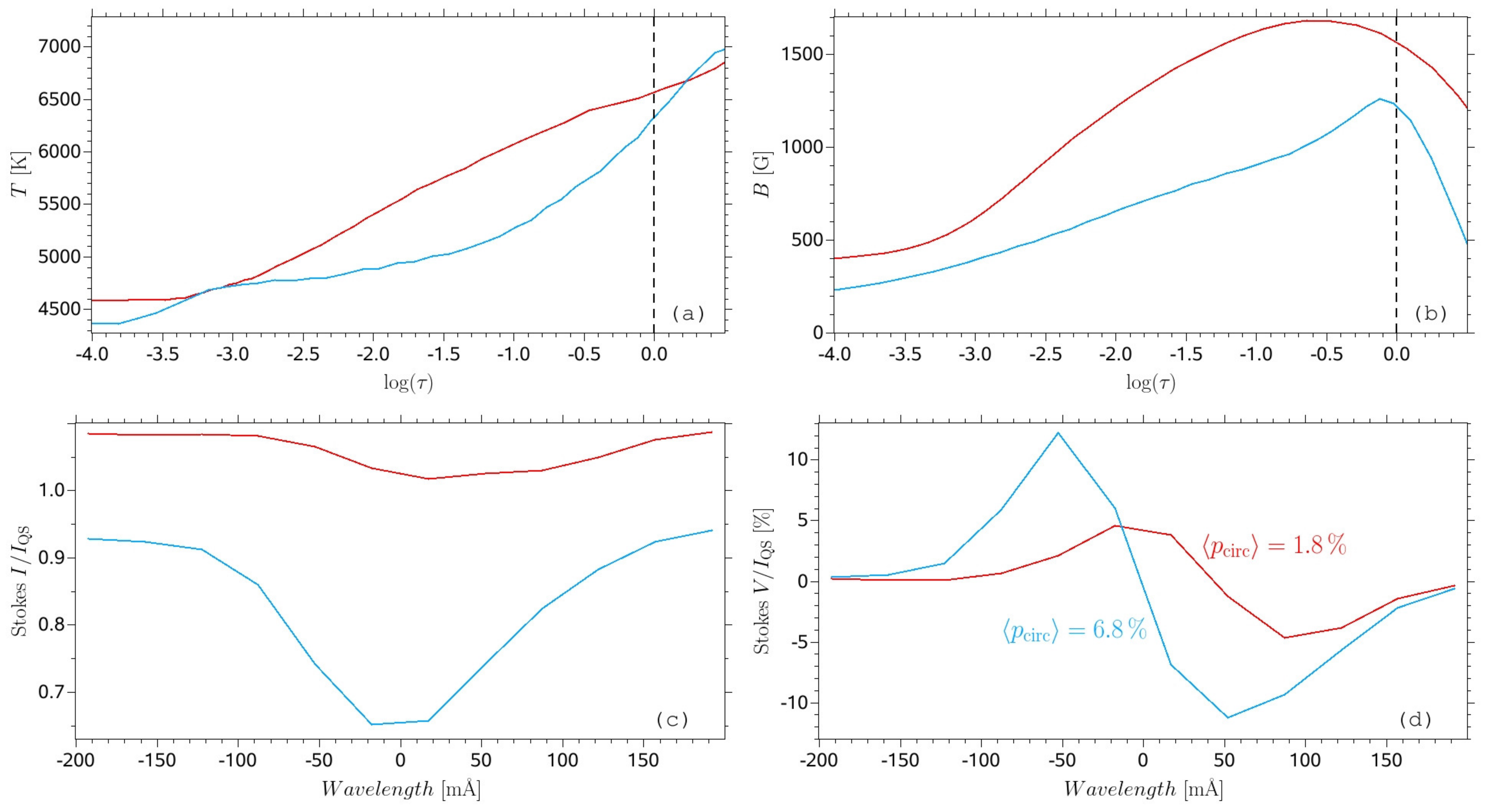}
   \caption{Demonstration of a weak polarization signal owing to the strong temperature sensitivity of the
   Fe\,{\sc i} line at 5250.2\,\AA{} by a comparison of a pixel in a particularly bright BP (red lines)
   with a pixel located in a much less bright BP (blue lines). Panel~(a) exhibits the vertical temperature
   stratification as a function of $\log(\tau)$ at 5000\,\AA{}, panel~(b) the magnetic field strength
   stratification. Panels~(c) and (d) depict the Stokes~$I/I_{\rm{QS}}$ and $V/I_{\rm{QS}}$ signals from
   the original MHD data at the twelve IMaX L12-2 scan positions in the line.}
   \label{FigTempEffect1}
   \end{figure*}

   Even at original resolution of the simulations the average $\langle p_{\rm{circ}} \rangle$ is only
   6.66\,\% and only 2.6\,\% of all BPs reach the 10.71\,\% retrieved from the 1\,kG HSRASP atmosphere.
   This remaining discrepancy can be explained by the high temperature sensitivity of the Fe\,{\sc i}
   line at 5250.2\,\AA{}, as illustrated in Fig.~\ref{FigTempEffect1}, where two pixels taken from the
   simulation data are compared. The pixel colored in blue belongs to a faint BP with a low brightness
   in the CN band, but one that is still identified as a BP, while the pixel colored in red is part of
   a BP with a high contrast. Panel~(a) of Fig.~\ref{FigTempEffect1} shows the vertical temperature
   stratification where the atmospheric height is given in logarithmic units of $\tau$, which is the
   optical depth of the continuum at the wavelength of 5000\,\AA{}. In the middle photosphere (where
   the spectral line is mainly formed), the red pixel's temperature is about 800\,K higher than the
   temperature of the blue pixel, but in the lower and upper photosphere both temperatures are almost
   the same. The vertical stratification of the magnetic field strength is displayed in panel~(b).
   In the middle photosphere, e.g. at $\log(\tau)=-2$, the blue pixel reveals a field strength of
   roughly 650\,G, while the red pixel has a field strength of 1200\,G. As typical for the \Index{flux tube}
   model of BPs, at $\log(\tau)=0$ both pixels show a magnetic field stronger than 1\,kG that is
   nearly vertically oriented (the mean field inclinations of the two pixels are $8.9^{\circ}$ and
   $7.6^{\circ}$). The original Stokes~$I/I_{\rm{QS}}$ signals of the twelve scan positions of the
   5250.2\,\AA{} line are shown in panel (c). The intensity of the red pixel is higher than
   $I_{\rm{QS}}$ while it is lower in the blue pixel. The most striking feature of the figure is the
   minimal line depth of the 5250.2\,\AA{} line in the red pixel. This is partly due to the large Zeeman
   splitting caused by the large field strength in this pixel. The strong temperature sensitivity also
   contributes just as much. The temperature sensitivity originates not just in the increased ionization
   of iron as the temperature is raised, but also from the \Index{excitation potential} of the lower level of
   this line of only 0.12\,eV. This weakening is also conspicuous in the Stokes~$V/I_{\rm{QS}}$ profiles
   of panel~(d). Although the red pixel has a higher magnetic field strength, its circular polarization
   degree of 1.8\,\% is much smaller than that of the blue pixel, 6.8\,\%. The ratio 1.8\,\%/6.8\,\% is
   larger than the ratio of the line depths, due to the contribution of the magnetic field which is
   stronger for the red pixel \citep[obviously \Index{Zeeman saturation},][is not complete, so that a residual
   Zeeman sensitivity is present]{Stenflo1973}.

\begin{figure}
   \centering
   \includegraphics[width=100mm]{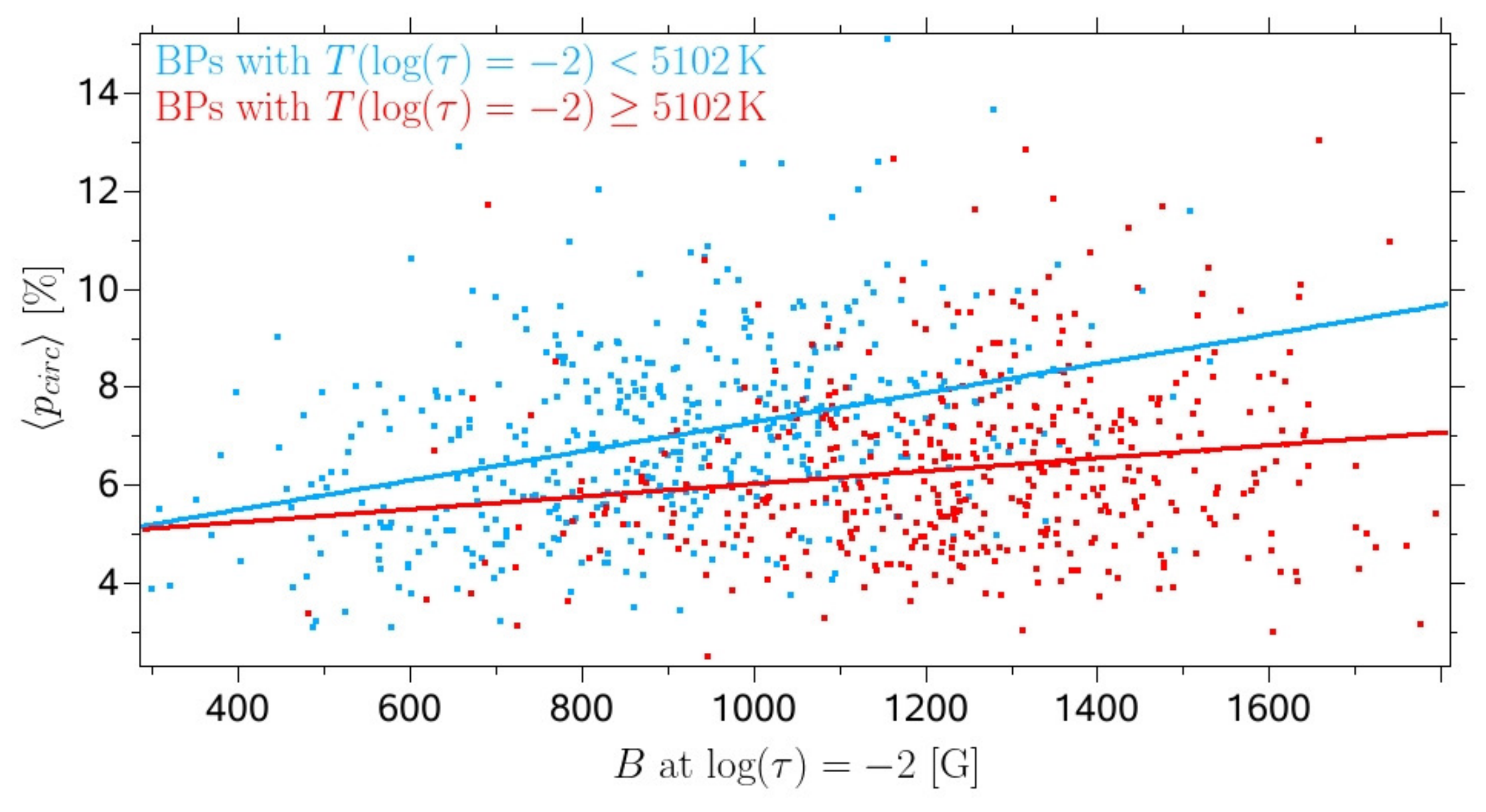}
   \caption{Circular polarization degree versus magnetic field strength at $\log(\tau)=-2$ as
   retrieved for the 898 BPs from the original 30\,G MHD data. Pixels with $T \ge 5102$\,K are colored
   in red, cooler pixels in blue. The solid lines are linear regressions.}
   \label{FigTempEffect2}
   \end{figure}

   The temperature effect is statistically relevant for the BPs in general. To show this, in
   Fig.~\ref{FigTempEffect2} we plot the \Index{circular polarization degree} versus the magnetic field
   strength at $\log(\tau)=-2$ (peak values for both quantities) for all 898 BPs in the original
   30\,G MHD data. At $\log(\tau)=-2$ these BPs have a mean temperature of 5102\,K, and all BPs
   hotter than the mean temperature are colored in red, the cooler ones in blue. The solid lines
   are the linear regressions of the two BP classes. We recognize a clear trend that for a given
   field strength, the BPs having a higher temperature show a weaker $\langle p_{\rm{circ}} \rangle$
   than cooler BPs, although the scatter is large.

   Since the Stokes~$V$ normalization to $I_i$, chosen by \citet{Riethmueller2010}, could enhance
   the effect of the line weakening of $\langle p_{\rm{circ}} \rangle$ we also considered other
   formulas to analyze the BP polarization or field strength, respectively: 

   \begin{equation}\label{Eq_Mean_CPc}
   \langle p_{\rm{circ}}^{\rm{QS}} \rangle = \frac{1}{12 I_{\rm{QS}}} \sum_{i=1}^{12}{\left| V_i \right|}~,
   \end{equation}

   \begin{equation}\label{Eq_Mean_Blos}
   \langle B_{\rm{LOS}} \rangle = \frac{1}{N} \sum_{i=1}^{N}{\frac{4 \pi c m_e}{e \lambda_0^2 g} \left| \frac{V_i}{(\rm{d}I/\rm{d}\lambda)_i} \right|}~.
   \end{equation}

   The scatterplot for $\langle p_{\rm{circ}}^{\rm{QS}} \rangle$ (not shown) exhibits the same
   qualitative behavior than that for $\langle p_{\rm{circ}} \rangle$ (shown in Fig.~\ref{FigTempEffect2}),
   so that the temperature effect cannot be reduced by this widely used normalization for the circular
   polarization calculation.

\begin{figure}
   \centering
   \includegraphics[width=100mm]{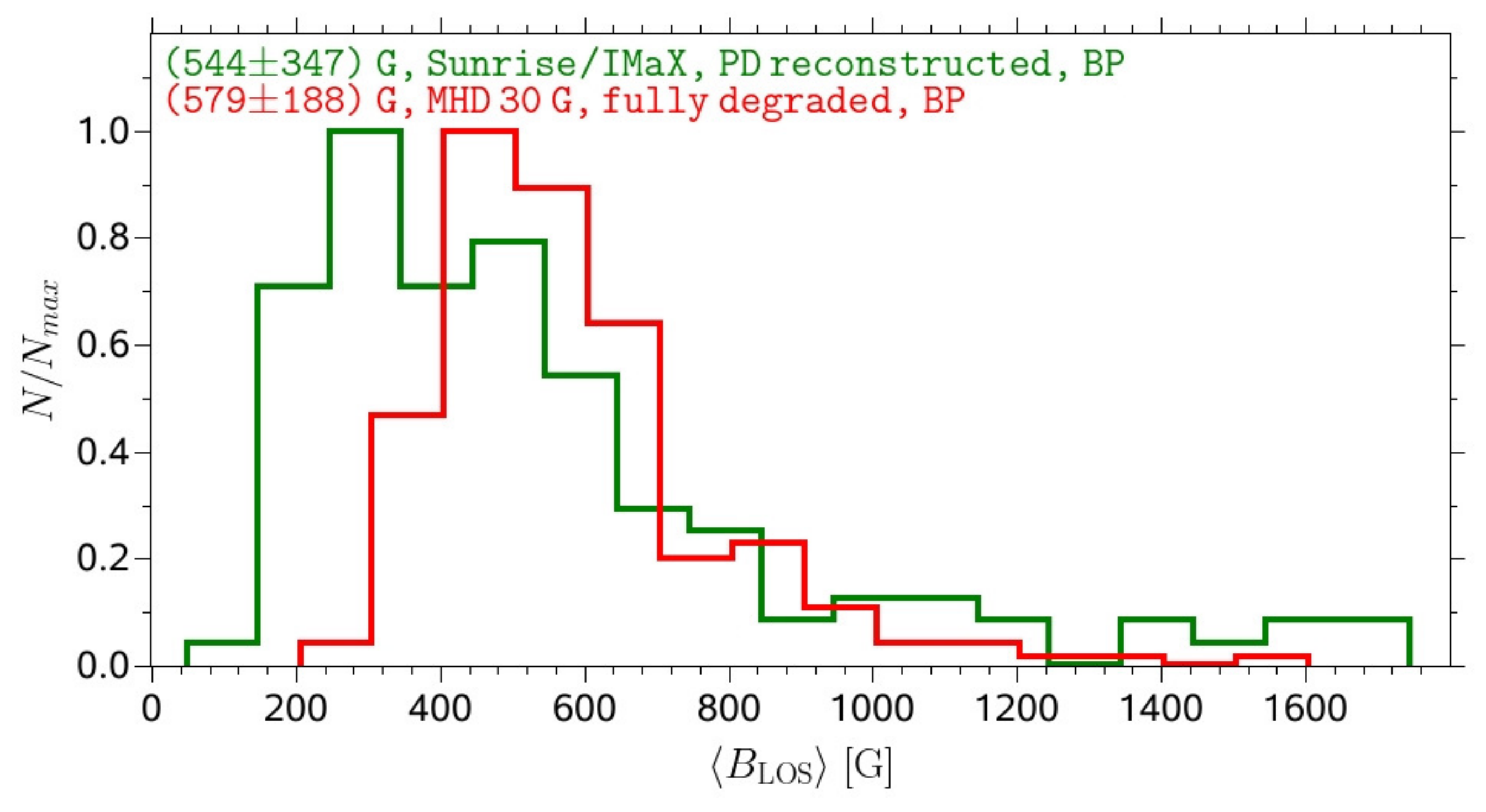}
   \caption{Histograms of the peak value of the BP field strength calculated via the weak field
   approximation (see main text). The green line shows the observed BPs, the red line displays the BPs
   from the degraded 30\,G MHD data.}
   \label{FigBPMean_BlosHist}
   \end{figure}

   Eq.~(\ref{Eq_Mean_Blos}) is based on the weak field approximation which holds if the Zeeman
   splitting is much smaller than the line width \citep{LandiDeglInnocenti2004}. The derivative of
   Stokes~$I$ was determined from the Gaussian fit of the Stokes~$I$ profile. $c$, $m_e$,
   and $e$ have the usual meaning, $\lambda_0 = 5250.2$\,\AA{} is the reference wavelength, and $g = 3$
   is the effective \Index{Land\'e factor} of the line. To avoid division by zero, the sum in Eq.~(\ref{Eq_Mean_Blos})
   had only been calculated over the $N$ scan positions with $\left| \rm{d}I/\rm{d}\lambda \right| > 3 \sigma$,
   where $\sigma$ is the ratio of the Stokes~$I$ noise level and the scanning step size. A scatterplot
   of $\langle B_{\rm{LOS}} \rangle$ versus the magnetic field strength at $\log(\tau)=-2$ (not shown)
   revealed that Eq.~(\ref{Eq_Mean_Blos}) very significantly reduces the effect of the line weakening.
   The $\langle B_{\rm{LOS}} \rangle$ values obtained from applying Eq.~(\ref{Eq_Mean_Blos}) on
   observational and degraded synthetic data are plotted in  Fig.~\ref{FigBPMean_BlosHist}. Similar
   to $\langle p_{\rm{circ}} \rangle$ (Fig.~\ref{FigBPMean_CPHist}), $\langle B_{\rm{LOS}} \rangle$
   shows a long tail of stronger fields which is more pronounced for the observational data. Only
   9.9\,\% of the observed BPs and 3.6\,\% of the synthetic BPs yielded $\langle B_{\rm{LOS}} \rangle$
   values higher than 1\,kG.

\subsection{Properties of simulated bright points}
\begin{figure}
   \centering
   \includegraphics[width=100mm]{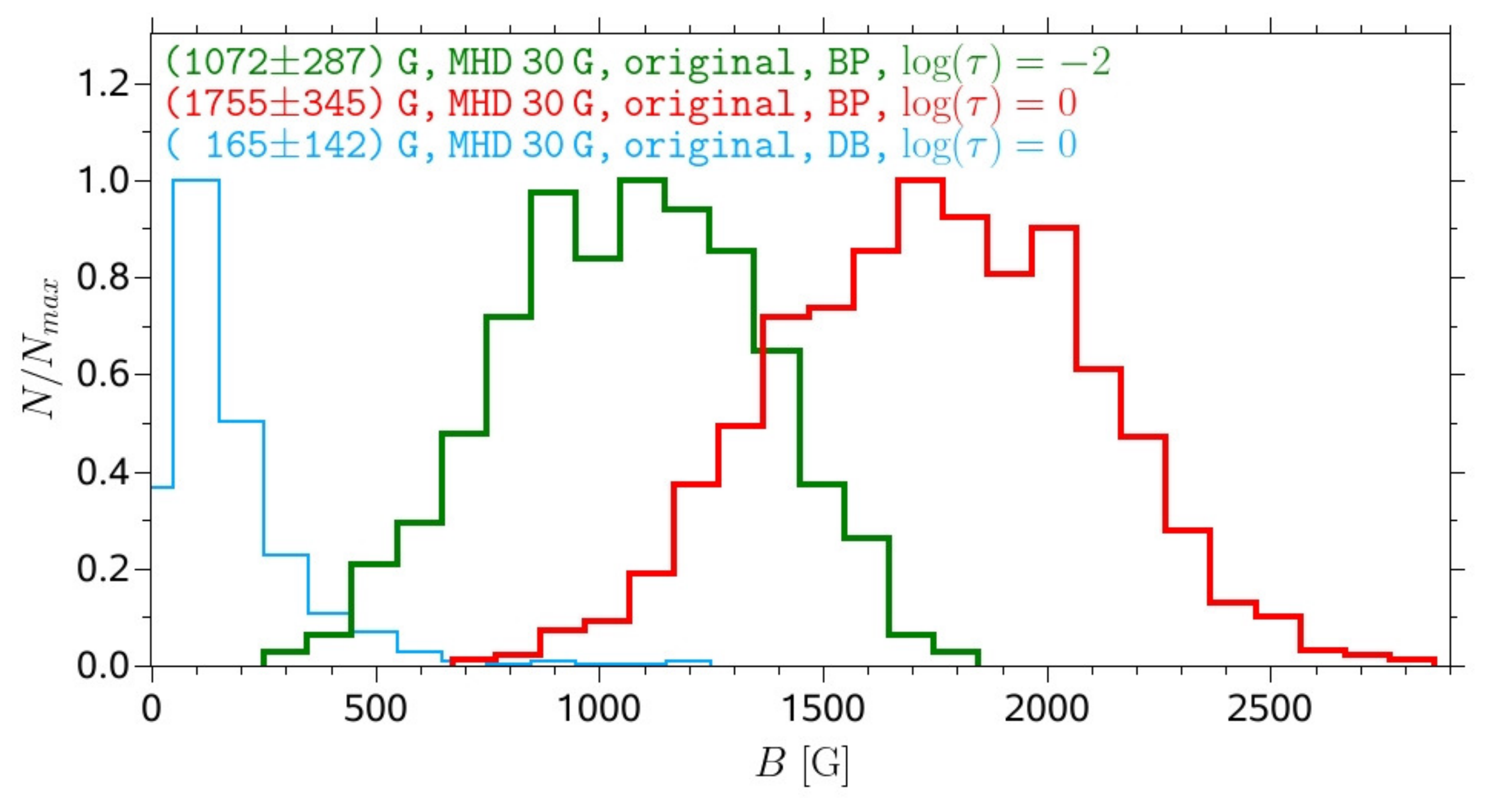}
   \caption{Histograms of the magnetic field strength of the BPs in the middle photosphere (green line)
   and of the BPs (red line) and the DB (blue line) in the lower photosphere as retrieved from the undegraded
   30\,G MHD data. Mean values and their standard deviations are given in the text labels.}
   \label{FigBPBfieldHist}
   \end{figure}

\begin{figure}
   \centering
   \includegraphics[width=100mm]{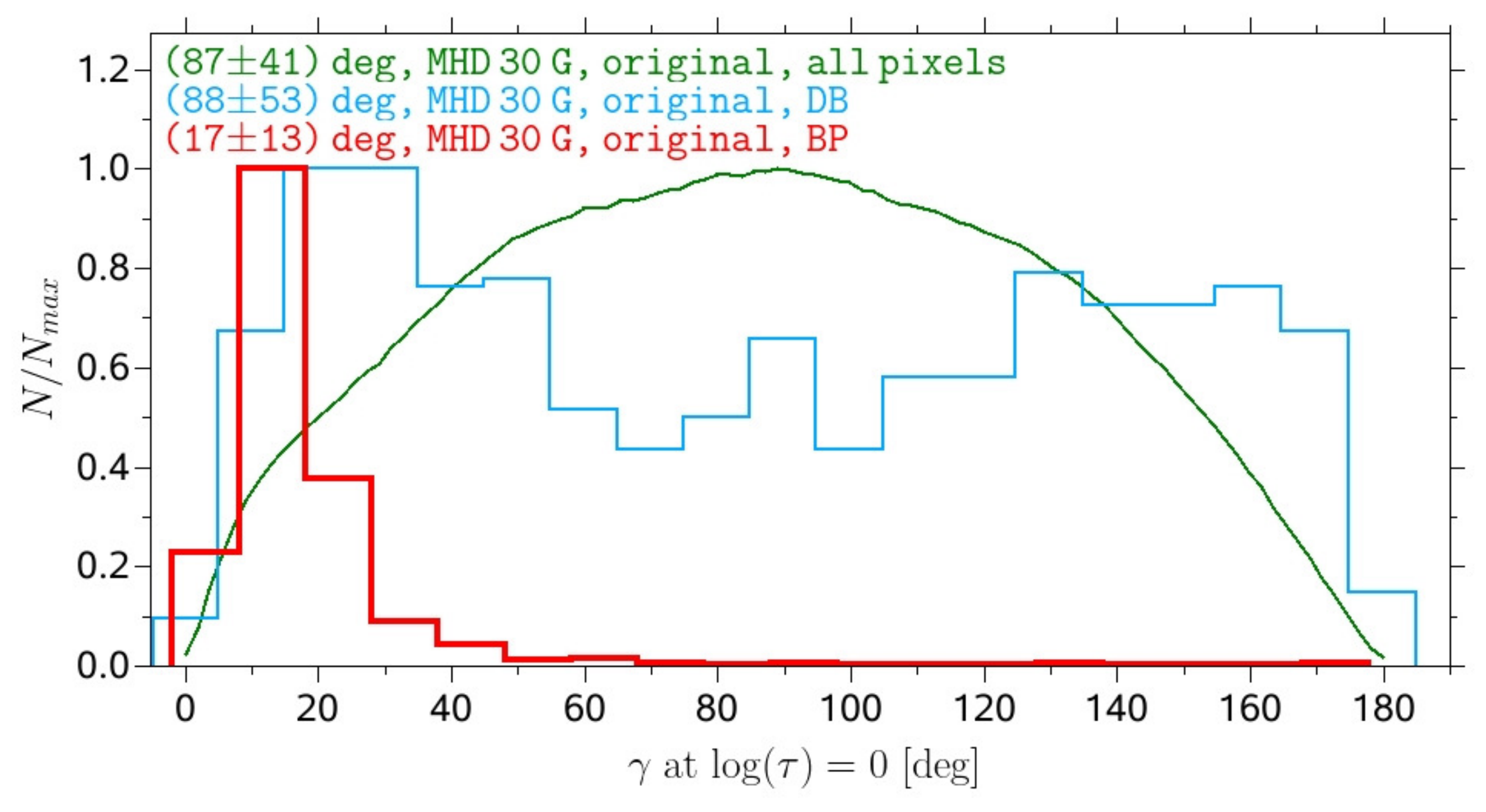}
   \caption{Histograms of the magnetic field inclination at optical depth unity of BPs, DB, and of all
   pixels. The color coding is the same as in Fig.~\ref{FigBPOHHist}.}
   \label{FigBPGammaHist}
   \end{figure}

\begin{figure}
   \centering
   \includegraphics[width=100mm]{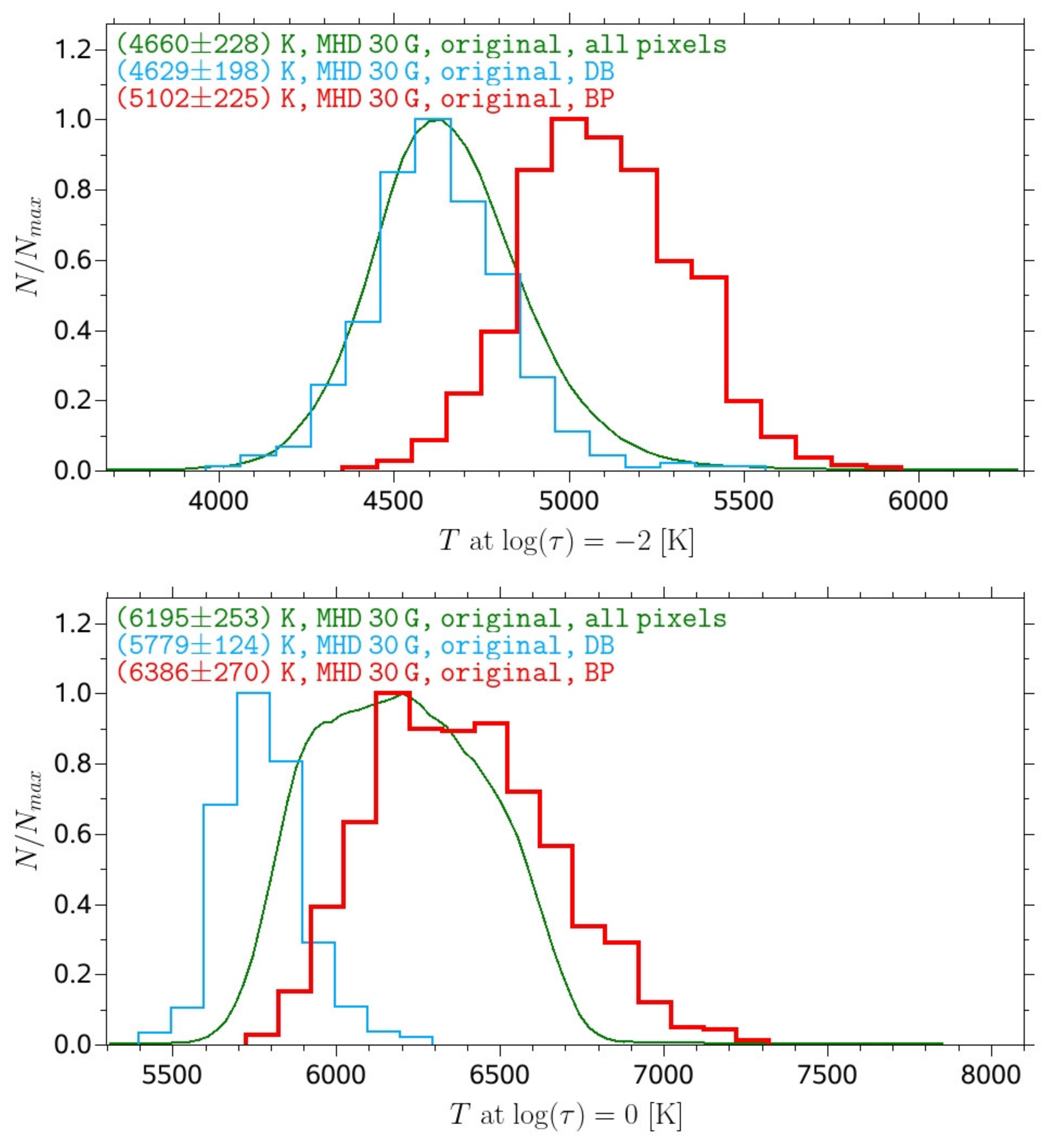}
   \caption{Histograms of the temperature for the simulated BPs (red lines), for the BPs' dark
   background (blue lines), and for all pixels in all frames (green lines). Mean values and their
   standard deviations are given in the text labels. The top panel shows the temperature at $\log(\tau)=-2$,
   the bottom panel at optical depth unity (note the different $T$ scales).}
   \label{FigBPTempHist}
   \end{figure}

\begin{figure}
   \centering
   \includegraphics[width=100mm]{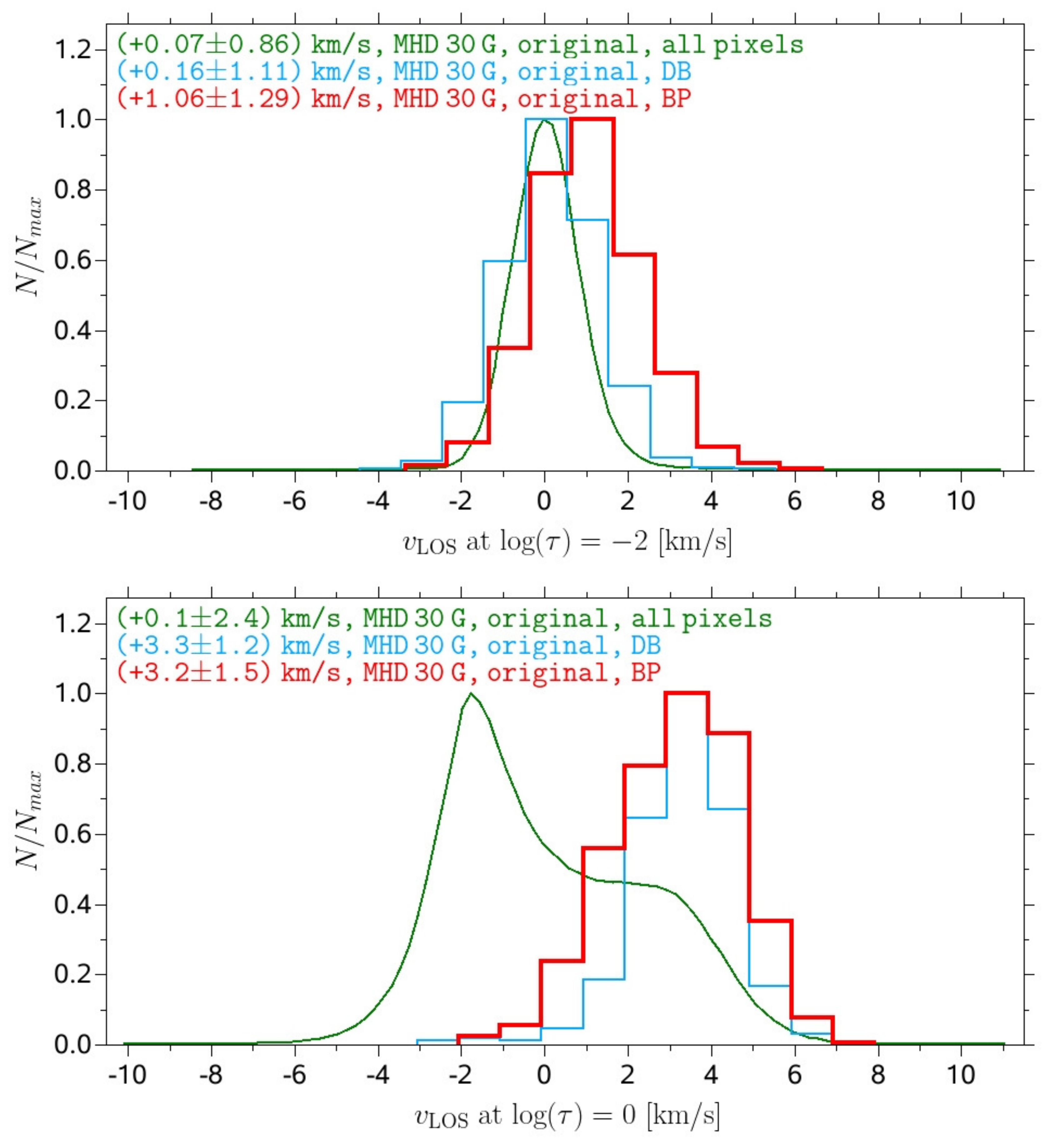}
   \caption{Same as Fig.~\ref{FigBPTempHist} for the LOS velocities (vertical component of the velocity
   vector taken directly from the MHD output).}
   \label{FigBPV_LOSHist}
   \end{figure}

   In sections \ref{CompBPs} \& \ref{ExplWeakBPs} we showed that the \Index{MURaM} MHD simulations reproduce
   the properties of the observed BPs reasonable well. Therefore we can obtain a better understanding
   of the physical phenomena underlying BPs by analyzing BP properties in undegraded simulations.
   This we now proceed to do. In the following we used the 898 BPs detected in the undegraded CN images.
   It is this set of BPs that underlies Figs.~\ref{FigBPBfieldHist}~to~\ref{FigBPV_LOSHist} and
   Fig.~\ref{FigBPI388vsBm0Scat}.

   Fig.~\ref{FigBPBfieldHist} shows histograms of the BP peak magnetic field strength taken directly
   from the 30\,G MHD simulations. The peak values were determined as the maximum field strength at a given optical depth
   within a BP. Owing to lateral force balance, the decreasing gas pressure with height and the need
   for magnetic flux conservation, the \Index{flux tube}s expand, so that the BP field strength drops from an
   average of 1755\,G to 1072\,G at $\log(\tau)=-2$. At optical depth unity, only 14 of the considered
   898 BPs had a field strength lower than 1000\,G, i.e. 98\,\% of the BPs were in the kilo-Gauss range,
   the strongest BP field strength was found to be 2825\,G, the weakest one was 721\,G. On average, the
   DB field strength is an order of magnitude lower than in the BPs.

   Fig.~\ref{FigBPGammaHist} displays the field inclinations (angle between the field vector and the
   surface normal) of the 30\,G MHD BPs at optical depth unity. The inclinations were averaged over all
   pixels covered by a BP at optical depth unity. The mean inclination\footnote{The undegraded data were
   noise-free and hence a magnetic field was found at every pixel, so that a field strength and inclination
   could also be given for each pixel.} of the BPs is $17^{\circ}$, i.e. the BPs are almost vertical as
   expected from buoyancy considerations \citep{Schuessler1986}. 4 of the 898 BPs showed inclinations
   greater than $90^{\circ}$, i.e. their magnetic field direction had been reversed from the initial
   condition of a homogeneous unipolar field. The field in the DB, in contrast, displays all possible
   inclination values, where vertical fields of either polarity are slightly preferred. The distribution
   of the field inclinations of all pixels is quite similar to the distribution of an isotropic field.

   In Fig.~\ref{FigBPTempHist} we compared the BP temperatures between the middle photosphere and the
   lower photosphere. The temperatures were averaged over all pixels of a BP as determined by the MLT
   algorithm applied to the CN maps. The mean BP temperature in the middle photosphere was 442\,K higher
   than the mean quiet-Sun temperature. The mean DB temperature was 31\,K lower than the mean quiet-Sun value.
   In the lower photosphere, the mean BP temperature was 191\,K higher than the mean quiet-Sun temperature,
   while the dark background was 416\,K colder than the quiet Sun. The temperature gradient in the
   BP and in the DB was significantly lower than on average.

   Fig.~\ref{FigBPV_LOSHist} displays histograms of the LOS velocity obtained directly from the MHD
   calculations. The LOS velocities were averaged over all pixels of a BP as determined by the MLT
   algorithm applied to the 5250.4\,\AA{} continuum intensity images. At an optical depth of
   $\log(\tau)=-2$ (top panel) the mean BP velocity was a moderate downflow of 1.06\,km~s$^{-1}$.
   With it the BPs revealed a stronger downflow than the DB with on average 0.16\,km~s$^{-1}$.
   In the lower photosphere at $\log(\tau)=0$ (bottom panel), the LOS velocities of the BPs showed
   a strong downflow of 3.17\,km~s$^{-1}$. This time the DB downflow was on average slightly stronger
   with 3.35\,km~s$^{-1}$. The histogram of all pixels (green line) clearly shows a superposition of
   two populations. The pixels from the interior of the granules formed the first population with
   an upflow of around $-2$\,km~s$^{-1}$ as the most numerous velocity. The second population corresponds
   to intergranular lanes with a 3\,km~s$^{-1}$ downflow as the most numerous velocity.

\begin{figure}
   \centering
   \includegraphics[width=100mm]{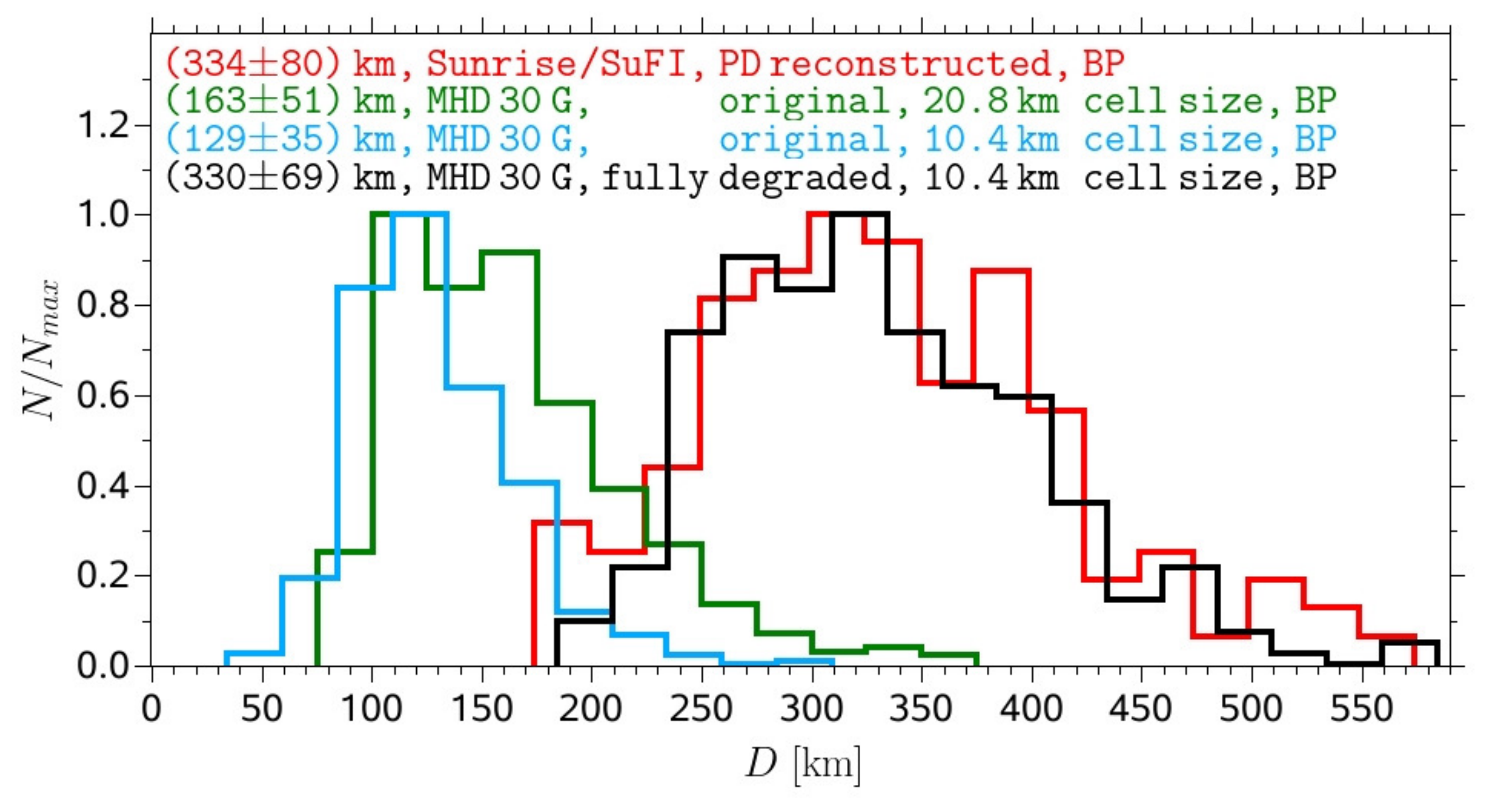}
   \caption{Histograms of the effective BP diameter for the observed BPs (red line), for the undegraded
   MHD BPs simulated with a cell size of 20.8\,km (green line), for the undegraded MHD BPs simulated
   with a cell size of 10.4\,km (blue line), and for the degraded MHD BPs simulated with a cell size
   of 10.4\,km (black line).}
   \label{FigBPDiameterHist}
   \end{figure}

\begin{figure}
   \centering
   \includegraphics[width=100mm]{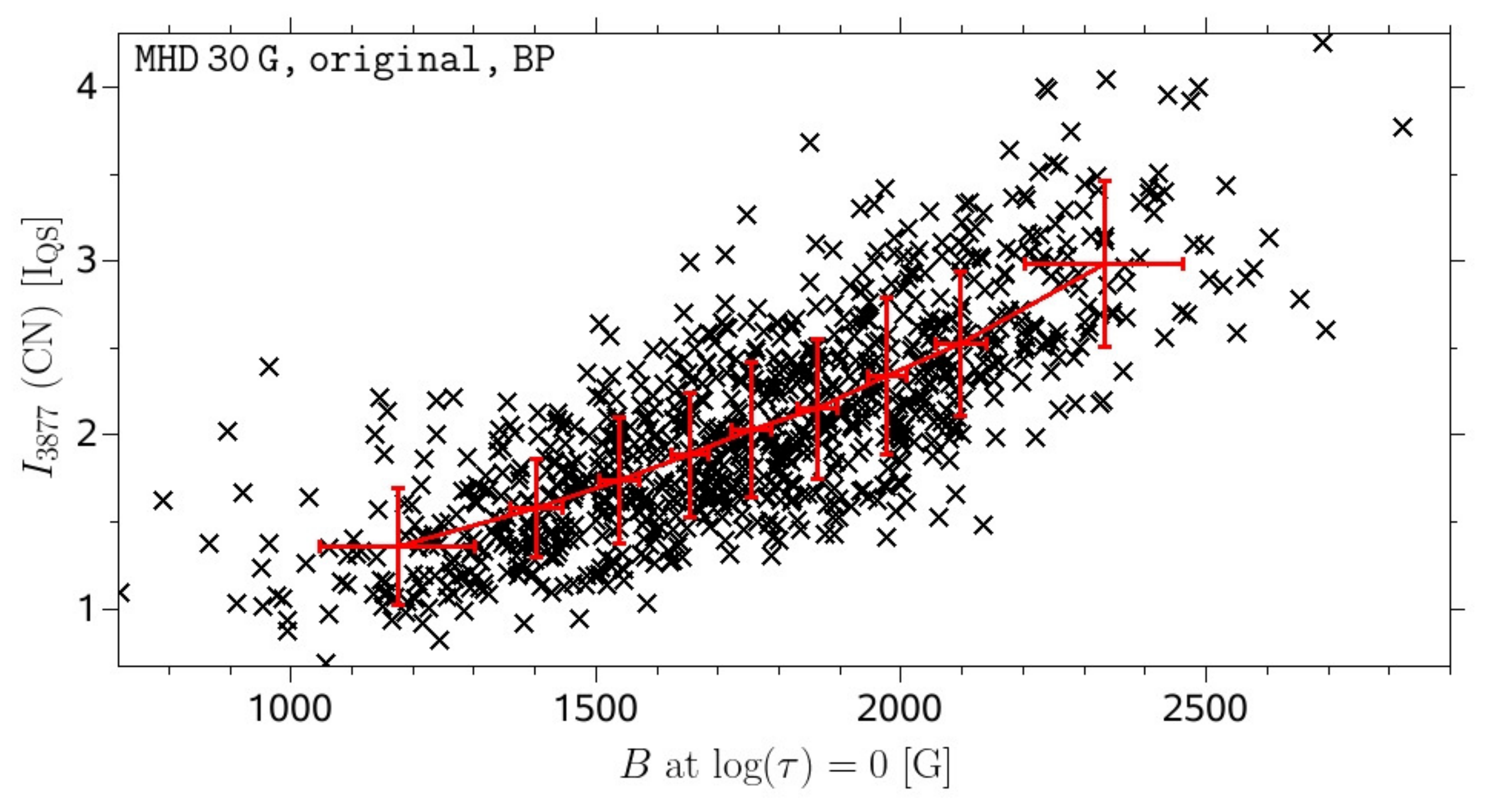}
   \caption{Scatterplot of the BP peak intensity at 3877\,\AA{} versus the magnetic field strength at optical
   depth unity (black crosses). The BPs were detected from the original 30\,G MHD data with a cell size of
   10.4\,km. The solid red line connects binned values.}
   \label{FigBPI388vsBm0Scat}
   \end{figure}

   Fig.~\ref{FigBPDiameterHist} compares histograms of the effective BP diameter between the \sunrise{}
   observations and various 30\,G MHD simulations. The diameters were calculated from the BP boundaries
   as determined from the CN intensity images. The observed mean BP diameter of 334\,km (red line) was fairly
   similar to the mean diameter of the degraded synthetic BPs, 330\,km (black line). The
   influence of the degradation can be seen by comparing this black line with the blue line, showing the
   histogram of the undegraded MHD simulations. The degradation increased the mean BP diameter from
   129\,km to 330\,km. The cell size of the MHD grid may also influence the BP diameters. We repeated
   the 30\,G MHD simulation, the full spectral line synthesis of the CN band, and the BP detection for
   a cell size of 20.8\,km while all other parameters were unmodified. 475 BPs were found in the 30 snapshots
   of $ 6 \times 6\,\mathrm{Mm}^2$ size (number density was 0.44 BPs per $\rm{Mm}^2$) with a mean BP diameter
   of 163\,km (green line). This dependence of BP diameter and number density on the grid size of the
   simulations is interesting in the sense that it indicates that an even smaller grid size could
   result in smaller BPs, which casts doubts on the claims of \citet{Crockett2010} that they resolved
   essentially all BPs, from the comparison of BP area distributions from observations and from
   simulations. However, the employed simulations had a grid size of 25\,km, so that true BP sizes are
   almost certainly smaller than was claimed. Note that the green line of Fig.~\ref{FigBPDiameterHist}
   is the only one result in this study that corresponds to an MHD cell size of 20.8\,km, all other MHD
   results stem from 10.4\,km simulations.

   Finally, we were interested in correlations between certain BP properties by looking at scatterplots.
   The field strength at optical depth unity, the CN intensity, as well as the 5250.4\,\AA{} continuum
   intensity were only weakly correlated with the BP diameter. The scatterplots (not shown) exhibited a
   slight increase of the three quantities for large BP diameters. A decrease in the intensities for
   very large diameters, that we expect for micropores and pores, could not be found by our study owing to
   the BP detection method, which looked for bright features and not for strongly magnetized ones.
   Also, the scatterplot of the CN intensity versus the magnetic field inclination at optical depth unity
   (not shown) revealed nearly no correlation. We found a slight increase in intensity only for the most
   common inclinations around $17^{\circ}$. A strong correlation was found between the CN intensity
   and the field strength at optical depth unity, see Fig.~\ref{FigBPI388vsBm0Scat}. The binning graph
   is monotonic over the whole field strength range. Again, a decrease of the intensity for very strong
   fields could not be found owing to the BP detection method which excluded micropores.

\section{Discussion}
   The aim of this study is to learn more about magnetic BPs in the solar atmosphere. To achieve
   this we combine high-resolution observations with realistic radiation-MHD simulations. First we
   compare a row of parameters deduced from the observations (intensity, LOS velocity, line width,
   \Index{circular polarization degree}) with their counterparts obtained from the simulations under conditions
   matching those of the observations as closely as possible (noise, spatial and spectral resolution
   and sampling, straylight, etc.). After establishing that the simulations give a reasonable
   description of the data, both in general and for the BPs in particular, we have employed the
   simulations to deduce more about the BPs than can be gleaned from the observational data alone.

   We observed quiet-Sun regions at disk center with the balloon-borne observatory \sunrise{}.
   Photometric data at 3118\,\AA{} and 3877\,\AA{} as well as spectropolarimetric data at twelve
   wavelengths in and around the Fe\,{\sc i} line at 5250.2\,\AA{} were acquired quasi simultaneously.
   Compared to high-resolution observations with other telescopes, we benefited from the following
   advantages of \sunrise{}: a) The Sun was not only observed in the visible but also in the near UV.
   b) The PSF was measured during the observations, so that the influence of the central obscuration
   by the secondary mirror, the spiders, and the low-order aberrations like \Index{defocus}, \Index{coma},
   astigmatism were known and we did not have to rely on a theoretical PSF. c) A \Index{stray light} analysis
   was possible owing to observations of the solar limb. d) The data were practically free of seeing
   effects. By means of that, it was possible to carefully determine the degradations that were acting
   during the observations and to apply them to the MHD simulations. In contrast to other studies, which
   worked with opacity distribution functions \citep[ODFs, e.g.][]{Danilovic2008,Afram2011}, we applied
   full spectral line syntheses, for the first time also for the OH band at 3118\,\AA{}. Our study
   concentrated on a comparison of the following BP properties at disk center: intensity in the visible
   and near UV, LOS velocity, and polarization degree, while other studies focussed on center to limb
   variations of only intensity histograms \citep{WedemeyerBoehm2009,Afram2011}.

   A reasonable match between the observations and the degraded simulations was found for the intensity
   histograms of all pixels for all three considered wavelength ranges as well as for the LOS velocities
   and the circular polarizations. The intensity histograms of the undegraded 30\,G simulations showed a
   superposition of two populations which was greatly weakened by the degradation and which was
   not found in the observations. The two populations were also clearly present in comparable simulations
   with the Stagger and \Index{CO$^5$BOLD} code \citep{WedemeyerBoehm2009,Beeck2012}. \citet{Afram2010}
   used also the \Index{MURaM} code and found the same superposition for their undegraded 0\,G and 50\,G
   simulations but not for their 200\,G data. We can confirm this result by our intensity histograms
   of the degraded MHD data for varying mean flux densities.

   The degradation of our MHD data reduced the rms contrast at 3118\,\AA{} from 32.4\,\% to 21.1\,\%,
   at 3877\,\AA{} from 30.8\,\% to 20.5\,\% and at 5250.4\,\AA{} from 22.1\,\% to 12.1\,\%. With the help
   of ODFs, \citet{Hirzberger2010} calculated an rms contrast of 28.3\,\% at 3118\,\AA{} and of 23.9\,\%
   at 3877\,\AA{}. In both cases, complete spectral line syntheses led to significantly higher contrasts
   than with the ODF method, although the smaller grid size of the simulation employed by us may also
   have contributed. For the CN band at 3877\,\AA{}, \citet{Hirzberger2010} applied also a full spectral
   line synthesis but they used an older version of \Index{SPINOR} whose number of wavelength points was limited.
   They converted from geometrical heights into optical depths by a different opacity package and
   they also used a different horizontal grid resolution of 20.8\,km. They report an rms contrast of
   25.3\,\% for the undegraded CN data while we found a value of 30.8\,\% with the improved \Index{SPINOR} code.
   The deviation of the rms contrasts between degraded simulations and observations was 0.9\,\% at
   3118\,\AA{}, it was negligible at 5250.4\,\AA{}, at 3877\,\AA{} it was somewhat higher, but at 1.7\,\%
   the agreement was still remarkably good, given the various sources of uncertainty, e.g. slightly varying
   image quality owing to the remaining \Index{pointing jitter} of the gondola, incomplete spectral line lists,
   inaccurate atomic or molecular data or neglected \Index{non-LTE} effects. Given that the computed CN lines are
   distinctly too strong even for the average quiet Sun (see Fig.~\ref{FigSynthSpectra}), we expect these
   last points to play a significant role in producing the discrepancy. Even a comparison between several
   \Index{MHD code}s by \citet{Beeck2012} led to a variation in the bolometric intensity contrast of up to 1\,\%.
   \citet{Hirzberger2010} found rms contrasts of 18.3\,\% and 20.1\,\% for two different \sunrise{}
   observations at 3877\,\AA{}, while our observation showed 18.8\,\%. Depending on the ratio of the
   jitter frequency to the exposure time, different image qualities are possible for observations at
   different wavelengths. The exposure time of the 3877\,\AA{} images was by far the shortest one
   compared to the other wavelengths.

   We found that the various degradation steps considerably influenced the shape of the histograms
   of various parameters. In particular, we want to emphasize the \Index{stray light} since it influenced
   all types of histograms and often it provided the most important contribution to the degradation.
   This was also found by \citet{WedemeyerBoehm2009}. Additionally, for the line width histogram a
   good knowledge of the spectral PSF of the instruments is needed. The circular polarization histograms
   depended strongly on the noise level of the Stokes~$V$ images, while the noise level of the Stokes~$I$
   images as well as the other degradation steps played only a minor role.

   Stray light may also contribute to the last remaining discrepancy between the data and the simulations,
   the larger scatter of the observed 5250.2\,\AA{} line widths than in the simulations. We used the
   continuum images of the IMaX limb observations, but the IMaX stray light may have been somewhat
   dependent on the wavelength within the spectral line. Due to the sensitivity of the histogram shapes
   to stray light, we believe that such a wavelength dependence could be a good candidate to explain
   the larger scatter in the observed line widths. Another possibility is the influence of the
   evolution of the granulation during the relatively long acquisition time of IMaX of almost 32\,s
   which may also change the line width histogram.

   The good match between observation and simulation (only a moderate deviation remained for the scatter
   of line widths) enhanced our trust in the simulations, so that we used them to probe BP properties.
   We found a BP number density of 0.05 BPs per $\rm{Mm}^2$ in our observations which is at the lower end
   of the wide spread of values found in the literature: \citet{Muller1984} reported 0.04 BPs per $\rm{Mm}^2$,
   0.12 BPs per $\rm{Mm}^2$ were found by \citet{Bovelet2008}, 0.3 BPs per $\rm{Mm}^2$ by
   \citet{SanchezAlmeida2004} -- all these were quiet-Sun studies, and \citet{Berger1995} analyzed
   active-region data and resulted in 0.37 BPs per $\rm{Mm}^2$. Our BP number density
   is larger than the 0.03 BPs per $\rm{Mm}^2$ obtained by \citet{Jafarzadeh2013} from
   Ca\,{\sc ii}~H observations made by \sunrise{}, but this is likely due to the fact that they restricted
   their study to features narrower than 0\carcsec{}3. On the other hand, the effective BP Diameter was
   on average 334\,km in our observations while most other studies reported somewhat smaller BP diameters,
   e.g.: \citet{Berger1995} found a mean FWHM intensity diameter of 250\,km by taking the smallest dimension
   across the BP features. \citet{SanchezAlmeida2004} fitted the minor and major axes of the BPs with average
   values of 135\,km and 230\,km, respectively. \citet{Utz2009} reported a decrease in the mean BP diameter
   from 218\,km to 166\,km by doubling the spatial sampling of data from the Hinode telescope.
   \citet{Crockett2010} found a distribution which peaked at an effective BP diameter of 240\,km.

   A direct comparison of the BP number density and BP diameter with the results of other studies, in
   particular from other telescopes is rather difficult because both quantities depend strongly on the
   image degradation, on the mean vertical flux density of the observed region, and probably on the
   method employed to identify the BPs. For our simulated BPs we found that the degradation of the data
   decreased the BP number density by a factor of 3.2 while the BP diameter was reduced by a factor of 2.6.
   An increase of the mean vertical flux density from 30\,G to 200\,G led to a 3.1 times higher BP number
   density. Also the BP detection method can influence the two quantities since, e.g., almost all methods
   depend on several thresholds. Possibly we found a lower BP number density and a higher BP diameter (which
   results in an BP area coverage somewhat closer to the results of the above mentioned studies) because our
   manual detection method tend to consider such larger bright patches (associated with larger concentrations
   of magnetic flux) as single features whereas other techniques may identify them as chains or clusters
   of multiple BPs. However, there still remains a discrepancy between our observations and the degraded
   simulations since we found a very good match in the mean BP diameter but a BP number density that is
   lower in the observations by a factor of 5. This implies that a smaller fraction of the magnetic
   field in the observations is in the form of kilo-Gauss magnetic elements than in these simulations.
   At the same time, the observations show some large and strongly polarized BPs which were not present
   in our simulations. Remember that the average magnetic field strength of 30\,G was determined by
   comparing the distribution of $\langle p_{\rm{circ}} \rangle$ in all pixels, whereas the BPs cover
   a very small fraction of the surface area. This conclusion agrees with the results of
   \citet{Danilovic2010}, who got very good agreement between both the circular and linear polarization
   measured by Hinode and obtained from a local turbulent dynamo simulation \citep{Schuessler2008},
   which have more weak and horizontal fields than the simulations considered here, which, however,
   we find to be better suited to provide the correct properties of BPs, which was our main aim.
   One reason for the relatively low number of BPs is that \sunrise{} observed a very quiet region
   in the deepest part of the last activity minimum. \citet{Foukal1991} and \citet{Meunier2003} found
   evidence that the number of quiet-Sun BPs is correlated with the solar cycle.

   For a deeper understanding of the BP phenomenon, we analyzed further properties of the simulated BPs.
   The histogram of the inclination of the magnetic field vector exhibited a nearly vertical magnetic
   field for most of the BPs, which is to be expected for kilo-Gauss fields and not too strong horizontal
   flows \citep{Schuessler1986}. A significant difference between the middle and lower photosphere was
   found for the vertical velocities of the simulated BPs. While the BPs showed on average a downflow of
   3.2\,km~s$^{-1}$ in the lower photosphere, the downflow was reduced to 1.06\,km~s$^{-1}$ in the
   middle photosphere. A decrease of velocity with height was also found by \citet{BellotRubio1997} from
   the inversion of asymmetric Stokes~$V$ profiles observed in plage regions. The deeper we look into
   the atmosphere the stronger were the downflows of the BPs, so that it would be interesting to observe
   BPs spectropolarimetrically at a spectral line that is formed deep in the photosphere, e.g. the
   C\,{\sc i} line at 5380\,\AA{}. For a line formed so deep in the atmosphere we expect BP downflows
   that are stronger than the ones found with \sunrise{}. These strong flows in the deep layers of
   photospheric magnetic elements raise the question of the origin of the mass. Either it diffuses
   into the magnetic features across field lines, which runs counter to the estimates of \citet{Hasan1985},
   or the lifetimes of BPs, i.e. kilo-Gauss features are very short (i.e. we are generally catching them
   during their convective collapse phase), or the plasma with the strong field is continually mixing with
   relatively field-free plasma in the immediate surroundings of the magnetic elements. This last process
   may be related to the vortices found in the simulations around magnetic elements by \citet{Moll2011}.

\section{Conclusions}\label{Bp2Conclusions}
   We have compared high-resolution \sunrise{} data in three spectral bands with three-dimensional
   radiative MHD simulations and found that the two agree in most areas remarkably well, as long as
   all instrumental effects that degrade the data are properly introduced into the simulations as
   well. This represents a stringent test of the simulations, since we consider many more parameters
   than just intensities and also consider BPs separately.

   We showed that although the majority of BPs is weakly polarized in the observational data
   \citep[see also][]{Riethmueller2010} they correspond to magnetic elements with kilo-Gauss fields.
   The small signals can be partly explained by spatial smearing due to residual \Index{pointing jitter} and
   the instrumental \Index{stray light}. The large temperature sensitivity of the Fe\,{\sc i} line at
   5250.2\,\AA{} also contributed to reducing the polarization signal for the generally hot BPs.
   In the original simulations 98\,\% of the BPs are almost vertically oriented magnetic fields in
   the kilo-Gauss range, which confirms the physical model of magnetic flux concentrations as evacuated
   and laterally heated structures \citep{Spruit1976,Deinzer1984}.

   MHD simulations with a horizontal cell size of 20\,km or higher are widely used in the literature
   \citep{Afram2010,Afram2011,Hirzberger2010,OrozcoSuarez2008,Roehrbein2011,WedemeyerBoehm2009}.
   We found a reduction from 163\,km to 129\,km for the effective BP diameter and an increase for the
   BP number density from 0.44 BPs per $\rm{Mm}^2$ to 0.83 BPs per $\rm{Mm}^2$ by doubling the
   horizontal grid resolution. We cannot rule out that in order to obtain true BP properties by MHD
   simulations a horizontal cell size lower than 10\,km is needed.

   The observations, in particular when taken together with the simulations also indicate that
   phenomena are present in the BPs that are not yet understood. One of these is the downflow velocity
   of on average 0.6\,km~s$^{-1}$ in the observations and 1.25\,km~s$^{-1}$ in the undegraded synthesized
   line profiles. Such high universal downflows would lead to an evacuation by an order of magnitude
   (i.e. 10 times reduced gas pressure) within the time it takes the gas to flow down to scale heights,
   i.e. roughly 200\,km. At 1.25\,km~s$^{-1}$ this will take place within 160\,s. Comparing with the
   mean lifetime of Ca\,{\sc ii}~H BPs of 612\,s in similar \sunrise{} data sets \citep{Jafarzadeh2013},
   this implies that the magnetic elements would be almost completely evacuated within
   a fraction of their lifetimes unless the gas is continuously replenished (i.e. this downflow is not
   a sign of a continuous convective collapse).

   The plasma can be replenished either by gas flowing up along the opposite footpoints of loops that
   end in the BPs. As shown by \citet{Wiegelmann2010} at the \sunrise{} resolution in the quiet Sun,
   strong-field regions, such as BPs, are mainly connected to weak-field regions. Also, most of these
   loops are rather low-lying, i.e. not reaching above the chromosphere. Along such loops a siphon
   flow from the footpoint with weaker field to that with the stronger field can take place
   \citep{Meyer1968,Montesinos1989}. Such a flow would produce a downflow in the BPs and has been
   observed along the neutral line of an active region \citep{Rueedi1992,Degenhardt1993}.
   However, the simulations also show such downflows and these have a closed upper boundary through
   which the magnetic field of the BPs passes, but no flow can go through. Hence the flow must be
   replenished locally. Note that the observed downflows of on average 0.6\,km~s$^{-1}$ are larger
   than the simulated ones, 0.27\,km~s$^{-1}$. The drift may be due to siphon flows, or just to lower
   Reynolds number of the simulations.

   The \sunrise{} images contain a few large and strongly polarized BPs which possibly were part of
   network elements. Such network elements were not present in the simulations. Usually, they are
   observed at the boundaries of supergranules and are possibly formed deeper in the convective zone.
   In a future study we will provide MHD simulations with a computational box large and deep enough
   to contain one or more supergranules.

\chapter{Outlook}\label{Outlook}

In this last chapter an outlook is given how the presented research on small-scale magnetic features can be
reasonably continued, in particular taking into account modern telescopes and instruments that were either
recently put into operation or whose commissioning is planned in the near future.

\section{Dark lanes and downflow channels of umbral dots}

The synthetic UDs of the MHD simulations of \citet{Schuessler2006} often show a horizontally elongated
form and almost all UDs exhibit a central \Index{dark lane}, some cases of larger UDs display a threefold dark lane.
In the simulations, the endpoints of the dark lanes often show narrow downflow channels. In fact, dark
lanes inside UDs were found in the observations of \citet[50~cm \hinode{}/SOT,][]{Bharti2007} and
\citet[76~cm \Index{Dunn Solar Telescope} (\Index{DST}),][]{Rimmele2008}, but some doubts still remain if these are really the
phenomena seen in the simulations. The observed UDs \citep[effective diameter roughly 900~km,][]{Bharti2007}
are significantly larger than the synthetic UDs \citep[diameter roughly 300~km,][]{Schuessler2006}.
Either these large observed structures are bona fide UDs, or they are related, but distinct features
(they look like fragments of a \Index{light bridge}) or the impression of a \Index{dark lane} only exists
because two UDs without a visible internal structure came very close to each other (see
section~\ref{QualResults}). Although downflow patches at the edges of UDs were found by
\citet[1-m SST, bisector method,][]{Ortiz2010} as well as in chapter~\ref{Ud3Chapter} (50-cm SOT, 2D inversion\index{inversiontwodim@2D inversion}),
concentrated narrow downflow channels at the endpoints of dark lanes could not yet be observed beyond doubt.
A verification of the MHD predictions is hence desirable with spectropolarimetric data of telescopes
having higher spatial resolution than \hinode{}/SOT or \Index{DST}, respectively.

One promising opportunity for such a verification is the second science flight of the \sunrise{}
observatory planned for 2013. The main argument for a second flight is the prospect of active region
observations (and hence of UDs) which were not possible in 2009 due to the solar activity minimum.
At least equally important is a considerable improvement of the gondola's \Index{pointing stability}, so that
the \Index{optical performance} of the telescope can be fully exploited and longer time series as well as a
significantly increased spatial resolution for the UV channels can be reached. The dynamics of
small-scale magnetic features can only be analyzed with preferably long time series of constantly good
image quality. At the present time, \hinode{}/SOT provides the highest resolution data of this type,
but the \Index{NFI} instrument does not work properly. The clear aperture of \sunrise{} is twice as large as
that of \hinode{}/SOT and the shortest wavelength, 214~nm, is almost half of the shortest \hinode{}
wavelength, 388~nm, so that \sunrise{} has the potential of nearly four times higher \Index{spatial resolution},
which possibly allows completely new insights into the fine structure of small-scale features. Note,
however, the long integration times of 30~s required to achieve a good signal-to-noise level in the
quiet Sun at 214~nm, which makes reaching the \Index{diffraction limit} more challenging. The SuFI filters
envisaged for the reflight concentrate more on chromospheric observations. There will be three filters
for chromospheric observations, a Mg\,{\sc ii}~K 2796~\AA{} filter (FWHM 4.8~\AA{}), the Ca\,{\sc ii}~H
3968~\AA{} filter of the 2009 flight (FWHM 1.8~\AA{}), and a more narrow Ca\,{\sc ii}~H 3968~\AA{} filter
(FWHM 1.1~\AA{}). The 214~nm filter, used for observations of the upper photosphere, is now 22~nm wide
instead of only 10~nm, which keeps the exposure time of umbral images roughly constant (on average the
UDs are nearly half as bright as the quiet Sun, see panel (c) of Fig~\ref{FigHistOverview}) and reduces
the integration time of quiet-Sun images. A 300~nm filter (FWHM 4.4~\AA{}) is planned again
for observations of the lower photosphere which is the proper filter choice for the search for dark lanes
in UDs (expected exposure time is around 500~ms), while downflow channels at the endpoints of \Index{dark lane}s
can be searched in the IMaX data at 525.02~nm.

The experiences with the \Index{IMaX} data of the first \sunrise{} flight revealed that inversions with only
four scan positions within the spectral line (plus one in the continuum) contain some uncertainties
\citep{Borrero2010}, so that IMaX modes with more than four scan positions are recommended for the
second flight. Additionally, the data shall be inverted with the noticeably improved 2D inversion\index{inversiontwodim@2D inversion} method,
which takes into account the instrumental effects responsible for the spatial and spectral degradation
of the observational data, so that the inversion results correspond to spatially deconvolved values, but
without the added noise usually introduced by deconvolution \citep{vanNoort2012}. Note, that the
weak downflows surrounding the UDs could only be detected with the new 2D inversion\index{inversiontwodim@2D inversion} method used for the
study in chapter~\ref{Ud3Chapter}, while this was not yet possible with the classical inversion method used
in chapter~\ref{Ud2Chapter}. The 2D inversion\index{inversiontwodim@2D inversion} technique in combination with the 1-m \sunrise{} telescope
can possibly allow a safe detection of the narrow downflow channels expected from the simulations.

\sunrise{} makes quasi simultaneous observations of different layers of the solar atmosphere possible and
hence potentially allows a verification of the MHD predictions that \Index{dark lane}s in UDs are only visible in the
deeper layers of the photosphere \citep{Schuessler2006}. The acquisition of at least 20~min time series enables
analyses of the temporal evolution of UDs that allow deciding if a dark lane is really a property of a unique
UD or only the result of a temporarily vicinity of two UDs. It should be reviewed if the UD magnetic field
weakenings of roughly 500-700~G found in chapters~\ref{Ud2Chapter} and \ref{Ud3Chapter} are more pronounced
in the \sunrise{} data due to the higher spatial resolution.

On the other hand, a new UD study would also be interesting with the two ground-based solar telescopes of the
1.5~m class that were recently inaugurated. The \Index{New Solar Telescope} (\Index{NST}) at the Big Bear Solar Observatory
is an off-axis Gregory telescope with an aperture of 1.6~m \citep{Cao2010,Goode2010}. The \Index{GREGOR solar telescope}
at the Observatorio del Teide on Tenerife is an on-axis Gregory telescope \citep{Schmidt2012a,Schmidt2012b}.
It has a clear aperture of 1.44~m and complements the NST's capabilities\footnote{E.g., the NST Spectro-Polarimeter
for Infrared and Optical Regions \citep{SocasNavarro2006} allows infrared observations up to
1.6~$\mathrm{\mu{}m}$, while the GREGOR Infrared Spectrograph \citep[\Index{GRIS},][]{Collados2012} allows
spectroscopical observations of up to 2.2~$\mathrm{\mu{}m}$.}. A first photometric UD study with broadband 
NST data at 705.68~nm was already done by \citet{Andic2011} and revealed UDs without any fine structure with
a mostly roundish appearance. Since unfavorable seeing conditions, a possibly low \Index{optical performance} of the
telescope, and the use of a spectral range with an improper formation height range could not be excluded, one
cannot conclude from this study that UDs generally do not show a \Index{dark lane} at a \Index{spatial resolution} of 80~km.
A new UD study with the GREGOR Fabry-P\'erot Interferometer \citep[\Index{GFPI},][]{Puschmann2012} or the GREGOR Broad-Band
Imager \citep[\Index{BBI},][]{vonderLuehe2012} is still missing.

NST and GREGOR serve as precursors for the next generation of solar
telescopes: The Americans are building the \Index{Advanced Technology Solar Telescope} \citep[\Index{ATST},][]{Keil2011}
and the Europeans are planning the \Index{European Solar Telescope} \citep[\Index{EST},][]{Collados2010}, so that two
telescopes of the 4~m class should be available in the future whose \Index{diffraction limit} at 500~nm corresponds
to a \Index{spatial resolution} of only 20~km on the solar surface. Such large apertures allow only for \Index{diffraction limit}ed
observations, if significant progress can be achieved on the \Index{adaptive optics} side. One approach
currently under development is Multi Conjugate Adaptive Optics \citep[\Index{MCAO},][]{Berkefeld2006,Berkefeld2010,Kellerer2012},
which uses multiple \Index{wavefront sensor}s in order to measure the turbulence in different layers of the
terrestrial atmosphere. For a 4~m telescope, the theoretical \Index{diffraction limit} of 20~km is on the order of
the horizontal \Index{mean free path} of the photons in the lower photosphere and concerns were expressed that such
small structures are not resolvable because of \Index{photon diffusion}, i.e. due to smoothing \Index{radiative transfer}
effects (in particular scattering). The 2D \Index{non-LTE} radiative transfer computations for thin \Index{flux sheet}s of
\citet{Bruhls2001} and \citet{Holzreuter2012} as well as the dynamical \Index{flux tube} simulations of \citet{Stein2006}
showed that the size limit where the photospheric structures are not resolvable due to smoothing radiative transfer
effects must lie well below 10~km.

\section{Comparison of \Index{plage} bright points and quiet-Sun bright points}
\citet{GrossmannDoerth1994} compared MHD \Index{flux sheet} models with observations (recorded with the
McMath telescope close to the disk center) and determined a flux sheet width of roughly 200~km
in the network but 300-350~km in plage regions. \citet{Kobel2012} analyzed \hinode{}/SP data,
also near disk center, and found that the continuum brightness as well as the dispersion of the
LOS velocities (retrieved from \Index{Milne-Eddington inversion}s) of magnetic elements in active regions
are lower than in the quiet Sun. The presence of a magnetic field partly suppresses the convection
in the surroundings of the magnetic elements and leads finally to a reduction in their heating.
The total radiation flux of the MHD simulations of \citet{Voegler2005a} showed a similar behavior.

If both, plage and quiet-Sun regions can be observed during the reflight of \sunrise{}, BPs can be
detected in both regions and the comparison of their properties can be revisited at the higher spatial
resolution of the \sunrise{} telescope. It is also possible to determine the mean flux density of both
regions with the method described in chapter~\ref{Bp2Chapter}, to run MHD simulations with the retrieved
flux densities, and to look if the match between observation and simulation is equally good for both regions.

\section{Height dependence of the mean downflows in bright points}
The MHD simulations of chapter~\ref{Bp2Chapter} revealed that on average the BPs exhibit downflows that
increase with optical depth (see Fig.~\ref{FigBPV_LOSHist}). This prediction from the simulations
should be verified with observations. One possibility to determine the height dependence of the LOS velocities
is the application of the \Index{bisector method} \citep{Dravins1981} to the Stokes~$I$ profiles of the IMaX L12 data
of the first \sunrise{} flight. It is also possible to alternately scan two spectral lines with preferably
different formation heights (e.g. the Fe\,{\sc i} 5250.2~\AA{} line formed in the mid photosphere and the
C\,{\sc i} 5380\,\AA{} line formed in the deep photosphere) with a spectropolarimeter like SST/\Index{CRISP}
\citep{Scharmer2006}. The BP velocities can then be determined, e.g., by a Gaussian fit of the Stokes~$I$
profiles as done in chapter~\ref{Bp2Chapter}. A possibly more auspicious method to retrieve the velocity
stratification of BPs is the 2D inversion\index{inversiontwodim@2D inversion} \citep{vanNoort2012} of \hinode{} data, since the current
studies of \Index{plage} regions \citep{Buehler2013} and penumbrae \citep{vanNoort2013} give some hints that the
velocity information in deep layers can also be retrieved by this method from \hinode{}/SP data at 6302~\AA{}.

\section{Detailed investigations of synthetic bright points}
Several BP properties were already studied by investigations described in chapter~\ref{Bp2Chapter} but
should be extended by further analyses. A deeper insight into the nature of quiet-Sun BPs can be gained,
amongst others, by the search for correlations between the various BP properties, e.g. between
magnetic field strength and inclination, between field strength and temperature at different
optical depths, or between field strength and BP size. Likewise interesting is the question which
quantities correlate with the BP downflows, e.g. the field strength or its spatial or temporal
derivative. Additionally, an in-depth analysis of a few typical BPs can contribute to answering
the question what is the origin of the continuous plasma flow needed to maintain the detected BP
downflows (see discussion in section~\ref{Bp2Conclusions}).

Furthermore, our knowledge about BPs should be improved by a study on their dynamics. For such a study
not only the detection of BPs in single images is needed, but also the tracking of the features is
required to determine the BPs' trajectories as done, e.g. in Chapter~\ref{Ud1Chapter} for umbral dots
or in \citet{Jafarzadeh2013} for Ca\,{\sc ii}~H BPs in the low \Index{chromosphere}. Thus, questions regarding the
formation, evolution, lifetimes, interaction, and destruction of BPs can be answered. Since synthetic data
are practically free of noise and have high \Index{spatial resolution}, the realization of such a dynamics study
is significantly easier than with observational data.

\section{MHD simulations of network \Index{bright point}s}
In spite of the match between the properties of synthetic and observed BPs found in chapter~\ref{Bp2Chapter},
some discrepancies were also found. The most remarkable disagreement is the fact that the observations
show a few relatively large and strongly polarized BPs, probably network BPs, that were not found in
the simulations. One should check if MHD simulations with a significantly larger and deeper computational box
than the one used in chapter~\ref{Bp2Chapter} is able to reproduce such network BPs and hence lead to
more realistic simulations of quiet-Sun phenomena.

\appendix
\chapter{Operator identities}\label{Appendix_A}

If $s$ is any scalar field, $\vec{v}$ any vector field, $\nabla$ denotes the Nabla operator, and $\Delta$ is the
Laplace operator, then:

\begin{equation}\label{Eq_OpIdent2}
\nabla \times (\nabla s) = 0
\end{equation}

\begin{equation}\label{Eq_OpIdent4}
\nabla \cdot (\nabla \times \vec{v}) = 0
\end{equation}

\begin{equation}\label{Eq_OpIdent6}
\nabla \cdot (\nabla s) = \Delta s
\end{equation}

\begin{equation}\label{Eq_OpIdent8}
\nabla (\nabla \cdot \vec{v}) = \Delta \vec{v} + \nabla \times (\nabla \times \vec{v})
\end{equation}

\begin{equation}\label{Eq_OpIdent9}
(\nabla \times \vec{v}) \times \vec{v} = (\vec{v} \cdot \nabla)\vec{v} - \nabla \left( \frac{v^2}{2} \right)
\end{equation}


\printindex

\chapter*{Acknowledgments\markboth{Acknowledgments}{Acknowledgments}}
\addcontentsline{toc}{chapter}{Acknowledgments}

I would particularly like to thank my doctoral adviser Prof. Sami Solanki, who has bravely ventured on the
supervision of a non-astrophysicist. His constructive criticism always improved my work considerably.
I also thank Prof. K.-H. Gla\ss{}meier for his support, in particular for his valuable advice during the
writing of this thesis and for his willingness to act as a referee.

Important for me were also the brilliant introductions into solar physics which I was given by Johann Hirzberger,
among others during the long nights of observation campaigns on La Palma. I thank Vasily Zakharov for
recording and reconstructing the data used in chapter~\ref{Ud1Chapter} and Achim Gandorfer for the
supervision in the beginning phase of my Ph.D. time.

During a stay at the National Astronomical Observatory of Japan in Tokyo, Luis Bellot Rubio made it easy
for me to learn everything about the \hinode{} data handling. Andreas Lagg provided an always kindly
assistance to me, in particular with respect to the SPINOR inversion code, and he enabled me two stays
at the VTT on Tenerife with Alex Feller and Michiel van Noort, where Godehard Monecke taught me how to
influence the seeing via the Ronmiel technique.

I'm grateful to Manfred Sch\"ussler and Robert Cameron for helping me to handle and run the MHD simulations.
I would also like to thank Valentin Mart\'{\i}nez Pillet and Jose Carlos del Toro Iniesta for the careful
review of my \sunrise{}-based papers.

Roughly half of this thesis concentrates on \sunrise{} observations which were not possible without the
untiring efforts of a large number of colleagues. In place of all, only a few examples can be mentioned
here. In a friendly manner, Peter Barthol managed even the hot phases of the project with his
usual aplomb and has proved a recipe for an always pleasant climate between the people. The inherent
insistence of Reinhard Meller contributed crucially to the success of \sunrise{}. Georg Tomasch is
really proficient in reanimating the SuFI camera, several times presumed dead. Rafael Morales from IAA
is positively stuck in my memory not only because of the unforgettable Tapas tour through Granada.
The teamwork with Dietmar Germerott and Martin Kolleck led finally to the enormous improvement of my
software engineering skills that I experienced in the last 10 years. In my opinion, Alice Lecinsky
and Jack Fox from HAO accomplished much more for the \sunrise{} project than one could expect. 

\chapter*{Lebenslauf\markboth{Lebenslauf}{Lebenslauf}}
\addcontentsline{toc}{chapter}{Lebenslauf}

\begin{tabular}{lll}                                  
   Name:                  & \multicolumn{2}{l}{Tino L. Riethm\"uller} \\[2ex]
   Geburtstag:            & \multicolumn{2}{l}{02. 03. 1971} \\[2ex]
   Geburtsort:            & \multicolumn{2}{l}{Karlsburg} \\[2ex]
   Staatsb\"urgerschaft:  & \multicolumn{2}{l}{deutsch} \\[2ex]
   Ausbildung:            & 1977-1987             & Polytechnische Oberschule 1 Dingelst\"adt \\[2ex]
                          & 1987-1989             & Spezialklassen f\"ur Mathematik und Physik der \\ && Martin-Luther-Universit\"at Halle/Wittenberg, Abitur \\[2ex]
                          & 1990-1995             & Humboldt-Universit\"at zu Berlin, \\ && Diplomstudium Physik \\[2ex]
                          & 2008-2013             & Technische Universit\"at Braunschweig, \\ && Promotionsstudium Physik \\[2ex]
   Wehrdienst:            & 1989-1990             & Mot.-Sch\"utze in Brandenburg, \\ && Bausoldat in Potsdam \\[2ex]
   Beruflicher Werdegang: & 1995-1996             & Daimler-Benz AG, Forschung Systemtechnik, \\ && Berlin, Wissenschaftlicher Mitarbeiter \\[2ex]
                          & 1996-2002             & LMS Instruments GmbH, G\"ottingen,\\ &&  Software-Ingenieur \\[2ex]
                          & 2002-jetzt            & Max-Planck-Institut f\"ur Sonnensystemforschung,\\ &&  Lindau, Software-Ingenieur \& Wissenschaftler \\
\end{tabular}

\end{sloppypar}
\end{document}